\documentclass[traditabstract,longauth]{aa} 
\usepackage[english]{babel} 
\usepackage[utf8]{inputenc} 
\usepackage[T1]{fontenc}
\pdfoutput=1
\usepackage{txfonts} 

 \usepackage{longtable} 
 \usepackage{supertabular}
 
\usepackage{graphicx}

\graphicspath{ {./figures_HFI_DPC_astroph/} }   

\usepackage[mathscr]{eucal}

\usepackage{natbib} \bibpunct{(}{)}{;}{a}{}{,} 
\usepackage[breaklinks, colorlinks, citecolor=blue]{hyperref}

\usepackage{geometry}
\usepackage{array}
\usepackage{setspace}

\usepackage{listings}

\newcommand\lab[1]{\label{#1}\end{equation}}

\newcommand\beq{\begin{equation}}\newcommand\eeq{\end{equation}}	
\newcommand\bma{\begin{displaymath}}\newcommand\ema{\end{displaymath}}
\newcommand{\bce}[0]{\begin{center}}\newcommand{\ece}[0]{\end{center}}
\newcommand{\ben}[0]{\begin{enumerate}}\newcommand{\een}[0]{\end{enumerate}}
\newcommand{\bit}[0]{\begin{itemize}}\newcommand{\eit}[0]{\end{itemize}}

\newcommand{\hide}[1]{}

\def\spose#1{\hbox to 0pt{#1\hss}}

\def\simpropto{\mathrel{\spose{\lower 3pt\hbox{$\mathchar"218$}}
     \raise 2.0pt\hbox{$\propto$}}}
\newcommand\ltsima{$\; \buildrel < \over \sim \;$}
\newcommand\simlt{\lower.5ex\hbox{\ltsima}}
\newcommand\gtsima{$\; \buildrel > \over \sim \;$}
\newcommand\simgt{\lower.5ex\hbox{\gtsima}}
\newcommand\simprop{\lower.5ex\hbox{$\; \buildrel \propto \over \sim \;$}}

\newcommand{\sqr}[2]{{\vcenter{\vbox{\hrule height.#2pt
             \hbox{\vrule width.#2pt height#1pt \kern#1pt
                   \vrule width.#2pt}
             \hrule height.#2pt}}}}

\newcommand{\GHz}{{\rm GHz}}

\newcommand{\K}{\,{\rm K}}	
		
\newcommand{\microK}{\,$\mu \mathrm{K}$}
\newcommand{\muK}{\mu {\rm K}}

\newcommand{\eg}{\textit{e.g.,  }} 
\newcommand{\ie}{\textit{i.e., }}
 
\newcommand\etc{\textit{etc.}}	
\newcommand{\rms}{\emph{rms}}

\newcommand{\cf}{\textit{cf.\ }}

\newcommand{\FIRAS}{\textit{FIRAS\/}}
\newcommand{\COBE}{\textit{COBE\/}}

\newcommand{\DMC}{\textsc{DMC}} 
\newcommand{\HFIDMC}{\textsc{HFI-DMC}} 
\newcommand{\IMO}{\textsc{IMO}} 
\newcommand{\IMORealisation}{\textsc{IMORealisation}} 
\newcommand{\healpix}{HEALPix}
\newcommand{\polkapix}{\texttt{polkapix}}
\newcommand{\thinc}{\textsc{ThinC}} 
\newcommand{\MagiqueIII}{\textsc{Magique-III}}

\newcommand{\degree}{\ensuremath{^\circ}}

\def\microm{\ifmmode \,\mu$m$\else \,$\mu$\hbox{m}\fi}

\newcommand{\Planck}{\textit{Planck\/}}
\newcommand{\WMAP}{\textit{WMAP\/}}
\def\GHz{\ifmmode $GHz$\else \,GHz\fi}
\def\MHz{\ifmmode $MHz$\else \,MHz\fi}
\def\Hz{\ifmmode $Hz$\else \,Hz\fi}

\def\muK{\ifmmode \,\mu$K$\else \,$\mu$\hbox{K}\fi}
\def\microK{\ifmmode \,\mu$K$\else \,$\mu$\hbox{K}\fi}
\def\muW{\ifmmode \,\mu$W$\else \,$\mu$\hbox{W}\fi}
\def\kms{\ifmmode $\,km\,s$^{-1}\else \,km\,s$^{-1}$\fi}

\def\allearlypapers{\nocite{planck2011-1.1, planck2011-1.3, planck2011-1.4, planck2011-1.5, planck2011-1.6, planck2011-1.7, planck2011-1.10, planck2011-1.10sup, planck2011-5.1a, planck2011-5.1b, planck2011-5.2a, planck2011-5.2b, planck2011-5.2c, planck2011-6.1, planck2011-6.2, planck2011-6.3a, planck2011-6.4a, planck2011-6.4b, planck2011-6.6, planck2011-7.0, planck2011-7.2, planck2011-7.3, planck2011-7.7a, planck2011-7.7b, planck2011-7.12, planck2011-7.13}}

\begin{document}
\titlerunning{AA 536, A6 (2011)}
\authorrunning{Planck HFI Core Team: Planck early results. VI. HFI data processing}

\title{\textit{Planck\/} early results. VI. The High Frequency Instrument \\ data processing}

 
\author{\texorpdfstring{\small
Planck-HFI Core Team\thanks{This list includes those who contributed substantially over the years to the development of the HFI instrument, its operation and its data processing, as well as all the members of the \textit{Planck\/} science team.}:
P.~A.~R.~Ade\inst{52}
\and
N.~Aghanim\inst{29}
\and
R.~Ansari\inst{44}
\and
M.~Arnaud\inst{40}
\and
M.~Ashdown\inst{38, 3}
\and
J.~Aumont\inst{29}
\and
A.~J.~Banday\inst{57, 6, 46}
\and
M.~Bartelmann\inst{56, 46}
\and
J.~G.~Bartlett\inst{2, 36}
\and
E.~Battaner\inst{59}
% ...
}{} }
\institute{\small
Agenzia Spaziale Italiana Science Data Center, c/o ESRIN, via Galileo Galilei, Frascati, Italy\\
\and
Astroparticule et Cosmologie, CNRS (UMR7164), Universit\'{e} Denis Diderot Paris 7, B\^{a}timent Condorcet, 10 rue A. Domon et L\'{e}onie Duquet, Paris, France\\
\and
Astrophysics Group, Cavendish Laboratory, University of Cambridge, J J Thomson Avenue, Cambridge CB3 0HE, U.K.\\
\and
Atacama Large Millimeter/submillimeter Array, ALMA Santiago Central Offices, Alonso de Cordova 3107, Vitacura, Casilla 763 0355, Santiago, Chile\\
\and
CITA, University of Toronto, 60 St. George St., Toronto, ON M5S 3H8, Canada\\
\and
CNRS, IRAP, 9 Av. colonel Roche, BP 44346, F-31028 Toulouse cedex 4, France\\
\and
California Institute of Technology, Pasadena, California, U.S.A.\\
\and
DAMTP, University of Cambridge, Centre for Mathematical Sciences, Wilberforce Road, Cambridge CB3 0WA, U.K.\\
\and
DSM/Irfu/SPP, CEA-Saclay, F-91191 Gif-sur-Yvette Cedex, France\\
\and
DTU Space, National Space Institute, Juliane Mariesvej 30, Copenhagen, Denmark\\
\and
D\'{e}partement de Physique Th\'{e}orique, Universit\'{e} de Gen\`{e}ve, 24, Quai E. Ansermet,1211 Gen\`{e}ve 4, Switzerland\\
\and
Department of Astronomy and Astrophysics, University of Toronto, 50 Saint George Street, Toronto, Ontario, Canada\\
\and
Department of Physics, Princeton University, Princeton, New Jersey, U.S.A.\\
\and
Department of Physics, University of California, One Shields Avenue, Davis, California, U.S.A.\\
\and
Department of Physics, University of Illinois at Urbana-Champaign, 1110 West Green Street, Urbana, Illinois, U.S.A.\\
\and
Department of Physics, University of Oxford, 1 Keble Road, Oxford, U.K.\\
\and
Dipartimento di Fisica, Universit\`{a} La Sapienza, P. le A. Moro 2, Roma, Italy\\
\and
Dipartimento di Fisica, Universit\`{a} degli Studi di Milano, Via Celoria, 16, Milano, Italy\\
\and
European Southern Observatory, ESO Vitacura, Alonso de Cordova 3107, Vitacura, Casilla 19001, Santiago, Chile\\
\and
European Space Agency, ESTEC, Keplerlaan 1, 2201 AZ Noordwijk, The Netherlands\\
\and
INAF - Osservatorio Astronomico di Trieste, Via G.B. Tiepolo 11, Trieste, Italy\\
\and
INAF/IASF Bologna, Via Gobetti 101, Bologna, Italy\\
\and
INAF/IASF Milano, Via E. Bassini 15, Milano, Italy\\
\and
INSU, Institut des sciences de l'univers, CNRS, 3, rue Michel-Ange, 75794 Paris Cedex 16, France\\
\and
IPAG: Institut de Plan\'{e}tologie et d'Astrophysique de Grenoble, Universit\'{e} Joseph Fourier, Grenoble 1 / CNRS-INSU, UMR 5274, Grenoble, F-38041, France\\
\and
Imperial College London, Astrophysics group, Blackett Laboratory, Prince Consort Road, London, SW7 2AZ, U.K.\\
\and
Infrared Processing and Analysis Center, California Institute of Technology, Pasadena, CA 91125, U.S.A.\\
\and
Institut N\'{e}el, CNRS, Universit\'{e} Joseph Fourier Grenoble I, 25 rue des Martyrs, Grenoble, France\\
\and
Institut d'Astrophysique Spatiale, CNRS (UMR8617) Universit\'{e} Paris-Sud 11, B\^{a}timent 121, Orsay, France\\
\and
Institut d'Astrophysique de Paris, CNRS UMR7095, Universit\'{e} Pierre \& Marie Curie, 98 bis boulevard Arago, Paris, France\\
\and
Institut de Ci\`{e}ncies de l'Espai, CSIC/IEEC, Facultat de Ci\`{e}ncies, Campus UAB, Torre C5 par-2, Bellaterra 08193, Spain\\
\and
Institut de Radioastronomie Millim\'{e}trique (IRAM), Avenida Divina Pastora 7, Local 20, 18012 Granada, Spain\\
\and
Institut de Radioastronomie Millim\'{e}trique (IRAM), Domaine Universitaire de Grenoble, 300 rue de la Piscine, 38406, Grenoble, France\\
\and
Institute of Astronomy, University of Cambridge, Madingley Road, Cambridge CB3 0HA, U.K.\\
\and
Instituto de Astrof\'{\i}sica de Canarias, C/V\'{\i}a L\'{a}ctea s/n, La Laguna, Tenerife, Spain\\
\and
Jet Propulsion Laboratory, California Institute of Technology, 4800 Oak Grove Drive, Pasadena, California, U.S.A.\\
\and
Jodrell Bank Centre for Astrophysics, Alan Turing Building, School of Physics and Astronomy, The University of Manchester, Oxford Road, Manchester, M13 9PL, U.K.\\
\and
Kavli Institute for Cosmology Cambridge, Madingley Road, Cambridge, CB3 0HA, U.K.\\
\and
LERMA, CNRS, Observatoire de Paris, 61 Avenue de l'Observatoire, Paris, France\\
\and
Laboratoire AIM, IRFU/Service d'Astrophysique - CEA/DSM - CNRS - Universit\'{e} Paris Diderot, B\^{a}t. 709, CEA-Saclay, F-91191 Gif-sur-Yvette Cedex, France\\
\and
Laboratoire Traitement et Communication de l'Information, CNRS (UMR 5141) and T\'{e}l\'{e}com ParisTech, 46 rue Barrault F-75634 Paris Cedex 13, France\\
\and
Laboratoire d'Astrophysique de Marseille, 38 rue Fr\'{e}d\'{e}ric Joliot-Curie, 13388, Marseille Cedex 13, France\\
\and
Laboratoire de Physique Subatomique et de Cosmologie, CNRS/IN2P3, Universit\'{e} Joseph Fourier Grenoble I, Institut National Polytechnique de Grenoble, 53 rue des Martyrs, 38026 Grenoble cedex, France\\
\and
Laboratoire de l'Acc\'{e}l\'{e}rateur Lin\'{e}aire, Universit\'{e} Paris-Sud 11, CNRS/IN2P3, Orsay, France\\
\and
Lawrence Berkeley National Laboratory, Berkeley, California, U.S.A.\\
\and
Max-Planck-Institut f\"{u}r Astrophysik, Karl-Schwarzschild-Str. 1, 85741 Garching, Germany\\
\and
National University of Ireland, Department of Experimental Physics, Maynooth, Co. Kildare, Ireland\\
\and
Observational Cosmology, Mail Stop 367-17, California Institute of Technology, Pasadena, CA, 91125, U.S.A.\\
\and
Optical Science Laboratory, University College London, Gower Street, London, U.K.\\
\and
Rutherford Appleton Laboratory, Chilton, Didcot, U.K.\\
\and
SUPA, Institute for Astronomy, University of Edinburgh, Royal Observatory, Blackford Hill, Edinburgh EH9 3HJ, U.K.\\
\and
School of Physics and Astronomy, Cardiff University, Queens Buildings, The Parade, Cardiff, CF24 3AA, U.K.\\
\and
Space Research Institute (IKI), Russian Academy of Sciences, Profsoyuznaya Str, 84/32, Moscow, 117997, Russia\\
\and
Space Sciences Laboratory, University of California, Berkeley, California, U.S.A.\\
\and
Stanford University, Dept of Physics, Varian Physics Bldg, 382 Via Pueblo Mall, Stanford, California, U.S.A.\\
\and
Universit\"{a}t Heidelberg, Institut f\"{u}r Theoretische Astrophysik, Albert-\"{U}berle-Str. 2, 69120, Heidelberg, Germany\\
\and
Universit\'{e} de Toulouse, UPS-OMP, IRAP, F-31028 Toulouse cedex 4, France\\
\and
Universities Space Research Association, Stratospheric Observatory for Infrared Astronomy, MS 211-3, Moffett Field, CA 94035, U.S.A.\\
\and
University of Granada, Departamento de F\'{\i}sica Te\'{o}rica y del Cosmos, Facultad de Ciencias, Granada, Spain\\
% ...
}


\abstract{We describe the processing of the 336 billion raw data samples from the High Frequency Instrument (hereafter HFI) which we performed to produce 
	six temperature maps from
   the first 295 days of \Planck-HFI survey data. These maps
   provide an accurate rendition of the sky emission at 100, 143, 217, 353, 545 
   and 857\GHz\ with an angular resolution ranging from 9.9 to 4.4\arcmin. 
   The white noise level is around 1.5 $\mu$K\ degree or less in the 3 main CMB 
   channels (100-217\GHz). The photometric accuracy is better than 2\,\% at
   frequencies between 100 and 353 \GHz and around 7\,\% at the two highest frequencies. 
   The maps created by the HFI Data Processing Centre reach our goals in terms 
   of sensitivity, resolution, and photometric accuracy. They are  already sufficiently 
   accurate and well-characterised to allow scientific analyses which are presented 
   in an  accompanying series of early papers. At this stage,
   HFI data appears to be of high quality and we expect that with further
    refinements of the data processing we should be able to achieve, or exceed,  the science goals
    of the \Planck\ project.
}

\keywords{Cosmology: cosmic background radiation -- Surveys -- Methods: data analysis}  

 \maketitle


 \section{Introduction}

\Planck\footnote{\Planck\ (\url{http://www.esa.int/Planck} ) 
is a project of the European Space
Agency (ESA) with instruments provided by two scientific consortia funded by ESA member
states (in particular the lead countries France and Italy), with contributions from NASA
(USA) and telescope reflectors provided by a collaboration between ESA and a scientific
consortium led and funded by Denmark.}
\citep{tauber2010a, planck2011-1.1}
is the third-generation space mission to measure the anisotropy of the cosmic microwave background (CMB).  It observes the sky in nine frequency bands covering 30--857\,GHz with high sensitivity and angular resolution from 31\arcmin\ to 5\arcmin.  The Low Frequency Instrument (LFI; \citealt{mandolesi2010, bersanelli2010, planck2011-1.4}) covers the 30, 44, and 70\,GHz bands with amplifiers cooled to 20\,\hbox{K}.  The High Frequency Instrument (HFI; \citealt{lamarre2010, planck2011-1.5}) covers the 100, 143, 217, 353, 545, and 857\,GHz bands with bolometers cooled to 0.1\,\hbox{K}.  

 Early science papers using HFI data are based on the products of the
 HFI Data Processing Centre (hereafter DPC). This paper describes the
 steps taken to transform the packets sent by the satellite into
 frequency maps, with the help of ancillary data, for example, from
 ground calibration.  At this early stage in the analysis, the LFI and
 HFI DPCs agreed to focus the analyses on temperature maps alone, as
 obtained from the beginning of the first light survey on the
 13$^{th}$ August 2009, to the 7$^{th}$ June 2010. These nearly ten
 months of survey data provide complete coverage of the sky by all
 detectors (by roughly 3 days more than the minimum duration needed),
 but only limited redundancy. Indeed the overlap between the two
 consecutive six-month surveys is only about $60\%$.

 The products of the early processing phase were made available to the
 joint HFI plus LFI DPC team in charge of producing the \Planck\ Early Compact
 Source Catalogue (hereafter ERCSC), and also to the CMB removal team,
 on the 17$^{th}$ July 2010, \ie only 5 weeks after the data
 acquisition was complete. Indeed, the early papers are based only on non-CMB products, so the DPCs produced a 
reference set of maps with the CMB removed.  
The CMB removed versions of the maps were
 made available to all of the \Planck\ collaboration on the 2$^{nd}$ 
 August, \ie three weeks later. The speed in completing this relatively
 complex chain of tasks was made possible thanks to the many years of
 preparation within the DPC, and to the excellent performance of the
 HFI in flight (described in the companion paper by the
 \cite{planck2011-1.5}). Of course, the accuracy of the processing
 presented here is not yet at the level required for precision CMB
 cosmology, but as we shall see, it has already reached an accuracy
 that is sufficient for many early science projects.

 The HFI DPC is the organisation in charge of analysing HFI data and
 ensuring strict traceability and reproducibility of the results. It
 is in charge of preparing all the formal deliverables of HFI at the six
 frequencies of the instrument. In addition to the concomitant ERCSC, the
 deliverables include maps of the sky, and their error properties,
 for all the HFI channels for the nominal survey duration of 15.5
 months. Early in the mission design, it was decided to have 
separate  DPCs for
 each the HFI and LFI  to make efficient use of the hardware expertise, 
 which is critical in generating science quality
 data from such complex instruments.  Nevertheless,
 wherever possible common tools have been developed jointly by the two
 DPCs, for example, the non-instrument specific parts of the \Planck\
 simulation pipeline. We have also exchanged calibrated processed
 data between the two DPCs, every alternate month since the start of
 regular observations, to allow cross-checks between frequency bands
 and various other tests.

 The HFI DPC relies on a centralised data base and common software
 infrastructures  operating on a reference hardware platform. This
 centralized backbone serves  geographically distributed groups of
 scientists in charge of various aspects of  code development and of validating
 results. The infrastructure also serves dedicated science  groups of the
 HFI core team. Parts of that infrastructure (and data) are
 replicated at various DPC supporting computer centres to avoid
 excessive data transfers. The combined computing power at these centres
 has proved important for some DPC activities, in particular for 
 end-to-end Monte-Carlo simulations. Appendix~\ref{Annex:infra} gives
 an overview of the 
 infrastructure that we have developed. This infrastructure
evolved from the initial implementation of a breadboard model (used
 as a prototype in designing the ground segment), a development model
(completed in January 2007 to validate the basic infrastructures), 
and a flight model designed to be ready before
launch\footnote{This flight readiness was checked through end-to-end
   tests on simulated data, based on knowledge of the
   instrument derived from the ground calibration campaigns.} 
to process the data and create clean calibrated maps
for the full duration of the nominal mission. Our main tasks after launch 
were therefore to compare the in-flight performance of the HFI to the
data derived from ground calibration, to identify, understand and model 
unanticipated systematic effects, and
to quantify uncertainties in the data as accurately as possible.

The next section provides an overview of HFI data processing.
Section~\ref{sec:L1} describes the preliminary part of the processing
which assembles telemetry packets in Time Ordered Information objects
(hereafter TOIs). Section~\ref{sec:toip} is devoted to the processing
of the detector TOIs to produce cleaned timelines which are used to
estimate the temporal noise properties in Sect.~\ref{sec:detnoise} and
to determine the beams and the focal plane geometry in
Sect.~\ref{sec:fpg}.  Sect.~\ref{sec:map} discusses the creation of
maps and their photometric calibration, and Sect.~\ref{sec:cmbrm}
describes the CMB removal.  Sect.~\ref{sec:sum} concludes
with a summary of the characteristics of the data,  as currently
processed.
 
 Appendix~\ref{Annex:infra} to \ref{Annex:CO} provides details on the DPC infrastructure, 
the temporal noise determination, focal plane measurements, the determination of the 
effective beams or band-pass measurements at CO lines frequencies. 

 \section{HFI Data Processing Overview}

 The overall data flow follows a succession of distinct steps which we
 refer to as `levels'. The first, L1, begins with formatting the data
 received from the satellite to build a database of raw data. The temporal
 sequences of measurements are grouped into objects generically called
 Time Ordered Information, or TOI. The second level of processing, L2,
 produces maps of the sky from the raw TOIs, their characterisation,  and
 a model of the instrument. This is the core of the DPC analysis, when
 instrumental effects are identified and corrected. The next
 level, L3, involves separating the sky emission maps into
 astrophysical components (including point sources) and their
 scientific characterization.

 Functionally, the data analysis may be thought of as using data, $y$,
 related to an unknown source, $x$, through some function, $F$, 
 describing the measurement process, to infer information on
 $x$ from $y$. In addition, we need to estimate the covariance matrix
 of the errors on the estimate of $x$. Different choices of $x$ and $y$
 describe different steps of the data analysis. For L2, the unknown
 $x$ is the sky emission in different bands. The function $F$ is then
 a description of how the instrument relates the sky to the data,
 where $y$ is the TOI and associated calibration data. The difficulty
 is, of course, that $F$ must be partly determined (or at least
 verified) from the flight data itself, which requires an iterative
 procedure involving a succession of analysis and synthesis steps.
 
 In the analysis phase, a parametric model is constructed of the
 relationship linking the TOIs to the unknowns, together with
 estimates of the current values of the parameters.  In the synthesis
 phase, this model of the measurements is used in conjunction with the
 TOIs to process the TOIs globally and to determine the unknowns.  The
 results may then be used to  improve the parametric model of the
 measurement to be 
 used at the next iteration (\eg\ by sampling the resulting sky map
 to obtain an improved signal predictor to be removed from the raw
 TOIs).  In practice,  most of the L2 tasks are
 devoted to improving the instrument model (instrument temporal
 transfer function, noise properties,  focal
 plane geometry, main beams, far side lobes, gains, 
 identification of systematic effects, \etc).

 Each new pass through the data results in new processed data and an
 improved understanding of the measurements. This understanding is
 retained through an update of the instrument model which therefore
 has a central (conceptual and practical) role. The Instrument Model,
 hereafter the \IMO, is a synthesis of the information (from ground
 test and flight data, simulations and theoretical considerations)
 required to perform the data processing. Implementation details may
 be found in the appendix~\ref{sec:IMO}. The initial version of the \IMO\ was
 derived pre-launch from an analysis of ground calibration data and
 was used for processing the First Light Survey, \ie data acquired
 during the last two weeks of August 2009. The analysis of the First
 Light Survey verified that the satellite, instruments and their
 ground segments were all fully operational.  The next \IMO\ version
 included improved models derived from a detailed analysis of the
 Calibration and Performance Verification phase (hereafter CPV) data
 acquired during the previous 1.5 months. Further versions of the
 \IMO, and of the pipelines themselves were then derived after the
 completion of successive passes through the data. The data made available
 to the \Planck\ consortium for the early science papers 
resulted from the fourth pass through the data.

 \begin{figure*}
	 \includegraphics[width=1\textwidth]{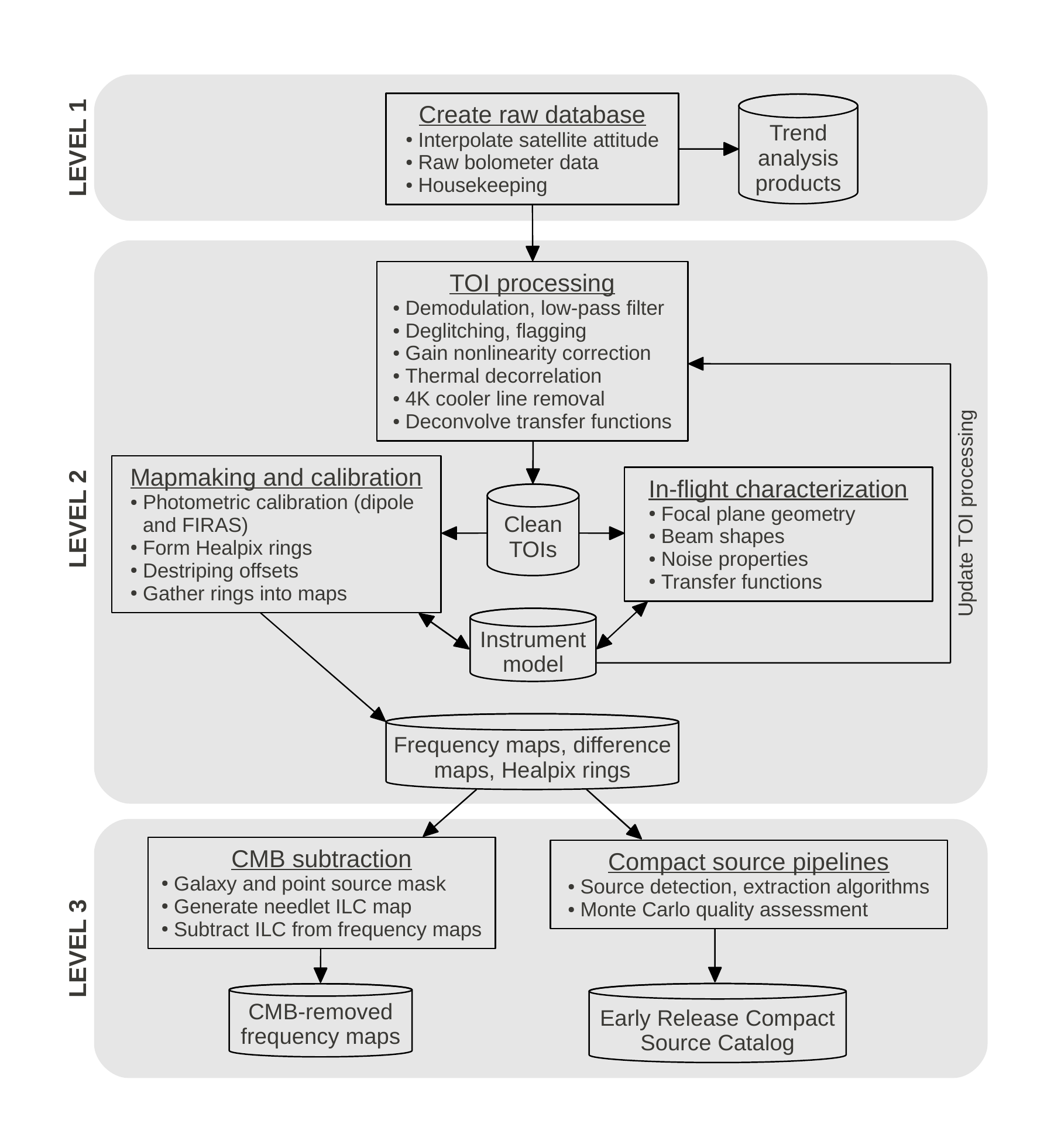}
	 \caption{Overview of the data flow and main
   		functional tasks of the DPC.  This flow
   		diagram illustrates the crucial role of the Instrument
                Model (\texorpdfstring{\IMO}{IMO})
   		which is both an input and an output of many tasks, and is updated
   		iteratively during successive passes of the data }
   	\label{fig:pipes} 
 \end{figure*}

 We now provide an overview of the main steps involved in the data
 processing, which can be visualised with the help of
 Fig.~\ref{fig:pipes}. First,  the L1 software fills the database and
 updates,  daily,  the various TOI objects as described in
 Section~\ref{sec:L1}. The satellite attitude data, which we receive 
 sampled at 8\Hz\ during science data acquisition and at  4\Hz\ otherwise, are 
densely sampled by interpolation to the 180.4\Hz\
 acquisition frequency of the detectors. Raw TOIs and housekeeping
 data are then processed to compensate for the instrumental response
 and to remove estimates of  known artefacts. To do so, the raw
 TOIs in Volts, are demodulated, filtered, deglitched, corrected for
 gain non-linearity on strong sources like Jupiter, and for temperature
 fluctuations of the environment using 
 correlations with the signal TOIs from the two dark bolometers.
 Narrow spectral lines caused by the 4K cooler are also removed before
 decorrelating the temporal response of the instrument. Finally,
 various flags are set to mark unusable samples,  or samples where the
 signal from planets or other strong sources is important.  The setting of
 flags is described further in Sect.~\ref{sec:toip}.

The \Planck\ satellite spins around an axis pointing towards a 
fixed direction on the sky, repeatedly scanning the same circle. 
As the spin axis follows the Sun, the observed circle sweeps 
through the sky at a rate of $\simeq 1$ degree/day. Assuming a focal plane 
geometry, \ie a set of relations between the
 satellite pointing and that of each of the detectors, one can proceed
 to build `rings' which are used to derive a new version of the
 \IMO. A ring corresponds to a sky signal estimate along a circle
 which can be derived by analysing all the data acquired by a detector
 during each stable pointing period,  when the spin axis is pointing
 towards an essentially fixed direction in the sky and the detectors
 repeatedly scan the same circle on the sky (up to a negligible
 wobbling of the spin axis). This redundancy permits averaging of the
 data on rings to reduce the instrument noise. The resulting estimate
 of the sky signal can then be removed from the TOI (by unrolling it
 periodically) to estimate the temporal noise power spectral density,
 which is a useful characterisation  of the detector data after
 TOI processing. As described in Sect.~\ref{sec:detnoise}, this noise
 may be described as a white noise component, dominating at
 intermediate temporal frequencies, plus additional low and high
 frequency noise (when passed by our high frequency filtering).
 
 The effect on maps of the low frequency part of the noise can be partially mitigated by
 determining an offset for each ring. These so-called `destriping' offsets 
 are obtained by requiring that the difference
 between intersecting rings be minimised.  
 Once the offsets are removed from each ring, the rings
 can be simply co-added to produce sky maps.  As explained in the
 map-making and calibration section (Sect.~\ref{sec:map}), a
 complication arises from the fact that the detector data include both
 the contribution from the \emph{Solar} dipole induced by the motion
 of the Solar System through the CMB (sometimes referred to as the 
`cosmological' dipole), and the \emph{orbital} dipole induced by the
 motion of the satellite within the Solar System.  The orbital dipole
 contribution must, of course, be removed from the rings before
 creating the sky map.  While the orbital dipole can in principle be used as
 an absolute calibrator for the CMB detectors, it is rather weak and
 needs a long baseline to break degeneracies with other signals. For
 the time being, we use the Solar dipole as determined by \WMAP, or a
 comparison with \COBE/\FIRAS\ at higher frequencies, as our
 calibrators at the map level. Since we need this calibration to
 remove the orbital dipole contribution to create the maps themselves,
 the maps and their calibrations are obtained iteratively. The dipoles are computed  
 in the non-relativistic approximation. 
 The resulting calibration coefficients are  also stored  in the \IMO,
 which can then be used, for instance, to express noise spectra in
 Noise Equivalent Temperatures (NETs).
 
 The `destriping' offsets, once obtained through a global solution, are also used to create
 local maps around planets which, as described in
 Section~\ref{sec:fpg} can be used to improve knowledge of the focal
 plane geometry stored in the previous version of the \IMO\ and to
 improve  measurements of the `scanning' beam (defined as the beam
 measured from the response to a point source of the full optical and
 electronic system, \emph{after} the filtering done during the TOI
 processing step).

 The ring and map-making stages allow us to generate many different
 maps, \eg using disjoint sets of detectors, the first or second
 halves of the data in each ring, or from different sky surveys
 (defined, by convention, as the data acquired over exactly six
 months).  The creation of `jackknife' maps, in particular, has proved
 extremely useful in characterising the map residuals.

 Section~\ref{sec:cmbrm} describes a further step, performed jointly
 with the LFI DPC, to estimate and remove the CMB contribution over
 the full sky in all of the channel maps. This was used by some of the
 scientific analyses reported in the \Planck\ early release papers
 (though other analyses used specific CMB removal techniques, described in the
 relevant papers, \eg using ancillary data available in their specific
 areas of the sky). The specific processing for extracting the
 \Planck\ ERCSC from the HFI and LFI maps is described in a companion
 paper \citep{planck2011-1.10} and additional details can be found in
 the ERCSC explanatory supplement \citep{planck2011-1.10sup}.
 
 The properties of the data are summarised in the concluding section~\ref{sec:sum}.  
 
 As already stated, this paper does not address polarisation
 measurements from the polarisation sensitive bolometers (hereafter
 PSBs) on the HFI at the $100$, $143$, $217$ and $353$ \GHz\
 channels. The accuracy of their properties derived from ground
 calibration is sufficient for our current purposes, which is
limited to using  PSBs as contributors to the total intensity
 measurements.  Ground
 measurements showed that the polariser orientations are within about
 1 degree of their nominal values.  Similarly, ground measurements
 showed levels of cross-polarization leakage of a few percent, which is good enough
 to use PSB TOIs to create the temperature maps discussed in this
 paper. Additionally, we have verified for Tau A, a bright polarised
 calibration source with a S/N $>$ 400 in a single (unpolarised) detector scan,
 that in-flight determinations are consistent with pre-launch data.

\section{From packets to Time Ordered Information} \label{sec:L1}

L1 is a collection of hardware and software items used to retrieve and
gather data from the satellite and the ground segment. Functionally,
L1 has two objectives.  The first is to provide the instrument
operation team with the tools required to monitor the HFI instrument
with a typical response time of a few seconds. Those tools are the
SCOS\footnote{SCOS-2000 is the generic mission control system software
  of ESA.}  visualisation tool and the HFI Quick Look Analysis tool
(hereafter QLA) which enables the visualisation of lower level
parameters, as well as command checking and acknowledgement. The Trend
Analysis (TA) tool provides quasi-automated reports on the evolution
of various instrument parameters.  The second L1 objective is to build
Time Ordered Information objects (hereafter TOI) and ingest them into the
DPC database for the next level of data processing.

The data set handled by the L1 consists of: \begin{itemize}

\item the telemetry packets which are, 	
     either coming from the satellite in near real time during the
    Daily Tele-Communication Period  (hereafter DTCP), 
    or dumped daily from the satellite, consolidated and made
      available by ESA's Mission Operating Centre (the MOC); 
 
 \item auxiliary files such as 
      pointing lists, orbit files, attitude data, 
     On-Board-Time \& Universal Time (OBT-UTC) time correlation data,
      telecommands history file as built by the MOC.

\end{itemize}

Regarding the data received  from the satellite,  L1 handles the satellite, 
sorption cooler and the HFI housekeeping data and the HFI `scientific' data,
each with  its own sampling rate. HFI housekeeping data include the 7
setup parameters for each detector (bias, voltage, gains, compression) and the HFI general
configuration (thermometry, temperature control, power supply, state of coolers).
From the housekeeping telemetry packets
the parameters are extracted and stored in the data base.

Signals in a telemetry packet or house keeping parameters, usually come in ADU
(analogue-to-digital unit). They are stored in the form in which they
are received and often need to be combined in various ways and
converted to  appropriate 
physical units, which is done on-the-fly using  a dedicated library
(\cf Appendix~\ref{sec:comlib} describing the so-called transfer
functions).

 \subsection{Building the HFI science TOIs} \label{sec:L1fill}

\begin{figure}[htbp] \centering
	\includegraphics[width=0.5\textwidth]{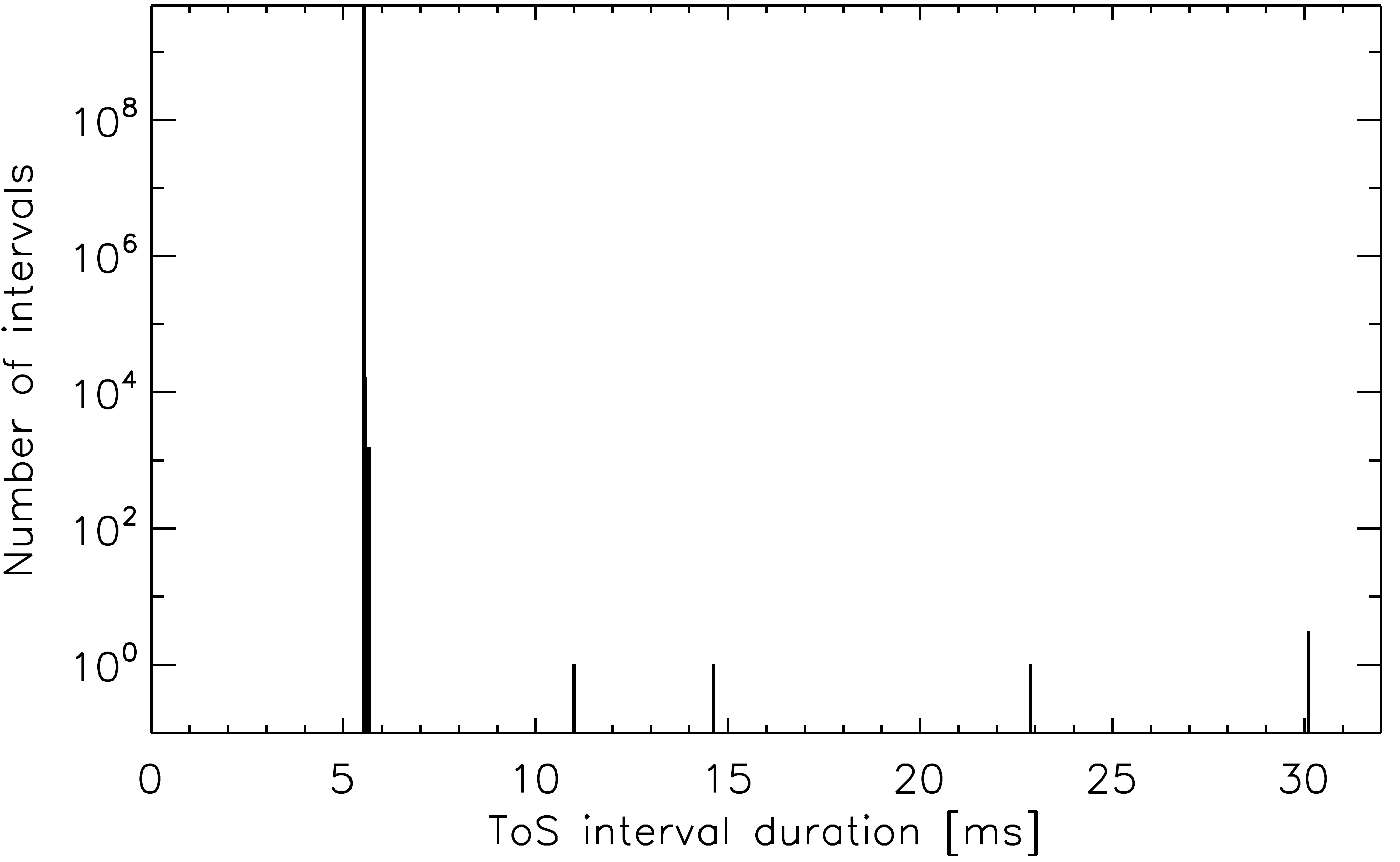}
	\caption{Histogram of time differences between two successive samples. 
		The time quantization step is $15\,\mu{\rm s}$.}
  	\label{fig:DeltaTos} 
\end{figure}

 On board, the signal from the 72 HFI channels is sampled at 180.4\,Hz
 by the Read-Out Electronic Unit (hereafter REU). 254 samples per
 channel are grouped into a {\it compression slice\,}. The Data
 Processing Unit (hereafter DPU) then builds a set of several
 telemetry packets containing the compression slice data and adds to
 the first packet the start time of the compression slice. When
 receiving this set of telemetry packets, the L1 software extracts the
 $72 \times 254$ samples and computes the time of each sample based on
 the compression slice start time (digitized with 15\,$\mu$s
 quantization steps) and a {\it mean sample time\,} between
 samples. For the nominal instrument configuration, the sample
 integration time is measured to be $T_{\rm samp} = 5.54404\,{\rm
   ms}$.  At one rotation per minute, this corresponds to an arc of
 $2.0\, \rm arcmin$ on the sky. This interval is quantized in units of
 $2^{-16}\, {\rm s} = 15.26 \,\mu{\rm s}$, so that occasionally the
 time intervals differ by that amount, as can be seen from
 Fig.~\ref{fig:DeltaTos}.

It should be noted that the time of sample (ToS) depends on the REU
sampling time and the DPU clock.  Although the REU sampling time was
fixed once at the beginning of the mission by telecommand, it relies
on the REU clock quartz frequency which depends slightly on its
temperature.  From the satellite Thermal Balance/Thermal Vacuum tests
in Li\`ege, this dependency has been computed to be about 7.1
nanoseconds for a change of the REU temperature by one
Kelvin. Given the observed temperature stability of the REU (a drift
by less than 0.6\,K for the mission duration so far), this correction
does not currently need to be taken into account in the ToS
computation.

\begin{figure}[htbp] \centering
     	\includegraphics[width=\columnwidth]{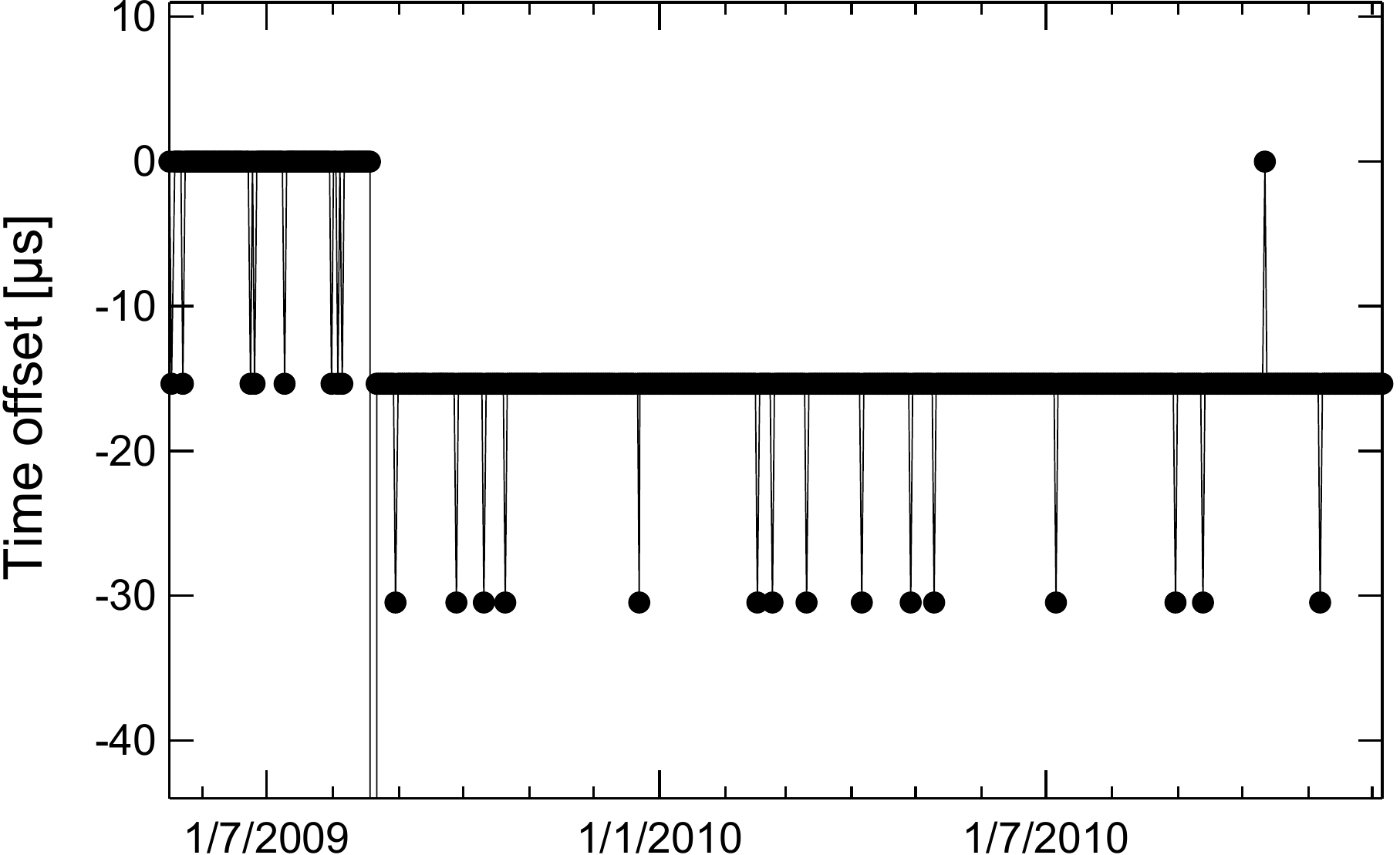}
     	\caption{Time offsets showing that the HFI DPU clock is in
                synchronisation with the satellite  
     		on-board time within a 15 microsecond quantization step.} 
       	\label{fig:DeltaOff} 
\end{figure}

 The ToS also depends on the DPU clock through the compression slice
 time stamping. Figure~\ref{fig:DeltaOff} shows that the DPU clock
 remains synchronised,  within a 15 microseconds quantization step, 
 with the satellite clock, with the largest exception so far being the
 large step which can be seen on the plot around the end of August
 2009. This was caused by  a software patch of the satellite Command and
 Data Management Unit. This local de-synchronization is small enough
 that it is ignored in the computation of the ToS for this analysis.

 \subsection{Statistics about the data gathered by  L1} \label{sec:L1stat}
 
During the first year of survey (which excludes the commissioning and CPV
phase, when the HFI was in specific modes of data production) the L1 handled
$4.0\times 10^{8}$ telemetry packets; $29 \%$ for satellite housekeeping; $6
\%$ for spacecraft housekeeping; $4 \%$ for HFI housekeeping; and $61 \%$ for
HFI science data. The total number of housekeeping parameters stored in the
database is 25,425.

Regarding HFI science packets, only 20 packets (\ie $8.1\times 10^{-6}\, \%$)
have been lost at the satellite level for well understood technical
reasons. None has 
been lost during transmission from space to the ground nor at the ground
segment level.

A total number of $4.66\times 10^{9}$ time samples have been stored in the
database for each of the 72 detectors, leading to a total of $3.36\times 10^{11}$ 
science data samples. 

 \subsection{Numerical compression of science data} \label{sec:comp}

 The quantization performed during the compression process was tuned
 during the CPV and the early phase of the survey. The compression
 performance complies with the requirements and is detailed in
 \cite{lamarre2010}.

 Because of the finite telemetry bandpass between the spacecraft and
 the ground station, the signal cannot be transmitted in some rare
 circumstances,  specifically, large amplitude glitches (which, in any
 case, are flagged out during the deglitching process) and the
 Galactic centre region for the 857\GHz\ detectors (where less than
 200 samples were lost during the March 2010 crossing). Because of the
 redundancies inherent in the \Planck\ scanning pattern, and the
 irregular angular distribution of the small number of compression
 errors, no pixels are missing in the maps of the
 Galactic centre regions.

 \subsection{Pointing processing}\label{sec:ptgproc}
 
 As is described in the mission overview paper \citep{planck2011-1.1}, 
the DPCs retrieve the
 reconstructed pointing information of the satellite every day from
 ESA's Mission Operating Centre. 

 The main data consists of the attitude as a function of time,
 represented by a normalized quaternion.  The attitude is the rotation
 linking a fixed reference frame on the satellite, in this case that
 formed by the telescope nominal line-of-sight and the nominal
 spin axis, with a sky reference frame in ecliptic J2000 coordinates.
 Both the raw attitude data (\ie as provided by the on-board satellite
 pointing system) and a processed version of this data, are
 distributed to the DPCs.  The main processing consists of 
 filtering high frequency noise during the stable pointing periods
 (\ie\ between slews). The order of magnitude of this high frequency
 noise can be roughly evaluated by measuring the mean quaternion
 distance $\Omega = \arccos(q_r\ q_f^{-1})$ between the raw quaternion
 $q_r$ and the filtered one $q_f$ (see below), which is of the order
 of 7\,arcsec. 

 Other ancillary data are also retrieved which are either related to
 the performance of the pointing system (operating mode, quality of
 the attitude determination, time of firing of the thrusters) or are
 derived quantities (in particular the nutation, mean and
 instantaneous direction of the spin axis).

 This data is retrieved by L1 in the form of two files: the Raw
 Attitude File (RAF) and the Attitude History File (AHF) which
 contains the raw and ground processed daily pointing
 information. Both are stored in the HFI database, and their content
 is used to compute the pointing at each data sample for each
 bolometer.

 At this stage of the mission, we assume that the pointing
 reconstruction,  once processed,  is perfect and thus the
 computation of the pointing at each data sample is very simple.  The
 filtered attitude is interpolated to the time of sampling of the
 bolometers using the spherical linear interpolation algorithm
 \citep{shoemake1985}, \ie the attitude quaternion $q(t)$ at time
 $t_0<t<t_1$ is computed from the AHF quaternions,  $q_0 = q(t_0)$ and
 $q_1 = q(t_1)$,  from a linear interpolation on the sphere which is given by
 $q(t) = \left(q_1 q_0^{-1} \right)^{\Delta t} q_0$, where $\Delta
 t=(t-t_0)/(t_1-t_0)$. This formula can
 be rewritten as
 \begin{equation}
  q(t) = \frac{\sin(\Omega(1-\Delta t))}{\sin\Omega} q_0 + 
  \frac{\sin(\Omega\Delta t)}{\sin\Omega} q_1 ,
 \end{equation}
 with $\Omega$ the quaternion distance between $q_0$ and $q_1$ defined above.
 
 The interpolation is done once (since all bolometers are sampled at
 the same time) and then submitted to the database.  This has a cost in
 disk space but, in the future, may be used to correct for possible
 time dependent deformations of either the star tracker assembly or
 telescope assembly. No such correction has been applied at this
 stage, given the low level of deformation which has been measured so
 far (see 
 Section~\ref{sec:fpg}).

 The pointing for each bolometer is computed on the fly\footnote{This is done
 	using the HFI  pointing library, which can compute the pointing in different
	sky reference frames, using different pointing representations
	(Cartesian, spherical) and polarisation angle conventions (IAU or
 	\healpix). } 
by combining
 the attitude with the location of the bolometer in the focal plane,
 as computed by the focal plane geometry pipeline (see Sect.~\ref{sec:fpg}) and stored in the \IMO.  This means that a single pointing 
timeline is kept in the
 database, rather than one per bolometer, producing substantial
 savings of disk space (a pointing timeline requires eight
 times the storage space of a signal timeline: four double precision
 values for the quaternions compared to one single precision value for
 the data).

\begin{figure*}[!htbp]
	\includegraphics[angle=180,width=1\textwidth,totalheight=.35\textheight]
		 {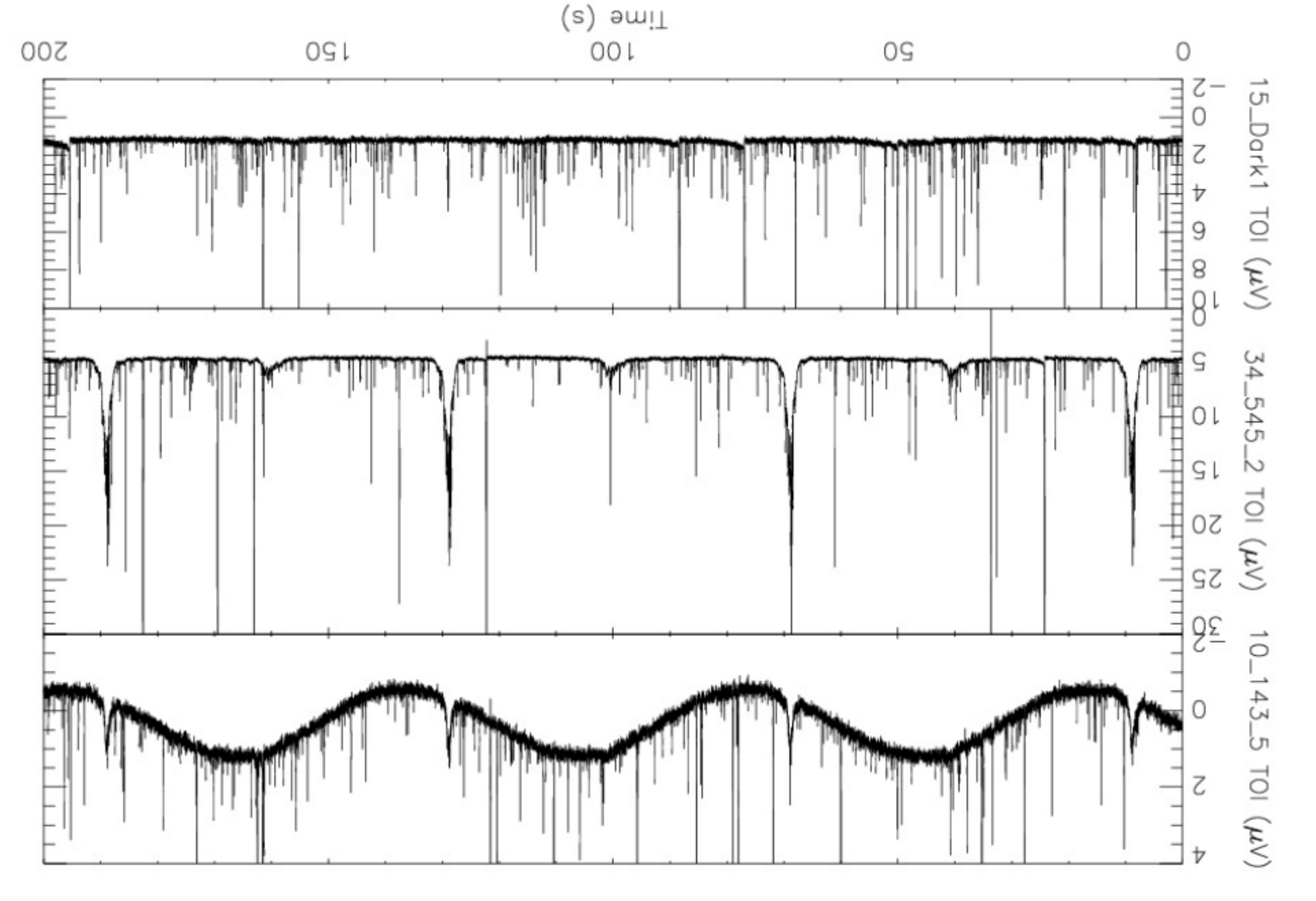}
	\caption{Raw TOIs for three bolometers,  the `143-5' (top),
          `545-2' (middle), and `Dark1' (bottom) illustrating the typical
	behaviour of a detector at 143\GHz, 545\GHz, and  a blind
        detector over the course of  three rotations of the spacecraft at 1 rpm.  At
	143\GHz, one clearly sees the CMB dipole with a  60\,s  period. 
	The 143 and 545\GHz\ bolometers  show vividly the two
	Galactic Plane crossings, also with 60\,s periodicity. The
        dark bolometer exhibits a nearly constant baseline together with  a
        population of glitches from cosmic rays similar to those seen in the 
	two upper panels.}
	\label{fig:TOIExample1} 
\end{figure*}

 The AHF and RAF are also used to obtain the reference times used to
 segment the data streams into rings (or pointing periods)
 which correspond to the intervals between successive depointing
 manoeuvres.  These rings are used both as an aid to index data and
 as a basis for some of the data processing tasks\footnote{ The start
 and end time of rings are the actual times of thruster firings.}.
  
 Finally a set of flags is created to identify the part of each ring that
 corresponds to the stable pointing period. Only this part of the data  
 is currently used in further processing.

 \section{TOI Processing}\label{sec:toip} 

The TOI processing pipeline produces cleaned TOIs and
associated flags for use in the subsequent data processing. The basic steps
involved are: (1) signal demodulation and filtering; (2) deglitching, which 
flags the strongest part of the glitches and subtracts a model of
their tails; (3) 
conversion from instrumental units (Volts) to physical units (Watts of
absorbed power), including a correction for non-linearity; (4)
decorrelation of thermal fluctuations; (5) removal of the systematic
effects induced by the 4\,K cooler mechanical vibrations;  (6)
deconvolution of the bolometer time constant.  These steps are described in
more detail in the remainder of this section.

\subsection{Demodulation and initial filtering} 

The data are first demodulated from the AC bolometer modulation. A 
low-pass filter is then applied to the TOI in order to remove the 
modulation frequency carrier (at $90.19\,\mathrm{Hz}$). This low-pass 
filter (a finite width digital symmetrical filter) is adapted to the temporal 
transfer function (see \ref{sec:TF}) using data from Mars
crossings. Hence, the demodulated data $d(t_{i})$ is filtered with a
  centred filter $f_{j}$ of width $2N+1$:
$$TOI_{i}=\sum_{j=-N}^{+N}d(t_{i+j})f_{j}$$

The optimization criteria to derive the filter coefficients  include 
(1)~to have an exact zero at the modulation frequency, 
(2)~to minimize beam smearing in the in-scan direction (less than 20~\% increase of the
in-scan beam width), and (3)~to reduce the enhancement of high-frequency
noise after the temporal transfer function deconvolution.  It was found that a
Kaiser-like filter \citep{kaiser1974,walraven1984} with 13 points 
offers a good trade-off for the 100 and 143\GHz, while we use a
21-point filter for the 217 and 353\GHz, and a 3-point filter for the
545 and 857\GHz\ channels (this last filter is simply $\{\frac{1}{4},\ \frac{1}{2},\ \frac{1}{4}\}$). 

Figure~\ref{fig:TOIExample1} displays the signal for three typical bolometers
after demodulation. At 143\GHz, one clearly sees the CMB dipole
pattern with a 60 second period. Both the 143 and 545\GHz\ 
bolometers clearly show the two Galactic 
plane crossings, also with a 60\,s periodicity. The dark bolometer exhibits a
rather constant baseline, as well as a population of glitches similar to those
apparent in the 143 and 545\GHz\ detectors.

\subsection{Glitch Handling} \label{sec:deglitching}

Glitches result from the impact of cosmic rays in the vicinity of HFI
detectors. As can be seen in Fig.~\ref{fig:TOIExample1}, they are
quite conspicuous in the raw data. We start by describing the glitch
types we found in the data \cite[see][for a discussion of their
physical origin]{planck2011-1.5}, then we describe the method that we
have developed to process them, and then we show how this processing
modifies the data.  Finally we address the question of errors in the
data introduced by the deglitching process.

\subsubsection{`Glitchology'} \label{sec:glitchology}

\begin{figure}[htbp] 
	\centering
  	\includegraphics[width=\columnwidth]{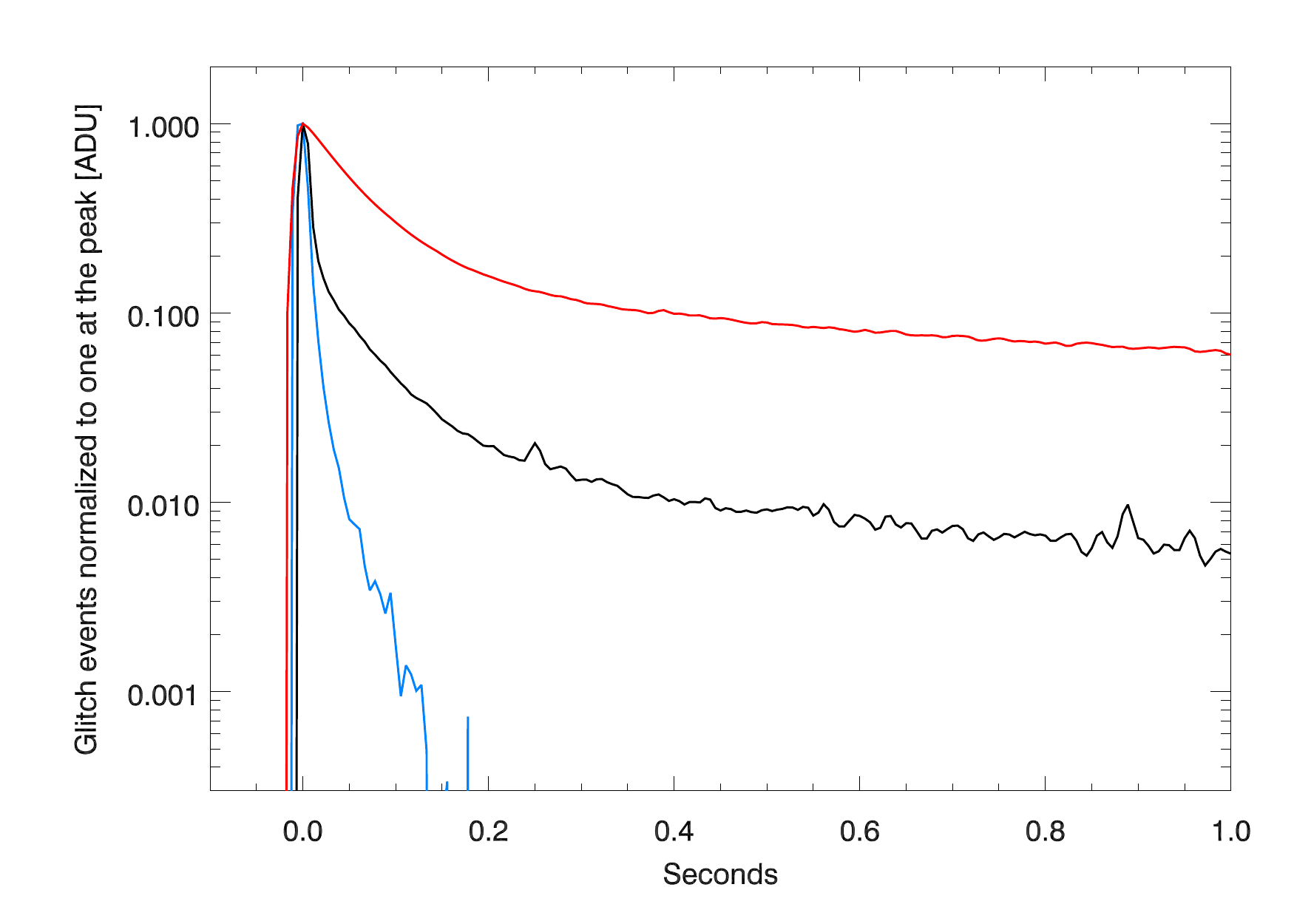}
  	\caption{Three examples of high amplitude glitch events for one bolometer. 
  		One event for each category of glitch is shown, normalized to 
unity at the peak.  
  		The blue curve shows  a short glitch, the black curve  a 
long one, 
  		and the red curve is for a very long glitch. Short and long glitches both display a fast  
  		bolometer time-constant decay; in the case of long glitches
this is  followed by an extended tail. Very long  
  		glitches are described by the same extended template as long glitches but do not show 
  		the fast bolometer time-constant decay.}
	\label{fig:GlitchTypes} 
\end{figure}

The shape of a typical glitch is characterised by a sharp rise
followed by a slow decay. The occurrence of glitches are not
correlated with each other (except between detectors belonging
to the  same PSB
pair) and they are independent of the sky signal. This first finding was
quantified by measuring the histogram of the time delay between two
glitches and  showing that it follows the expected exponential law
for uncorrelated events. Nevertheless, we were not able to check this
behaviour accurately below 20 sample intervals ($\approx 100~$ms) because
of confusion between events.  The `noise' in
the data produced by glitches is naturally dealt with in the time
domain, as its source is well localized in time. Three significant
populations of glitches have been identified and characterized:
\emph{short glitches}, which are characterised by a fast decay (of a
few milliseconds) followed by a low amplitude slower decay;
\emph{long glitches}, whose fast decay is followed by a long tail with
time constants of about 60\,ms and 2\,s; even \emph{longer glitches},
which show only a slow decay very similar to the second
category. Figure~\ref{fig:GlitchTypes} shows examples of these three
glitch types. The population of long glitches predominates at low and
intermediate amplitudes whereas the short glitches dominate at higher
amplitudes (see Fig.~\ref{fig:GLdist} for a quantitative assessment).

\begin{figure}[htbp] 
	\centering
  	\includegraphics[width=\columnwidth]{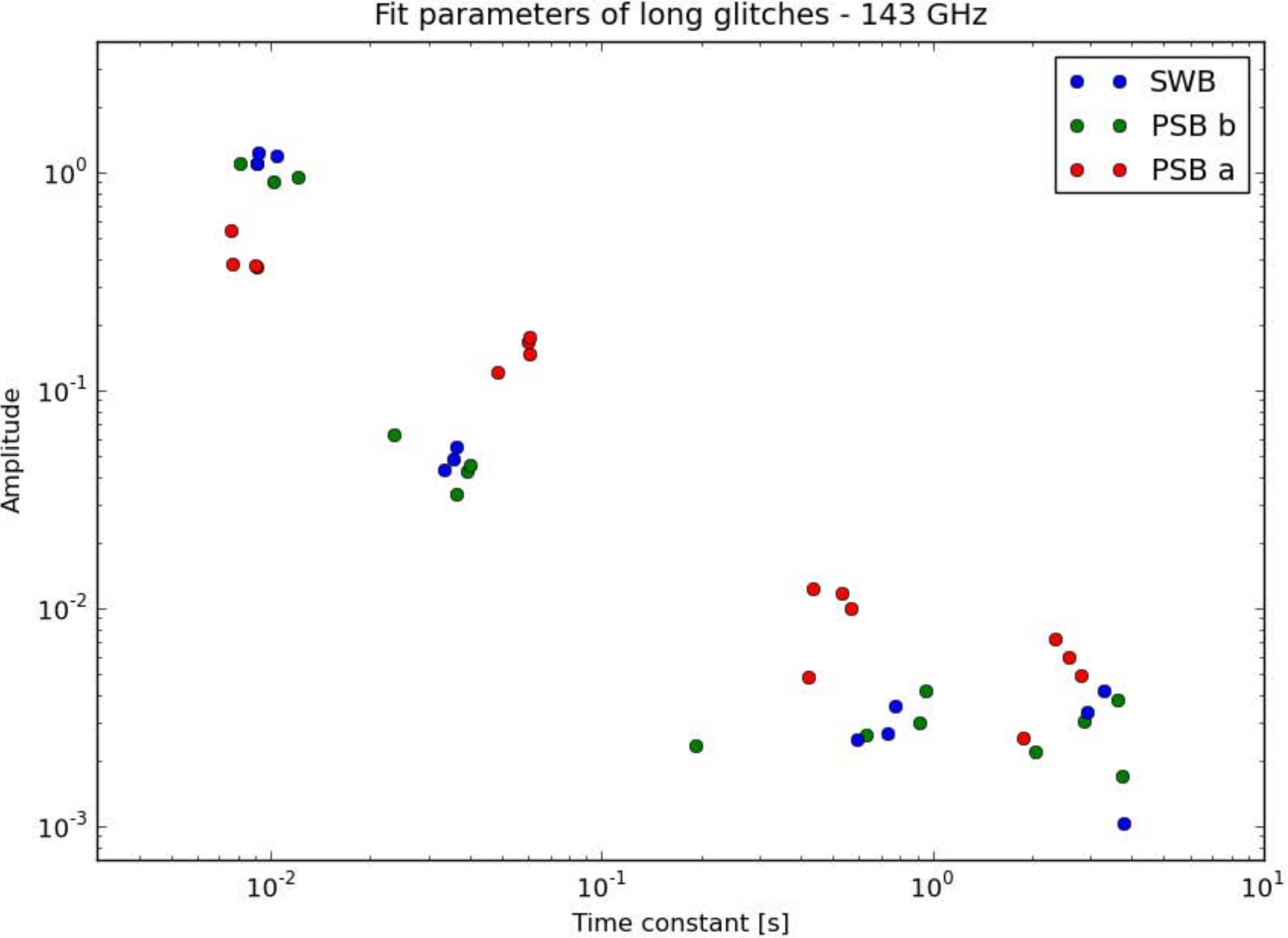}
  	\caption{Amplitudes of the components of the long glitch templates versus
          their exponential decay times. For each  
  	bolometer (here those at 143\GHz), the glitch template is fitted as a sum of four
    	exponentials, $\sum_{i=1}^4 A_i \exp{-t/\tau_i}$ (and the plot shows $A_i$ vs $\tau_i$). 
          We observe some homogeneity in the grouping of the long glitch characteristics of the 
         detectors, even more so within each detector type. }
	\label{fig:GlitchTimeConstant} 
\end{figure}

Precise templates have been obtained for each glitch population in
each bolometer by stacking many events \citep[see][for more
details]{planck2011-1.5}. Figure~\ref{fig:GlitchTimeConstant}
summarizes the properties of the long glitch templates for all the
143\GHz\ detectors; each template is a sum of four exponentials, and
the figure displays,  for each exponential,  its amplitude and decay
time. Note the long duration ($\sim 0.5-5$s) of the low-amplitude
components. By comparing the long glitch templates for different
bolometers, we observe that for PSB-a detectors long glitches have a
60\,ms exponential decay with an amplitude relative to the peak of
about  10 \%, whereas for PSB-b and SWB detectors this intermediate
time-constant decay is at a substantially lower amplitude. The long
time constant decay of long glitches of about two seconds produces noise
excess at frequencies approximately $ 0.01$~Hz, which is 
 well above instrumental and
photon noise for most of the detectors. The shape of the long glitch decay 
does not vary significantly over the time of the mission or as a function of amplitude.

Given a glitch rate of typically one per second,  and their long decay
times, we cannot simply flag them out, since that would lead to an
unacceptably large loss of data.  We therefore flag (and in practice
discard) only the initial part of the glitch and fit the tail with a
model which is subtracted from the TOI.  The temporal redundancy of
the sky observations is used to separate the glitch signal from the
sky signal. Since each ring is scanned 40 -- 70 times with
excellent pointing accuracy (see the paper by
\cite{planck2011-1.1}) the \Planck\ scanning pattern provides
a high degree of redundancy.
 This redundancy is of crucial importance as it
provides a way to separate from strong  signals, such as
point-sources, which will generally show similar temporal behaviour to the
glitches. The algorithm used for the glitch processing iterates
between the estimates of the signals from the sky and from the
glitches, and hence should not bias the estimation of the
final sky signal since, at convergence, the glitch detection is to first
order independent of real sky signals.

Many previous bolometric CMB experiments have deconvolved the
time streams prior to deglitching, which has the benefit of increasing
the signal-to-noise on glitches and minimizing the fraction of flagged
data. For a very good reason, this is not what is done in HFI; the
populations of glitches each exhibit distinct transfer functions, so
there is no single Fourier representation that would appropriately
treat them all.

\subsubsection{Glitch handling methodology}

An initial estimate of the sky signal is obtained ring-by-ring by
using only data within three times the dispersion around the median
value in each sky phase bin (ring pixel) in order to mitigate the effect of possible outliers. 
A third-order polynomial
interpolation within ring pixels is then performed to account for
sub-pixel variations of the signal (\eg large gradients close to the
Galactic plane). This is used to subtract the estimated sky signal
from the TOI at the exact time/position of the sample. The glitch detection
threshold is first set relatively high to prevent errors in the
reconstructed signal from inducing spurious detections.
Then the threshold is decreased by a constant ratio at each iteration
as the estimate of sky signal and glitches improves. Usually some five
iterations are needed for convergence.  After the first iteration, the
sky signal is re-estimated using a more elaborate method  described
later.

\begin{figure}[htbp] 
	\centering
  	\includegraphics[width=\columnwidth]{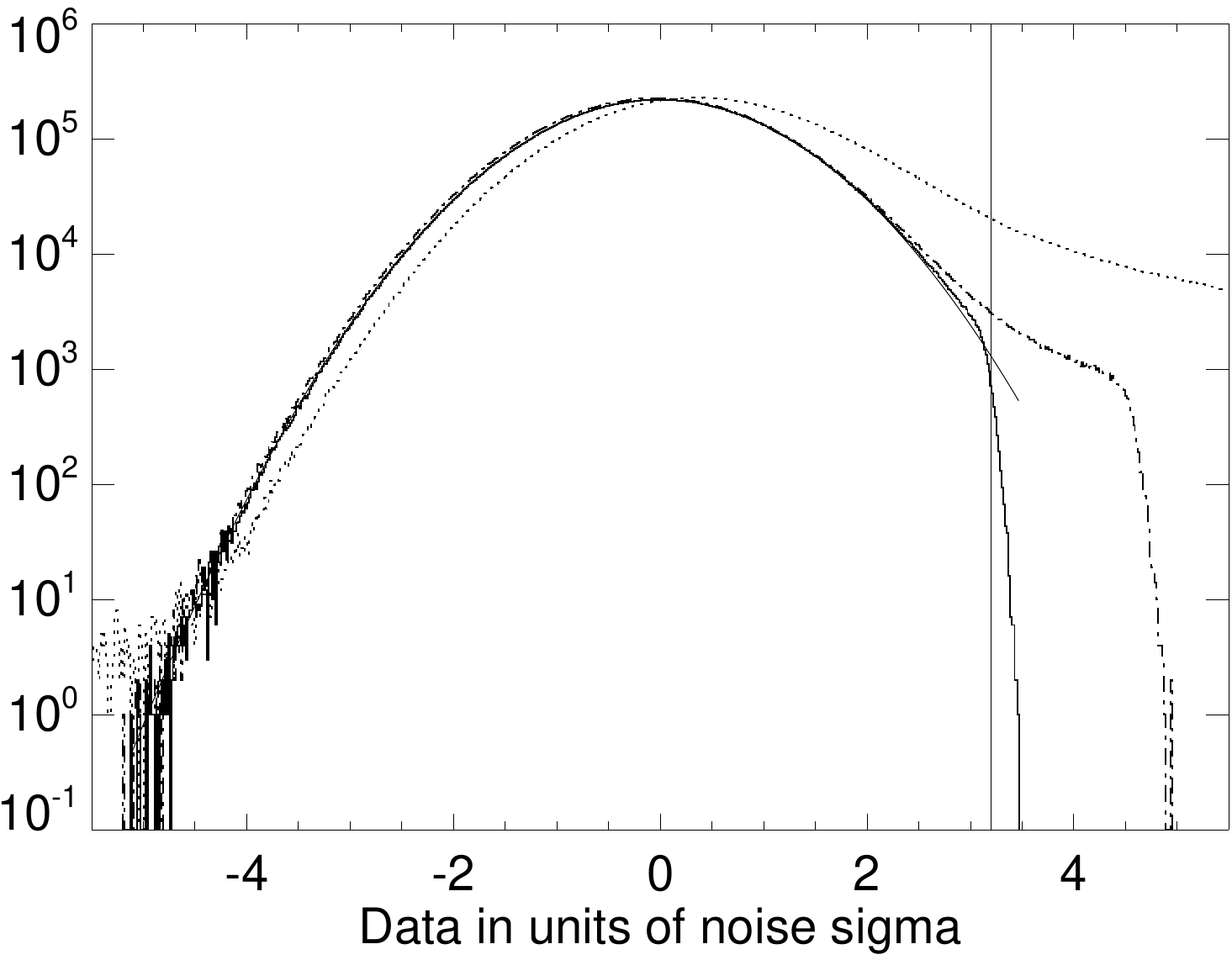} 
  	\caption{The solid curve is the histogram of a series of bolometer
    	values after deglitching with a $3.2 \sigma$ threshold, subtracting
    	estimates of the sky signal and glitch templates, as well as high pass filtering. A
    	fit with a Gaussian is also shown. We do not see a sharp cut at
    	$3.2 \sigma$ because of the low pass filter applied to display the
    	data. The dot-dashed curve is the same as before but after
    	deglitching with a $4.5 \sigma$ threshold. The dotted curve is for
    	data without deglitching. A $3.2 \sigma$ threshold offers a good
    	trade-off between data losses and remaining contaminants.}
	\label{fig:histoData} 
\end{figure}

The final detection threshold of glitches is set to $3.2 \sigma$, where
$\sigma$ is the local noise level.  This choice is a trade-off 
between the amount of data flagged and the number of small glitches
left in data, as shown in Fig.~\ref{fig:histoData}. Reconstructed
distributions of cosmic rays appear to show a break at lower
amplitudes indicating that only a small fraction of the small  glitches
remain in the data.

Events with an amplitude larger that $3.2 \sigma$ are detected in small
data windows by finding their maxima after 3-point filtering of
data. Very confused events are not distinctly detected using only
maxima, but many are detected by applying a matched filter tuned to
detect abrupt changes of slope. Data flagging and cleaning is performed
progressively window by window.  Each window is set with the
maximum event of the interval at the center,  provided it has not yet been flagged from the
processing of other windows .  Within a window, a
multi-parameter fit of the amplitudes of long templates is performed jointly for
all events which are  eight samples after the maxima. This is done in practice
by minimizing:
\begin{equation}
  \chi^2 = \sum_{t=0}^{N_s} (d_k(t) - \sum_i a_i T_l(t-t_i) + K)^2 M(t),
\end{equation}
with respect to the amplitude parameters $a_i$ for all the events $i$
in the window as well as an offset $K$, where $N_s$ (= 1000 in practice) is the
number of sample per window, $d_k(t)$ is the signal--removed data value
at sample $t$ and iteration $k$, $T_l(t-t_i)$ is the normalized long
glitch template at $t-t_i$ samples after the maximum located at $t_i$,
and $M(t)$ is a mask function (which is zero within eight samples after the
glitch maxima and also where data have been flagged at a previous iteration,
and unity otherwise).  All of the amplitude parameters are constrained to be
positive. Joint fitting is critical at this stage
because of the high level of confusion between events. Low frequency
drifts of noise are well represented by constants in intervals within
windows ($K$ in the expression above) and are fitted jointly with
the glitch templates.

If events in the centre of the window have maxima above $10 \sigma$,
and the fitted long templates are above $0.5$ times the amplitude
expected for long glitches, then the template is subtracted from the
data using the fitted amplitude as follows:
\begin{equation}
  d_{k+1}(t) = d_k(t) - \sum_j {\hat{a_j}} T_l(t-t_j),
\end{equation}
where $d_{k+1}(t)$ is the cleaned data used for the next window at time
$t$, $\hat{a_j}$ the estimated template amplitude for event $j$ such
that all $t_j$ are separated by less than eight samples starting from
the event at the centre of the window. Samples around glitch maxima
are flagged in intervals defined such that the expected remaining
glitch signal in its tail is well below the noise \rms. For glitches
below 10$\sigma$,  or for detected short glitches, nothing is subtracted from
the data and the glitch is entirely flagged.
The window is then moved to other events which have not yet
been cleaned. In total, between 9 to 16\,\% of the data are flagged,
depending on the rate of glitches (see Fig.~\ref{fig:gldataloss}
below).

Once the templates have been removed window-by-window and the data
flagged, a matched filter built from the long glitch template is
applied to data.  Rare events detected by the matched filter are
flagged in long intervals. Up to 2\% of the data are flagged in this
way for bolometers with the highest rate of glitches.

At the end of this procedure, the sky signal is re-estimated after removing the glitch
templates from the original data, using unflagged  data only, and a new
iteration of the whole detection procedure is performed.  In contrast
with the first iteration, we use for all ensuing ones a more elaborate method for this sky
estimate. We fit splines with nodes at the location of ring pixels.
This has proven to be very efficient at capturing sky variations
within ring pixels. Simulations show that errors of less than 0.1\%
are made in the signal reconstruction of a Gaussian with a FWHM of
three times the ring pixel size (close to the characteristic size of a
point source in \Planck\ data).  Errors in the signal reconstruction
by this method are dominated by the effect of small pointing drifts,
which are of order 10\,arcsec within a ring; this is significant only
for the higher frequency channels at  Galactic plane crossings or
for exceptionally strong sources.  We have increased the glitch detection
threshold to account for signal reconstruction errors by adding
quadratically a term in the noise \rms\ proportional to the signal
amplitude in each pixel. This correction is negligible for all the
data except for planet transits in CMB channels (100, 143, 217\GHz)
and is significant only within the Galactic plane and scanning strong
sources at higher frequencies.

We obtain convergence between signal and glitch template estimates for
99.99\,\% of the rings. Nevertheless, the very few rings for which 
convergence is not reached can be detected by checking when glitch tail templates
are subtracted systematically at exactly the same location in the sky for consecutive scan.  
These rings are discarded from the analysis.

\subsubsection{Glitch handling results} 

\begin{figure}[htbp] 
	\centering
  	\includegraphics[width=\columnwidth]  {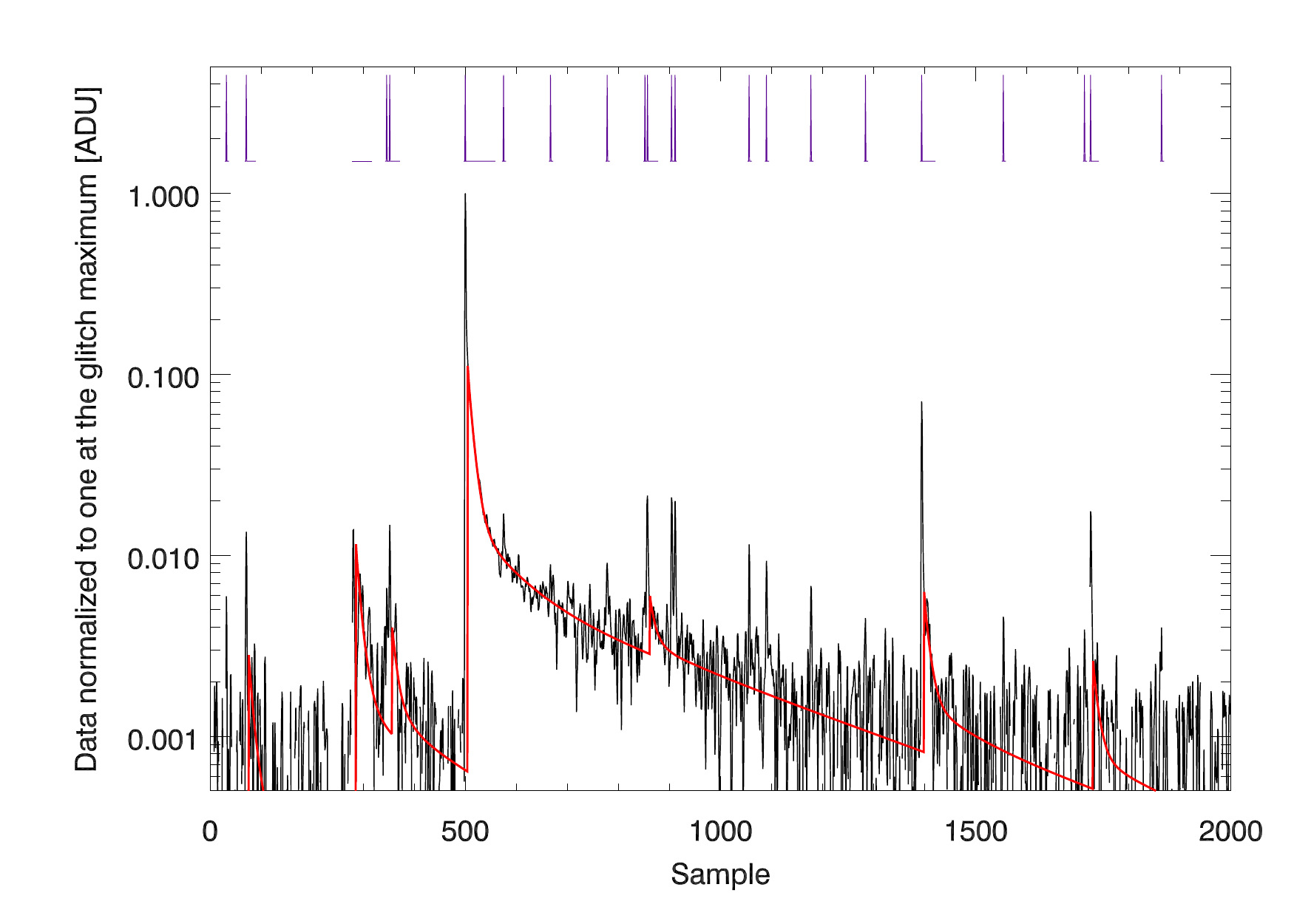} 
  	\includegraphics[width=\columnwidth]  {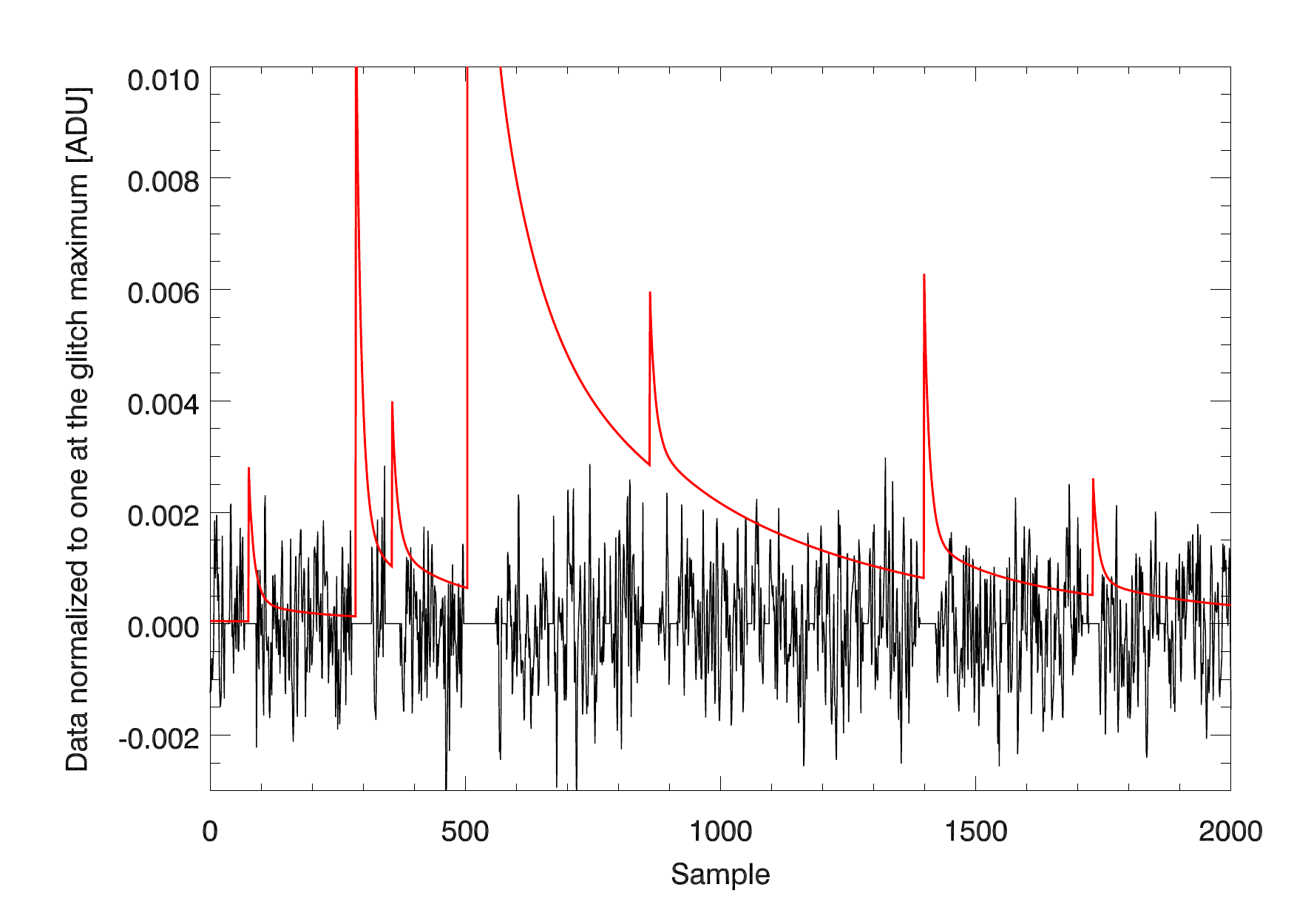} 
  	\caption{Top panel: Example of 2000 samples of sky subtracted
          data encompassing a large 
  	event in black and our best fit glitch templates (red). The
        purple ticks in the upper part of the 
  	figure show where data are flagged and indicate the detected
        position of glitches. Bottom 
  	panel: The cleaned residual in black (with the flagged areas
        set to zero) compared with the  
  	fitted  template in red. }
	\label{fig:residualTemplSub} 
\end{figure}

\begin{figure}[!htbp] 
	\centering
	\includegraphics[width=\columnwidth]  {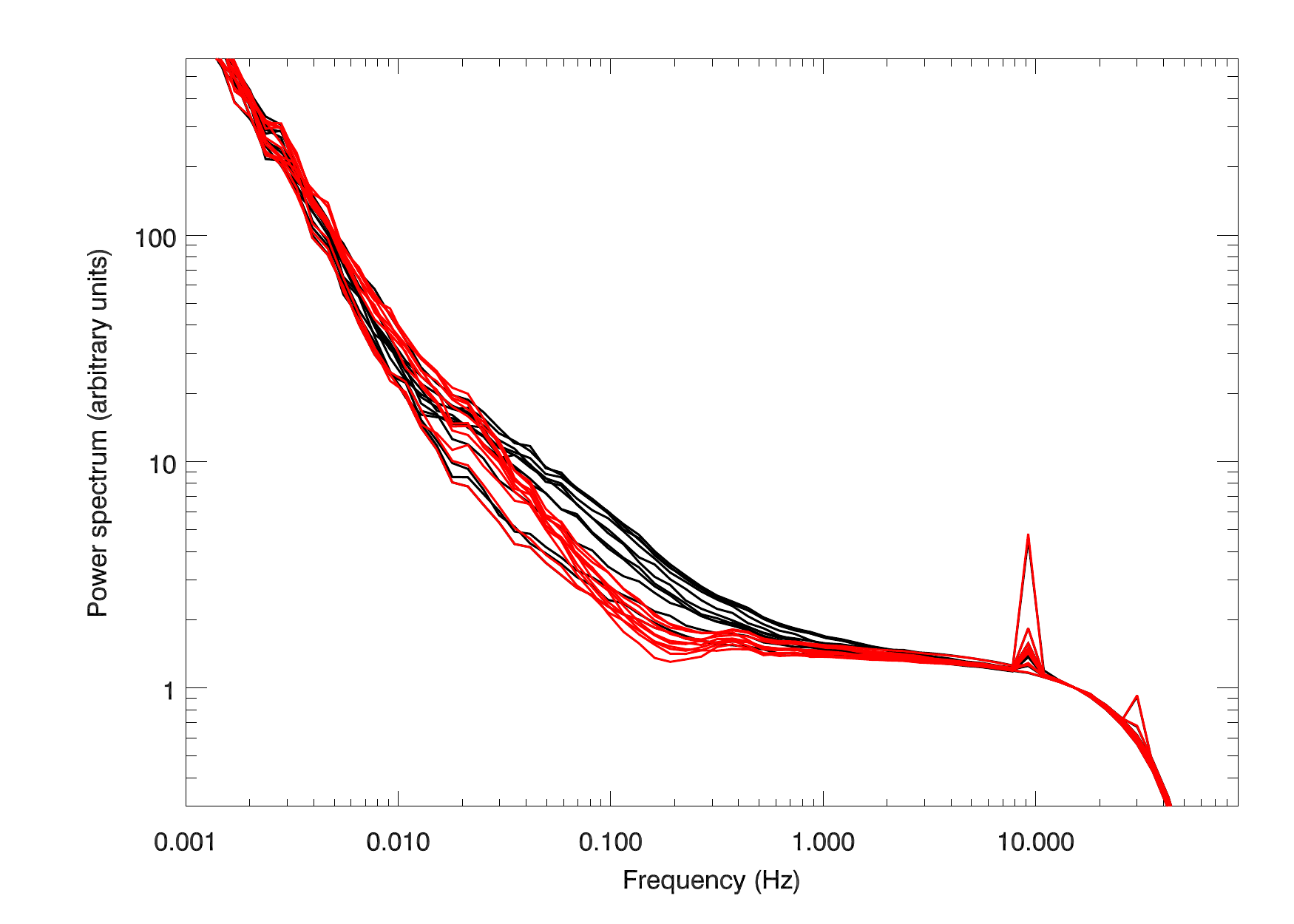} 
	\includegraphics[width=\columnwidth]  {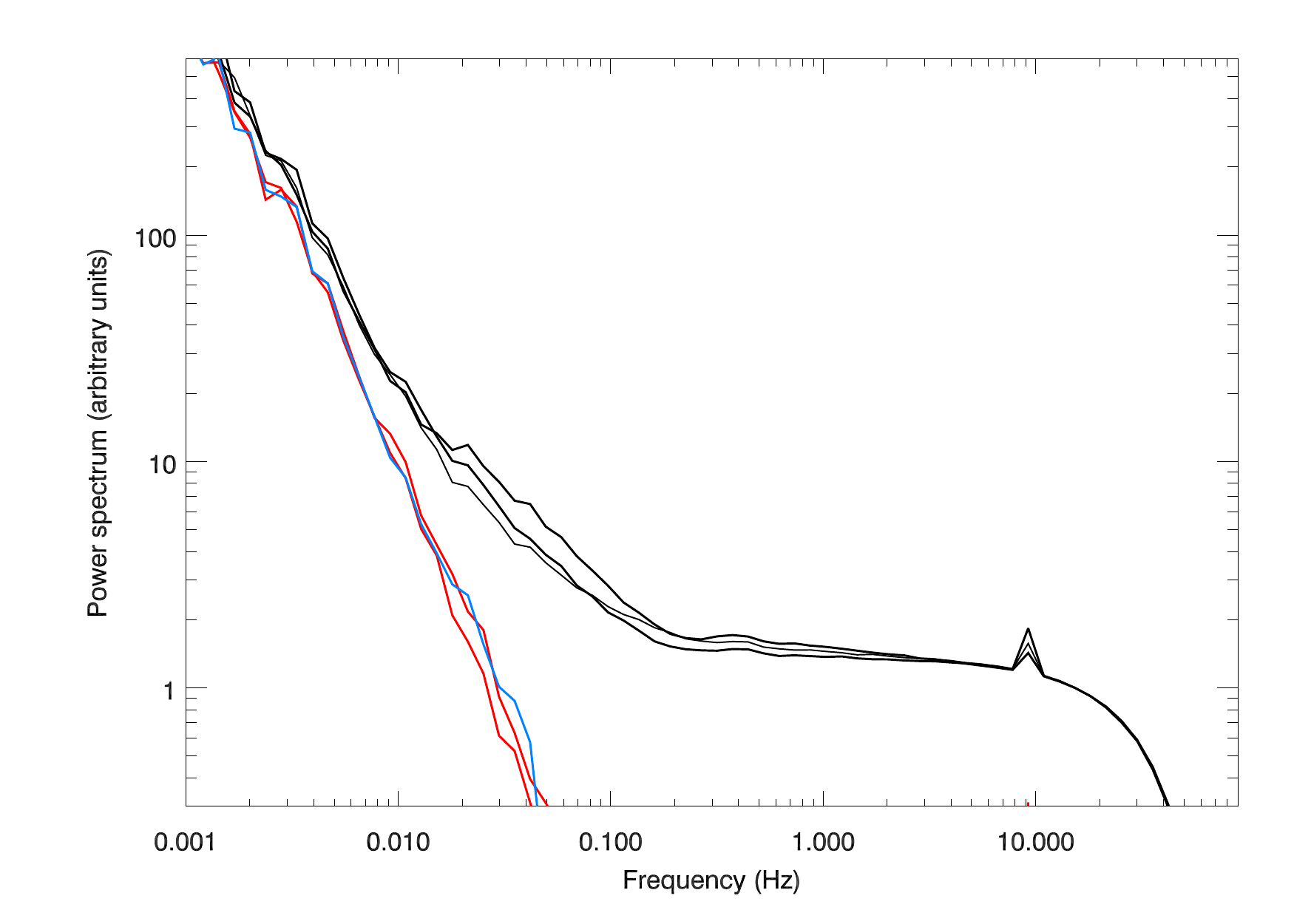} 
	\caption{a) top: power spectra of the noise in the TOIs for
          all bolometers at 143\GHz\ without 
	(black) and with (red) subtraction of long glitch templates. The data corresponds to 
	sky-subtracted TOIs, before any further processing; one can see,  for 
	instance, 4\,K 
	lines harmonics at 10 and 30\Hz\ which have not been removed at this stage
     (this is done later in the TOI processing pipeline). Two of the 143\GHz\  detectors have a rate of long glitches 
	about a factor of five below that seen in those with the highest glitch rate. The power spectra 
	for these two bolometers are those with the lowest amplitude
        at around $0.03$\Hz; as can be 
	seen from the figure, the template correction in these two
        cases has only a very small impact. 	 
	As a result, these bolometers can be used to test  for noise
        from sources other than glitches.  
	This plot illustrates the efficiency of the template
        subtraction method in reducing contaminant  
	noise in the frequency range 0.02\Hz\ to 2\Hz\ whilst simultaneously
        reducing the amount of flagged 
	data. b) bottom: the auto-spectra  after deglitching of 3 of the 143\GHz\
	 bolometers of the top panel,  in black, are compared with  
	their cross-spectra in blue and red. The blue curve
        corresponds to the cross-spectra of two  
	bolometers in a PSB pair. This suggests that the decorrelation
        from common mode thermal 
	fluctuations cannot have much of an effect on the noise at frequencies
        greater than the spin frequency at 0.02\Hz. }
	\label{fig:PowerSpectra143} 
\end{figure}

Figure~\ref{fig:residualTemplSub} shows examples of the
deglitching process on the sky-subtracted data. The glitches in this
segment are overlapping, a situation which arises quite often.  
Globally, glitch template correction is very effective,
despite the fact that the glitch rate varies appreciably from bolometer to
bolometer.  Figure~\ref{fig:PowerSpectra143} compares the power spectra of
noise in TOIs for all bolometers at 143\GHz\ (except 143-8, which is affected 
by Random Telegraphic Signal (RTS) noise , see Sect.~\ref{sec:rts}), 
with and without long glitch template subtraction. Flags have been
applied in both cases (flagged samples have been masked and filled using a
simple interpolation method). For this plot, the flags applied to estimate the
noise power spectra  without template subtraction are extended to mask the
data for which the long glitch template is above one sigma of the noise. This
has a non-negligible extra cost (up to 9\% of the data for some bolometers) in
the amount of data masked. We can see that the template correction reduces
the spurious noise power by a factor of around 4 for a large fraction
of the bolometers at frequencies in the range 0.02\Hz\ to 2\Hz, whilst
simultaneously reducing the amount of data flagged for exclusion in the
map-making stage.  Nevertheless some low level residual is still expected in the data
after template subtraction. We have performed simulations to assess the level of these residuals, see sec.~\ref{sec:glitchres}. 

\begin{figure}[htbp] \centering
	\includegraphics[width=\columnwidth]{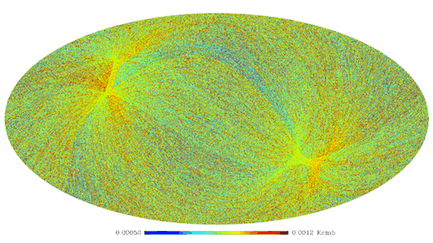}
    	\includegraphics[width=\columnwidth]{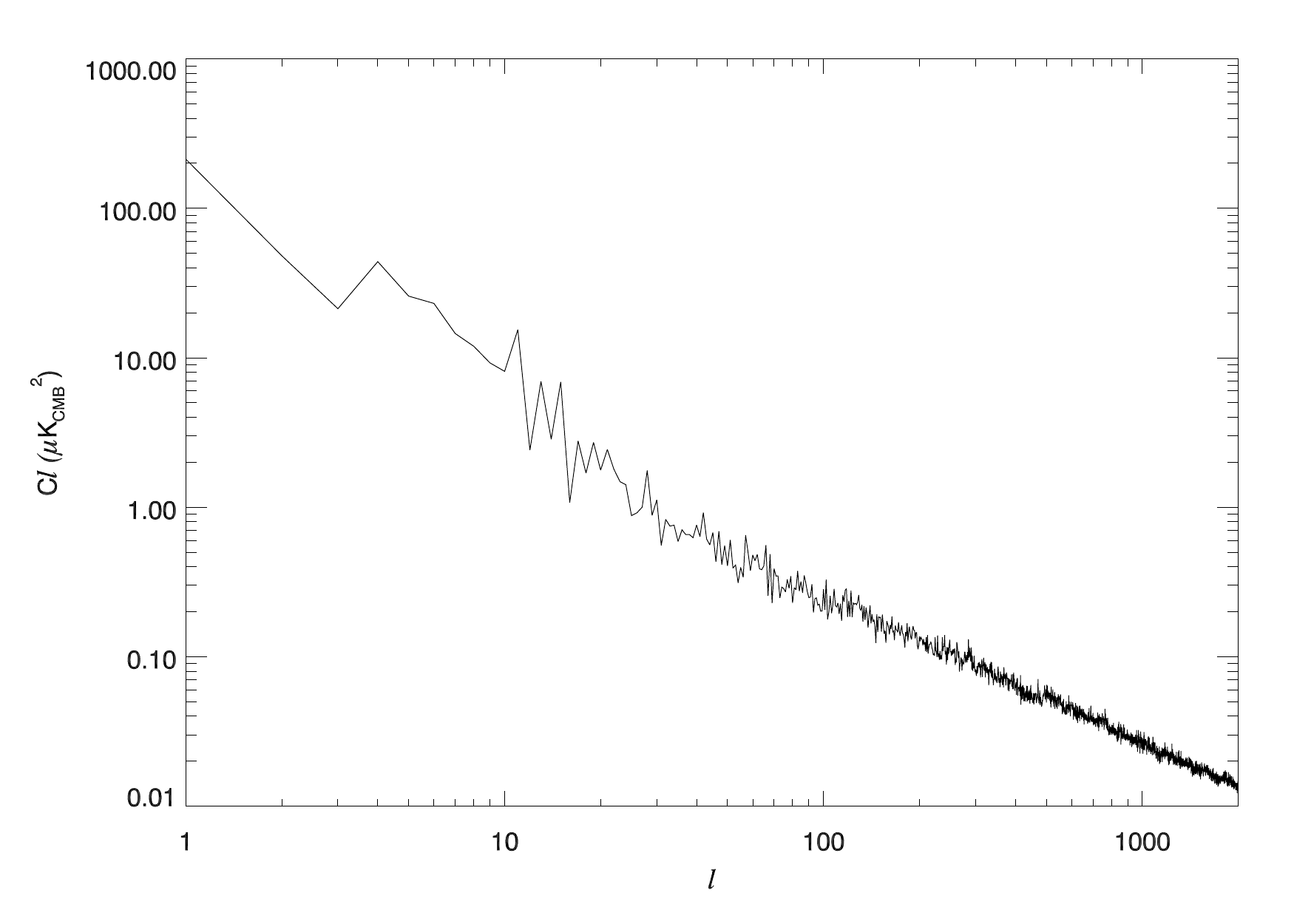}
        \caption{Top: map in Galactic coordinates built by projecting
          the estimated long glitch template 	for one bolometer at
          353\GHz. Comparison with the Jackknife map at the same frequency in 
	Fig.~\ref{fig:mapmaking_noise_maps} shows that, if uncorrected, that
	contribution would be the dominant part of the residual `noise'. 
	Bottom: power spectrum of the map shown in  the top panel.} 
	\label{fig:maptmpl} 
\end{figure}

To evaluate the effects of glitch removal on the final maps, we have
projected the correction TOI (the sum of the estimated glitch
templates) into maps and computed the corresponding power spectra. Note
that these correction time-line are generally non-zero nearly
everywhere \ie all samples are corrected by this `baseline removal'.
A map and associated power spectrum are shown in
Fig.~\ref{fig:maptmpl} for one bolometer at 353\GHz. There is no
obvious correlation of the glitch template estimation with the
Galactic signal at that frequency, but we have noted some
anti-correlation at 857\GHz, when the Galactic signal is
strongest. This is expected since the glitch detection threshold is
significantly increased for the highest frequencies in presence of
strong signal. The power spectrum of the glitch template map is
roughly proportional to $1/\ell$, as expected \citep{efstathiou2007, tristram2011}.
Conversely, we checked whether any CMB has leaked into the glitch time-stream, 
$g = g_0 + \alpha\ CMB$, by finding limits on $\alpha$ through cross-correlating our glitch 
maps with the 7 years WMAP ILC map. We then estimate $\alpha$ as the ratio of the cross-spectra 
(glitch X ILC)/(ILC X ILC) for $\ell < 300$;  we found no evidence for leakage of the CMB 
into the 100, 143, and 217\GHz\ glitch time-streams. 
Indeed this leakage was constrained to be smaller than $0.01\%$.

\begin{figure*}[!htbp] %
  	\includegraphics[angle=180,width=1\textwidth, totalheight=1.05\columnwidth]
  		{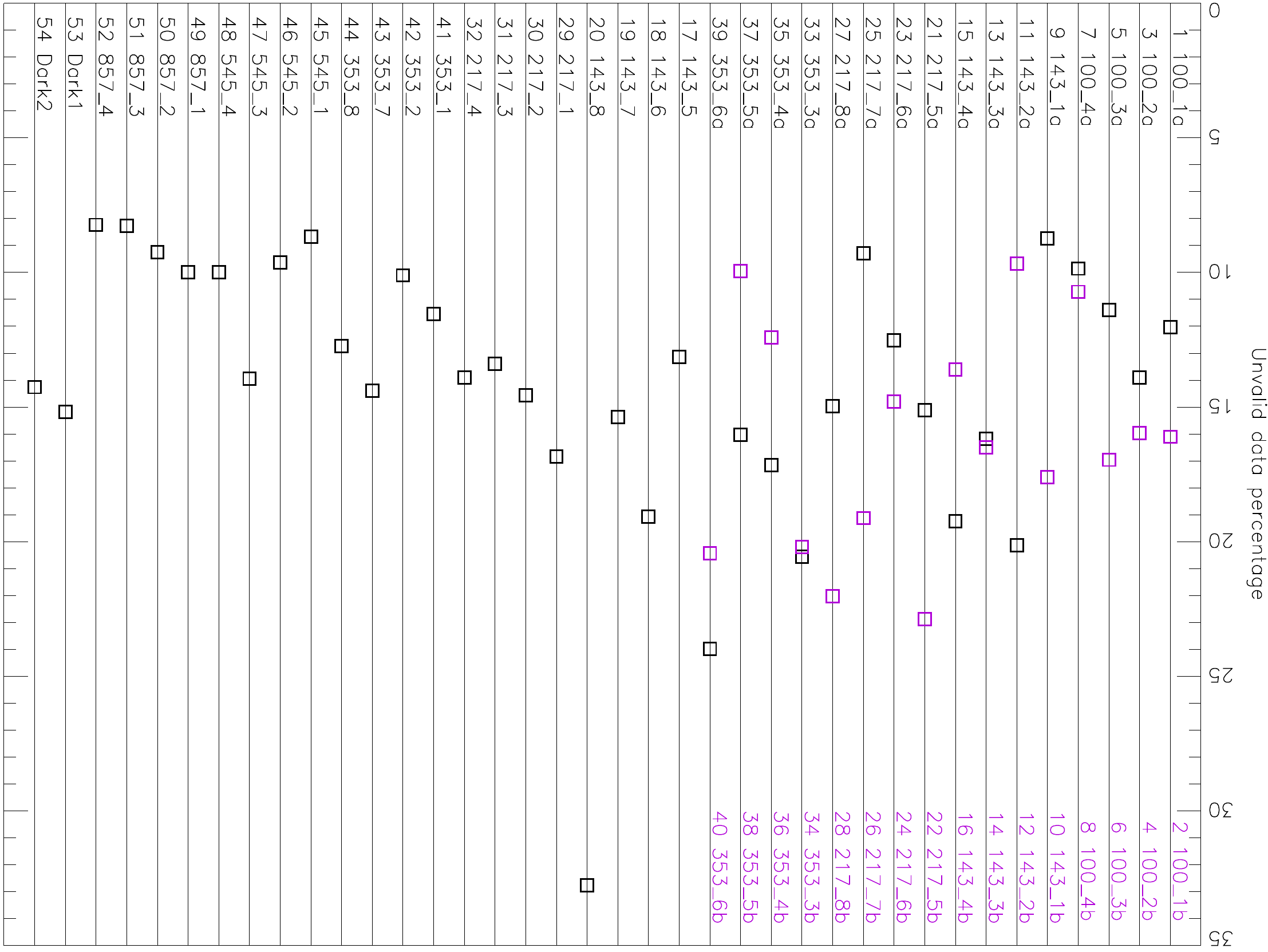}
  	\caption{Average fraction of data lost in the first sky survey
          for each bolometer as a result of glitches. The number for the 143\_8 is not representative since it is corrupted by RTS. }
  	\label{fig:gldataloss} 
\end{figure*}

We have also tested whether significant glitch residuals are
left in the data  near the event peaks after template subtraction, by
multiplying the flag extents after each glitch peak by a factor of two. We applied
this for all detectors at 143\GHz\  and computed the noise
power spectrum of the combined 143\GHz\ map using jackknife analysis
as described in Sect.~\ref{mapmaking:jackknife}. We found that the noise
at small scales increased by a factor comparable to the square root
of the inverse ratio between the number of hits with and without flag
extension, as expected in the limit of no residual glitch
contamination. The noise increased by a  larger factor at
the lowest multipoles. We  interpret this as being due to the
loss of accuracy of the destriping map making algorithm 
as a result of the extra data 
flagging.

Overall, 10 to 15\% of the data are flagged as unusable due to the presence
of glitches. Figure~\ref{fig:gldataloss} summarizes the data loss caused by cosmic
rays for each individual bolometer, while Table~\ref{tab:dataloss} gives the
average over detectors within the same frequency channel. This is to be compared
with other data losses, in particular the phases of depointing which are not
used at present. A full ring has a mean duration of 46.5\,minutes. On average, 3.8\,minutes
(8.3~\%) of data are taken during re-pointing manoeuvres between
stable pointings. They are currently not used, so at present we lose about
20\% of the data due to depointing and glitches. Data
compression in the high frequency channels leads to 
negligible amounts of lost data. In addition,  the deglitching processing
produces anomalous results in some rings. Statistical inspection of all rings
showed that 0.5~\%  are anomalous and these are  flagged and discarded
from further analysis.

\begin{table}[!htbp] 
  	\caption{ Summary of the average data lost because of glitches per bolometer frequency 
  		(the number in the first line correspond to the frequency of the channel). 
  		$N_{b}$ is the number of bolometers used, \ie
          excluding bolometers affected by 
  		RTS noise.} 
	\label{tab:dataloss} 
	\centering  
	\scalebox{0.975}{ 
	\begin{tabular}{lrrrrrrr}  \hline \hline 
         &  100 & 143 & 217 & 353 & 545 & 857 & Dark \\ \hline
	$N_{b}$ & 8 & 11 & 12 & 12 & 3 & 3 & 2 \\
	\%  &   13.4 &  15.4  & 15.8 & 15.8  & 9.4 & 9.2 & 14.7 \\  \hline 
	\end{tabular}  
	} 
\end{table}

\subsubsection{Glitch simulations and residuals}  \label{sec:glitchres}

\begin{figure}
	\centering
	\includegraphics[width= 0.5\textwidth,angle=0] {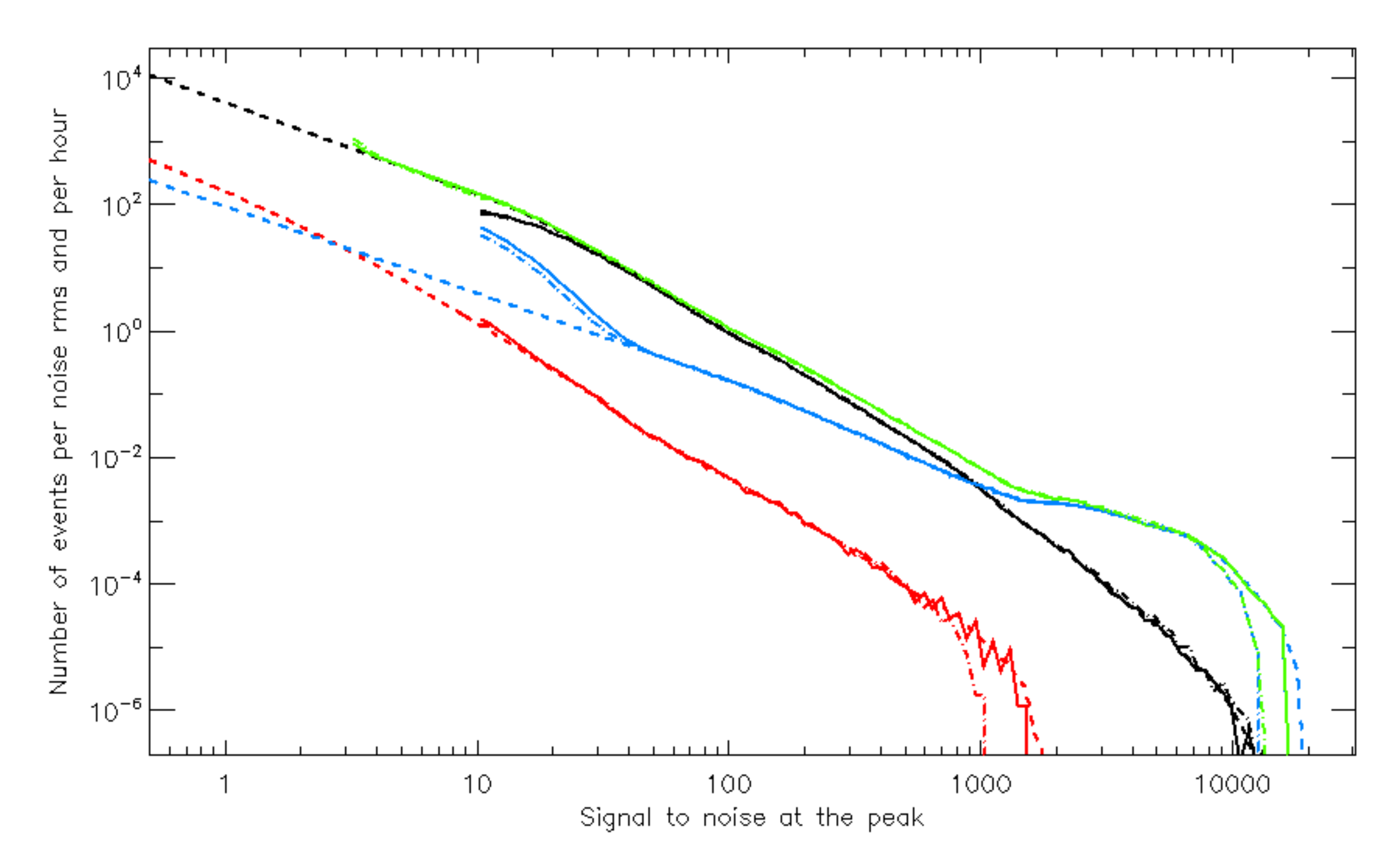}
	\caption{Distributions of the three populations of glitches
          for one PSB-a bolometer at 143\GHz; black 
	 is for long glitches, blue is for short glitches, and red is
         for the very long ones. Green is for the 
	 total. The measured distributions from real data are
         displayed with solid curves, the model with 
	 dashed curves and the measured distributions in simulation
         with dot-dashed curves. One can 
	 see the excellent agreement between the estimated
         distributions from real data and from  
	 the simulations. The total distribution is well measured down to $3.2
	 \sigma$. The separation of glitches into different
         populations is not attempted below $10 \sigma$ ($\sigma$ stands for the \rms\ of the noise), 
         and is quite secure above  
	 $40 \sigma$ for PSB-a, and above $200 \sigma$ for PSB-b and SWB.  A 
	 two-power law model with breaks has been used to represent
         the glitch populations at the faint 
	 end. For this bolometer, the breaks are fixed to $\approx
         100 \sigma$ for the short glitches, 
	 $\approx 20 \sigma$ for the long and $\approx 3 \sigma$ for the very long glitches.}
	\label{fig:GLdist}
\end{figure}

To assess the effect on the data of our glitch removal procedure, we
have simulated TOIs of noise with glitches representative of the three
populations described in Sect.~\ref{sec:glitchology}.  We have used,
as a reference, the glitch amplitude distributions measured for one of
the 143\GHz\ bolometers (cf. Fig.~\ref{fig:GLdist}). Since the
populations are not well separated in the data for glitch peak
amplitudes below $\approx 40 \sigma$ (where $\sigma$ is the \rms\ of
the noise) the simulated distributions for all populations have been
extrapolated using two-power law models at the faint end, keeping the
sum of the distributions identical to the data down to about $5 \sigma$.
Figure \ref{fig:GLdist} compares the measured and the model
distributions, as well as the recovered distributions after applying
the glitch analysis on simulations. Distributions for other bolometers
are rescaled in glitch amplitudes by a factor specific to each
population. We have observed that this factor varies strongly from
bolometer to bolometer for the long glitch distributions, and is about
constant for the short one.

\begin{figure}[!htbp] 
	\centering
	\includegraphics[width=\columnwidth]  {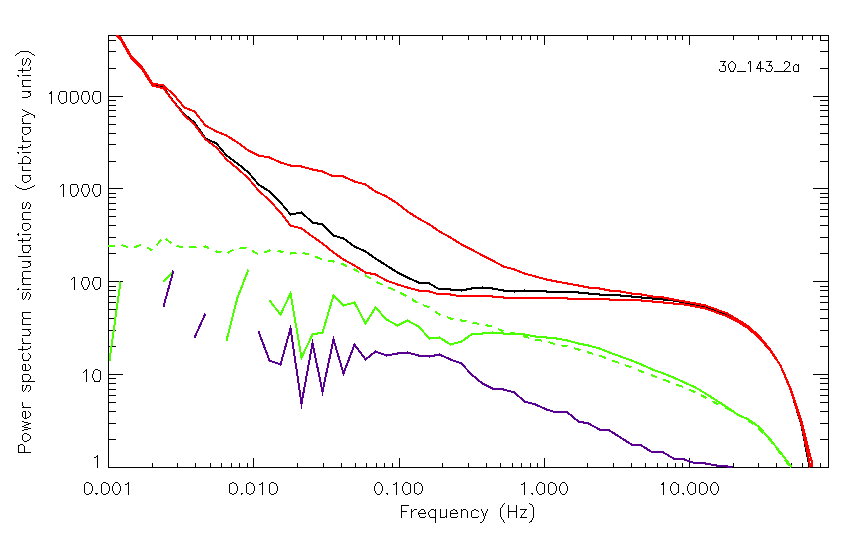} 
	\includegraphics[width=\columnwidth]  {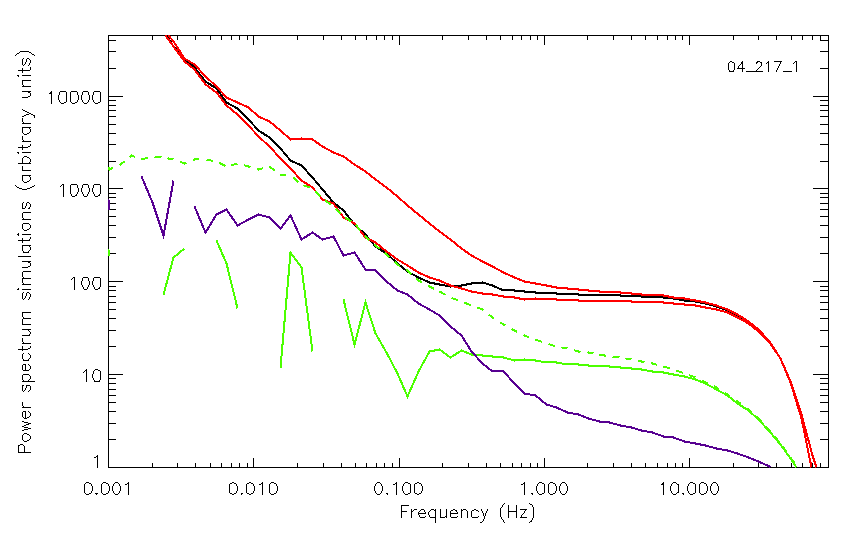} 
	\caption{Power spectra of the residual noise in simulations after template
  	subtraction and flagging (in black) for one bolometer (PSB-a) at 143\GHz\ 
  	(top panel) and one SWB at 217\GHz\ (bottom panel). Both bolometers have high rates
  	of long glitches. No other processing has been applied to the
        simulations, except a 3-point  
  	filter (as applied to real data before deglitching), the
        effect of which can be seen at frequencies 
  	above 10\Hz. This can be compared with the surrounding red
        curves. The top red curves  
  	in both plots correspond to the power spectra of the simulated
        data (pure noise+glitches)  
  	before template subtraction 
  	The lower red curves correspond to the power spectra of the pure noise TOIs used in
  	generating the simulations. The green dashed curves are the
        power spectra of the residual 
  	glitches 
  	remaining after flagging and template subtraction, which are measured from 
  	the difference between the estimated and the input TOIs of glitches. Those curves are
  	not exactly equivalent to the difference between the black and
        the bottom red curves because  
  	some noise is subtracted by the template subtraction process. 
  	The green curves are the cross-spectra between the input  TOIs
        of glitches and the cleaned  
  	data, and the blue curves show the cross-spectra between the recovered
  	 TOIs of glitches (which are subtracted from data) and the input noise TOIs.}
	\label{fig:SPSimus1432a} 
\end{figure}

The (sum of) four--exponential glitch templates used for the detection were also 
used to simulate the glitch tail
profiles. The fast part with a decay given by the bolometer time-constant (not
included in the tail templates) is added. Glitch amplitudes and 
sub-sample arrival times are set randomly according to  Poisson
statistics. Glitch profiles are integrated within samples. This gives
some dispersion in the peak-to-template amplitude ratio depending on
the arrival time within the sample integration duration, as observed in data. A non-linearity
coefficient is also applied to the amplitude of the long glitch template
relative to the fast part of the glitch, also as observed in data. 

The excellent agreement between the recovered distributions in data and simulations displayed in Fig.~\ref{fig:GLdist} gives us confidence in our glitch modelling. 

\begin{figure}[htbp] 
	\centering
	\includegraphics[width=\columnwidth]  {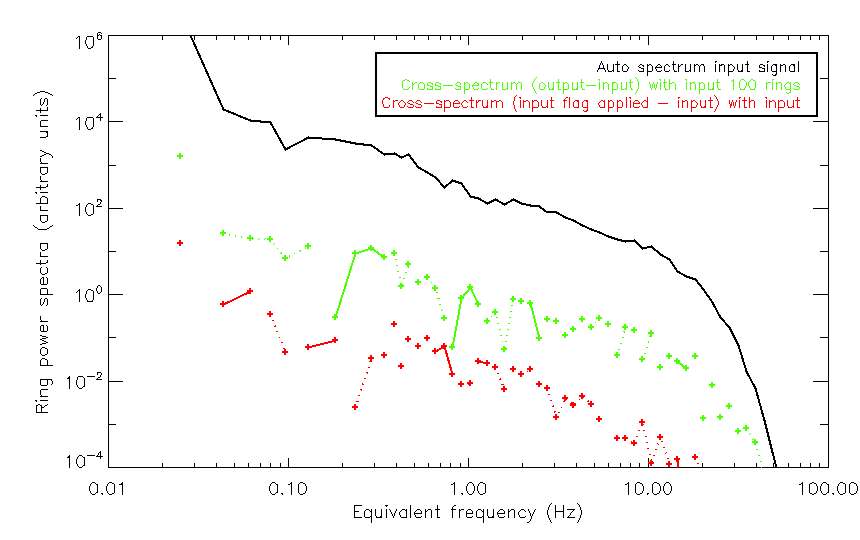} 
	\caption{In black: power spectrum of pure simulated signal at 143\GHz\ projected on 
		1.5 arcmin rings. In green: cross-power spectrum
                between (1) pure signal minus 
		recovered signal and (2) pure signal. Solid lines join
                positive values and dotted lines 
		negative values. In red: cross-power spectrum between
                (1) pure signal minus pure  
		signal flags applied, and (2) pure signal. We see no apparent bias in signal 
		estimation from either template subtraction or flagging.}
	\label{fig:1dSignalPower} 
\end{figure}

Figure~\ref{fig:SPSimus1432a} shows the noise power spectra 
after deglitching and template subtraction in simulations of two bolometers
(one PSB-a and one SWB) with a high rate of glitches. Power spectra of the
outputs are compared to input noise power spectra. We can see a small residual
contamination from glitches remaining at frequencies higher than 0.01\Hz\ for
the PSB-a, which is at most at the level of the input noise around
0.04\Hz. For the SWB, the contamination is stronger and of the order of the
noise between 0.01\Hz\ and 0.1\Hz. Residual contamination is at the 10--20\%
level above 0.2\Hz\ in both cases. The stronger residual contamination for a
SWB (and PSB-b) as compared to a PSB-a, is explained by the fact that the
intermediate 60\,ms decay, which is a trigger for the template fitting, is
lower in the SWB, hence the $\approx 2$ second tail (mostly responsible for
the noise excess) is removed less efficiently. Also, the knee frequency for
the SWB at 217\GHz\ is higher than that for the PSB-a at 143\GHz. Nevertheless,
at all frequencies the residual contamination  does not exceed the input noise level.

Fig.~\ref{fig:SPSimus1432a} shows the cross--spectrum
between the template subtracted data and the input glitch TOI
in simulations (green curve). The low level of relative correlations
below 0.1\Hz\, as seen by comparing the green dashed and solid curves,
indicates that the excess noise residuals associated with glitches are
dominated by the errors in the glitch template fit, whereas above
0.1\Hz\ the high level of relative correlations show that the excess
noise is due to unidentified (or unsubtracted) glitches below the
detection threshold. This has been confirmed by inspection of the
difference between input and recovered glitch TOIs. The measured
cross-correlation between the input noise in simulations and the
recovered glitch TOI (blue curve in Fig.~\ref{fig:SPSimus1432a})
indicates that by removing the latter from the data, some detector
noise is also removed. This is because errors in the glitch fit are
naturally correlated with the noise in the data. This effect is at about the
20\% level at 0.1\Hz.

To evaluate potential biases in the recovered sky signal which would
be introduced by the deglitching procedure, we have computed, using
simulations, the cross--power spectrum between the recovered signal
residual, estimated from the difference of pure signal minus recovered
signal phase binned rings, and pure signal phase binned rings for one
detector at 143\GHz. Results are shown in
Figure~\ref{fig:1dSignalPower} including and excluding the glitch
flags to project the pure sky signal TOI. The recovered signal phase
binned rings is a product of the deglitching procedure at the last
iteration as already described. We do not see any significant
correlations at 143\GHz\, nevertheless we see some small effect at the
level of a tenth of a percent for the two highest frequencies. This is
caused by the strong Galactic signal, which leads to a higher glitch
detection threshold and hence less complete subtraction of glitches.
We have also constrained potential biases on the signal by
computing the cross-spectrum of the estimated sky signal from real
data on rings with the estimated glitch templates projected on the
same rings for one detector at 143 \GHz. We have not found any
significant correlation at any ring frequency after averaging the
cross-spectra of a thousand rings. The same test applied at 545\GHz\
have shown a very small anti-correlation between the reprojected
glitch templates and the galactic signal of the order of 0.04\% of the
signal. This is due to the fact that less glitches are subtracted on
average while observing strong sources, as the detection threshold is
increased depending on the intensity of the signal. This leads to a
negligible bias after template subtraction.

\subsection{Gain} \label{sec:gaincor}

Bolometers are non-linear devices. The voltage output must be
converted to absorbed power by using a gain model.  This is discussed
in the companion paper by the \citet{planck2011-1.5}. The slowly
varying gain is corrected by using parameters measured during the
ground-based tests and in-flight calibration periods. The time response is
dealt with in Sect.~\ref{sec:TF}. The gain factor is typically of the
order of $10^9\,\mathrm{V/W}$.  The gain non-linearity amplitude for
typical signals (the dipole or the Galaxy) is of the order of a few parts in 
$10^{4}$.

\subsection{Decorrelation of Thermal fluctuations} \label{sec:thermalfluct}

\begin{figure*}[!htbp]
  	\includegraphics[angle=180,width=1\textwidth, totalheight=.4\textheight]
  		{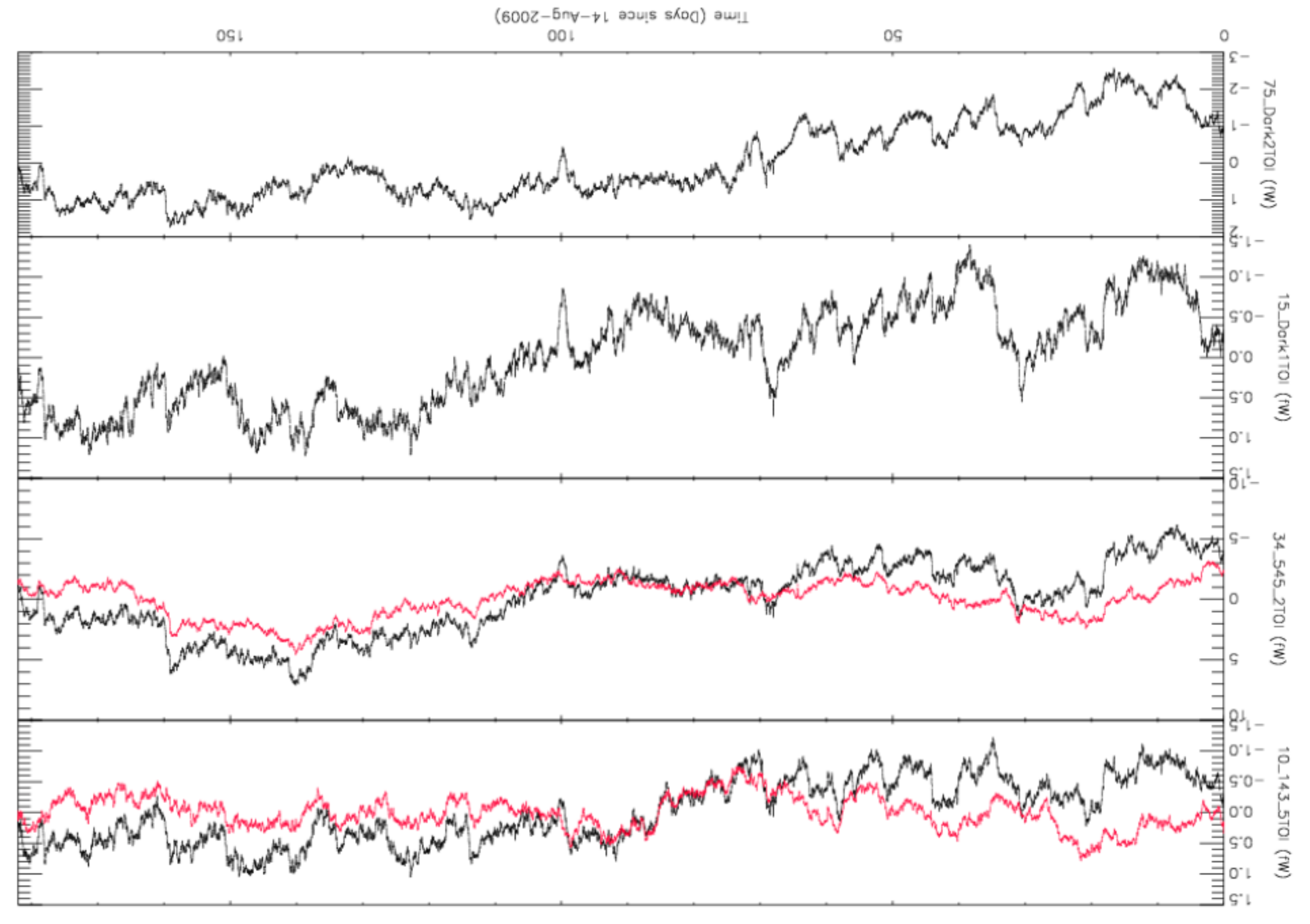}
  	\caption{Trend in processed TOIs of the 143-5 and 545-2 bolometers over the
  		first sky survey period (180 days), before (\emph{in black}) and after
  		(\emph{in red}) the 100\,mK fluctuations are removed
                (as monitored by the two dark 
  		bolometers, Dark1 and Dark2, in the two lower plots). These data
  		are averages over 1 minute of processed TOIs, smoothed to one
                hour. \textbf{ For the dark bolometers, one unit
                  ($1\,\mathrm{f W}$) is equivalent to about $10\,\mathrm{\mu K}$.}
                }
	\label{fig:TOIExample2} 
\end{figure*}

\begin{figure}[!htbp]
  	\includegraphics[width=1\columnwidth, totalheight=.7\columnwidth]
  		{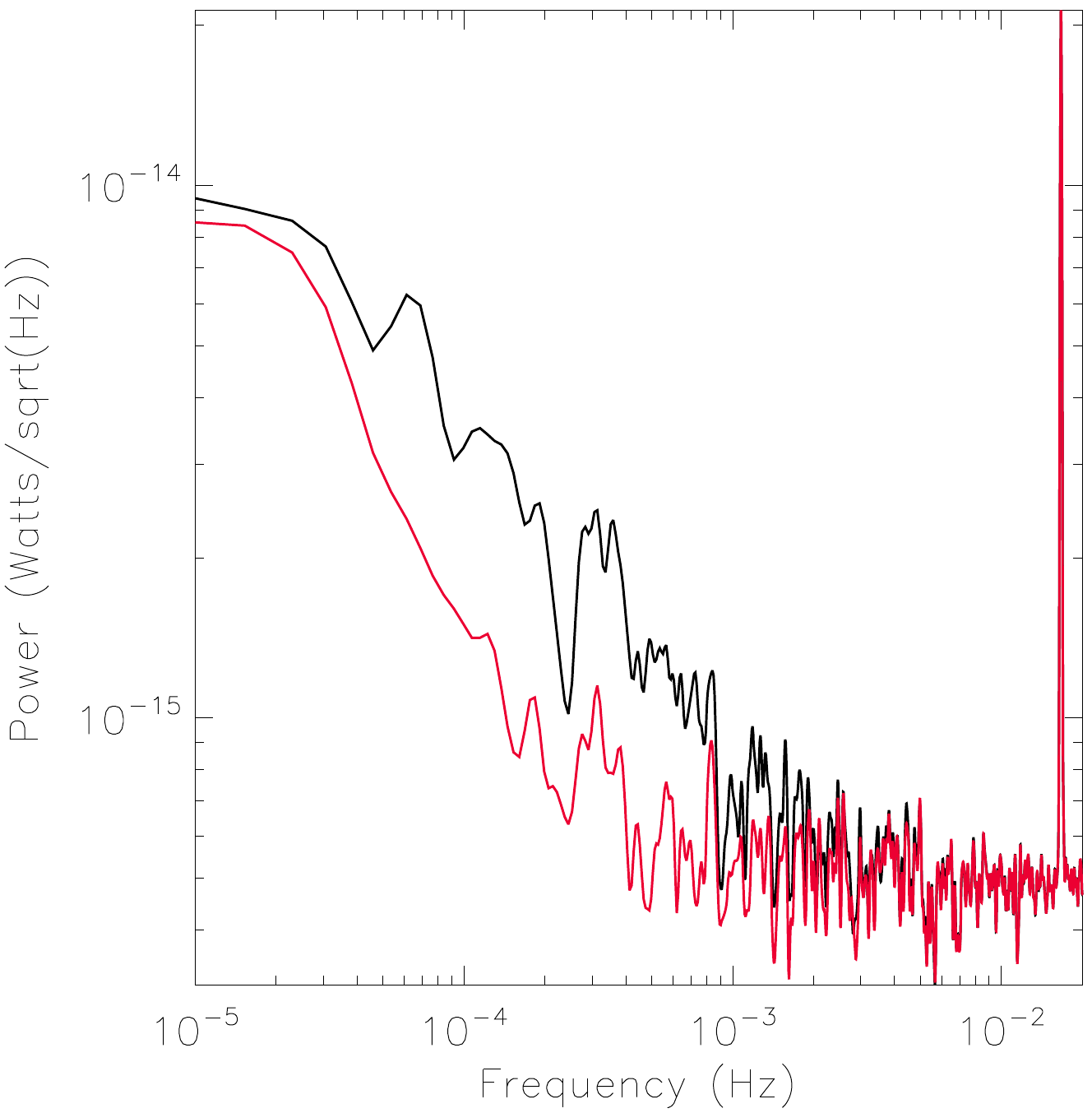}
	\caption{Power spectrum of the TOI of the 143-5 bolometer, before
		(\emph{in black}) and after
		(\emph{in red}) the 100\,mK fluctuations traced by the dark bolometers
		(smoothed by a 2-minutes box kernel) have
		been removed. The removal of 100\,mK fluctuations affects only the very low
		frequencies below the first sky harmonic at $f_{spin}\simeq 0.017$\Hz. }
	\label{fig:TOI_Tdecor} 
\end{figure}

The main thermal fluctuations identified originate from the bolometer plate at
103\,mK and are monitored by four thermometers, one of which is used for long
term (hour scale) regulation.  Unfortunately, the thermometer data is
seriously affected by glitches and proved to be unusable. We rely instead on
the deglitched signal from the two dark bolometers, which are smoothed with a
running window with a width of 2 minutes and combined to produce a template of
the thermal fluctuations.  Two correlation coefficients per bolometer, derived
from two weeks of flight data acquired before the start of the survey, are
used to decorrelate the bolometer signal using the template. This step reduces
the correlation coefficient from $\sim 0.5$ to zero (although these
coefficients have a rather large scatter of about 0.25 when computed on only
50 rings).  Fig.~\ref{fig:TOIExample2} displays the effect of the
decorrelation on a few TOIs, and Fig.~\ref{fig:TOI_Tdecor} shows the effect on
a TOI power spectrum.

\subsection{4K cooler line cleaning} \label{sec:4K}

The 4\,K mechanical cooler induces some noise  in the bolometer
signal via electromagnetic interference and coupling, as well as
microphonic effects.  The 4\,K coolers main operational frequency
($f_{4K}=40.0834\,$Hz) is locked-in with the signal modulation frequency
($f_{4K}=\frac{4}{9} f_{mod}$ with $f_{mod}=90.1876$\,Hz).  The 4\,K cooler
systematic effects therefore show up in power spectra of the signal TOIs as
narrow lines at predictable frequencies and with a power larger than the noise
by one or more order of magnitude.  Here we describe a simple method that we
have used to remove these lines.

Lines at nine frequencies ($\sim 10\times n$\,Hz with $n=1 \,\mathrm{to}\, 8$,
and at $17\,\mathrm{Hz}$) were systematically found in the power spectra that
could be traced to the 4\,K cooler.  They affect the 
72 detectors in different ways. Given a chunk of 54 data samples, taken at the acquisition
frequency $f_{acq}=2f_{mod}= 180.3751890$\,Hz for a given detector, the 4\,K
cooler lines show a Fourier pattern at a period of $3n$ and 5 samples (\ie at multiples of  $f_{acq}/54$) for the nine lines.

The removal method assumes that the lines correspond to single Fourier
components of the signal.  We compute the $a$ (cosine) and $b$
(sine) coefficients of the Fourier series at the line frequencies over
a given period of time, $L_{cut}$, from the input TOI (which is
interpolated over flagged data).  
We then construct a timeline by summing the 9 cosine and sine components of the Fourier
series and subtracting it from the initial data.

The cooler lines contribute coherently to the phase binned rings only when
their frequencies are close to a multiple of $f_{spin}$.  Given the roughly
random variation of the spin rate (3$\sigma$) between 59.95 and 60.05\,sec
\citep{planck2011-1.1}, this occurs in 1--2\% of the rings.  Indeed the
natural line width is $f_{spin}/N_{circle}$, where $N_{circle}\sim 30-40$ is
the number of circles in a ring, to be compared to the 
separation between sky-signal harmonics of $f_{spin}$.

\begin{figure*}[!htbp]
	\includegraphics[width=1\textwidth, height=.4\textheight] 
		{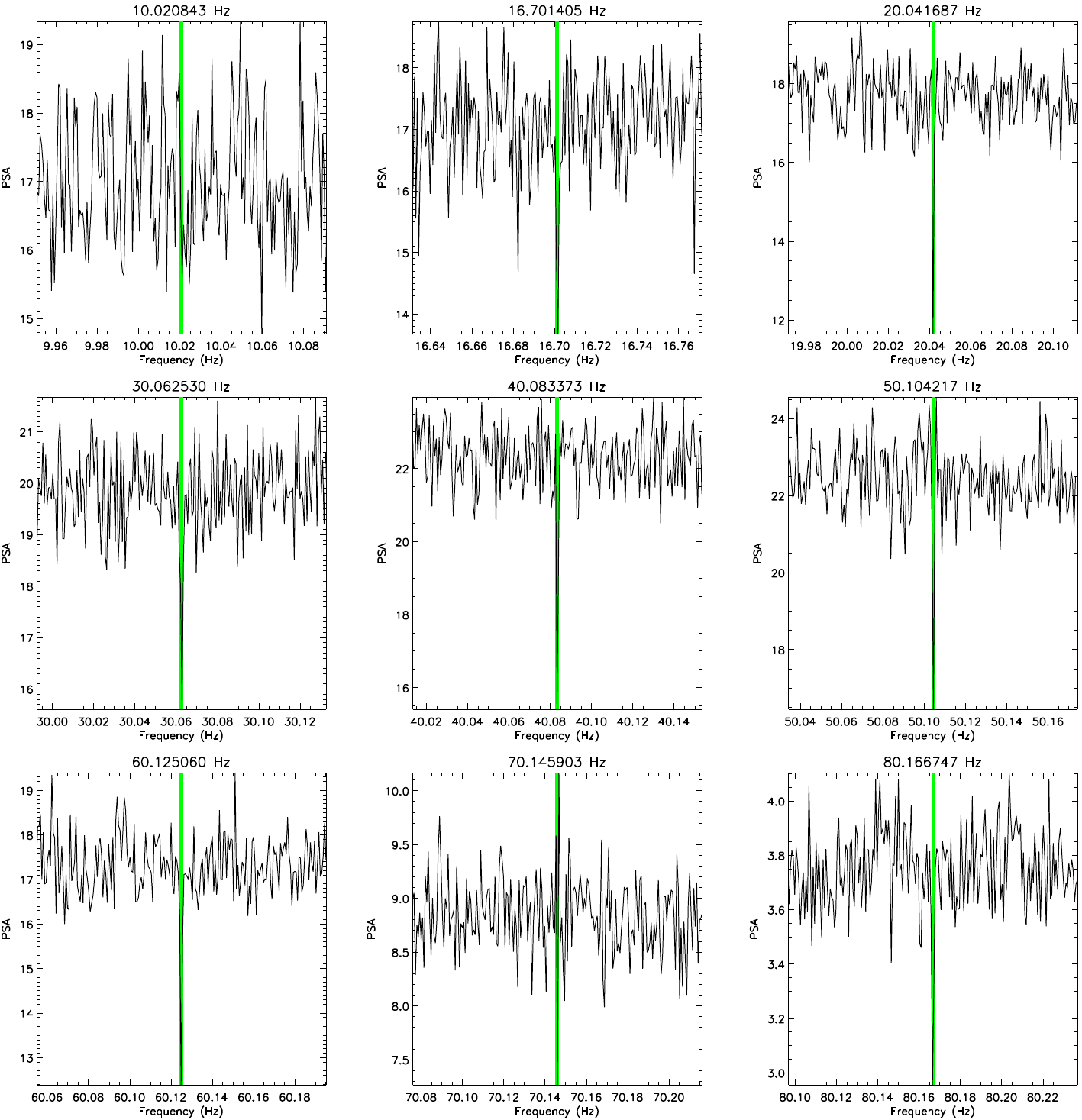}  
	\\ \\
	\includegraphics[width=1\textwidth, height=.4\textheight] 	
		{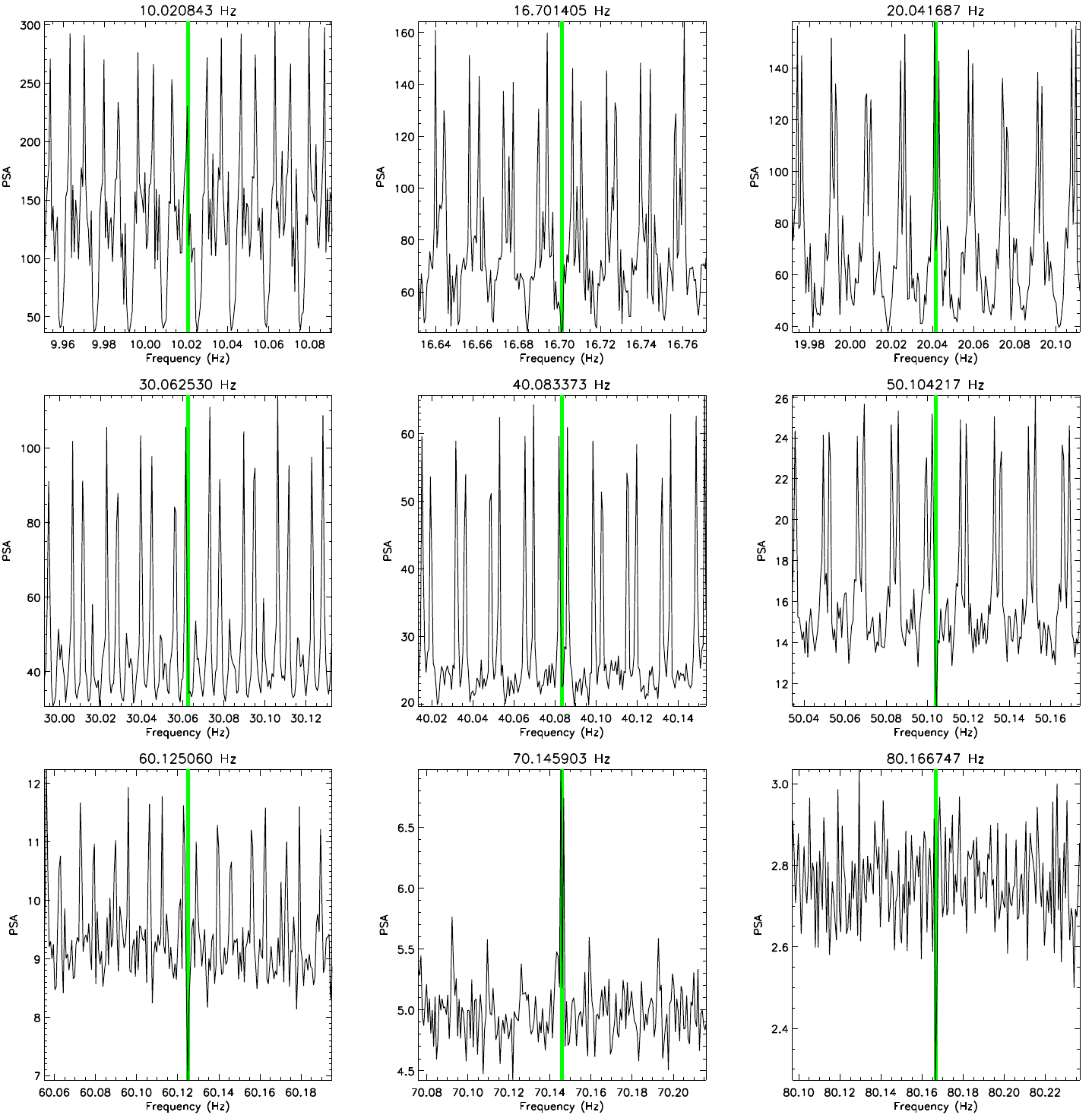}
     	\caption{The power spectrum amplitude around the nine
                  4\,K lines for the two bolometers 143-5 (upper panels) and 545-2
                  (lower panels). The
                   vertical bar shows the line frequency. The spectra
                  are obtained by averaging  200 individual
                  ring spectra (see  Sect.~\ref{sec:detnoise}) 
                  to reduce the noise. The residual lines are 
		at most comparable to the noise level. The signal shows up 
                  at the harmonics of the spin frequency $\simeq 0.16$\Hz.
		 In the CMB channel of the top panel, the high harmonics of the signal 
		do not stand out against the noise, but they are clearly visible in the 
		bottom panel. Note that the spin frequency variations between 
		rings yield the complex signal line pattern of interleaved combs. }
	\label{fig:4KlinePS} 
\end{figure*}

The sky signal may therefore impact on the measured cooler lines and
the computed $a$ and $b$ coefficients for those rings. In the
extreme case of strong sources this may induce ringing in the data.  The
effective frequency window width around each of the nine 4\,K cooler lines for
this analysis is defined by $\Delta f= {f_{acq}}/{L_{cut}}$. When a
multiple of $f_{spin}$ happens to fall into these windows, the $a$ and $b$
coefficients will be altered by the signal. Hence, the ringing increases
inversely to $L_{cut}$ requiring us to use larger values of $L_{cut}$.

On the other hand, the accuracy to which we remove the 4\,K cooler
lines from the TOI is limited by their time variability and therefore
pushes us to decrease $L_{cut}$.  
To mitigate the 
variability of the lines and the
ringing effect described above, a trade-off was chosen by taking
$L_{cut}= 345,600 $ samples (31\,min~56\,s) for all channels (a multiple
of 54 samples).

Fig.~\ref{fig:4KlinePS} presents the power spectrum around nine of the
4\,K lines. The 4\,K lines are subtracted reasonably well except for
the line at 70\Hz\ for the high frequency bolometers.

We have performed simulations to quantify the effect of the 4K-line removal
on the final HFI maps. The main adverse effect occurs when one of the removed
4\K\ lines happens to straddle one of the sky harmonics.  In such cases the
missing harmonics from the sky signal produces a `stitching' pattern on the
map, which can be pronounced if there are strong signal gradients along the
ring (\eg due to a very bright point source). Typically, a point
  source produces a `stitching' pattern with an amplitude of about $10^{-5}$ of
  the source signal. Troublesome artefacts are rare enough that, for now, we
  either remove such rings from the analysis or discard  `stitching'-related
  spurious sources from the ERCSC. The simulations show that a negligible
  number of spurious point sources remain.


\subsection{Temporal transfer function deconvolution}\label{sec:TF}

A simple FFT is used with the processed TOI to deconvolve the temporal
transfer function . This temporal transfer function is determined by using
Mars crossings (see Sect.~\ref{sec:fpg}) and optimising the symmetry of the
measured output beam profile \citep[see][]{planck2011-1.5}. A regularisation
filter (a low pass filter with a width of 5\,Hz, specifically
$\mathbf{ \sin^{2}\left[ \frac{\pi}{2} \left( \frac{f_{mod}-f}{5\,\mathrm{Hz} } \right) \right] }$ is
applied simultaneously to the inverse of the temporal transfer
function. Before applying the FFT, the TOI samples which are flagged as
invalid, are filled with a ring-based interpolation. The Jupiter signal is
replaced by a combination of linear interpolation and noise to avoid long
duration effects (ringing) of the deconvolution of this huge signal along the
scan. These interpolated samples are flagged and not projected onto maps.

To provide an assessment of how uncertainties in our current knowledge of the
transfer functions translate into processed data, we consider the statistical
uncertainties in the determination, and propagated these uncertainties to
source fluxes and angular power spectra. This was obtained by means of 50
Monte-Carlo simulations with the following elements: (i) sky simulation; (ii)
projection to time series for all the detectors within a frequency band, using one year of
flight pointing data; (iii) convolution of each time series with a time
response realization; (iv) deconvolution by the nominal time response
function; (v) projection to a single map per band. An overview of the
technical implementation of our simulations is given in
Sect.~\ref{sec:sims}. The time response realizations are estimated as transfer
functions in Fourier space, based on the model and error budget described in
\cite{planck2011-1.5}. The realizations include statistical errors, error
correlations, and the estimated level of systematic error. Correlations among
errors in transfer functions of different detectors are not included.
Convolution and deconvolution transfer functions are normalized at the
temporal frequency corresponding to the cosmic dipole signal observation, to
mimic dipole/large-scale calibration. Moreover the time response is such that
the point sources are not shifted in time.  For the point source flux
assessment, the simulated input sky included only point sources and diffuse
Galactic emission. For assessing the impact on angular power spectra, the
simulated sky was a realization of the CMB.

The results can be summarized as follows:\begin{itemize} 

\item aperture flux of point sources has an average error of 4\% at
  857\GHz\ and 0.5\% at 143\GHz;

\item the error on the angular power spectrum is order of 1\% above
  multipole $\ell = 100$ and reduces  to 0.1\% at larger scales.  
\end{itemize}
This sets the level of errors to which the results in the early papers 
 need to be immune to, provided  the errors on the transfer
functions assumed here are realistic. We are gratified to have achieved this
level of accuracy within only a month of the data being acquired. This rapid
delivery of accurate processed data was 
one of the key elements which allowed the early science analyses
 described in this special issue. Nevertheless, the accuracy of the data  is certainly not 
sufficient to reach the more demanding cosmological science goals of \Planck.
In the next two years, improving the fidelity of the data will be the major focus of  work 
by the HFI core team. This task will be greatly aided  by the increased level of 
redundancy offered by more data but we also expect to identify 
other systematic effects  that will need correction, in addition to those described here.
These will be described in detail in future papers.

\subsection{Other sources of noise} \label{sec:rts}

Three bolometers (143\_8, 545\_3 and 857\_4) have been identified as
being affected by Random Telegraphic Signal \citep[RTS,
see][]{planck2011-1.5}. These detectors are not used to
build frequency maps. There is no evidence for a significant
RTS contribution to the noise of other detectors. 

More quantitatively,
we have used simulations to assess the minimum jump between signal levels that can be
detected by a Viterbi-based algorithm \citep{viterbi1967}.  
We derive a best-fit distance
between two states ($D$, measured in units of the noise rms of that
bolometer), as well as the per-sample improvement in the fit from
using two states, $\delta L$. For bolometers at frequency
$\le\,353\,\mathrm{GHz}$, there are only a  few single rings which
show $D>1$ and $\delta L > 0.03$ in the real data whereas the
algorithm finds all RTS rings in a simulations with $D>0.5$. For
higher frequencies, the upper limit on any RTS is somewhat less
stringent. We conclude that the impact of any undetected RTS on the final
data products is negligible.

Finally, as discussed earlier, there is a low frequency excess noise which 
is seen in all detectors and as a result the amplitude at the spin frequency 
$f_{spin} =  0.016\Hz$ is higher by a factor of about three than the white noise level 
(as can be seen in Fig.~\ref{fig:PowerSpectra143}-a and \ref{fig:AmoPSD}), which 
as we shall see later  results in increased noise at large scales in the maps, 
though at a significantly lower level than the cosmic variance.
Given our study of glitch and RTS residuals, neither are  likely 
to contribute significantly to this excess.
One possible explanation mentioned in \cite{planck2011-1.3} 
may be thermal fluctuations linked to high energy cosmic ray showers 
\emph{which are not common to all detectors}, as indicated by the weak 
level of cross-correlation, at low-frequency, shown in Fig.~\ref{fig:PowerSpectra143}-b. 

\begin{figure}[!htbp]
	\includegraphics[width=1\columnwidth]{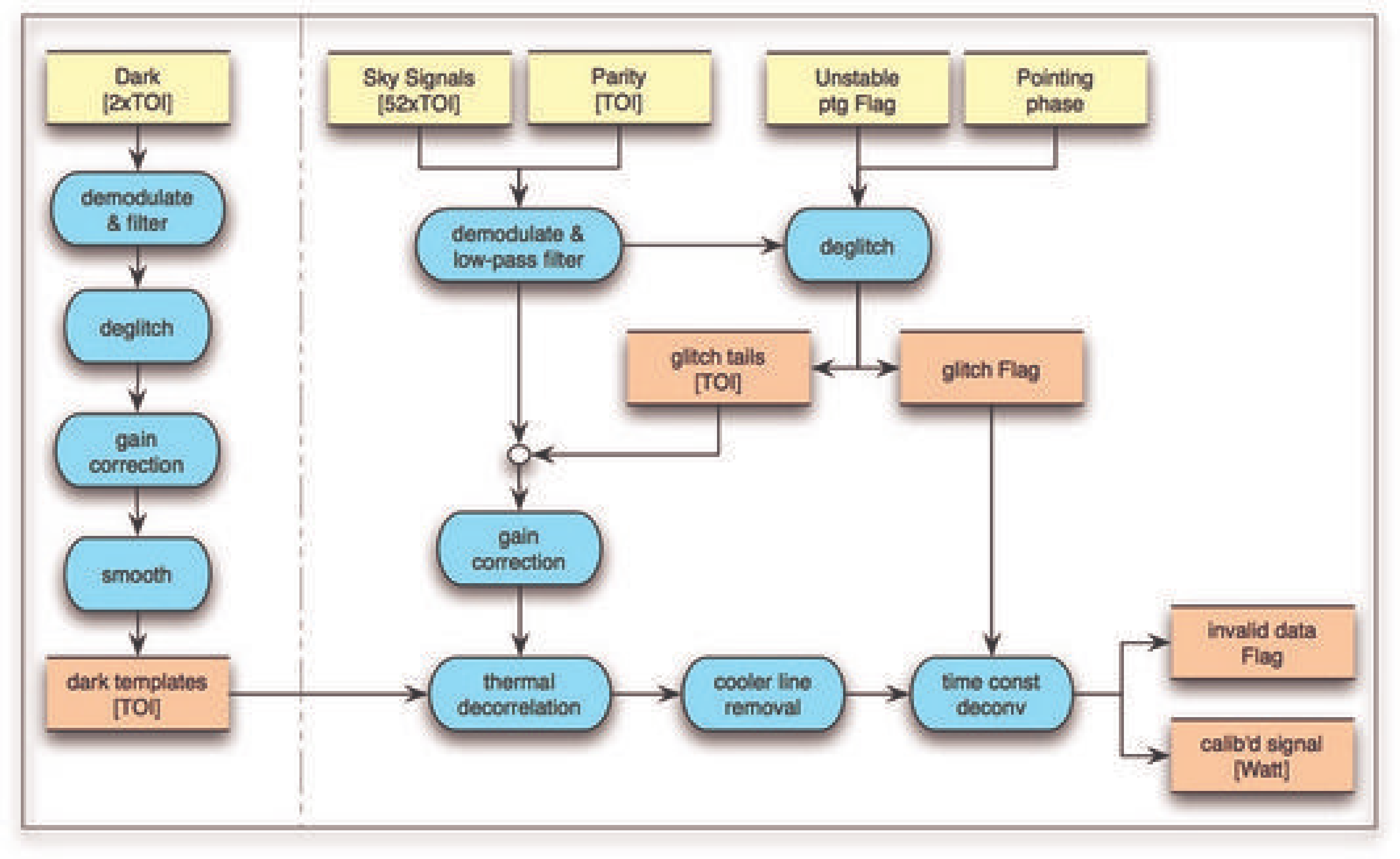}
  	\caption{TOI processing data flow. See text for details. }
  	\label{fig:TOIoverview} 
\end{figure}

\subsection{The overall TOI processing pipeline} 

\begin{figure*}[!htbp]
  	\includegraphics[angle=180,width=1\textwidth, totalheight=.35\textheight]
  		{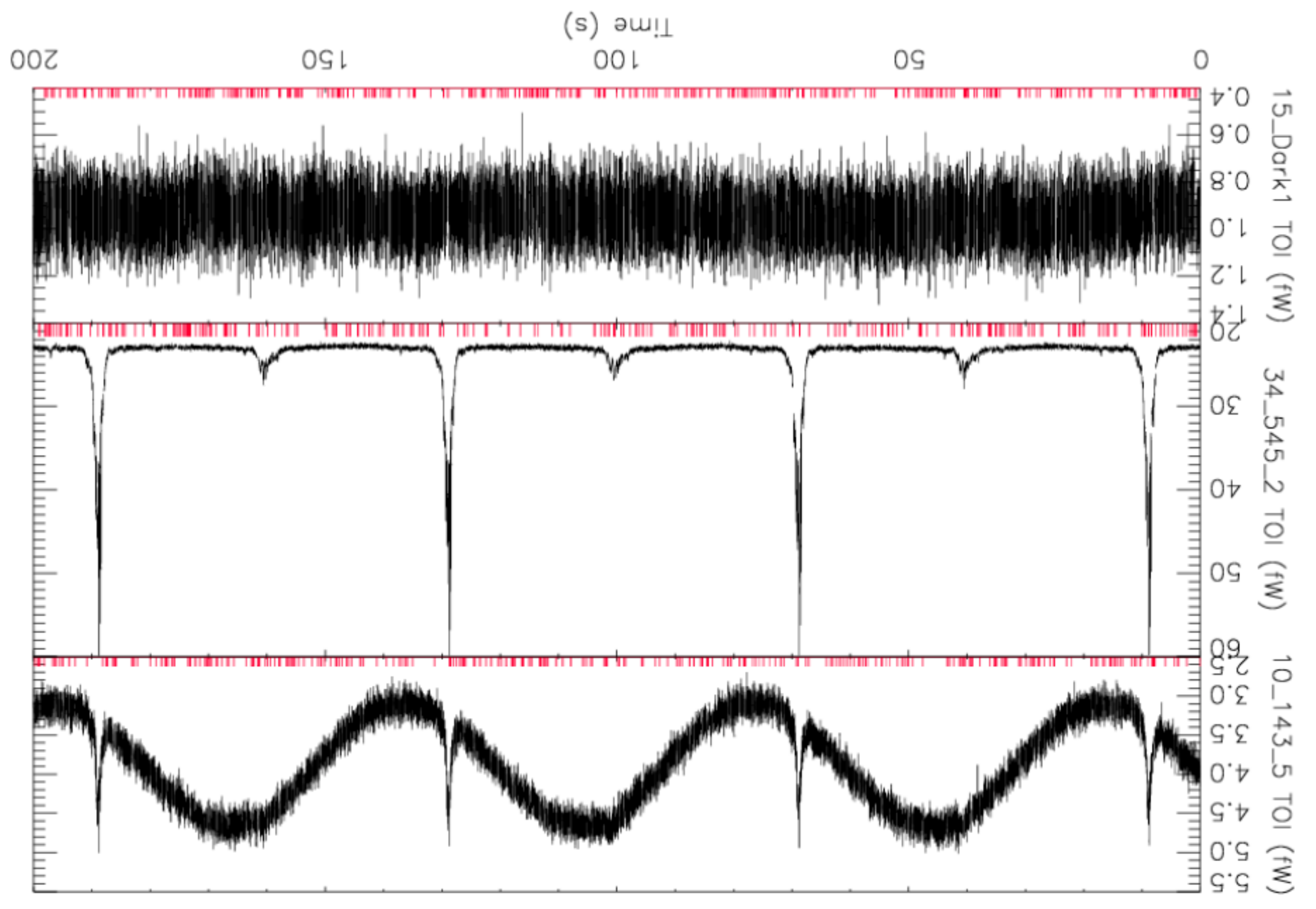}
  	\caption{Processed TOI for the same bolometers and time range as shown in 
  		Fig.~\ref{fig:TOIExample1}. Times where
  		data are flagged, are indicated by the red ticks at
                the bottom of each plot.} 
	\label{fig:TOIExample4} 
\end{figure*}

\begin{figure*}[!htbp]
  	\includegraphics[angle=180,width=1\textwidth,
          totalheight=0.35\textheight] {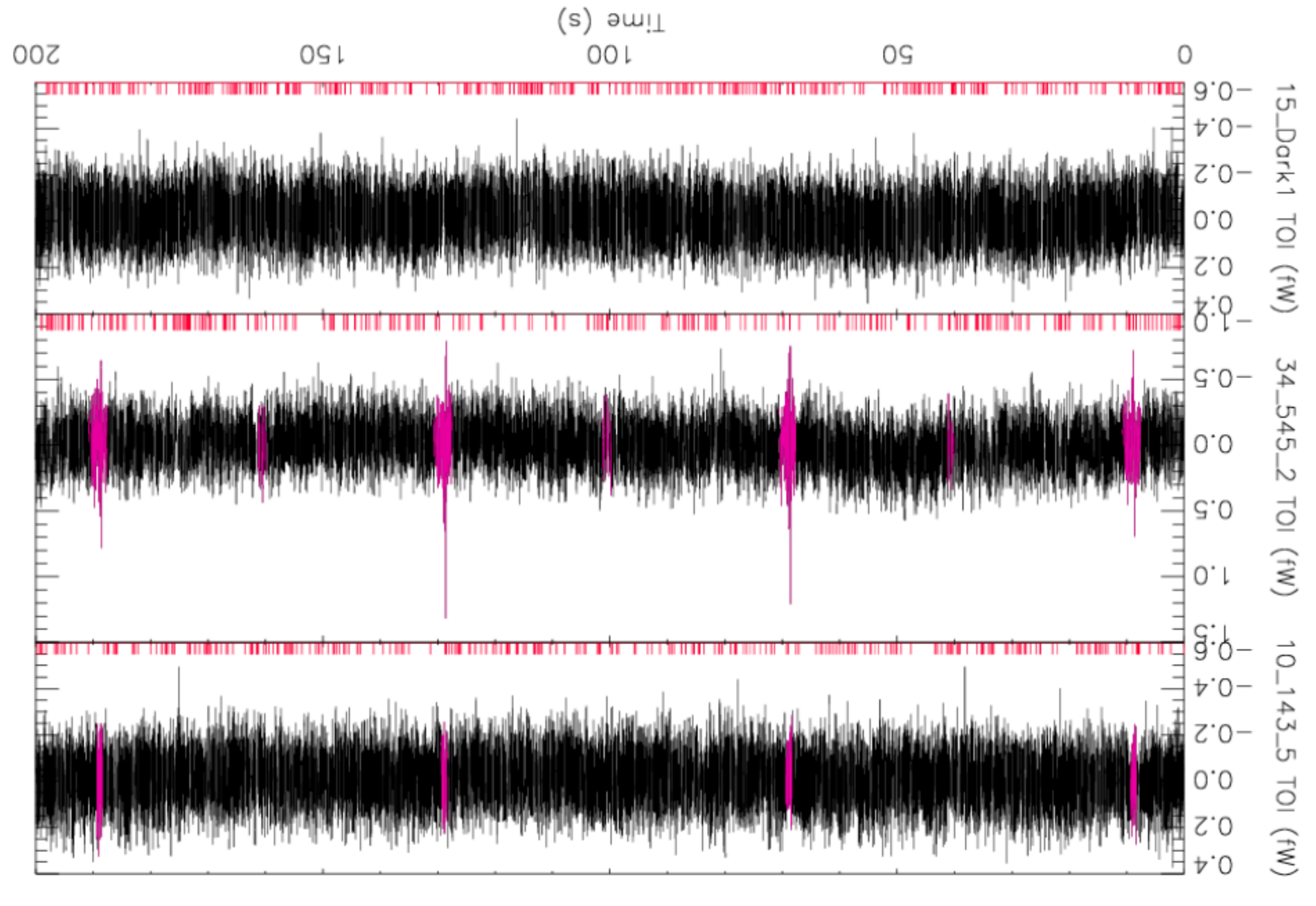}
  	\caption{Processed TOI as in Fig.~\ref{fig:TOIExample4}, but with
    		the ring averaged signal subtracted to enhance features near the
    		noise limit. Times where data are flagged, are shown by the
    		red lines at the bottom of each plot. Purple zones show where
                the strong signal flag is set and where
    		the phase-bin ring average subtraction is not expected to yield a perfect cancellation of the
                signal.}
	\label{fig:TOIExample5} 
\end{figure*}

Figure~\ref{fig:TOIoverview} provides an overview of the TOI
processing data flow.  The dark detectors are processed first to
prepare them for use as templates in the 100\,mK decorrelation of the
other detectors. Each bolometer is then processed independently of the
others. The independence of the processing enables various jackknife
tests to be constructed to assess errors. The TOI processing pipeline
chains the modules described above in a specific order. Namely, the
raw signal TOI in Volts is read from the database. It is demodulated
and filtered.  The deglitching step produces a flag TOI
and a new bolometer TOI in which glitch tails have been corrected for.
This new TOI is then gain-corrected.  A (small)
decorrelation from the smoothed dark templates is then applied. For
Fourier-domain processing we need to fill the gaps. Hence, for invalid
data samples, an interpolation is performed using the bolometer
average signal computed from phase-binned ring derived by
Fourier-Taylor expansion (see Sect.~\ref{sec:detnoise} and
Appendix~\ref{sec:FTR} for details). We then remove the 4K cooler lines
 and finally deconvolve the  temporal transfer function.  Two
additional flags are created for use in the map-making or noise
estimation. These are: (1) flags around planets, asteroids, and comets\footnote{For 
definiteness, here follows the list of solar system moving bodies flagged out. 26 Asteroids: 
Hygiea, Parthenope, Nemesis, Victoria, Egeria, Irene, Eunomia, Psyche,
Melpomene, Fortuna, Ceres, Massalia, Amphitrite, Pallas, Bamberga,
Juno, Daphne, Eugenia, Vesta, Davida, Europa, Interamnia,
Iris, Thisbe, Flora, Metis; 20 Comets:  Broughton, Cardinal,
Christensen, d'Arrest, Encke, Garradd, Gunn, Hartley2,
Holmes, Howell, Kopff, Kushida, LINEAR, Lulin, McNaught,
NEAT, Shoemaker-Levy4, SidingSpring, Tempel2, Wild2.}, as
moving bodies should not be projected onto the maps; (2) flags of strong
signals (mostly Galactic plane crossings) that are discarded from
noise estimation and destriper offset computation. Note that the
invalid data flag corresponds to data with the glitch flag set,
together with the very few missing data samples from the telemetry and
from compression saturation (for a huge glitch,  on average, once every 500 rings and 
when we crossed for  the Galactic centre at 857\GHz\ for the first time, see Sections~\ref{sec:L1stat}
and \ref{sec:comp}).

Examples of cleaned TOIs produced are shown
in Fig.~\ref{fig:TOIExample4}. The pipeline also produces TOI in which the
ring average signal has been removed (see Fig.~\ref{fig:TOIExample5}). This
provides the basis for estimating the noise as detailed in the next section.

 \section{Detector Noise estimation} \label{sec:detnoise} 

 In this section, we investigate the statistical properties of the
 detector noise timelines from flight data, both on raw and clean
 TOIs. Ground based measurement give only approximate 
 indications of the noise characteristics as many features (\eg\ long
 term drifts, microphonic noise) depend on the satellite environment
 and the instrument settings in-flight. The difficulty in interpreting
 flight data is to estimate the noise properties in the presence of
 signal, which is the goal of the approach described below.

\begin{figure}[!htbp]
	\includegraphics[width=1\columnwidth]{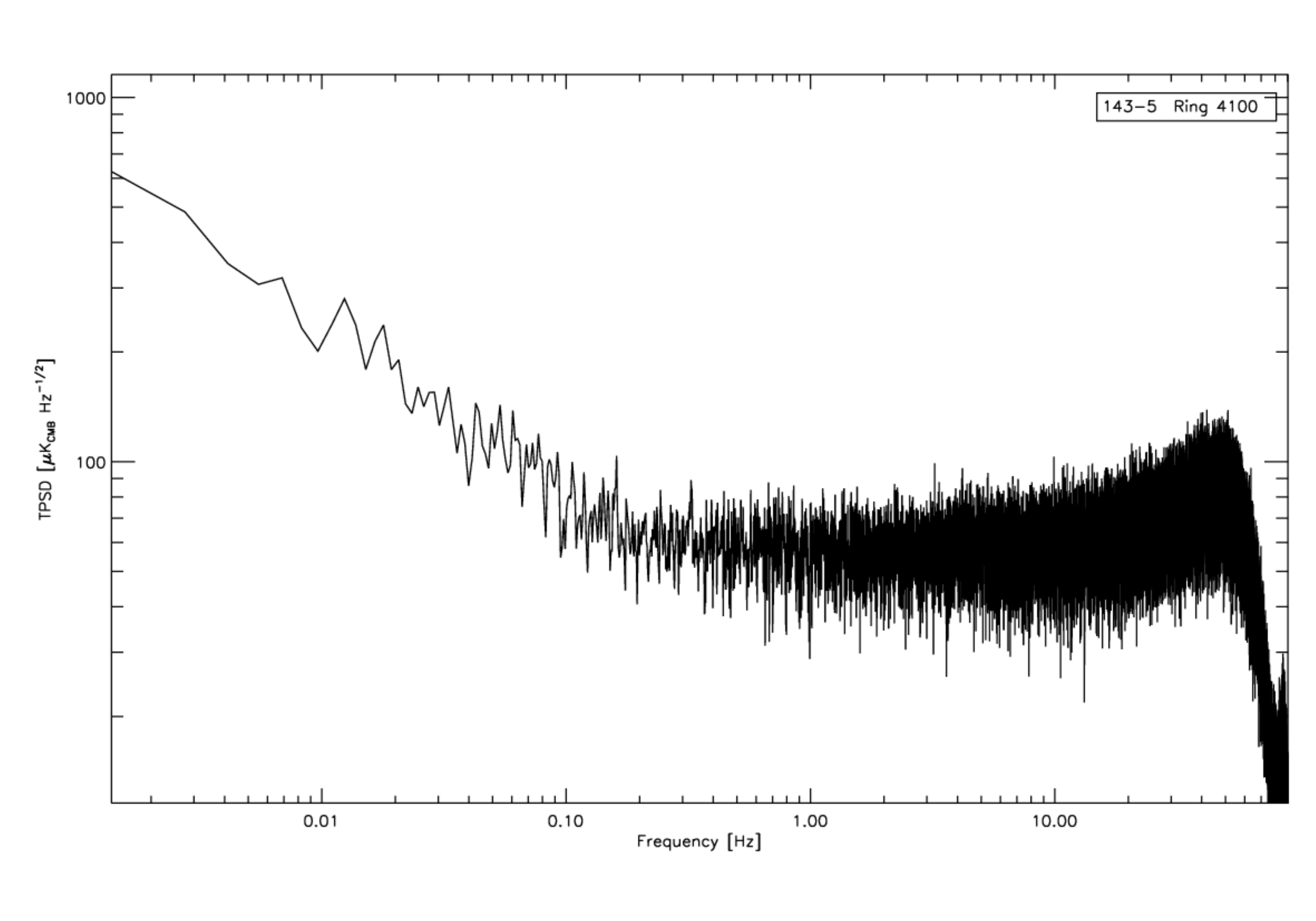}
	\includegraphics[width=1\columnwidth]{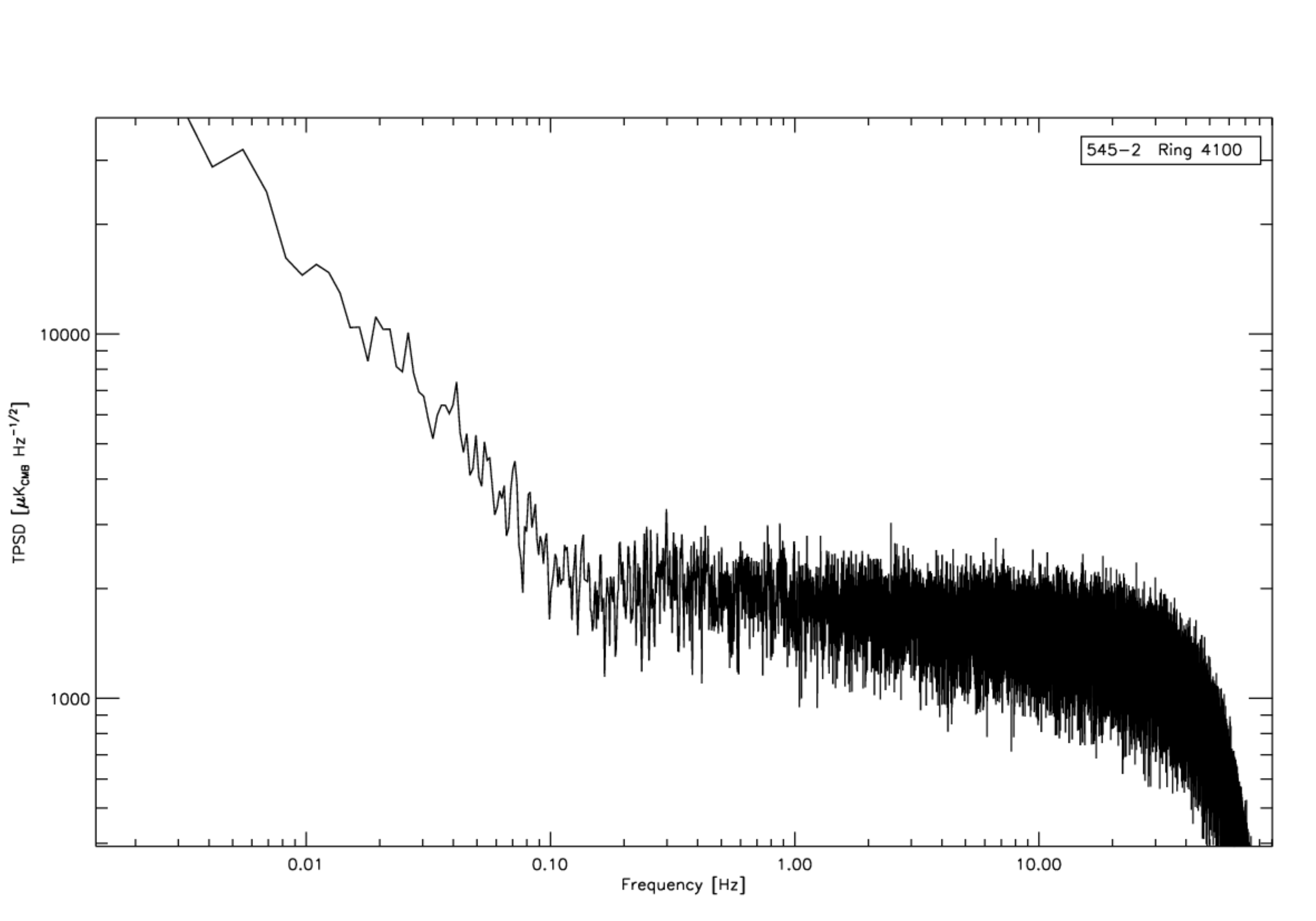}
	\includegraphics[width=1\columnwidth]{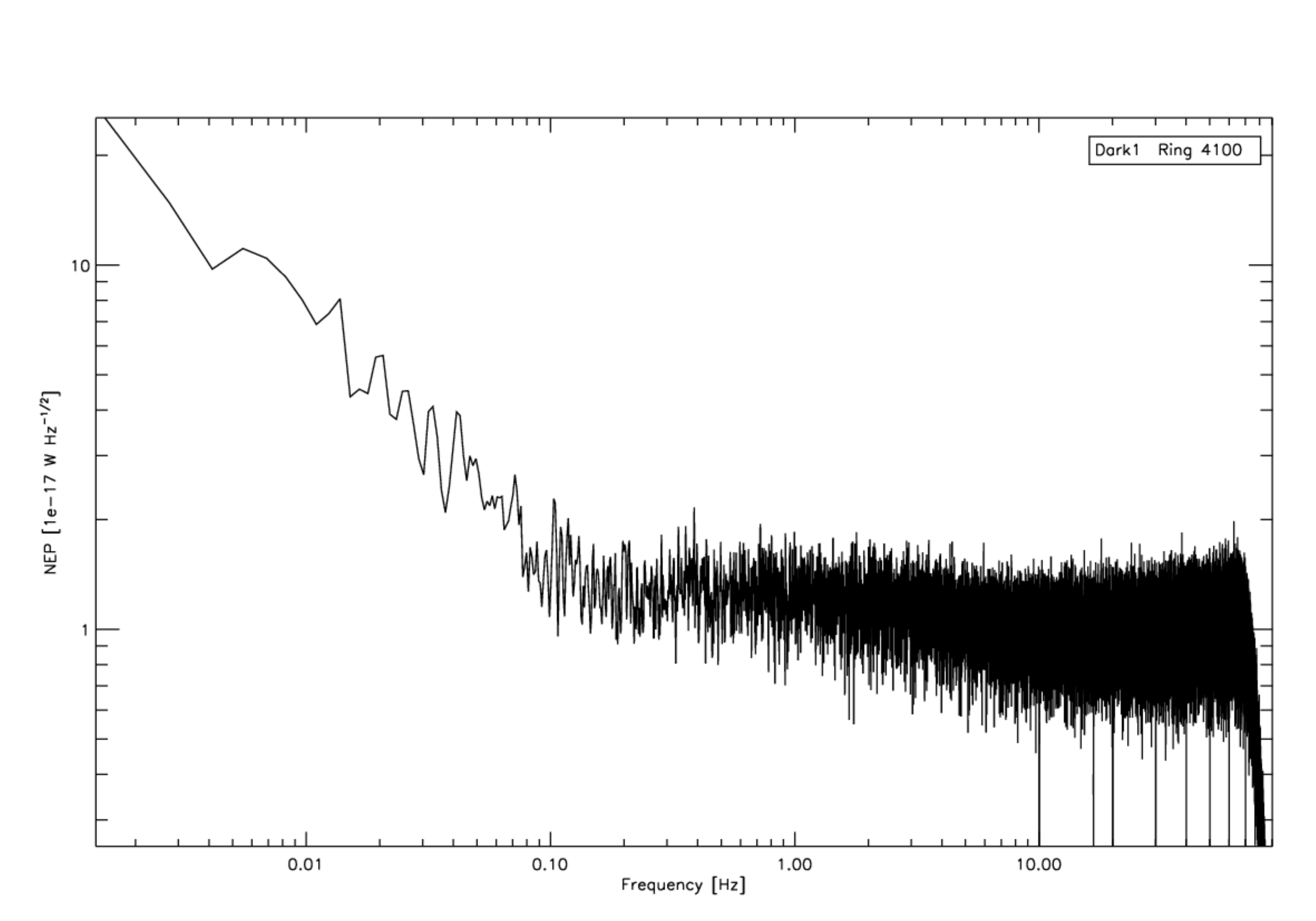}
  	\caption{Examples of noise power spectra for the
    		bolometers 143-5 (top), 545-2 (middle), and Dark1 (bottom).  The
    		first two have been calibrated in CMB temperature
                units, by using the calibration 
    		coefficients derived during the map making step. The
                last spectrum is in Watts. 
    		The central region shows a nearly white noise plateau,
    		with a low frequency `1/f' component, and a high frequency cut-off due to
    		the filtering of frequencies above the sampling frequency.  At 143\GHz, the
    		upturn due to the deconvolution of the (bolometer dependent) temporal transfer
    		function is clearly seen (see details in Sect.~\ref{sec:TF}).}
    \label{fig:AmoPSD}
\end{figure}

 It is shown in Appendix~\ref{Annex:detnoise} that the joint maximum
 likelihood estimate of the signal and the noise spectral parameters
 (taken here as flat frequency bin powers) can be achieved by 
 using the redundancy of the
 scanning strategy. In the case of the \Planck\ scanning
 strategy, a flat weighting in the signal estimation on rings is very
 close to optimal, so there is no need to iterate the signal
 and noise power estimation.

 The main difficulty in estimating the signal is to precisely sample
 the signal in phase on rings. This is achieved by making the
 assumption that the signal content is band-limited on the rings,
 together with an approximate (but arbitrarily accurate) irregular
 sampling method based on Fourier-Taylor expansions, leading to
 so-called Fourier-Taylor rings (hereafter FTR, see
 Appendix~\ref{sec:FTR}). Note that this signal removal approach is only 
possible when the estimation is done on a ring-by-ring basis 
 (map based signal estimation might replace it in the future).\\

The pipeline can thus be summarized as follows:
\begin{enumerate}

\item Estimate the signal content of each ring using the scanning
redundancy;

\item Subtract this estimate from the original data timeline to produce an
estimate of the noise content;

\item Compute edge-corrected, averaged periodograms of the estimated
noise timeline;

\item Optionally adjust a parametric model of the noise spectrum to
the periodograms to determine the noise parameters.

\end{enumerate}

Each step is described in further detail in  Appendix~\ref{Annex:detnoise}. 
Step 3 is repeated on all rings, though other zones can also be defined and
used for this purpose.   

The last step is implemented as a maximum likelihood estimate of the
spectral parameters (e.g. Noise Equivalent Power or NEP, knee
frequency and spectral index of low-frequency noise), where the
distribution of the averaged periodogram estimate is approximated as a
product of $\chi^2$ distributions with the appropriate number of
degrees of freedom.  The noise parameters are determined from the
spectra of all rings (or other zones), which is useful in monitoring
the evolution of these parameters with time. The results discussed below were determined by fitting a pure 
white noise model in the 0.6 -- 2.5\Hz\ frequency range.

Figure~\ref{fig:AmoPSD} shows the noise power spectrum estimates for
three bolometers: 143-5, one of the most sensitive CMB-dominated
channels (top); 545-2, operating at a frequency where the dust
emission of the Galaxy and IR galaxies dominate the signal (middle);
and the Dark1 bolometer (bottom).  Several features are apparent in 
this figure. The spectrum is flat at intermediate
frequencies, which gives an estimate of the bolometer's
NEP. At low frequencies, there is a rise of power that
begins at effective knee frequencies considerably higher than 
those measured during  ground based
calibration. The extra low-frequency power is believed to be mostly due
to residual  thermal fluctuations from cosmic ray hits (those which are not common to all
detectors, see Sect.~\ref{sec:toip}). The high frequency part of the spectrum shows a rise
of power due to noise amplification by the transfer function
deconvolution described in Sect.~\ref{sec:TF} and, in a few cases (not shown), weak
residuals of lines induced by the $4K$-cooler that have not been completely
removed. 

\begin{table}[htb] 
  	\caption{ Mean Noise Equivalent Temperatures for each channel. P and S after
    		the frequency value indicate polarised (PSB) and unpolarised (SWB)
    		bolometers. The last columns give the HFI sensitivity
                  goals, as can be found in 
    		\cite{lamarre2010}. }
  	\label{tab:NEP}  
  	\centering
  	\begin{tabular}{ c |  r  r }  \hline \hline
    		Frequency & NET & Goal \\ 
    		{[GHz]}  & \multicolumn{2}{c}{ [$\mu\mathrm{K_{CMB}\,s}^{1/2}$] } \\ \hline
          	100P & 65        &   100 \\
     		143P & 53         &     82 \\
     		143S & 41         &     62 \\
                   217P & 79         &   132  \\
                   217S & 68         &      91  \\
                    353P & 329   &    404  \\
                   353S & 220   &    277 \\
                   545S & 1410  &   1998  \\
                   857S & 41220 & 91000  \\ \hline
  	\end{tabular}
\end{table}

Table~\ref{tab:NEP} gives the average Noise Equivalent Temperatures per frequency channel (derived by using 
the calibration coefficients stored in the \IMO\ at the map-making stage) and,
for reference, the target values of the HFI  as given in the HFI pre-flight paper\footnote{The pre-flight paper provides goals for the polarised detectors in addition to the SWB only values of the Blue Book \cite{planck2005-bluebook}. It also sets the 100\GHz\ channels goal twice worse than in the Blue Book.} \citep{lamarre2010}. In 
the averaging per channel, some 
detectors showing pathological behaviour (Random Telegraphic Signal,
sometimes referred to as `pop-corn' noise) were removed, as they are
not used for map-making and other science analyses. The measured average NETs per channel
are, in all cases, much better than the goal values. As described in the Instrument performance companion paper \cite[see][and in particular Fig.~21]{planck2011-1.5}, the performance improvement as compared to goals comes from a lower background (and therefore of the corresponding photon noise) than initially assumed.

\section{Beams and Focal Plane geometry}\label{sec:fpg} 

\begin{figure*}[!htbp]\centering
	\includegraphics[width=\textwidth]{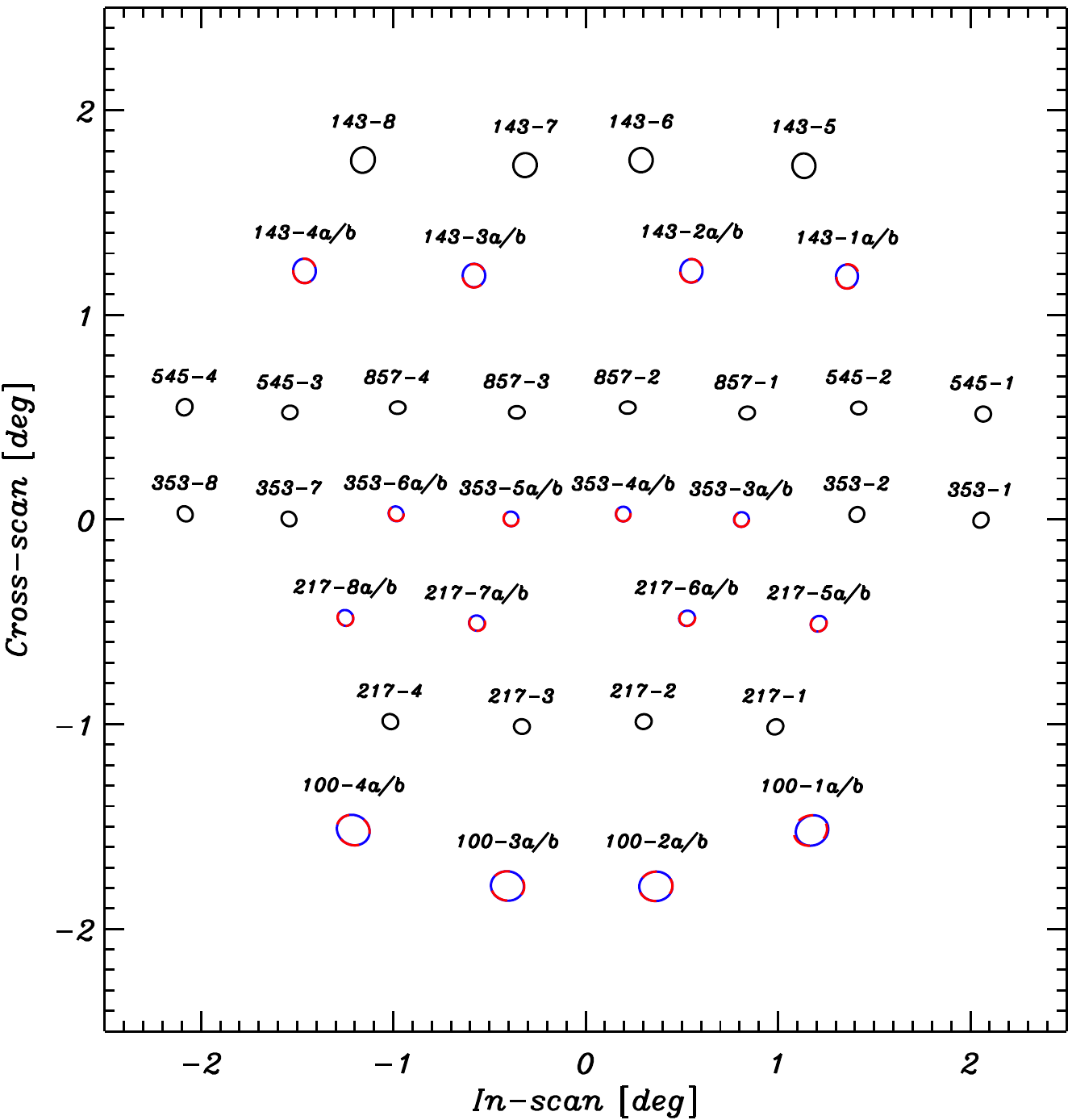}
	\caption{The HFI focal plane on the sky.  Individual detector locations are
		as measured, and beam shapes are represented by ellipses at the FWHM
		level. For polarized detectors, the beam shapes are over-plotted in red and
		blue and labelled (a/b) but are in practice
                indistinguishable by eye.  Fits are 
		on destriped data (except for 143-8) and represent the values from the
		Gaussian-only fits to the data.}
	\label{fig:fp_result_circle}
\end{figure*}


The pipeline for geometrical calibration determines,  simultaneously,  the
shape and location of the beam for each detector in the focal
plane. Detector locations are determined as three-dimensional
rotations with respect to the telescope attitude \citep{Shuster1993}. 
Beam shapes are parametrized in two ways: (1) as asymmetric Gaussians;
(2) with Gauss-Hermite polynomials of arbitrary order.  A visualization of the
focal plane is shown in Fig.~\ref{fig:fp_result_circle}.

The pipelines are capable of extracting measurements from planets as
well as from fixed sources. Planet ephemerides are calculated using
the Jet Propulsion Laboratory Horizons
software\footnote{\url{http://ssd.jpl.nasa.gov/?horizons}}
\citep{giorgini1996} which is programmed with \Planck's orbital
information. Movement over the course of a single planet
observation is significant in some cases and must be taken into
account.  Here, a \emph{single observation} refers to a set of
adjacent rings containing the object, typically lasting for about one week. 
Our planet observations are summarized in Table~\ref{tab:planets}.

\begin{table} [!htbp]
	\centering
    	\caption{Approximate dates and operational day (counted since launch), OD,
	 of HFI
          observation of planets.}      
    	\label{tab:planets}
        \begin{tabular}{l | c | c}
         \hline \hline
        Planet & Date & OD \\ 
         \hline
        Mars & 25/10/2009 & 165 \\
        Jupiter & 28/10/2009 & 168 \\
        Neptune & 03/11/2009 & 174 \\
        Uranus & 08/12/2009 & 209 \\
        Saturn & 05/01/2010 & 237 \\
        Mars & 13/04/2010 & 335 \\
        Neptune & 19/05/2010 & 371 \\
        Saturn & 14/06/2010 & 397 \\
        Uranus & 02/07/2010 & 415 \\
        Jupiter & 05/07/2010 & 418 \\
        \hline
        \end{tabular}
\end{table}

Single observations of Mars and Saturn provide high signal-to-noise
measurements of the beams and focal plane geometry.  Neptune and
Uranus also provide strong detections.  However, measurements of
extragalactic and Galactic sources are limited by CMB confusion and by
low signal-to-noise. Jupiter's very large flux drives some detectors
nonlinear and therefore results in artefacts which are removed from
the processed TOI. Mars and Saturn are therefore the primary
calibrators used in this section.

From a measurement of the position of each detector, we can compute an
overall rotation and scaling of the complete focal plane with respect
to a fiducial model (\eg\ the pre-flight optical model). These
parameters can be used to track mechanical and optical changes in the
telescope/detector system, although due to the phase shift induced by
the transfer function, we are insensitive to any  shifts in the
scan direction. 

Details of the algorithm for focal plane reconstruction, its pipeline
implementation, and validation on simulations, along with a separate
algorithm used to derive a preliminary estimate of the focal-plane
geometry, are given in Appendix~\ref{Annex:FPG}.

\subsection{Results} \label{sub:results_on_planck_data} 

\subsubsection{Focal-Plane Geometry} \label{ssub:FPG}

\begin{figure}[htbp] 
	\centering
	\includegraphics[width=0.49\columnwidth] {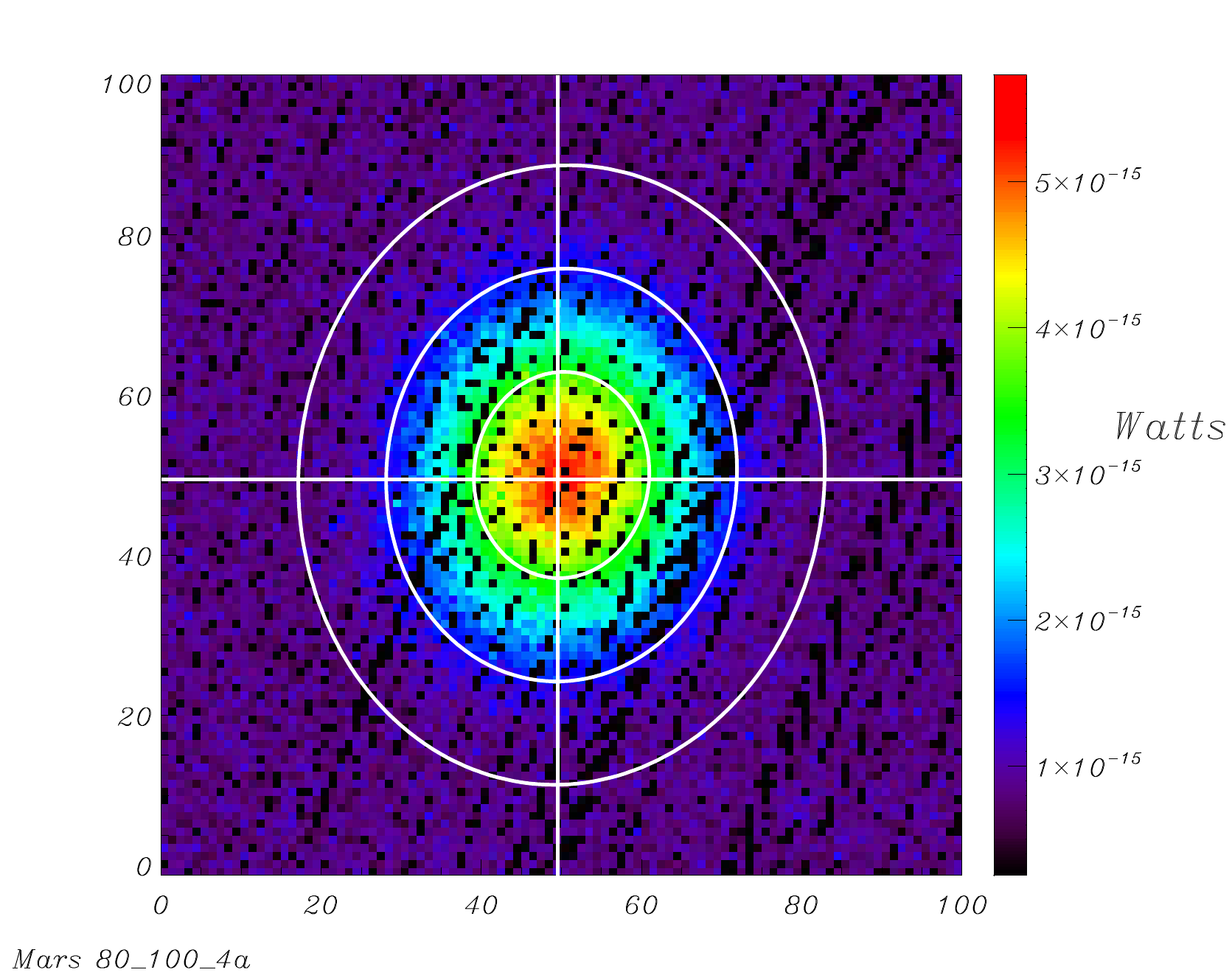}
	\includegraphics[width=0.49\columnwidth] {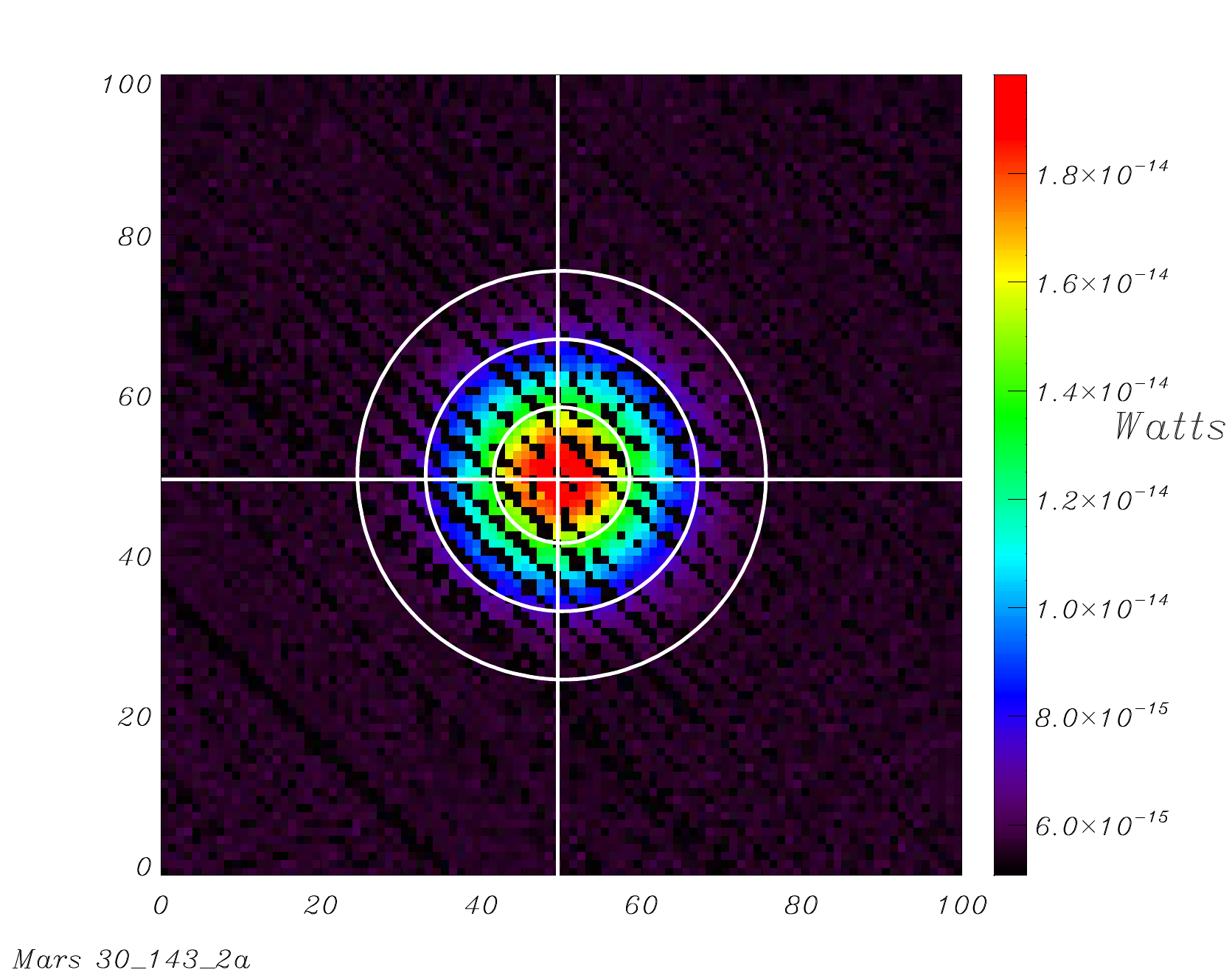}
	\includegraphics[width=0.49\columnwidth] {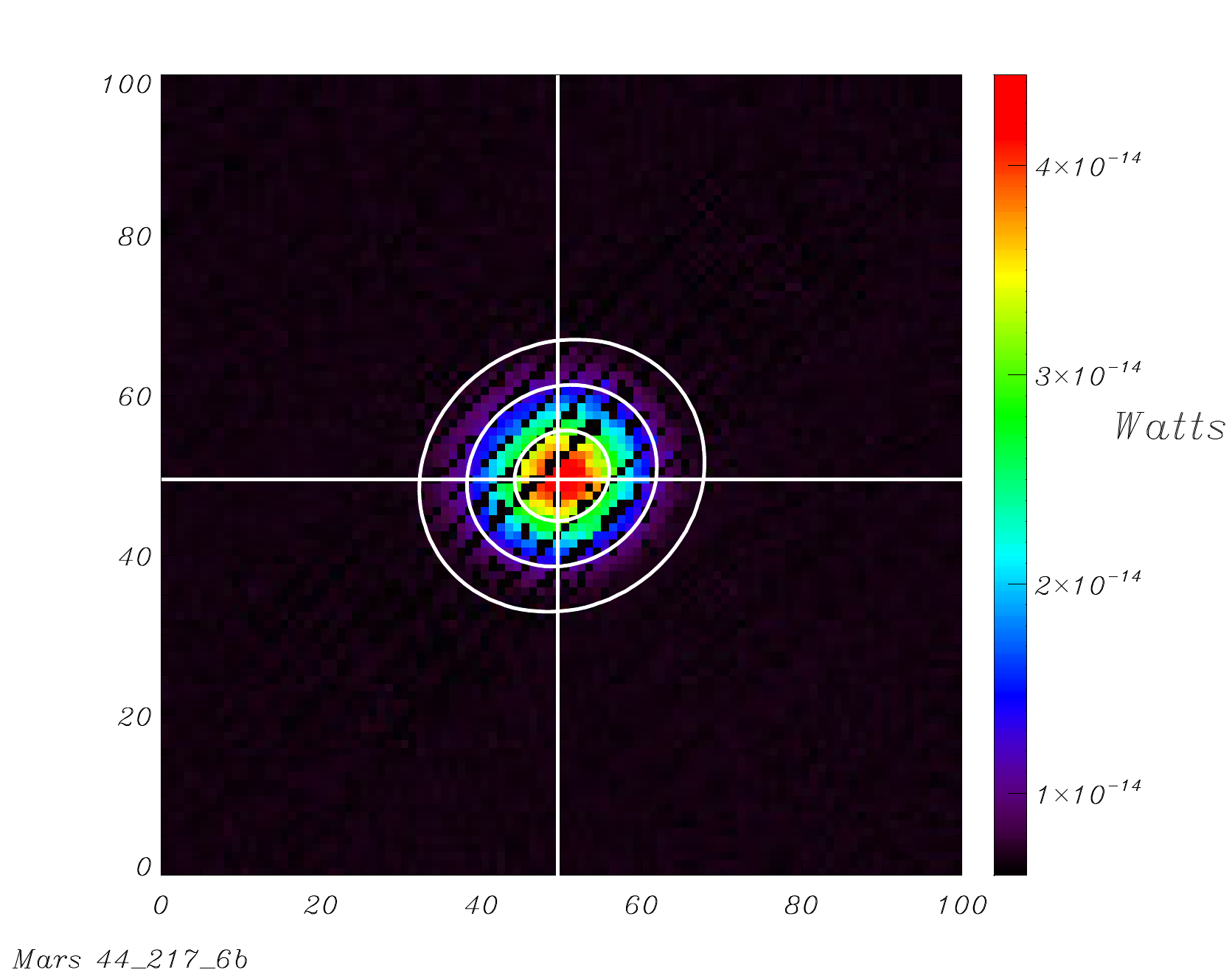}
	\includegraphics[width=0.49\columnwidth] {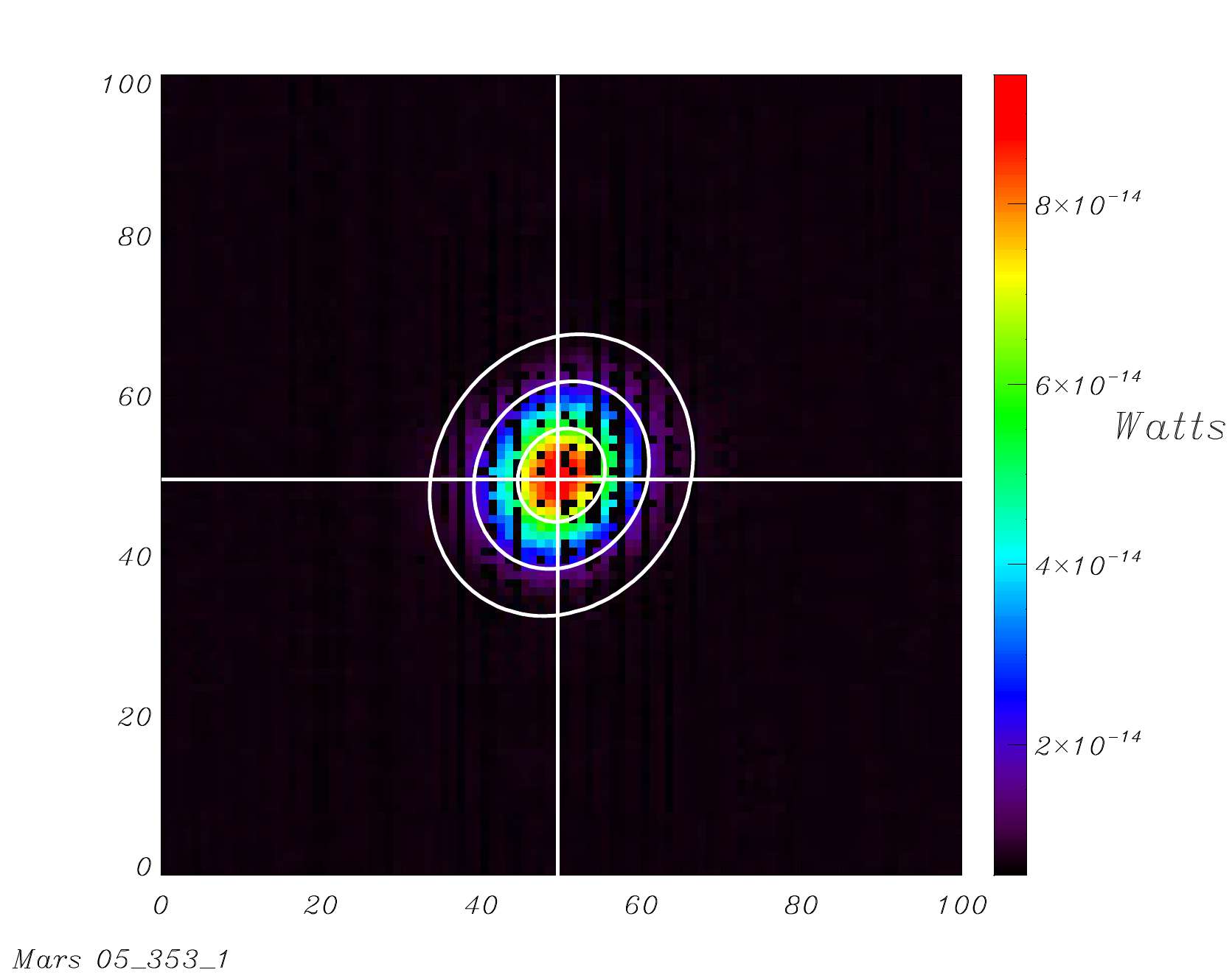}	
	\includegraphics[width=0.49\columnwidth] {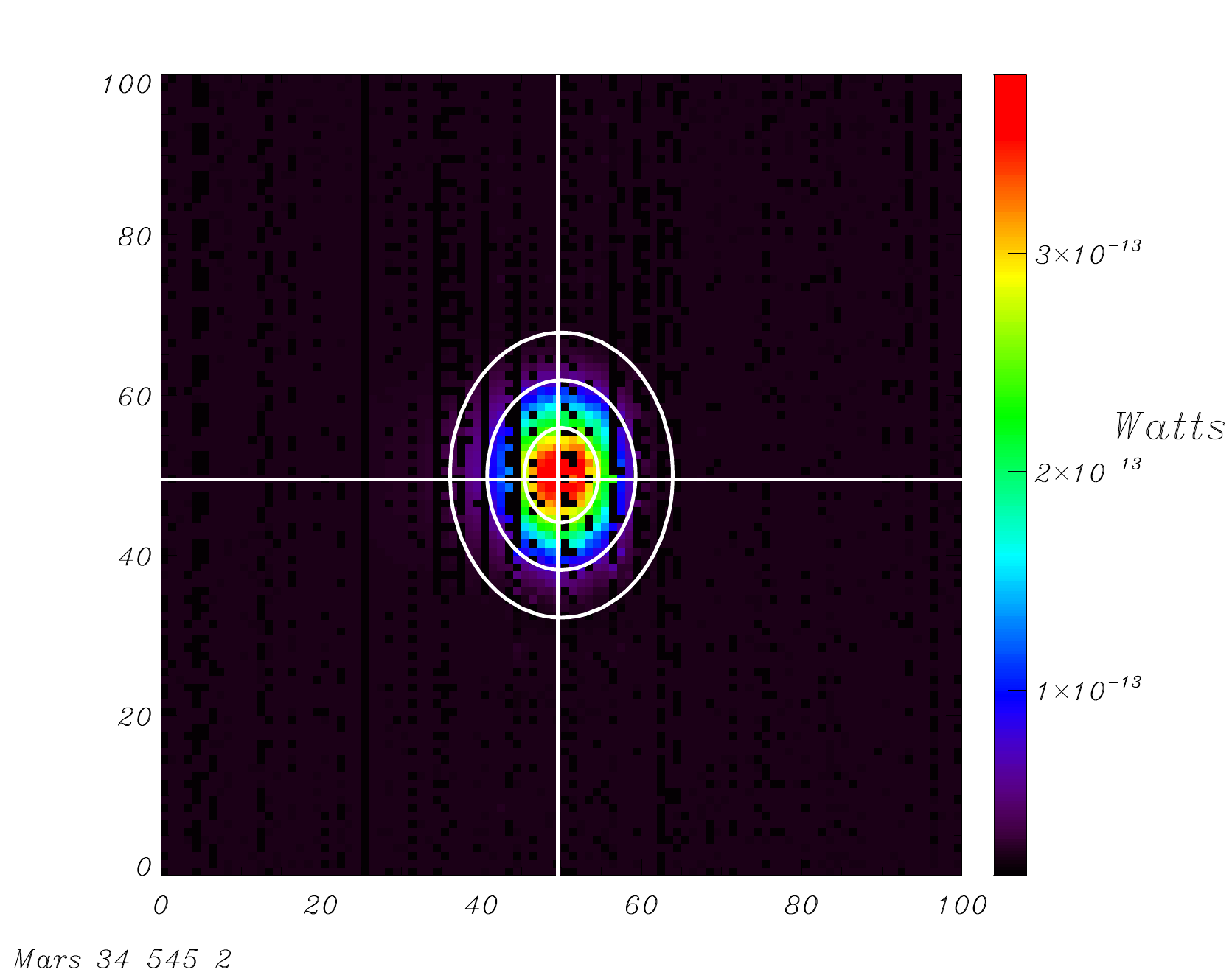}
	\includegraphics[width=0.49\columnwidth] {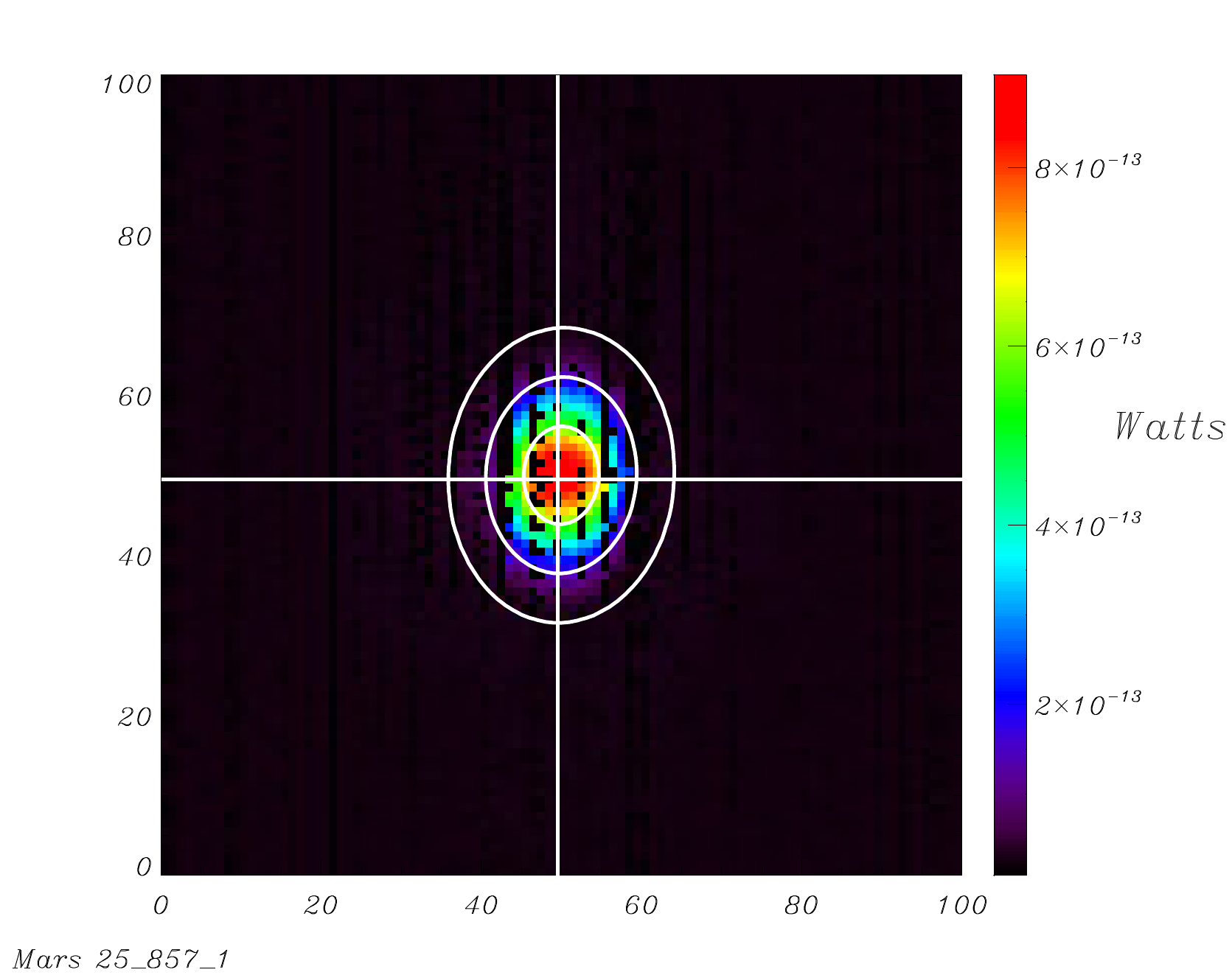}
	\caption{Maps of Mars, for the 100-4a, 143-2a,
 		217-6b, 353-1, 545-2 and 857-1 detectors, along with the best-fit
  		Gaussian beam. Note that the motion of Mars is taken
                into account. The stripes visible 
  		here are due to the spacing between successive rings,
                which is a significant fraction of 
  		the FWHM of the beam. }
   	\label{fig:Marsmaps} 
\end{figure}

In Fig.~\ref{fig:Marsmaps} we show maps made from the first crossing of Mars, 
which provides the highest
signal-to-noise measurement (that is not subject to nonlinear response) of
the focal-plane geometry and beam shapes.
Using the algorithm described in Appendix~\ref{Annex:FPG}, we fit the
data going into these `minimaps' to a Gaussian or Gauss-Hermite
model,  extracting the positions of the centre of the beams and
beam shape parameters.

\begin{figure}[htbp]
	\centering
	\includegraphics[width=\columnwidth]{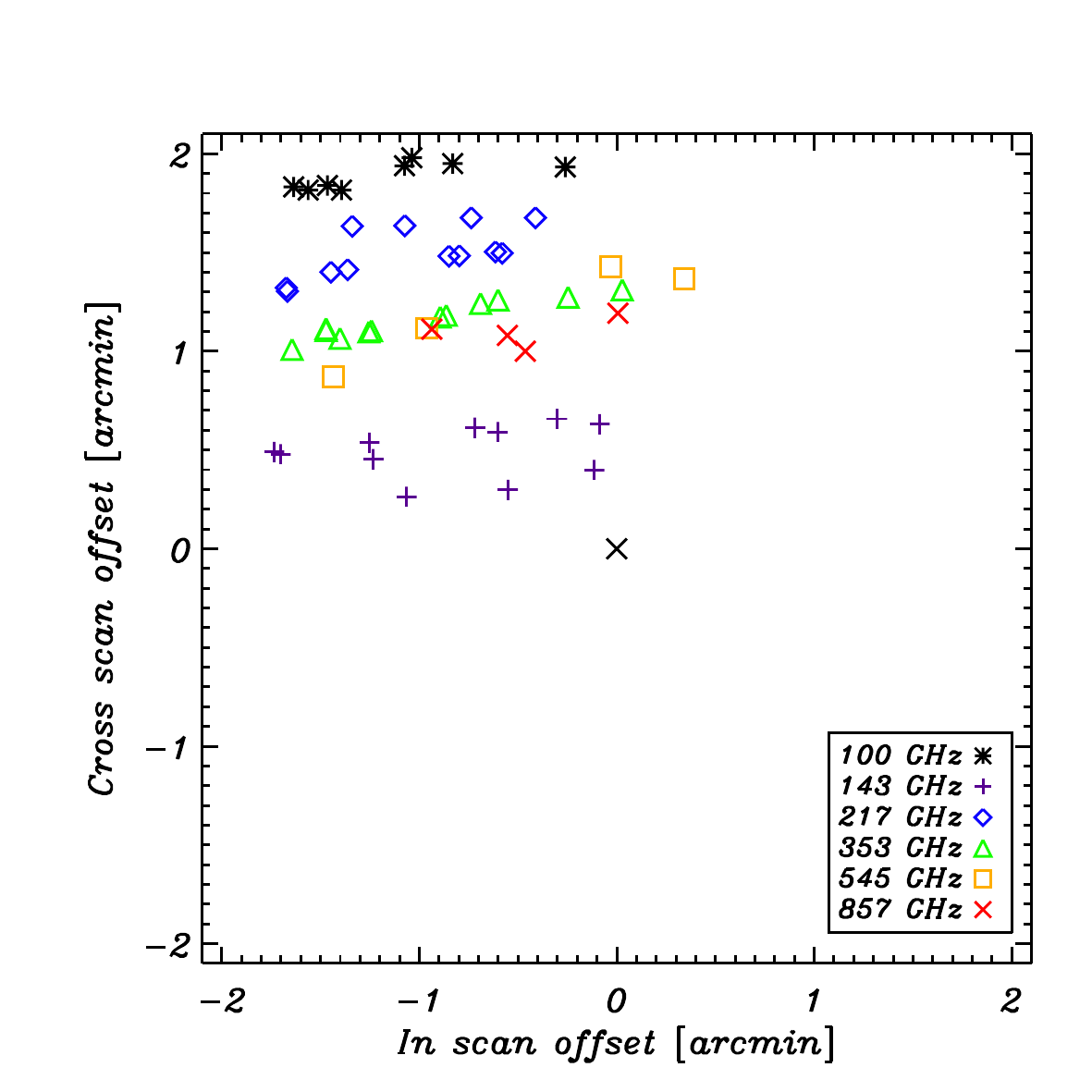}
	\caption{Detector positions on the HFI focal plane for the
          first observation of Mars with respect to the input RFFM
          Model.
          } 
	\label{fig:Mars1_v41}
\end{figure}

In Fig.~\ref{fig:Mars1_v41} we show the individual detector positions
with respect to the pre-launch Radio Frequency Flight Model (RFFM), in the scan
direction (in-scan, horizontal) and perpendicular to it (cross-scan, vertical).
We expect that some of the differences in the scan direction are due to
residual phase shifts after the deconvolution of the time-stream
filters \\  \citep[see Sect.~\ref{sec:toip} and][]{planck2011-1.5}. Although this
makes it difficult to measure optical or mechanical differences in that
direction, the measurement of these shifts and their incorporation into the
pointing model enable us to completely account for that effect in subsequent
analyses. 

In fact, the pattern of individual detector shifts in
Figure~\ref{fig:Mars1_v41} is a distorted and rotated image of the
focal plane itself, indicating errors in the initial modelling.  Using
these data, and attempting to account for the per-detector phase shift
introduced by transfer function deconvolution, we find an overall
rotation of 0.15 degree and a scaling by a factor of 1.007 in both the
in-scan and cross-scan directions. After such a scaling and
rotation, there is still a cross-scan rms scatter of 8 arcsec. As is
evident from Fig.~\ref{fig:Mars1_v41}, these results require 
an  overall rotation and scaling and are consistent
with subsequent optical models of the \Planck\ telescope.

\begin{figure}[htbp]
	\centering
		\includegraphics[width=\columnwidth]  {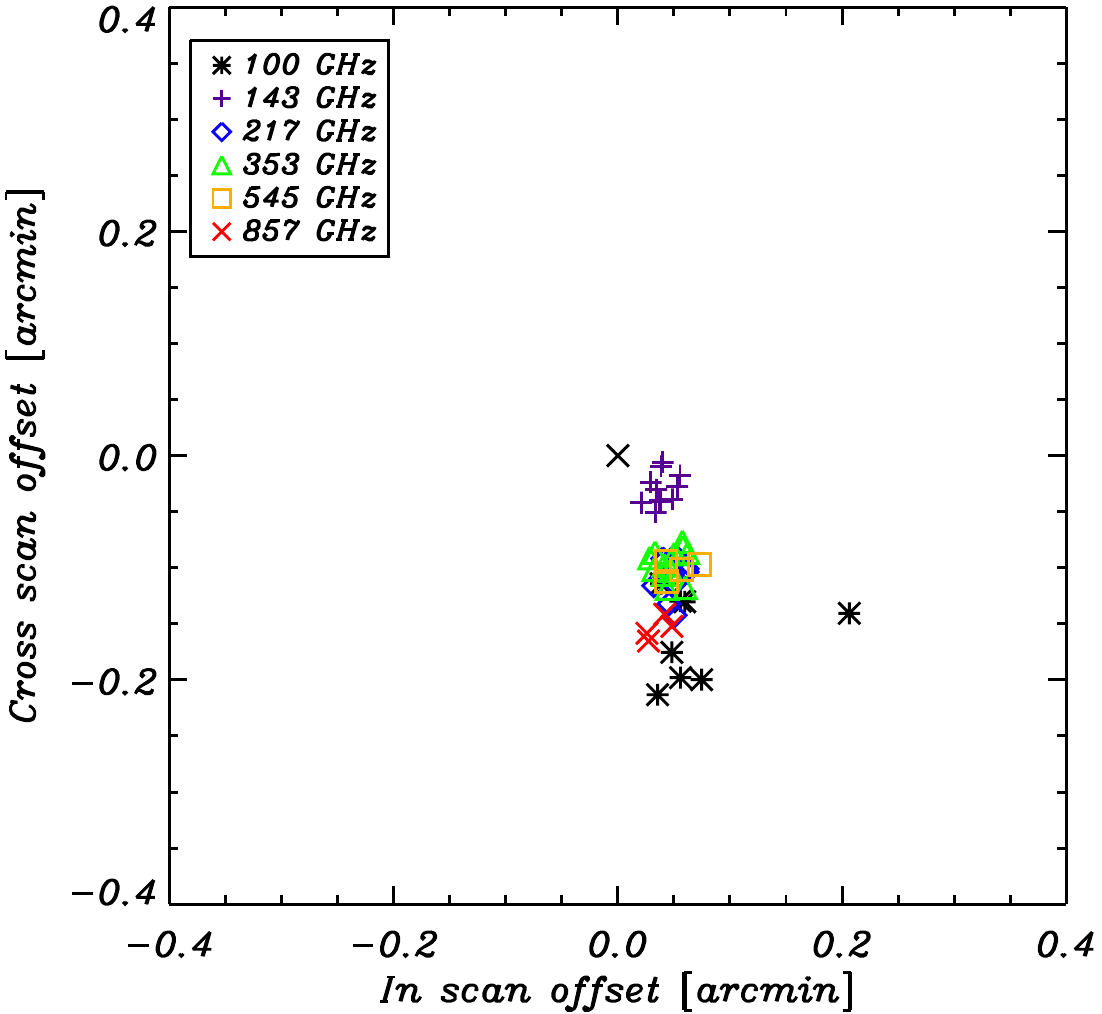} 
		\caption{Detector positions on the HFI focal plane for
                  the second observation of Mars with respect to the
                  first observation of Mars.} 
	\label{fig:Mars2-Mars1_v41}
\end{figure}

\begin{figure}[htbp]
	\centering
	\includegraphics[width=\columnwidth]  {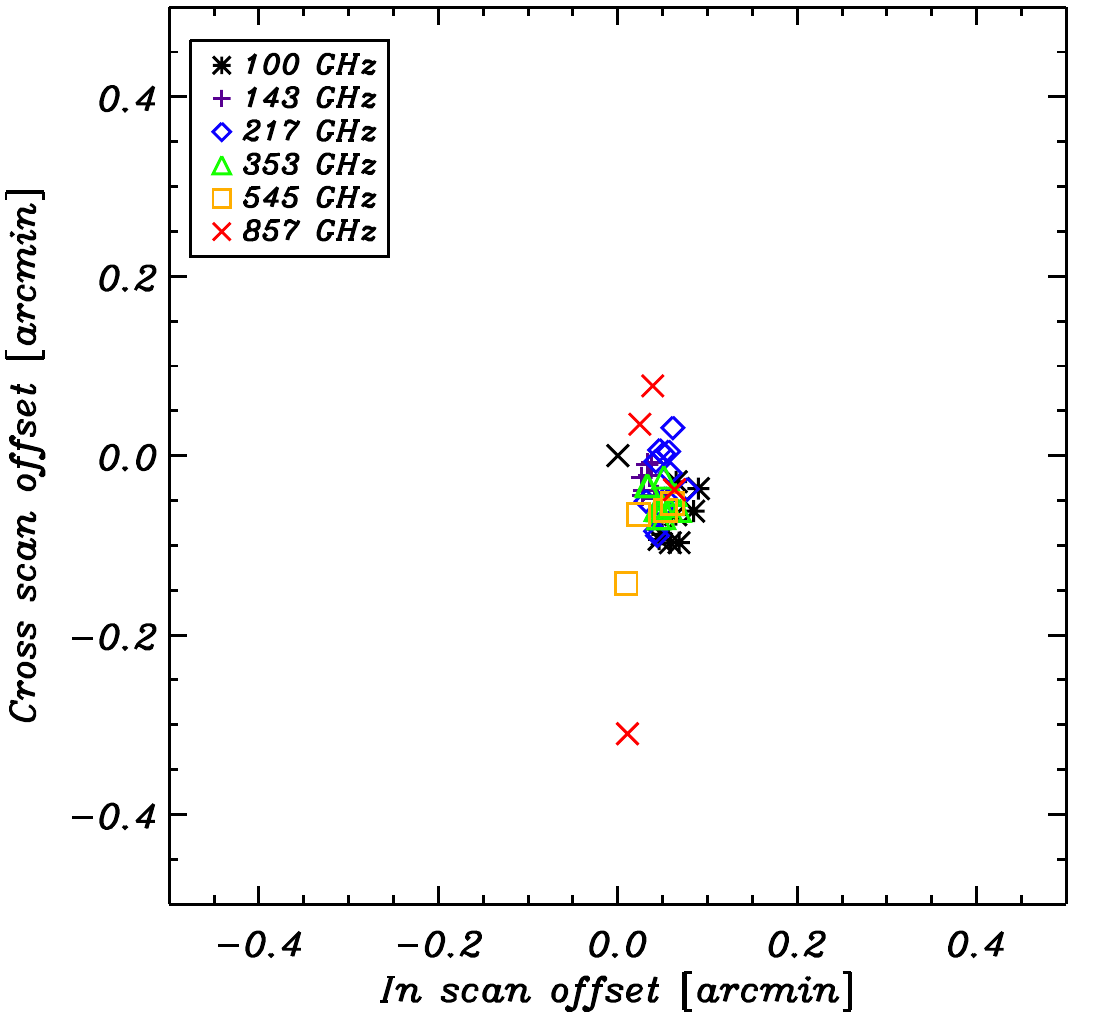} 
	\caption{Detector positions on the HFI focal plane for
                  the first observation of Saturn with respect to the
                  first observation of Mars.} 
	\label{fig:Saturn-Mars_v41}
\end{figure}

To determine any systematic sources of error and to monitor 
possible time evolution of the detector positions, we show in
Figs.~\ref{fig:Mars2-Mars1_v41} and
\ref{fig:Saturn-Mars_v41}  the residuals between
different  measurements of the focal plane.  Fig.~~\ref{fig:Mars2-Mars1_v41}
compares results from Mars transits in the first sky survey compared to
the second. Fig.~\ref{fig:Saturn-Mars_v41} compares results from
Saturn and Mars transits in the first survey. 
Comparison of the full set of planet observations yields an estimate of the remaining systematic error in the determination of the focal plane geometry to be approximately 20 arcsec cross-scan and 10 arcsec in-scan, considerably less than 10\% of the beam FWHM in all cases. 

The  effect on astrophysical sources is shown in Fig.~1 of
  \cite{planck2011-6.4a}, in which the locations of nearby galaxies as
  detected in the \Planck\ Early Release Compact Source Catalogue
  \citep{planck2011-1.10} are compared with measurements from
  IRAS. The rms difference, smaller than 1 arcmin, is dominated by
  confusion noise and pixelisation. We conclude that pointing errors
  are not significant for the HFI.

\subsubsection{Scanning Beams} \label{ssub:beams} 

\begin{figure}[htbp] 
	\centering 
 	\includegraphics[width=\columnwidth]  {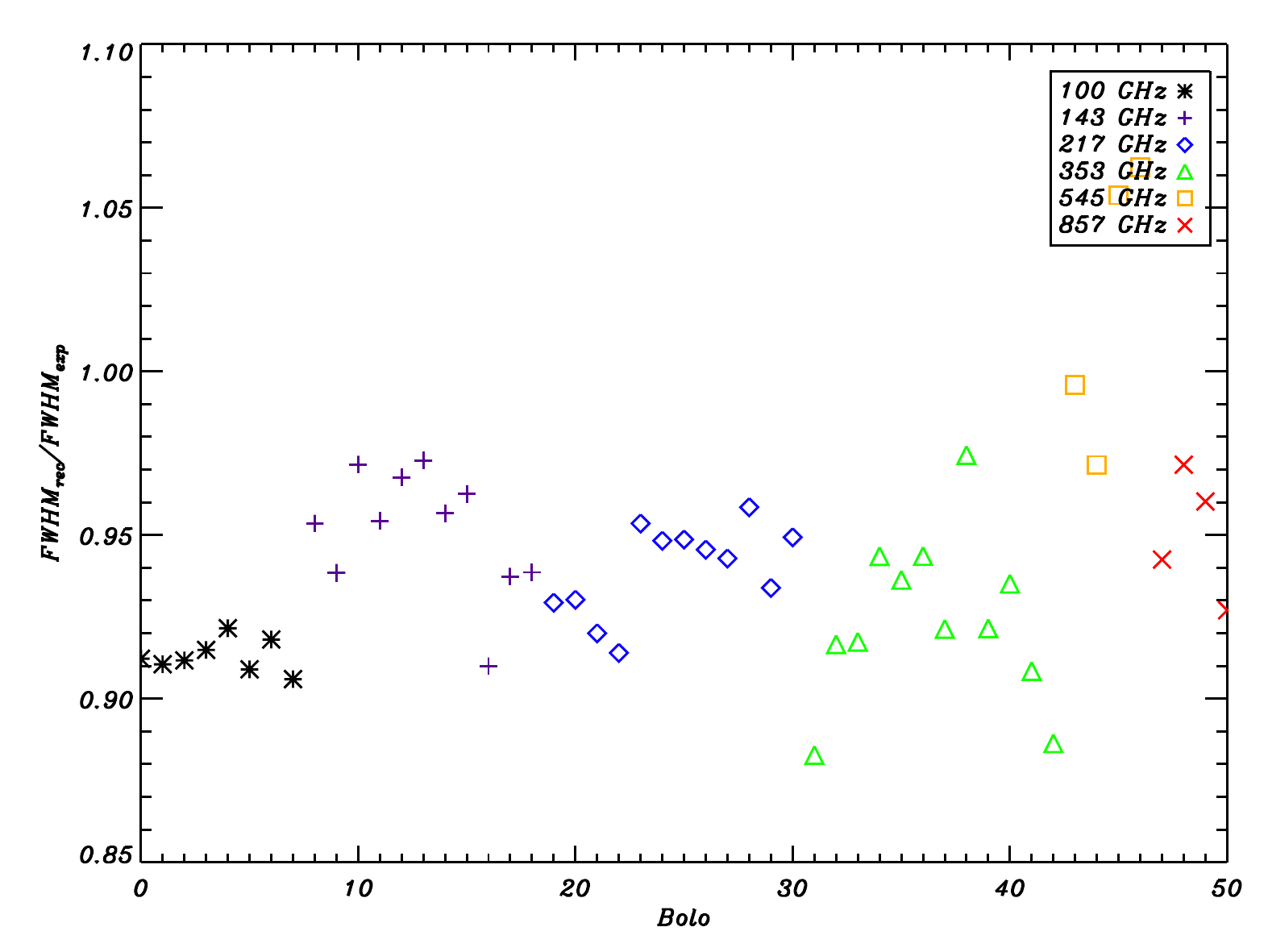} 
	\caption{The measured beam width (assuming a symmetrized
          Gaussian model) with respect to the RFFM model, for Mars.}  
	\label{fig:beam_comp_Mars1_v41} 
\end{figure}

\begin{figure}[htbp] 
	\centering 
	\includegraphics[width=\columnwidth]  {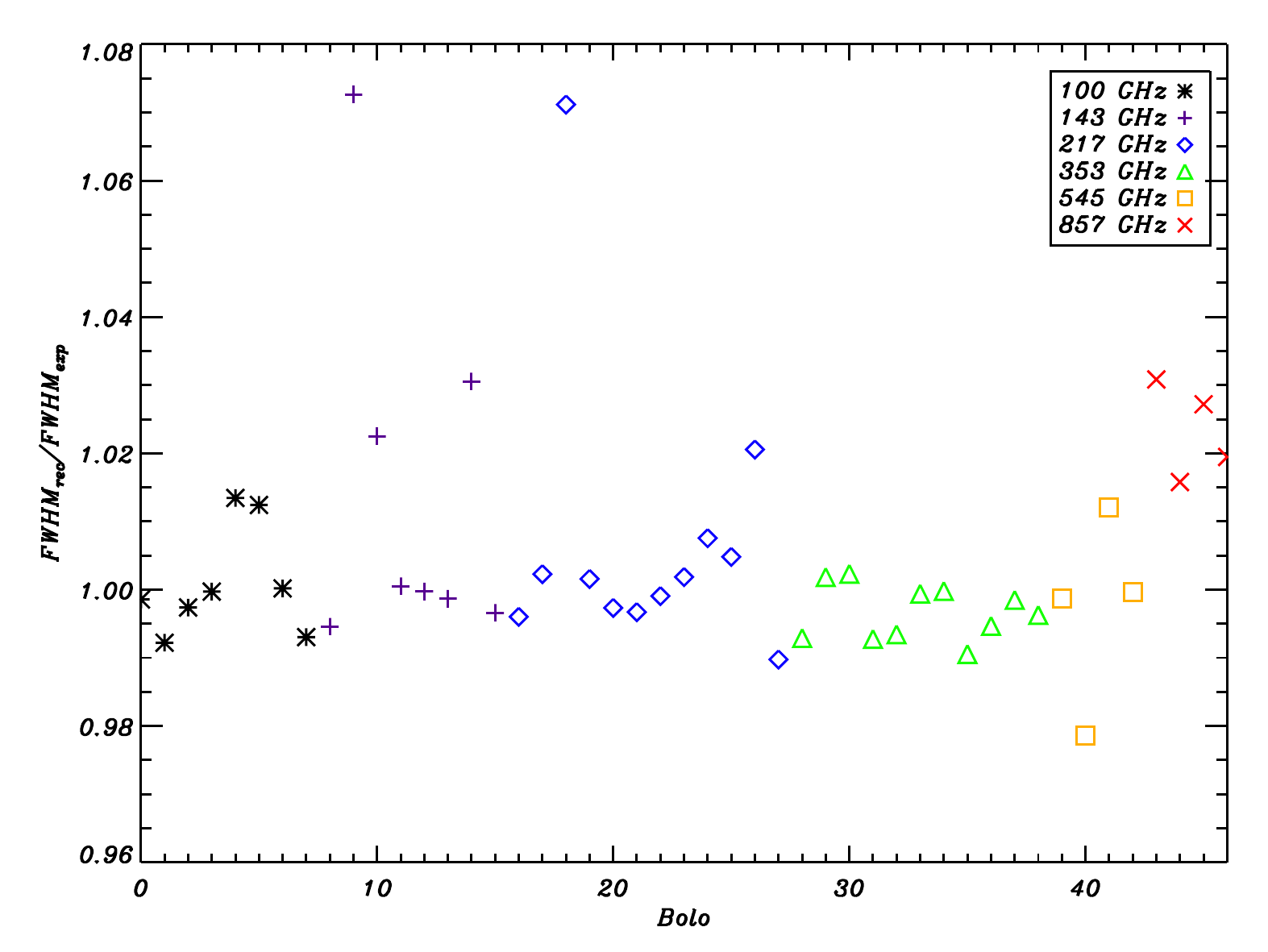} 
	\caption{The measured beam width (assuming a symmetrized
          Gaussian model) comparing the  
		first scan of Saturn to the first scan of Mars.}  
	\label{fig:beam_comp_Saturn_v41} 
\end{figure}

\begin{figure}[htbp] 
	\centering 
 	\includegraphics[width=\columnwidth] {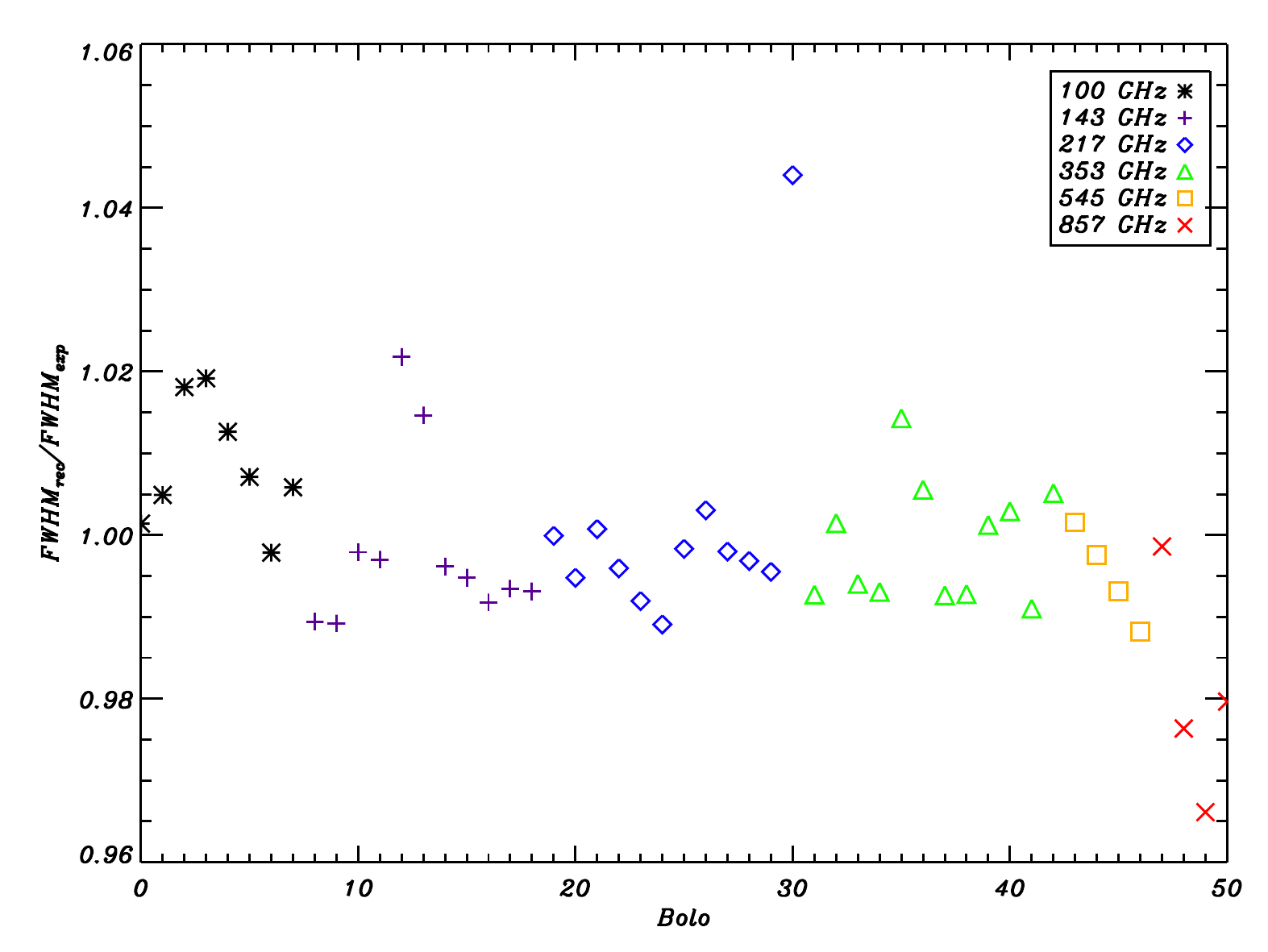}
  	\caption{The measured beam width (assuming a symmetrized Gaussian  model) comparing the 
  		two observations of  Mars.}  
    	\label{fig:beam_comp_Mars2-Mars1_v41}
\end{figure}

  In this section we discuss the
  measurement of HFI \emph{scanning beams}, defined as the beam
  measured from the response to a point source of the full optical and
  electronic system, \emph{after} the filtering described in
  Section~\ref{sec:toip} is applied. In the presence of low noise,
  perfect deconvolution, a perfect point source,  full sampling
and no filtering, 
  this would be equivalent to the optical beam, \ie\, the action of
  the telescope optics (mirrors and other optical elements) on the
  light path from infinity.  We first concentrate upon the
  well-measured elliptical Gaussian parametrization. We compare the
  symmetrized Gaussian beam ($\sigma^2 = \sigma_1 \sigma_2$ where
  $\sigma_1,\sigma_2$ are the major and minor beam widths) to the RFFM
  model as well as measurements of different planets.  In
  Fig.~\ref{fig:beam_comp_Mars1_v41} we show measurements using Mars
  with respect to the model. Except for a subset of the 545\GHz\
  detectors (which are highly non-Gaussian due to their multi-moded
  optics) the beams are typically narrower than the model.  (The 857
  GHz beams are also highly non-Gaussian, but happen to be better fit
  by a narrower beam). In Fig.~\ref{fig:beam_comp_Saturn_v41} we
  compare the measurement of Mars with that of Saturn and in
  Fig.~\ref{fig:beam_comp_Mars2-Mars1_v41} we compare the
  measurements of Mars between the first and second surveys. These
  few-percent changes are consistent with variations in the sampling
  of the beam-shape between different observations and are comparable
  to the systematic errors resulting from small algorithmic
  differences (\eg\ destriping). Finally, note that we expect that the fractional
  change in the size of Mars' disk due to the variation in observation
  angle and distance is expected to be a far-subdominant value of
$2\times10^{-5}$.

\begin{figure}[htbp] 
	\centering 
	\includegraphics[width=\columnwidth]{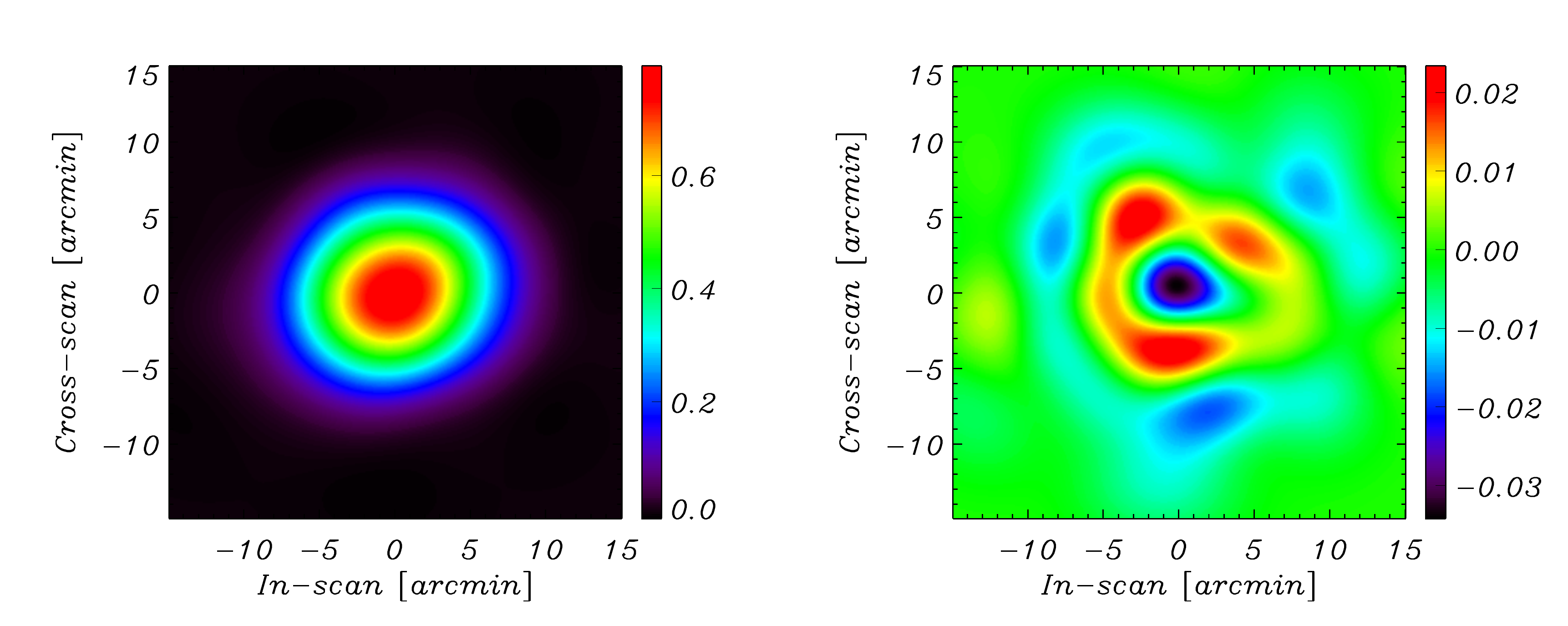} 
	\includegraphics[width=\columnwidth]{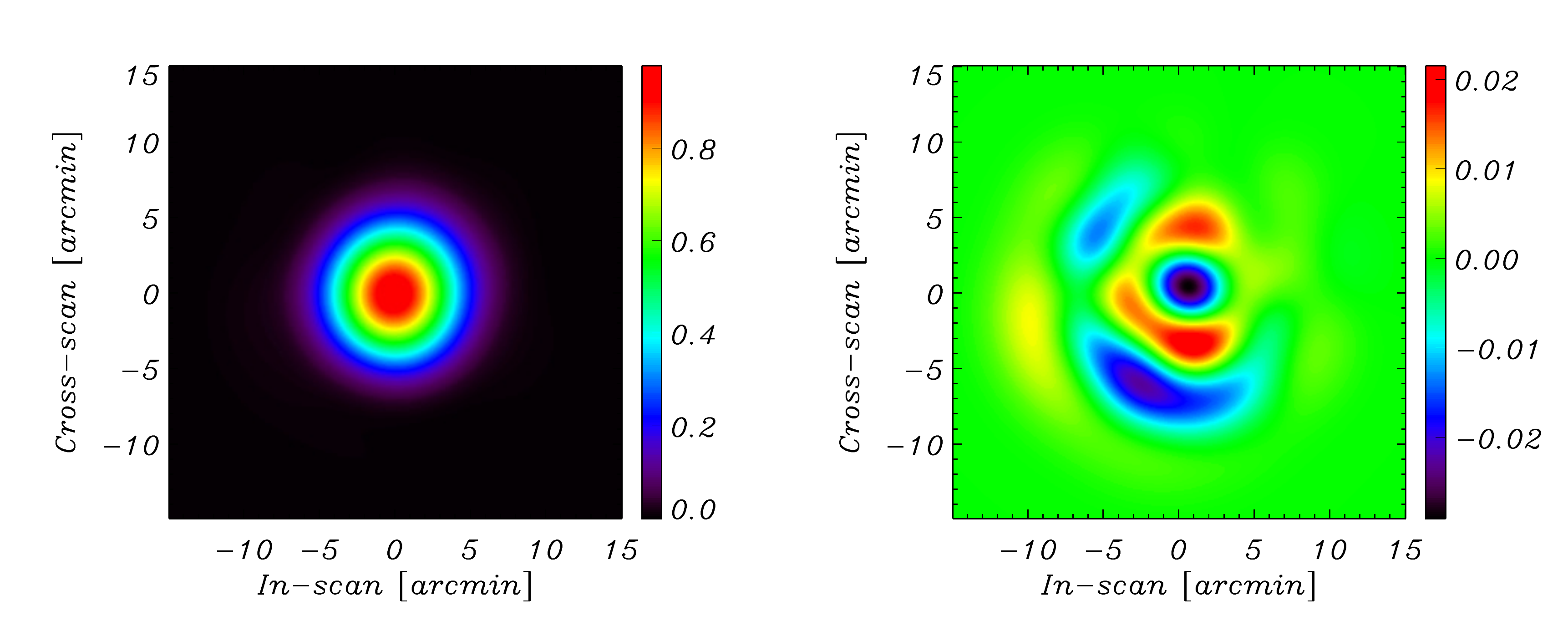} 
	\includegraphics[width=\columnwidth]{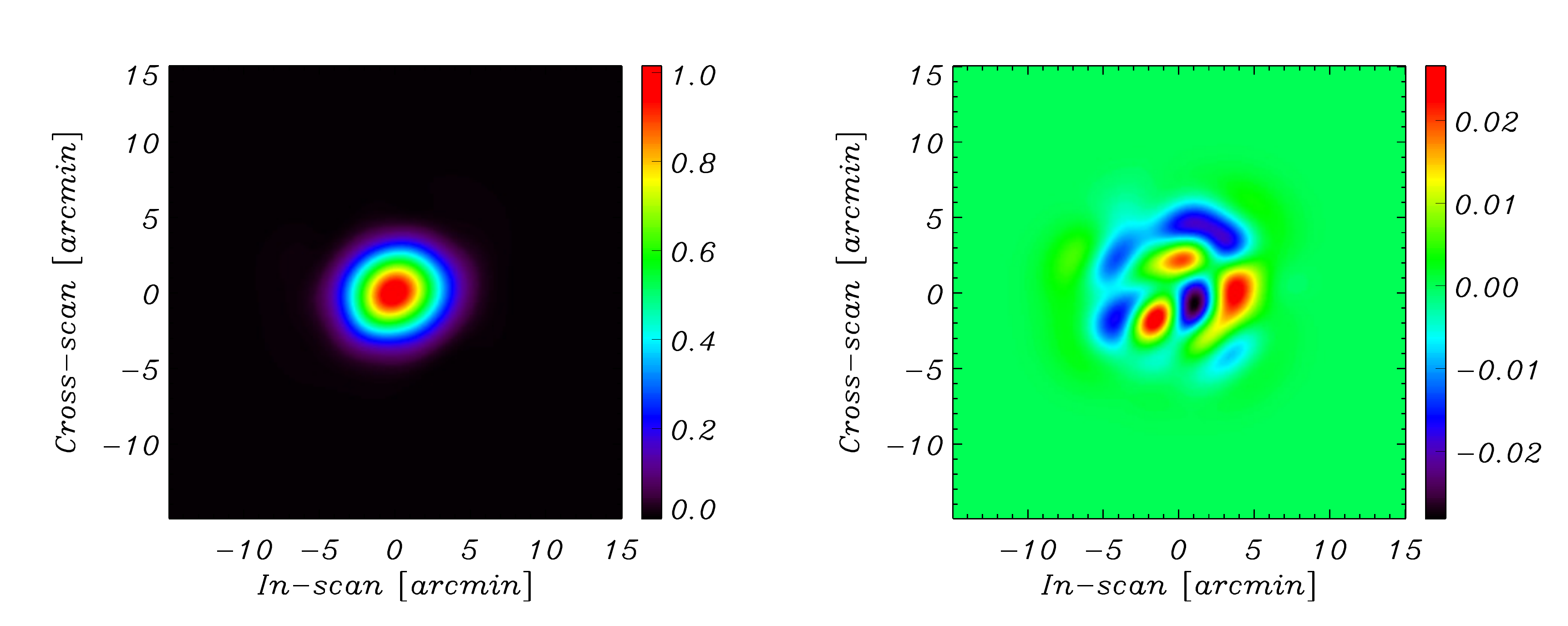} 
	\includegraphics[width=\columnwidth]{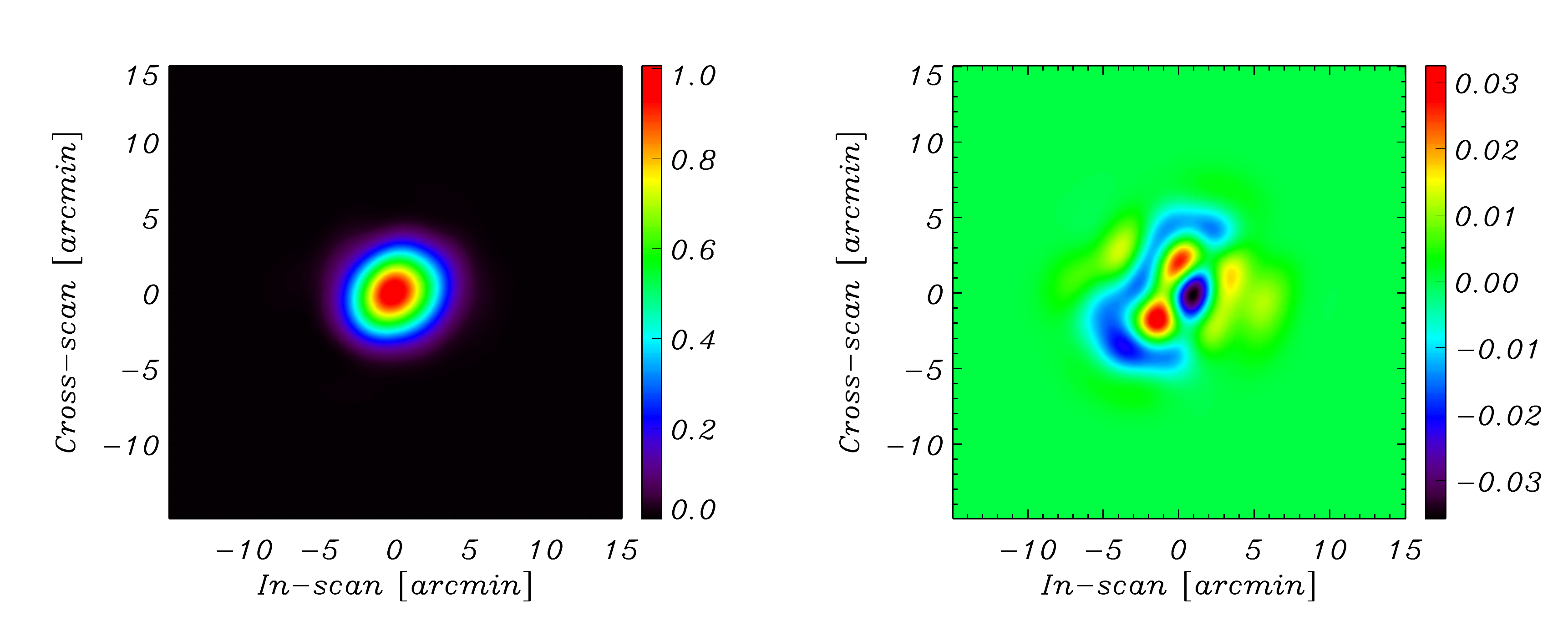} 
	\includegraphics[width=\columnwidth]{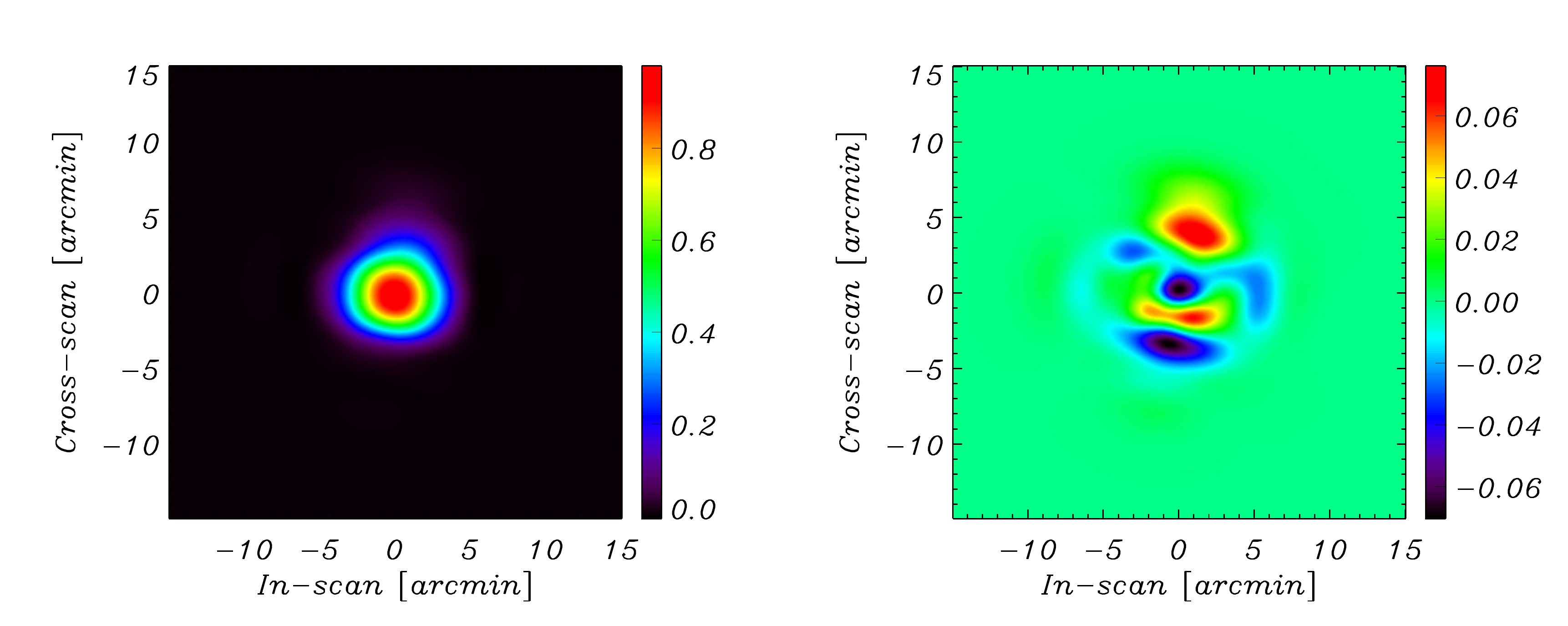} 
	\includegraphics[width=\columnwidth]{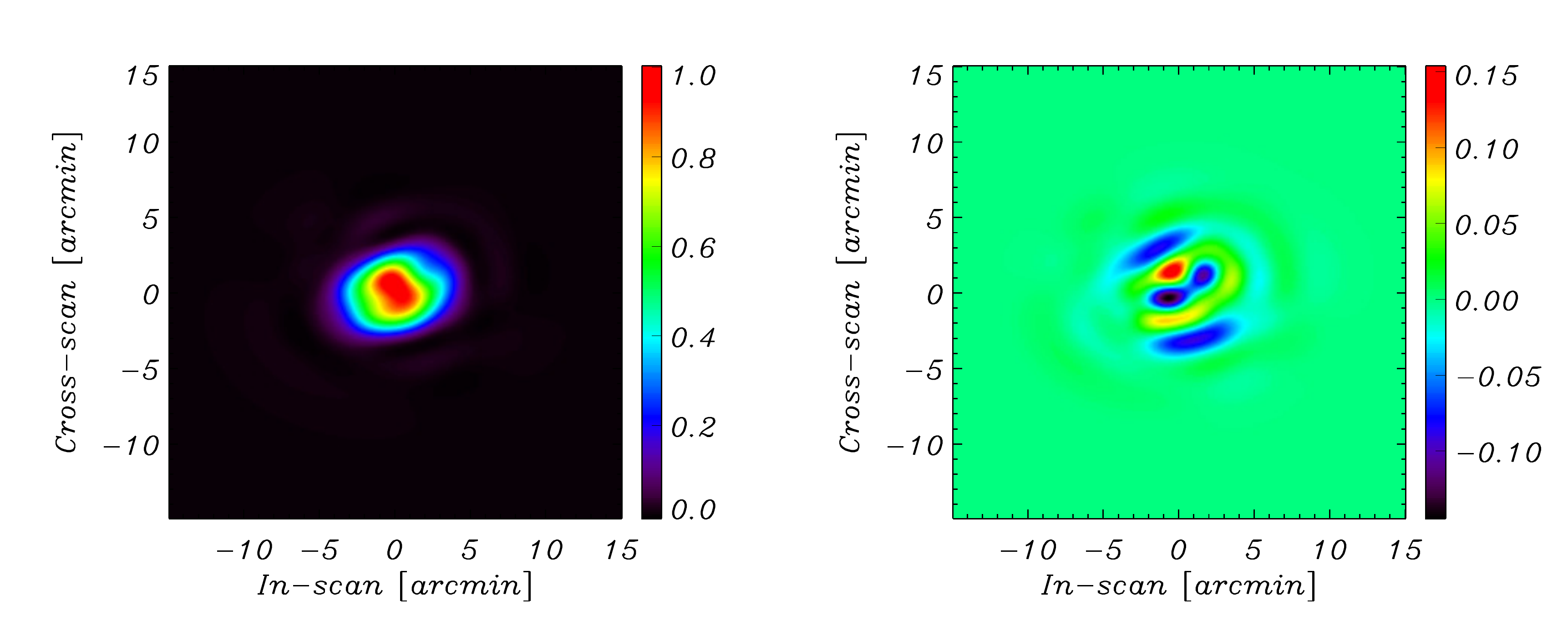} 
	\caption{Left Column: Full Gauss-Hermite fit to the detector
                scanning beam, for 100-1a, 143-1a,  
		217-1, 353-1, 545-1 and 857-1, from top to bottom.
                Right Column: Difference with an  
		elliptical Gaussian fit.}  
	\label{fig:MarsGH} 
\end{figure}

The more complex Gauss-Hermite model described in 
Appendix~\ref{Annex:FPG} allows us to take the full shape 
of the \Planck\ beams into account. In particular a pure 
Gaussian fit typically misestimates the effective solid 
angle of the beam, and hence affects the flux calibration from
 known sources. The effect is greatest at 545 and 857\GHz, for 
which the Gaussian approximation typically underestimates the 
solid angle by 5\%. At lower frequencies the effect is of 
order 1\% when averaged over a frequency channel.  In 
Fig.~\ref{fig:MarsGH}, we show the Gauss-Hermite 
fit for a number of detectors and the difference with an elliptical 
Gaussian fit. As 
expected, the effect of non-Gaussian beams is most 
significant for the two highest frequencies.  

The statistical error on the FWHM of a channel-averaged beam 
is typically of order 0.05--0.15 arcmin. In practice, this 
is dominated by various systematic effects. For example, 
differences in the destriping method can give a 
5-7\% difference for a few detectors (but 1-2\% is more typical).  
Other differences between analyses include pixelisation effects, 
the numerical likelihood maximization procedure, and the 
parametrization of the beam-shape. Overall differences of a 
similar magnitude in the final solid angle result from the use 
of these alternate pipelines.

\subsection{Effective beam per  channel} 

Let us define the \emph{effective beam} at the map level as the
overall angular response to the sky in a map pixel, which results from
the combined effect of the instrumental response, the scanning
strategy and the data processing.  

Scanning beams, described in the
previous section, are an expression of our knowledge of the combined
optical, electronic, and TOI processing induced response of each
\Planck\ detector to sky signals during the acquisition of each
individual sample, that is during the $\sim 5$\,ms period integration
on the sky.  Scanning beams are typically asymmetric, \ie not fully
rotationally symmetric around the pointing direction of a sample in 
the TOI.  

The \Planck\ scanning strategy is a combination
of several rotations: satellite spin, slow (6 monthly period)
precession of the spin axis around the anti-Solar direction in the L2
reference frame, and ecliptic motion of the anti-Solar axis of
\Planck.  These combined motions result in a complicated pattern of
pointing of each detector in the \Planck\ focal plane. Over the course
of the mission there is a build up of many observations in the areas
near the ecliptic poles, while the broad band around the ecliptic
equator is observed less frequently.  In the vicinity of the ecliptic
poles,  the asymmetric scanning beams are pointed at the sky with very
significant beam rotation, but the regions near the ecliptic plane see
very reduced local scanning beam rotation.  

Combination of all of the observations in a particular direction on the sky into 
 a  single pixel of a discretised sky map, results in the
addition of all scanning beams viewing this particular region of the
sky. We refer to the result of such addition as the effective beam.

We have developed two methods which can be used to estimate these
effects, FICSBell \citep{hivon2011} and FEBeCoP \citep{mitra2010}.
The first method uses
an approximate description of the scanning beam in  harmonic space
(retaining only the most relevant modes) while the second relies on an
approximate description in pixel space (retaining only the pixels
closest to the maximum). Both methods can be made precise to the
desired level by increasing the computational cost. They are further
described in the appendix~\ref{Annex:FEBECOP}. Both methods show that
the differences between the scanning beams and the effective beam are
small, at the few percent level. The Fig.~\ref{fig:febecop_GHstat}
in the appendix shows a dispersion of few percent  
in the FWHM and ellipticity around the sky and the mean 
characteristics are given in lines c1-c4 of the summary
table, Table~\ref{tab:summary} (Sect.~\ref{sec:sum}).   

Both of these beams have been used in our paper on the power spectrum
of CIB anisotropies with \Planck-HFI \citep{planck2011-6.6}. FEBECOP
was also used to estimate corrections to the derived flux of sources
in the \Planck\ ERCSC. One should note that the ERCSC production was
done shortly after maps had been produced by the DPCs and used the
best information on the scanning beams available at that time. This
paper provides the best information available at the time of writing,
\ie\ a few months later, but is  compatible with the
characteristics described  in the ERCSC paper and explanatory
supplement \citep{planck2011-1.10,planck2011-1.10sup}.

Observations of the outer planets suggest the presence of potentially
significant shoulder in the sub-mm beams, as would result from the
 diffuse scattering from random surface errors of order $5-$8$\,\mu m$ on scales
 between $10$ and $100$\,mm.
Surface errors at this level are
consistent with the estimates in Table 2 of \cite{stute2004}, and meet
the design specifications. The beam shoulder, which is typically evident at the
$-40$ dB level, is not described by either the Gaussian or Gauss-Hermite
 expressions of the main beam. The integrated throughput of this
 shoulder constitutes less than 0.5\% of the total solid angle of the
 scanning beam at frequencies below 353\GHz. However the contribution to
 the sub-mm channels is larger, with current estimates of 1.7, 4.8
 and 7.2\,\% of the total solid angle (as reported in Table 4, note b1) at
 353, 545 and 857\GHz, respectively. Beam solid angles derived
 exclusively from the main beam will correspondingly under-estimate the
 true instrumental solid angle. Therefore, flux estimates of point
 sources will generally be under-estimated by a comparable amount. We
 note that aperture photometry of sources with a signal-to-noise ratio
 less than about 40 dB will suffer similarly from this bias.

\subsection{Beam uncertainties}

We now turn to the effect of uncertainties in the scanning beams
on the knowledge of ERCSC point sources.  An approximate estimate 
is obtained by comparing simulations using the Gaussian elliptical and the
Gauss-Hermite description of the beams. In both cases, point-sources
were detected in the simulated 857-GHz map using a Mexican-hat wavelet
filter algorithm.  Outside the Galactic plane ($|b|> 20^{\circ}$),
$93.5\%$ of  input sources were detected using both beam
models.  Within the plane, this figure was $84.3\%$.  We estimated
point source fluxes by performing aperture photometry for the set of detected
input sources with Galactic latitude $|b|>20^{\circ}$.

\begin{figure}
	\centering
	\includegraphics[width= \columnwidth,angle=0]{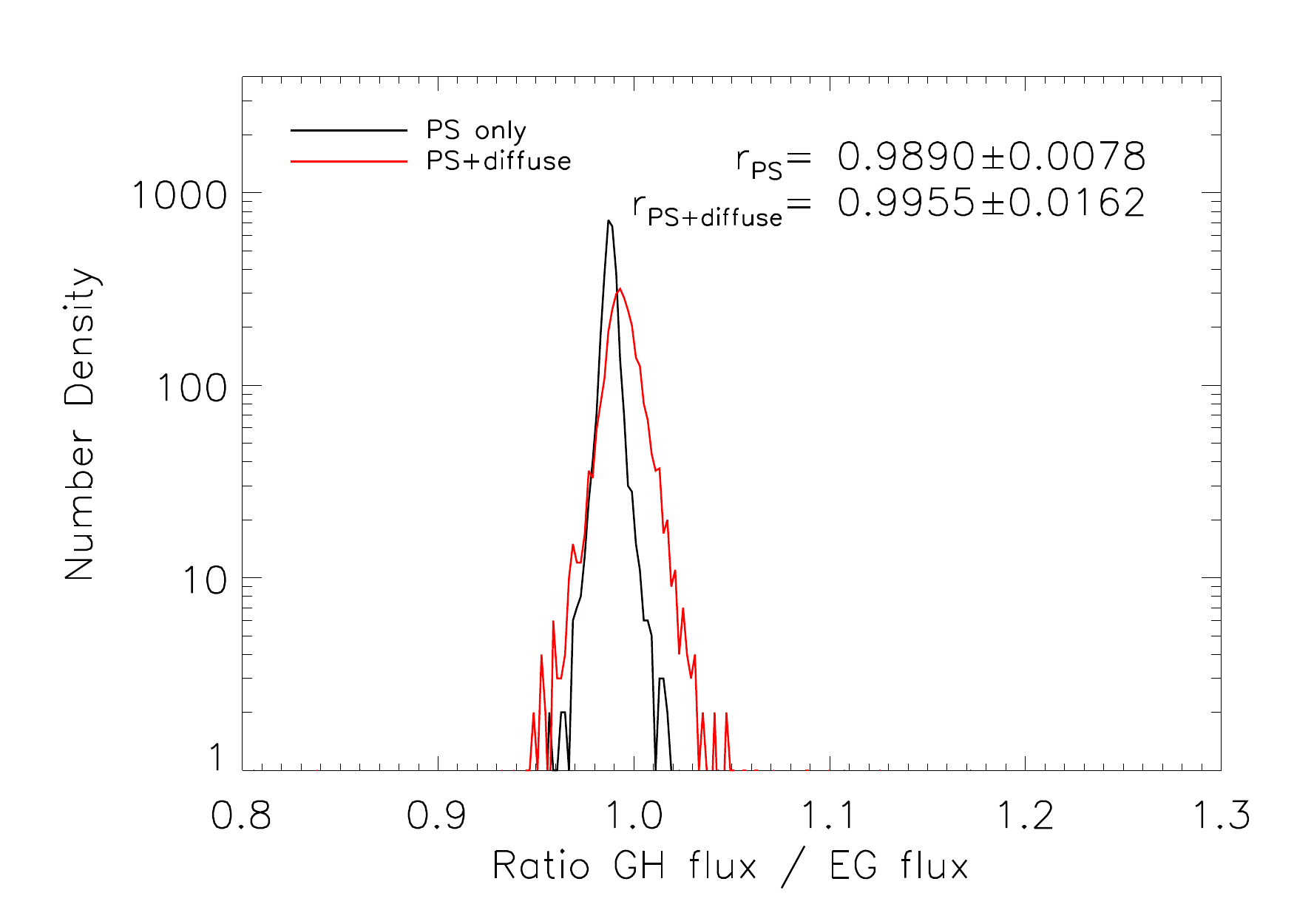}
	\caption{Histograms of the ratio of aperture fluxes from
                857\GHz\ simulations with elliptical 
		Gaussian and Gauss-Hermite beams. The quoted values of $r$ are the mean $\pm$ the 
		root-mean-square of the ratio across the source sample.}
	\label{fig:dwps857}
\end{figure}

Figure \ref{fig:dwps857} shows the histogram of the ratio of aperture
fluxes for the two beam models.   
The  black curve shows the distribution from a point source only
simulation, including only the beam  
difference of the point source signal.  The red curve shows the
distribution from a point source and  
diffuse component simulation, including point source signal
differences and differences in the 
background subtraction due to the beam.  From the latter we estimate
an upper limit on aperture flux 
errors due to beam uncertainties of 0.5\,\% at 143\GHz\ and 1.6\,\% at 857\GHz. 
This confirms that flux
error due to beam uncertainties  
are much smaller than the flux error due to source detection errors.  

For both simulations, the known input source centre was used for the aperture photometry.  
By design this ignores the effect of source detection errors (for
example,  mis-estimation of the centres).

We also computed the median absolute deviation of the flux error due
to detection errors as a function of absolute Galactic latitude and
found that it varies from 10-15\% in the Galactic plane to a few per cent at
$b>40\deg$.

\section{Map making and Photometric Calibration} \label{sec:map} 

The path from TOIs to maps follows 
two steps, ring-making and map-making.  The first step makes
use of the redundancy of observations provided by the \Planck\
spacecraft scanning strategy.  

As discussed in Sect.~\ref{sec:detnoise}, the in-flight
noise of the HFI detectors after TOI processing is mostly white with a `1/f' 
component at low frequency \citep[see section~\ref{sec:detnoise} and][]{planck2011-1.5}. 
To build sky maps and further reduce the
low-frequency noise, we adopted a destriping approach in which the
 $1/f$ component, is represented
by a uniform offset on a scan ring.  Using one offset per ring,
residuals of $1/f$ noise found in the cleaned map have been
shown to be significantly below the white noise level in the map
domain.

For a given channel, each measurement at the sample number $i$ may be
described as :  
\begin{equation}
 m_i \ = G\left( I_p + \frac{1-\eta}{1+\eta}\left( Q_p\ \cos{2\psi_i}
 + U_p\ \sin{2\psi_i} \right)\right)   + \ n_i  
 \label{eq:MM1}
\end{equation}
where $p$ denotes the sky pixel with Stokes parameters $I$, $Q$ and $U$; $n_i$
is the noise realization; $\eta$ is the cross-polarization parameter
(equal to 1 for an ideal SWB and 0 for an ideal PSB); $\psi$ is the
detector orientation; and $G$ is the detector's gain. 

Calibrating the detectors on an unpolarised source (with $Q=U=0$), \eg the
orbital dipole, determines $G$. The photometric calibration is 
performed either at ring level using the Solar dipole, for the lower
frequencies channels (see Sect.~\ref{mapmaking:calib_dipole}) or at map level
using \FIRAS\ data, for the 
channels at 545 and 857\GHz\ (see Sect.~\ref{mapmaking:calib_gal}). The polarization response
parameters (detectors orientations and cross-polarisation) have been
extracted from ground measurements as described in \citet{rosset2010}.  

\subsection{Ring making}

As an intermediate product, we average circles within a pointing
period to make rings with higher signal-to-noise ratio. 
To avoid introducing any additional binning of the data, we
choose a sky pixelisation as a basis for this ring making. 
We used the \healpix\ scheme \citep{gorski2005} with Nside$=2048$, corresponding to pixels
with 1.7\,arcmin on a side. Data are averaged in each scanned
\healpix\ pixel using the nearest grid point method.  The resulting
collection of \healpix\ pixels visited during a pointing period is
hereafter called a \healpix\ Ring or HPR.  

From the pointing direction and the satellite velocity at the time of each
sample, the component of the calculated orbital dipole signal is averaged as
described above and used for calibration. Once data are calibrated, it is
subtracted before projection.

The orientations of the detectors are propagated using the averaged  sine and
cosine values (Eq.~\ref{eq:MM1}) for the samples falling in each pixel of the HPR.  
Because of the low level of nutation of the satellite, 
the dispersion of the relative orientation within each bin is very low leading to a
negligible error in the averaged orientation.

Samples flagged by the TOI processing pipeline as invalid, observations
outside of the stable pointing periods,
and those near known moving objects (planets, asteroids and comets) are removed from
further analysis.

\subsection{Destriping}

Destriping  algorithms simplify the map making problem and 
require substantially fewer computational 
resources than those  for a full maximum likelihood solution.
They are very close to optimal when certain conditions
are satisfied (see, for example, \cite{Delabrouille1998,Revenu2000,
maino2002,keihanen2004,ashdown2007b})

In the destriping 
approach, the noise is divided into a low-frequency component
represented by the offsets $\mathbf{o}$, unfolded onto the time-ordered data
by the matrix $\mathbf{\Gamma}$, and a white noise part $\mathbf{n}$
which is uncorrelated with the low-frequency noise. The signal part in the TOI
is given by the projection of a pixelised sky map (or set of maps containing the 3 Stokes parameters, arranged as a single vector), $\mathbf{T}$ via a pointing matrix,  
$\mathbf{A}$,  leading to 
\begin{equation}
\mathbf{d} 
= 
\mathbf{A} \cdot \mathbf{T} + \mathbf{\Gamma}\cdot\mathbf{o} + \mathbf{n}, 
\end{equation}
The maximum-likelihood offset coefficients, $\mathbf{o}$, can be found from
the time-ordered data, $\mathbf{d}$, by solving 
\begin{equation}
\left( \mathbf{\Gamma}^T \mathbf{N}^{-1} \mathbf{Z} \mathbf{\Gamma} \right) 
\cdot \mathbf{o} 
= 
\mathbf{\Gamma}^T \mathbf{N}^{-1} \mathbf{Z} \cdot \mathbf{d},
\end{equation}
where
\begin{equation}
\mathbf{Z} 
= 
\mathbf{I} 
- 
\mathbf{A} \left(\mathbf{A}^T  \mathbf{N}^{-1} \mathbf{A}\right)^{-1} 
\cdot \mathbf{A}^T \mathbf{N}^{-1}.
\end{equation}

\begin{figure}[htbp]
   	\centering
	\includegraphics[width=0.5\textwidth]{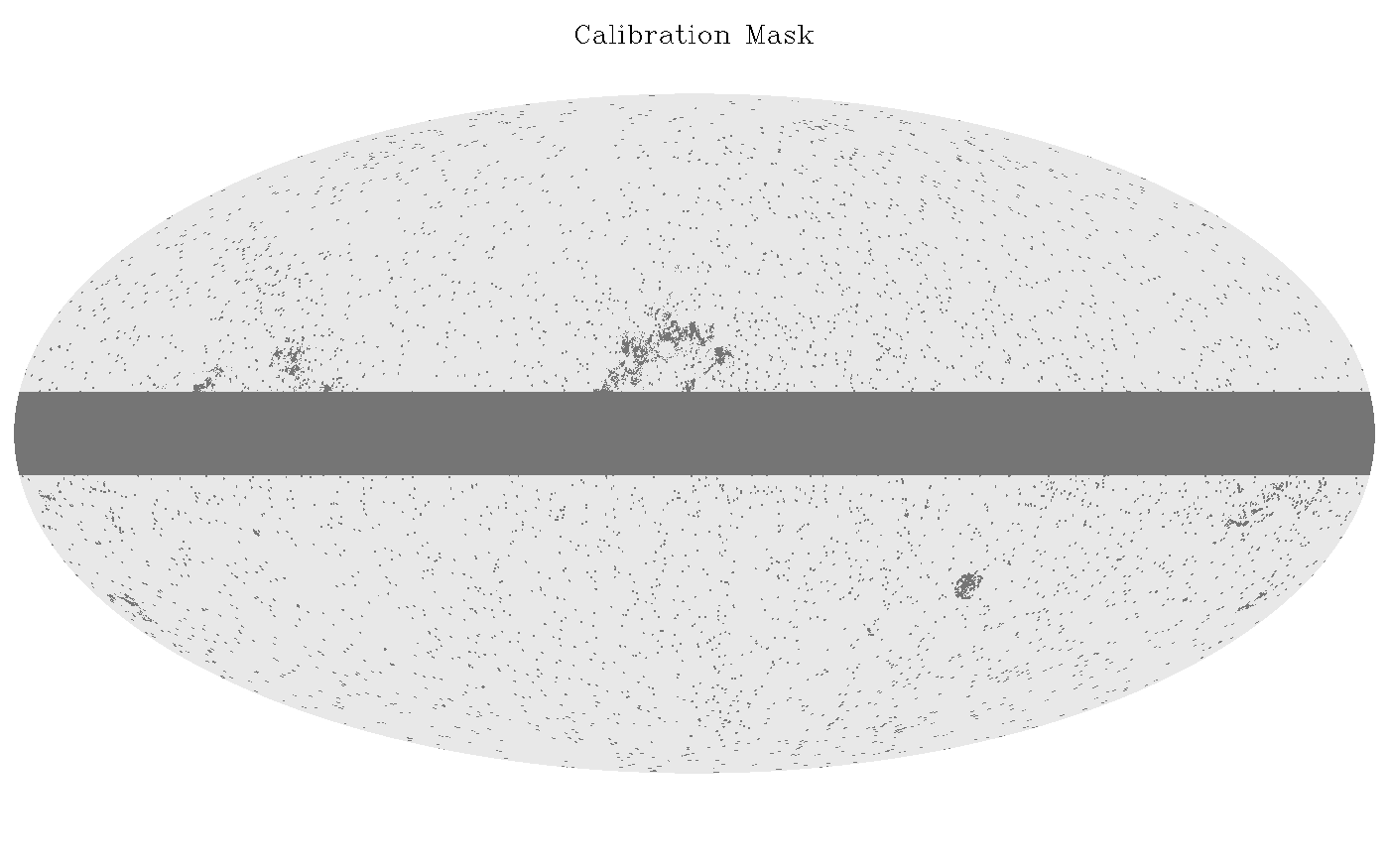}
	\includegraphics[width=0.5\textwidth]{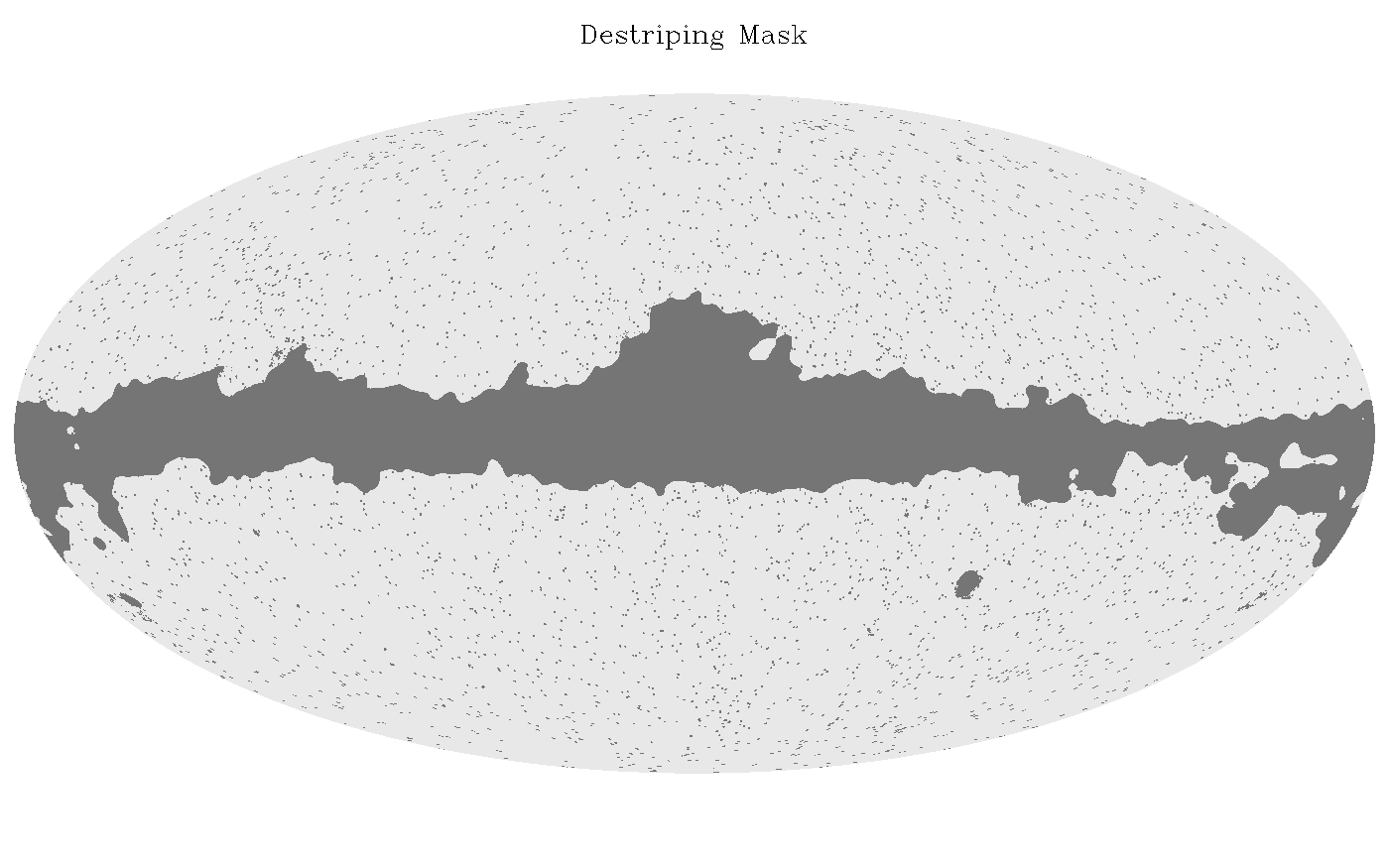}
    	\caption{Masks used for dipole calibration ({\it top}) and destriping ({\it bottom}). The former is 
    		less extended since the calibration methods include a Galactic template based on the 
    		\Planck\ Sky Model.}
    \label{fig:mapmaking_masks}
\end{figure}

The HFI destriper module, \polkapix, determines offsets (one
offset per ring) from a set of input HPR. The algorithm is based on  
underlying reference (\healpix) pixelisations of the I, Q and U sky
signals. These underlying sky maps may have different resolutions for
the temperature and polarization parts. To generate the
temperature maps for the early analyses, only the temperature signal has
been used to compute offsets, \ie polarization effects have been neglected. 
We chose an intermediate resolution of Nside$=128$  to
increase the signal to noise ratio without correlating the offsets of
too many neighbouring rings. At this resolution, it was necessary to
mask the inner part of the Galaxy where the gradients in the sky signal are strong.
The mask used is based
on a Galactic cut at 7\,${\rm MJy}{\rm sr}^{-1}$ on the IRAS 100$\mu$m map (removing
approximatively $19\%$ of the sky) and additionally includes some
bright sources (see Fig.~\ref{fig:mapmaking_masks}).

\subsection{Map projection}

\begin{figure*}[ht]
 	\centering
 	\includegraphics[width=0.49\textwidth]{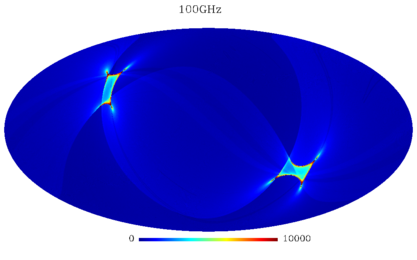}
 	\includegraphics[width=0.49\textwidth]{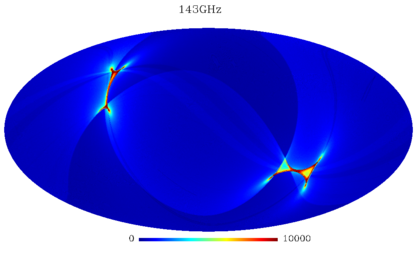}
 	\includegraphics[width=0.49\textwidth]{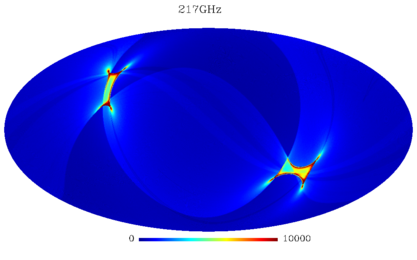}
 	\includegraphics[width=0.49\textwidth]{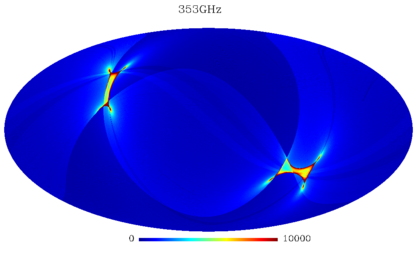}
 	\includegraphics[width=0.49\textwidth]{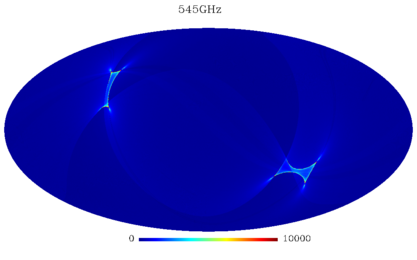}
 	\includegraphics[width=0.49\textwidth]{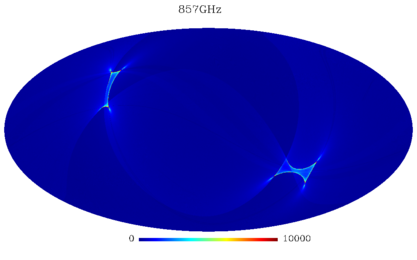}
 	\caption{Hit count  maps at each frequency ({\it from top left to bottom right :} 100, 143, 217,
 	 353, 545, 857\GHz) in 1.7\,arcmin pixels (Nside$=2048$).} 
 	\label{fig:mapmaking_hits_maps}
 \end{figure*}

Cleaned maps are produced using a simple co-addition of the \healpix\
Rings but taking into account the offsets estimated by the
destriper module. Maps are produced for each detector independently and per
frequency (combining all detectors of the same frequency channel). Polarization
maps are also constructed for polarization sensitive channels, \ie for
frequencies up to 353\GHz. 
We account for the different noise level in the combined maps,
weighting data from each detector using their NEP, as determined by the noise
estimation pipeline (\cf Sect.~\ref{sec:detnoise}). In each case, hit count maps
are also produced (shown in Fig.~\ref{fig:mapmaking_hits_maps}), as well as
maps of the 3x3 covariance matrix of I, Q and U in 
each pixel. Note that the ERCSC and the \Planck\ early results papers are based on
temperature maps only, even though polarization maps have been produced by
the pipeline. 

For jackknife tests and noise evaluation studies we produced several
other maps by using various subsets of data including: 
\begin{itemize}

\item two independent sets of detectors per frequency (so-called
  `half-focal plane' maps);

\item separate sky surveys (so-called `survey' maps, based on six months
  of data);

\item independent ring sets built using the first and second half of the
  stable pointing periods (so-called `half-ring'maps).

\end{itemize}
The last  set of maps can be used to estimate accurately the
high frequency noise (on time-scales less than the half-ring duration, \ie
about 20 minutes) in the rings, as well as in the maps.

Maps were also produced with \textsc{Springtide} \citep{ashdown2007b} and
\textsc{MADAM} \citep{keihanen2005} and lead to consistent results. 

\subsection{Absolute photometric calibration} 

The primary absolute calibration of the \Planck\ HFI instrument is based on extended
sources. At low frequencies (353\GHz\  and below), the
orbital dipole and the Solar dipole provide good absolute calibrators.  At high
frequencies (545\GHz\ and above), Galactic emission is used.  The best
available data,  in term of spectral coverage and absolute calibration accuracy,
are the \COBE/\FIRAS\ spectra. We used these data as an absolute photometric
calibrator for the 857 and 545\GHz\ channels.  Both calibration techniques
(dipoles and Galactic emission) have been applied at intermediate frequencies (353 and
545\GHz) for consistency checks.  \FIRAS\ data are also used at all frequencies
to set the zero level in the  HFI maps.

\subsubsection{Absolute photometric calibration using the Solar dipole} \label{mapmaking:calib_dipole}

Since we need to account for the orbital dipole, the dipole calibration 
is done at  the ring level rather than the map level. The Solar dipole
has been accurately measured by \WMAP\ \citep{hinshaw2009}.  
Gains are determined for each stable pointing period through a
$\chi^2$ minimization. Each input sample is modelled as :  
\begin{equation}
	m\ =\ g_d I_D + g_g T_g + C + \mathrm{noise}, 
	\label{eq:qdirt}
\end{equation}
where $g_d$ is the gain, $I_D$ is the dipole signal (including both
Solar and orbital dipoles), $C$ is a constant allowing for an
offset. The Galactic signal is modelled using a template $T_g$ derived
from the thermal dust maps in the Planck Sky Model (hereafter PSM,
version 1.6.3) at each central frequency smoothed to 
match the detector beam. 
For each ring we fit for the gain and the coefficient of the Galactic
template $g_g$ and the constant $C$. 

For calibration purposes, bright sources (detected by HFI or already
known a priori) are masked and we use a Galactic latitude cut
($b<9^\circ$) to avoid using the central part of the Galactic plane,
where the model used by the PSM (from \citet{finkbeiner1999}) is not
sufficiently accurate. The upper plot in
Fig.~\ref{fig:mapmaking_masks} shows the actual mask used for
calibration.

The main limitation of this approach is the contamination by Galactic
foregrounds, especially their polarised component which is poorly known. This
preferentially affects those  rings where, because of the orientation of  the spin axis, the
amplitude of the dipole is small compared to the Galactic signal.

An average over time of the gains estimated on the dipole is used for the
photometric calibration of each detector at frequencies of  353\GHz\ or less.
We have restricted this computation to a time interval of the first
survey, 
corresponding to rings 2000 to 6000, 
where the gain measurements appear less affected by systematic effects. 
On this interval, the \rms\ of the ring-by-ring gains is
of the order of 1\,\%, or less, for frequencies  up to 353\GHz. The same procedure
applied to the second survey gives comparable results, except for a few
bolometers that shows a yearly variation of less than 2\,\%.

Note that the photometric calibration has to be done prior to the
destriping so that the orbital dipole can be subtracted.

\subsubsection{Absolute photometric calibration using \FIRAS\ data:
  calibration factors and zero points} \label{mapmaking:calib_gal}

The scheme used for the photometric calibration of the \Planck\ HFI on
\FIRAS\ data is very similar to that adopted  by the {\sc Archeops}
collaboration \citep{macias2007}.  For calibration of
 the \Planck\ HFI, \FIRAS\ data was processed as follows:
\begin{itemize}

\item conversion of the data from their original \COBE\ quad-cube
pixelisation to the \healpix\ scheme by using a drizzling re-projection code 
\citep{paradis2011};

\item extrapolation of the data to obtain brightnesses at the nominal HFI frequencies.
The \FIRAS\ map at one selected frequency can be obtained by convolving
the \FIRAS\ spectra with the HFI bandpass filters.  However, this method
produces very noisy \FIRAS\ maps at high Galactic latitude (especially
for $\lambda>$ 700\,$\mu$m). Since we are interested in both the
Galactic plane and its surrounding, we have preferred to derive the
\FIRAS\ maps together with their errors using fits of \FIRAS\ spectra.  Each
individual \FIRAS\ spectrum is fitted with a black body,  modified
by a $\nu^{\beta}$ emissivity law. Since we are searching for the best
representation of the data and not for physical dust parameters, we
neglect the contribution of the cosmic infrared background. We
moreover restrict the fit to the frequency range of interest,  avoiding
the need for a second dust component as in \citet{finkbeiner1999}.

\end{itemize} 
The results of this processing consist of Nside$=16$ \healpix\ maps
for the sky signal extrapolated from \FIRAS\ data for each HFI detector,
together with associated errors.

The HFI data have to be convolved by the \FIRAS\ beam. This beam has been
measured on the Moon.  Due to imperfections in the sky horn antenna,
the effective beam shows both radial and azimuthal deviations from the
nominal 7$^\circ$ top hat beam profile. Since \COBE\ rotates about the
optical axis of the \FIRAS\ instrument, on average, the beam must
have cylindrical symmetry. However, the time it takes to collect a single
interferogram is less than a rotation period. Thus, a particular measurement
beam may be asymmetric. \citet{fixsen1997} estimate that the assumption
of beam symmetry may produce residual beam shape errors of order of
5\%, that are not taken into account in this analysis. We perform the beam
convolution in \healpix\ format.  To simulate the movement during an
integration of an interferogram, the HFI data were further convolved  
by a 2.6$^\circ$ top-hat in the direction perpendicular to the ecliptic
plane (which is roughly the \FIRAS\ scanning direction).  

\begin{figure*}[!htbp]
  	\centering
	\includegraphics[width=0.49\textwidth]{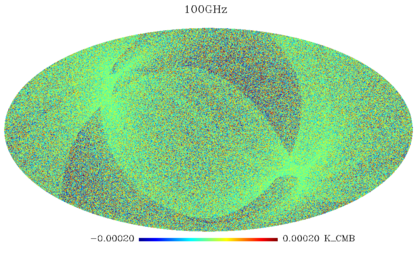}
         \includegraphics[width=0.49\textwidth]{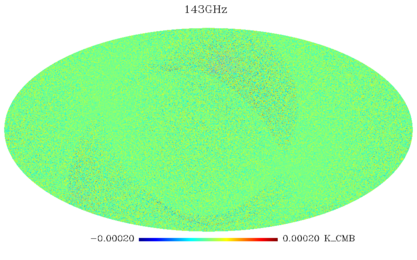}
	\includegraphics[width=0.49\textwidth]{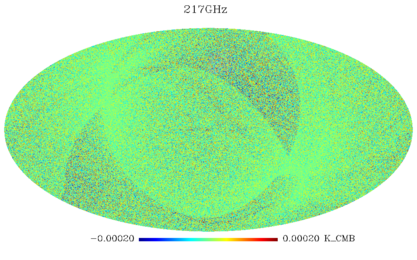}
	\includegraphics[width=0.49\textwidth]{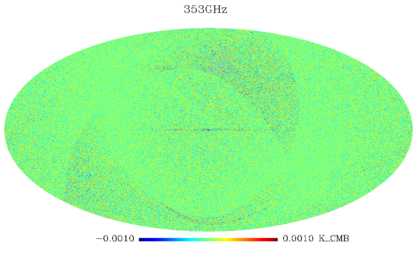}
	\includegraphics[width=0.49\textwidth]{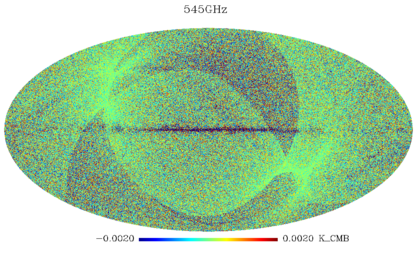}
	\includegraphics[width=0.49\textwidth]{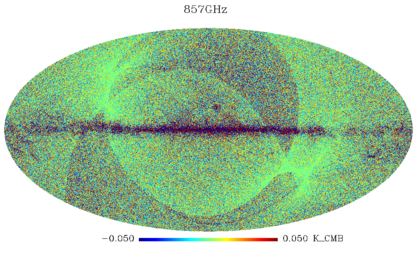}
    	\caption{Residual maps of the half differences
                between the maps made from the first  
    		and second half ring projection ({\it from top left
                to bottom right:} 100, 143, 217, 353, 
    		545, 857\GHz) in 1.7\,arcmin pixels (Nside$=2048$). Note that the
	 CMB channels at 100-217\GHz\ are all shown on the same color scale.  In
                addition to the noise pattern, 
    		which is well traced by the hit maps of
                Fig.~\ref{fig:mapmaking_hits_maps}, one also sees
    		small differences relative to the signal, when
                gradients of the signal are large 
    		(mostly in the Galactic plane) and sub-pixel effects
                become quite apparent. }  
    	\label{fig:mapmaking_noise_maps}
\end{figure*}

Following the IRAS convention, the spectral intensity data,
I$_{\nu}$, are expressed in MJy/sr at fixed nominal frequencies,
assuming the source spectrum is $\nu I_{\nu}$ = constant (\ie constant
intensity per logarithmic frequency interval). Since the source
spectrum is not a constant intensity per logarithmic frequency
interval, a colour correction has to be applied to obtain an accurate
intensity. The colour correction factor $cc$ is defined such that:
\begin{equation}
I_{\nu_0}({\rm actual}) = I_{\nu_0}({\rm quoted})/ cc, 
\end{equation}
where $I_{\nu_0}({\rm actual})$ is the actual specific intensity
of the sky at frequency $\nu_0$, $I_{\nu_0}({\rm quoted})$ is the corresponding
value given with the IRAS convention, and $\nu_0$ is the frequency
corresponding to the nominal wavelength of the considered band. With
these definitions: 
\begin{equation}
cc= \frac{\int (I_{\nu}/I_{\nu_0})_{\rm actual} R_{\nu} d \nu}{\int
  (\nu_0/\nu)  R_{\nu} d \nu} , 
\end{equation}
where $(I_{\nu}/I_{\nu_0})_{\rm actual}$ is the actual specific intensity of the 
sky normalised to the intensity at frequency $\nu_0$ and $R_{\nu}$
is the spectral response.
We derive the colour correction for each \FIRAS\ pixel using the 
HFI bandpass filters and the fits of \FIRAS\ spectra.  

Gains and zero levels are obtained for each detector by fitting:
\begin{equation}
  	I_{FIRAS}(\nu_0) \times cc = K \times I_{HFI}(7\mathrm{\,deg \ beam}) + zp, 
	\label{eq_FIRAS_calib}
\end{equation}
where $K$ is the calibration factor for the two high-frequency
channels and $zp$ is the zero level value ($I$ stands for the intensity). 
However, the fit is done 
for each detector, even for the low frequency channels,   to set
the zero levels of  our maps.  

In these low-frequency channels, we needed to add the CMB anisotropies
to the \FIRAS\ maps (or we could have removed them from the HFI
data) 
in order for Eq.~\ref{eq_FIRAS_calib} to be valid.
Currently, we are adding the \WMAP\ CMB map (convolved by the
\FIRAS\ beam). But we also have to deal with the fact that there is 
a substantial contribution in some wavebands from CO lines (\cf Sect.~\ref{sec:CO}). 
The \FIRAS\ dataset we are using does not contain the lines (but the CO(1-0) is
not detected in \FIRAS). At this stage, we simply reduce the fit to the 
regions outside a CO mask derived from the \cite{dame2001} survey, 
in order to minimize  the contamination of the
\Planck\ 217 and 100\GHz\ channels before computing the zero levels. 

Because of its high signal-to-noise ratio,  the Galactic plane is the best
place to perform the calibration of the HFI high-frequency data. 
Unfortunately, there are two problems that limit the accuracy in
 the Galactic plane: (1) the lack of precise knowledge of the
\FIRAS\ beam; (2) The fact that the colour corrections are applied at the
7$^\circ$ resolution, whereas they should be applied to each HFI pixel
prior to the convolution with the \FIRAS\ beam. We thus applied
Eq.~\ref{eq_FIRAS_calib} to bands slightly above the Galactic plane, \ie for
Galactic latitudes,  $10^\circ <  \vert b \vert <  60^\circ$, at
latitudes low enough to have a reasonable signal to noise ratio.   
The fit is done on one averaged
Galactic-latitude profile (i.e. one profile averaged over all
longitudes).  

Errors on $K$ and $zp$ take into account only statistical errors on
the \FIRAS\ data. There are mostly the same for $K$ and $zp$ and are
about 0.4\%, 0.8\%, 2.5\%, 5\%, 15\% and 20\% at 857, 545, 353, 217,
143 and 100\GHz, respectively. However, the systematic errors are
larger than the statistical errors at high frequencies. They have been
estimated using the dispersion of the fitted values in different parts
of the sky. They are about 7\% for $K$ at all frequencies and about
2\% for $zp$ (for 857, 545 and 353\GHz).  

Because of:  (i) the poor signal-to-noise of \FIRAS\ data at low frequency, and thus 
uncertainties in the \FIRAS\ spectra fits; (ii) the fact that we need to avoid
bright regions with CO;  (iii) the fact that we are calibrating outside
the Galactic plane (due to \FIRAS\ beam uncertainties), we compare the
dipole/Galaxy calibration only at 353\GHz\ and 545\GHz. We find
agreement within 2-3\% at 545\GHz, and within 5-6\% at 353\GHz. 

\subsection{Relative photometric calibration accuracy between channels} \label{sec:intercalib_cmb}

We evaluated the photometric calibration relative accuracy between
frequency channels up to 353\GHz\ using several methods.

In the first, we compare maps in CMB dominated sky areas. The first
step is to mask areas possibly contaminated by Galactic emission or
point sources.  We used a stringent mask, keeping only $\sim 6\,\%$ of
the sky, which we obtained by setting a threshold on the 857\GHz\
intensity as measured by HFI, combined with a point source mask built
using the HFI catalogue produced by the HFI pipeline (see the
description in Sect.~\ref{sec:masks}).  We then smoothed the frequency
maps to equivalent Gaussian FWHM of $1^\circ$.  The scatter plot of
the unmasked pixel values from each map is then fitted to a straight
line.  The slope of this line should be unity, as we calibrated both
maps in $K_{CMB}$ and selected sky areas where CMB dominates.  The
residual deviations from unity are below 1\,\% for the 100-143 and
143-217\GHz\ and about 4\,\% for 217-353\GHz\ comparison, where
foregrounds are more important even in the small unmasked area that we
used.  The same method, applied to compare the HFI 100\GHz\ and the
LFI 70\GHz\ maps, shows that their calibrations agree to within 0.5\%.

In the second method, we used a spectral-matching technique which jointly
fits all the spectra and cross-spectra of all HFI
channels~\citep{Delabrouille2003,cardoso2008}, except the 857\GHz\ channel (actually, that
channel is regressed out of the 5 lowest frequency channels before
fitting).  These auto- and cross-spectra are fitted via maximum
likelihood  to a model including contributions from the CMB, from
foregrounds and from noise. The calibration coefficients are
treated as additional parameters in the fits.  Since the
model includes a component capturing foreground emission, a relatively
large fraction of the sky can be used to estimate the spectra: we used
a mask excluding about 40\% of the sky, based on the emission at
857\GHz\ and on the ERCSC and ESZ catalogues.
The consistency of the estimates of the relative calibration
coefficients was checked using jackknifes: we compared the estimates
obtained using the whole masked sky or only the northern or southern
part of the sky.  We also compared the estimates using the first or
last part of the rings.  In all cases, the discrepancy between all
estimates was found to be well under 1\,\%.  Their mean value was found
to agree with the photometric calibration to better than 1\% for the
100, 143, 217 and 353 \GHz\ channels.  All those measurements were
obtained by fitting the spectra over the multipole range $100 \leq
\ell\leq 900$.  The robustness of the results against the choice of
multipole range was also investigated.  Fits performed over the sub-ranges
$[100,600]$, $[200,700]$, $[300,800]$, $[400,900]$ showed variations
of the calibration coefficients of the order of half a percent.

Finally, we also computed the relative calibration of HFI channels
using the CMB map \citep{delabrouille2009} derived by applying to the
WMAP 5 year data the needlet ILC method described later in
Sect.~\ref{sec:nilc}.  This was done on a selected range in $\ell$,
with a smooth rise of the spectral window between $\ell = 15$ and
$\ell=50$, and a cut at high $\ell$ set by the signal to noise ratio
of the CMB map derived from WMAP5. Different sets of masks based on
ancillary data (H$_\alpha$ from \cite{finkbeiner1999}, 408\MHz\ from
\cite{haslam1982}, 100\microm\ from \cite{schlegel1998}, or from our
own 857\GHz\ map and \Planck\ ERCSC sources) were used to perform the
calibration. The relative calibration factors were then computed
relative to our most sensitive 143\GHz\ channel for masks retaining
different fractions of the sky (from 30 to 70\,\%), and for maps
produced with different detector sets, or based on halves of the ring
data, or different hemispheres. The dispersion found (relative to the
143\GHz\ channel) was below about 1\,\% for the 100 and 217\GHz\
channels, and about 2\% for the 353\GHz\ channel.  

These three lines of investigation therefore consistently provide an
estimate better than 1\,\% for the relative calibration accuracy
between the 100, 143, and 217\GHz\ channels. For the 353\GHz\ channel,
we estimate the accuracy to be better than about 2\,\%, on the basis
of the two latter methods.  

\subsection{Jackknife tests and noise properties} \label{mapmaking:jackknife}

An important  characteristic of maps is obtained by analysing he
`half-ring' maps. For each frequency, we constructed the
half-difference between these two independent sets of half-ring maps; they are shown in
Fig.~\ref{fig:mapmaking_noise_maps}. As a result of the \Planck\ scanning strategy, the 
noise in these maps is  inhomogeneous; it is largely dominated by
white noise modulated by the hit count in each pixel, as can be seen from the hit maps 
shown in Figs.~\ref{fig:mapmaking_noise_maps}. 
In the high frequency difference maps (545 and 857\GHz), we also observe some additional 
residuals in the Galactic plane, where the gradients of the sky signal within the pixels are
strong. This can be understood by noting the poor sampling of 
those pixels, together with that of the beam FWHM for these channels; this
should be significantly reduced once more data is included.  

\begin{figure}[htbp]
   \centering
    \includegraphics[width=0.5\textwidth, height=.4\textwidth]
                    {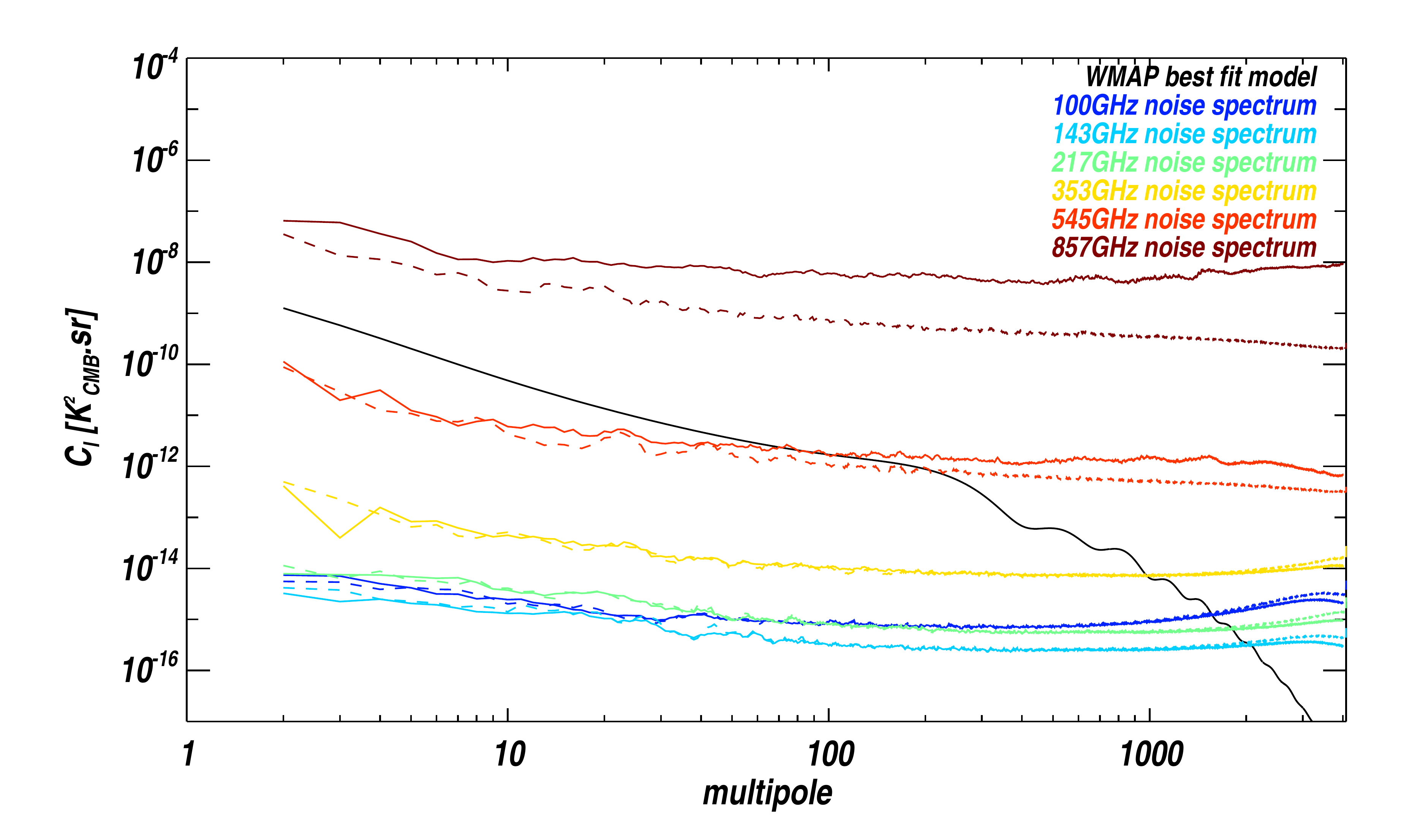}  
    \caption{Power spectra from the difference maps shown in 
    		Fig.~\ref{fig:mapmaking_noise_maps}, on the full sky (\textit{solid line}) and after
    		masking the Galactic plane (\textit{dashed line}). The sky coverage correction was done
    		according to \cite{tristram2005}. As expected, the difference is only substantial at high
    		frequency, when gradients of the Galactic signal are large.} 
    \label{fig:mapmaking_noise_spectra}
\end{figure}

In the harmonic domain (Fig.~\ref{fig:mapmaking_noise_spectra}), the
power spectra of these difference maps  are almost flat over a wide range of
multipoles. The effect of the time constant deconvolution is roughly
compensated by the low-pass filtering resulting in a relatively small
deviation from pure flat white noise at high multipoles ($\ell \sim
3000$).  The average level between $\ell =100$ and $\ell=1000$ of
these power spectra in the masked case provides an estimate of the 
white part of the noise in the maps which is given in line a3 of the summary
table, Table~\ref{tab:summary}. These levels are consistent with those 
obtained in clean region of the sky in the study of the cosmic infrared
 background fluctuations \citep{planck2011-6.6}.
  
 \section{CMB removal} \label{sec:cmbrm} 

This section was developed in common with LFI \citep{planck2011-1.6} and is reported 
identically in both papers.

In order to facilitate foreground studies with the frequency
maps, a set of maps was constructed with an estimate of the CMB
contribution subtracted from them.  The steps undertaken 
in determining that estimate of the CMB map,
subtracting it from the frequency maps, and characterising the errors
in the subtraction are described below.

\subsection{Masks} \label{sec:masks}

Point source masks were constructed from the source catalogues produced by the HFI 
pipeline for each of the HFI frequency channel maps.  The algorithm used 
in the pipeline to detect the sources was a Mexican-hat wavelet
filter.  All sources detected with a signal-to-noise ratio greater
than 5 were masked with a cut of radius $3\,\sigma \approx 1.27\,
\mathrm{FWHM}$ of the effective beam.  A similar process was applied
to the LFI frequency maps (\cite{planck2011-1.6}).

Galactic masks were constructed from the 30\,GHz and 353\,GHz frequency channel 
maps.  An estimate of the CMB was subtracted from the maps in order not to bias 
the construction.  The maps were smoothed to a common resolution of 5\degree.  
The pixels within each mask were chosen to be those with values above a threshold 
value. The threshold values were chosen to produce masks with the desired fraction 
of the sky remaining.
The point source and Galactic masks were provided as additional
inputs to the component separation algorithms.

\subsection{Selection of the CMB Template} \label{sec:CMBselection}

Six component separation or foreground removal algorithms were applied
to the HFI and LFI frequency channel maps to produce CMB maps. They
are, in alphabetical order:
\begin{itemize}
\item{AltICA}:      Internal linear combination (ILC) in the map domain;
\item{CCA}:         Bayesian component separation in the map domain;
\item{FastMEM}:     Bayesian component separation in the harmonic domain;
\item{Needlet ILC}: ILC in the needlet (wavelet) domain;
\item{SEVEM}:       Template fitting in map or wavelet domain;
\item{Wi-fit}:      Template fitting in wavelet domain.
\end{itemize}
Details of these methods may be found in \cite{leach2008}.  These six algorithms make
different assumptions about the data, and may use different combinations of frequency
channels used as input.  
Comparing results from these methods \textbf{(see Fig.~\ref{fig:cmbdispersion})} 
demonstrated the consistency of
the  CMB template and provided an estimate of the uncertainties
in the reconstruction.  A detailed comparison of the output of these
methods, largely based on the CMB angular power spectrum, was used to
select the CMB template that was removed from the frequency channel
maps.  The comparison was quantified using a jackknife procedure:
each algorithm was applied to two additional sets of frequency maps
made from the first half and second half of each pointing period.  A
residual map consisting of half the difference between the two
reconstructed CMB maps was taken to be indicative of the noise
level in the reconstruction from the full data set.  The Needlet ILC
(hereafter NILC) map was chosen as the CMB template because it
had the lowest noise level at small scales.

\begin{figure}
  	\includegraphics[width=\columnwidth]{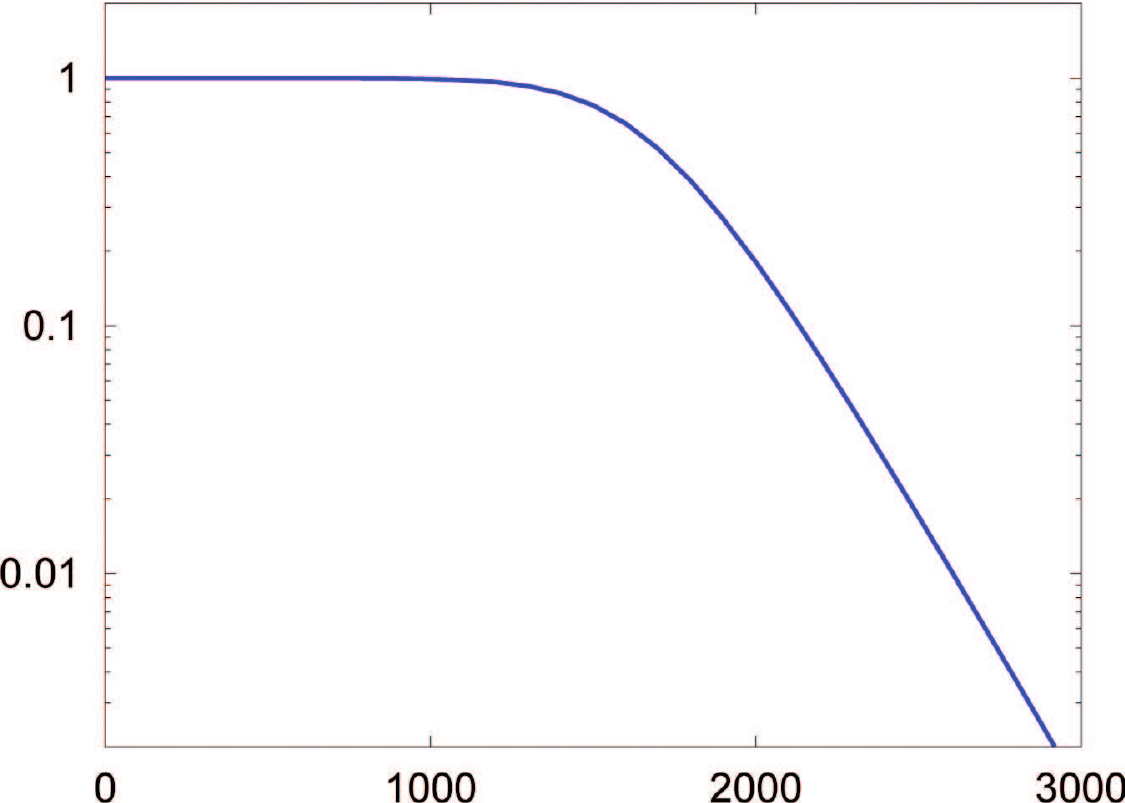}
  	\caption{Wiener-like filter function, plotted versus multipole, which was applied to produce the template for CMB removal.}
  	\label{fig:cmbremovalwienerfilter}
\end{figure}

The CMB template was removed from the frequency channel maps after
application of a filter in the spherical harmonic domain.  The filter
has a transfer function made of two factors.  The first
corresponds to the Gaussian beam of the channel to be cleaned; the
second is a transfer function attenuating the multipoles of the
CMB template that have low signal-to-noise ratio.  It is designed in
Wiener-like fashion, being close to unity up to
multipoles around $\ell = 1000$, then dropping smoothly to zero
with a cut-off frequency around $\ell=1700$ (see Fig.~\ref{fig:cmbremovalwienerfilter}).  All angular frequencies above $\ell=3900$ are completely suppressed.  
This procedure was adopted to avoid doing more harm than good to the
small scales of the frequency channel maps where the signal-to-noise
ratio of the CMB is low.

\subsection{Description of Needlet ILC} \label{sec:nilc}

\begin{figure}
  	\includegraphics[width=\columnwidth]{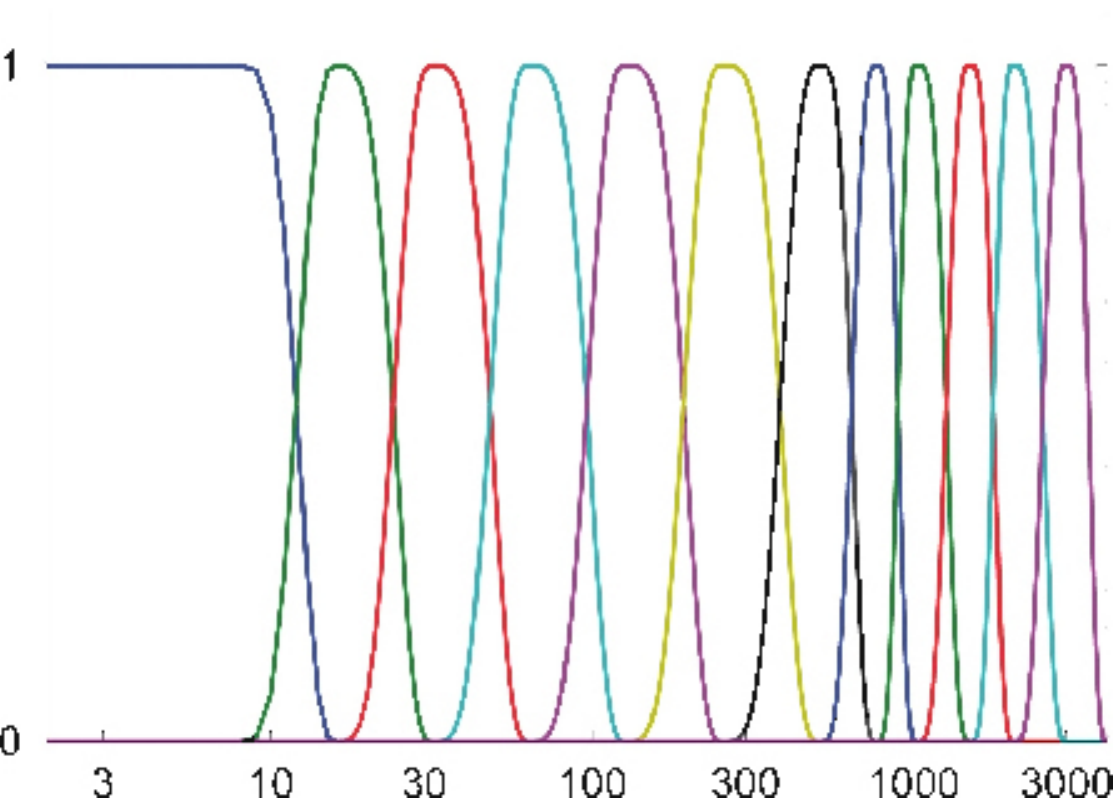}
  	\caption{The bandpass filters, plotted versus multipole, that define the spectral domains 
          used in the NILC.} 
  	\label{fig:needletbands}
\end{figure}

The NILC map was produced using the ILC method in the
``needlet'' domain.  Needlets are spherical wavelets that allow
localisation both in multipole and sky direction.  The input maps are
decomposed into twelve overlapping multipole domains (called
``scales''), using the bandpass filters shown in Fig.~\ref{fig:needletbands}
and further decomposed into regions of the sky.  Independent ILCs are
applied in each sky region at each needlet scale.  Large regions are used at 
large scales, while smaller regions are used at fine scales.

\begin{figure}
  	\includegraphics[width=\columnwidth]{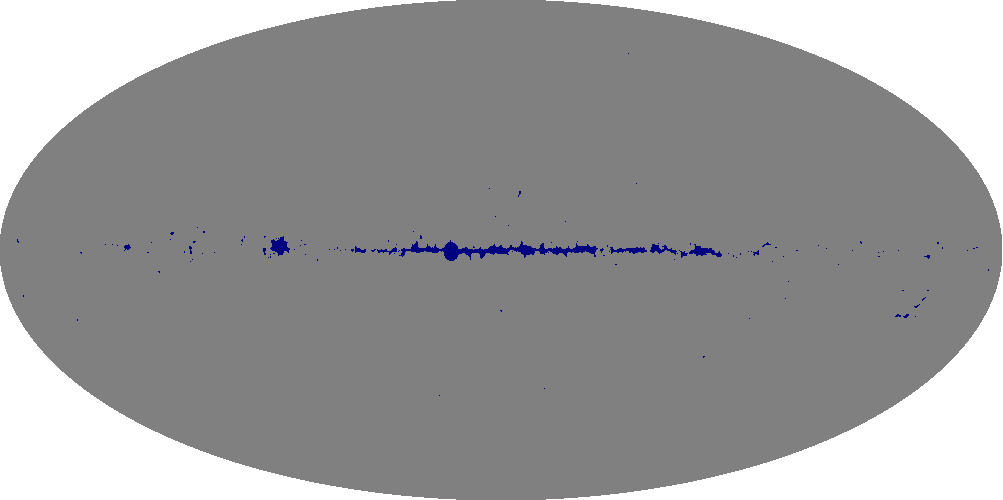}
  	\caption{Galactic mask used with NILC.}
  	\label{fig:nilcgalmask}
\end{figure}

The NILC template was produced from all six HFI channels, using the
tight Galactic mask shown in Fig.~\ref{fig:nilcgalmask}, which covers 99.36\% 
of the sky.  Additional areas are excluded on a
per-channel basis to mask point sources. Future inclusion of the 
LFI channels will improve cleaning of low-frequency foregrounds such as 
synchrotron emission from the CMB template.

Before applying NILC, pixels missing due to point source and Galactic 
masking are filled in by a ``diffusive inpainting'' technique, which consists of 
replacing each missing pixel by the average of its neighbours and iterating to
convergence.  This is similar to solving the heat diffusion equation in the
masked areas with boundary conditions given by the available pixel values at
the borders of the mask.  All maps are re-beamed to a common resolution of
5\arcmin.  Re-beaming blows up the noise in the less
resolved channels, but that effect is automatically taken
into account by the ILC filter.

The CMB template obtained after NILC processing is
filtered to have the `Wiener beam' shown in
Fig.~\ref{fig:cmbremovalwienerfilter}.  The ILC coefficients are saved
to be applied to the jackknife maps for performance evaluation as
described in Sect.~\ref{sec:NILCperf_internal}

\subsection{Uncertainties in the CMB Removal} \label{sec:cmb_uncertainty}

Uncertainties in the CMB removal have been gauged in two
ways, firstly by comparing the CMB maps produced by the different
algorithms and secondly by applying the NILC coefficients to jackknife maps.

\begin{figure}
  	\includegraphics[width=\columnwidth]{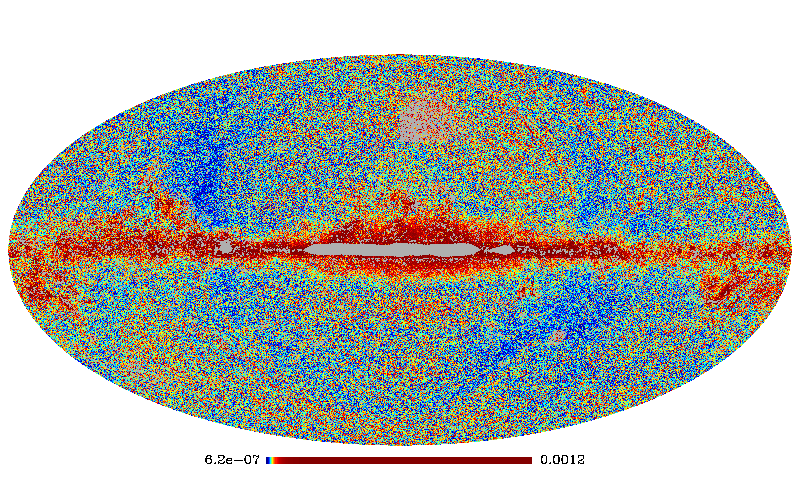}
  	\caption{Estimate of the \rms\ error in the CMB subtraction. The map is  
  		histogram-equalised to bring out the details.}
  	\label{fig:cmbdispersion}
\end{figure}

\subsubsection{Dispersion of the CMB maps produced by
  the various algorithms.} 

The methods that were used to produce the estimates of the CMB are
diverse.  They work by applying different algorithms (ILC, template
fitting, or Bayesian parameter estimation) in a variety of domains
(pixel space, Needlet/wavelet space, or spherical harmonic
coefficients).   Each method carries out its optimisation in a different way
and thus will respond to the foregrounds differently.  
Dispersion in the CMB rendition by different methods
provides an estimate of the uncertainties in
the determination of the CMB, and thus in the subtraction process.
The \rms\ difference between the NILC map and the other CMB estimates
is shown in Fig.~\ref{fig:cmbdispersion}.  As expected, the
uncertainties are largest in the Galactic plane where the foregrounds to remove are strongest, 
and smallest around the Ecliptic poles where the noise levels are lowest.

\subsubsection{CMB map uncertainties estimated by applying NILC filtering of jackknifes}
\label{sec:NILCperf_internal} 

\newcommand{\Var}[1]{\mathrm{Var}(#1)}

The cleanliness of the CMB template produced by the NILC filter
can be estimated using jackknives.  We apply the NILC filter to
the maps built from the first and last halves of the ring set.  
The power distribution of
the half-difference of the results provides us with a reliable
estimate of the power of the noise in the NILC CMB template,
(while previous results correspond to applying the NILC filter 
to the half-sum maps from which they can be derived).

The jackknives allow estimates of the relative contributions of sky signal and
noise to the total data power. Assume that the data are in the form $X = S + N$\
where $S$ is the sky signal and $N$ is the noise,
independent of $S$. The total data power $\Var{X}$ decomposes as
$\Var{X}  = \Var{S} + \Var{N}$. 
One can obtain $\Var{N}$ by
applying the NILC filter to half difference maps, and $\Var{S}$ follows from
$\Var{X} - \Var{N}$.  This procedure can be applied in
pixel space, in harmonic space, or in pixel space \emph{after} the
maps have been bandpass-filtered, as described next.

\begin{figure}
  	\centering
  	\includegraphics[width=\columnwidth]{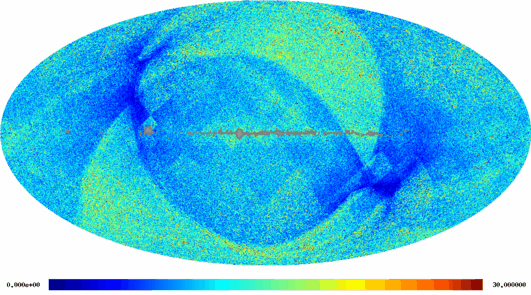}
  	\caption{Local \rms\ of the noise (estimated by jackknife) in the NILC
    		CMB map.  The colour scale is from 0 to 30 $\mu$K per pixel at
    		resolution $N_\mathrm{side}=2048$.}
  	\label{fig:nilcnoisespatial}
\end{figure}

\begin{figure}
  	\centering
  	\includegraphics[width=\columnwidth] {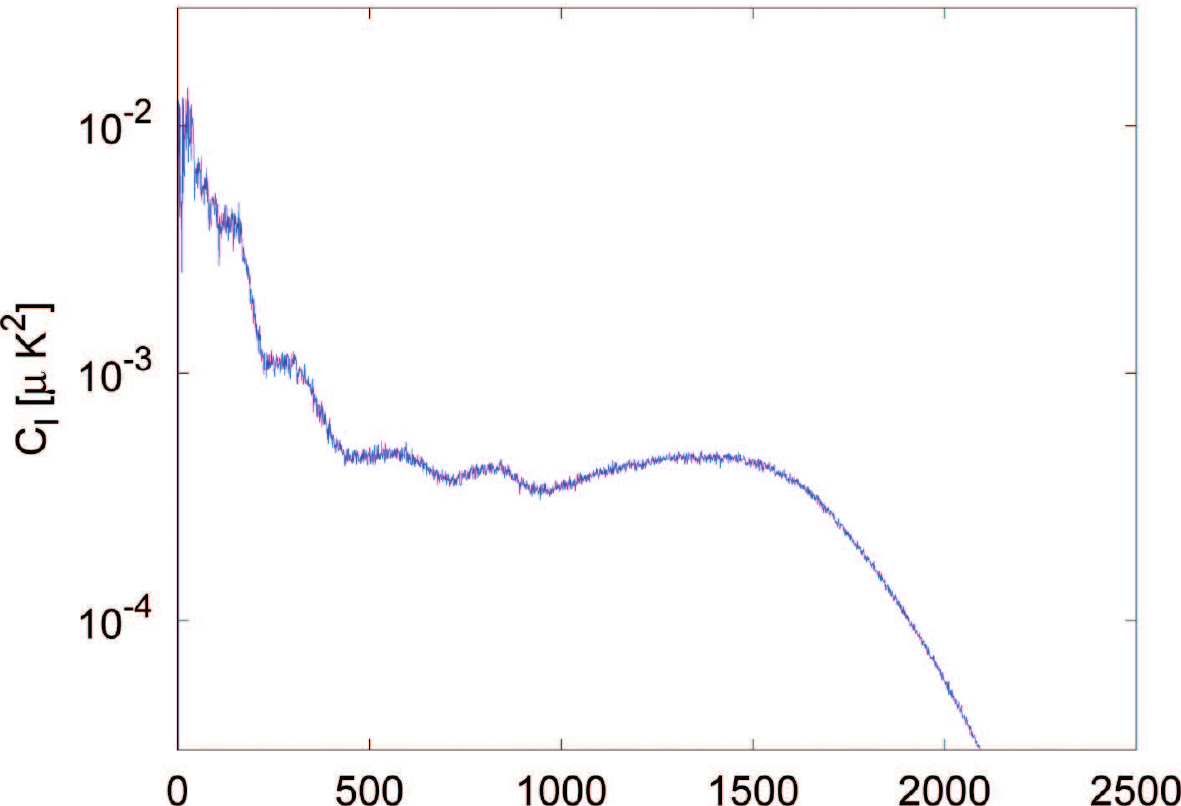}
  	\caption{Angular spectrum in $\mu\mathrm{K}^2$ of the noise (estimated by
    		jackknife) in the NILC CMB map.  It corresponds to 11
                $\mu\mathrm{K}$ per pixel.} 
  	\label{fig:nilcnoisespectral}
\end{figure}

We first used pixel space jackknifing to estimate the spatial
distribution of noise.  Figure~\ref{fig:nilcnoisespatial} shows a map
of the local \rms\ of the noise.  We applied the NILC filter to a
half-difference map and we display the square root of its smoothed
squared values, effectively resulting in an estimate of the local
noise \rms.
Using the same approach, we obtain an estimate of the angular spectrum of
the noise in the NILC map, shown in Fig~\ref{fig:nilcnoisespectral}.
That spectrum corresponds to an \rms\ $\left[(1/4\pi)\sum_\ell
(2\ell+1)C_\ell\right]^{1/2}$ of 11\,$\mu$K per pixel.
The ``features'' in the shape of the noise angular spectrum at large
scale are a consequence of the needlet-based filtering (such features
would not appear in a pixel-based ILC map).  Recall that the
coefficients of an ILC map are adjusted to minimize the total
contamination by both foregrounds \emph{and} noise.  The strength of
foregrounds relative to noise being larger at coarse scales, the
needlet-based ILC tends to let more noise in, with the benefit of 
better foreground rejection.

\begin{figure}
  	\centering
  	\includegraphics[width=\columnwidth]{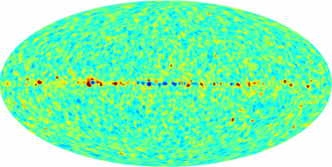}
  	\caption{Local power of the NILC CMB template in the range $\ell=500\pm200$.}
  	\label{fig:nilc_contamination}
\end{figure}

The half-difference maps offer simple access to the power
distribution of the residual noise in the estimated CMB template.
However,
it is more difficult to evaluate other residual contamination, since
all fixed sky emissions cancel in half difference maps.
Any such large-scale contamination is barely visible in the CMB
template, since it is dominated by the CMB itself.  However,
contamination is more conspicuous if one looks at intermediate scales. 
Figure~\ref{fig:nilc_contamination}
shows the local power of the CMB template after it is bandpassed to 
retain only multipoles in the range $\ell=500\pm200$. This smooth
version of the square of a bandpassed map clearly shows where the 
errors in the component separation become large
and so complicate some specific science analyses.

 \begin{figure*}[!htbp]
     \centering
     \includegraphics[width=0.49\textwidth] {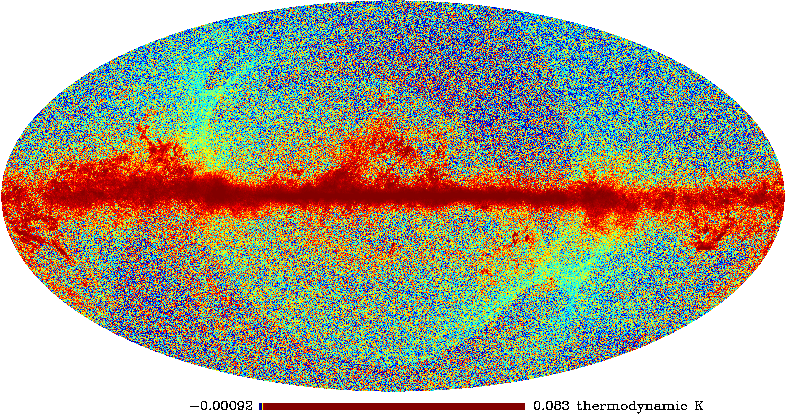} 
     \includegraphics[width=0.49\textwidth] {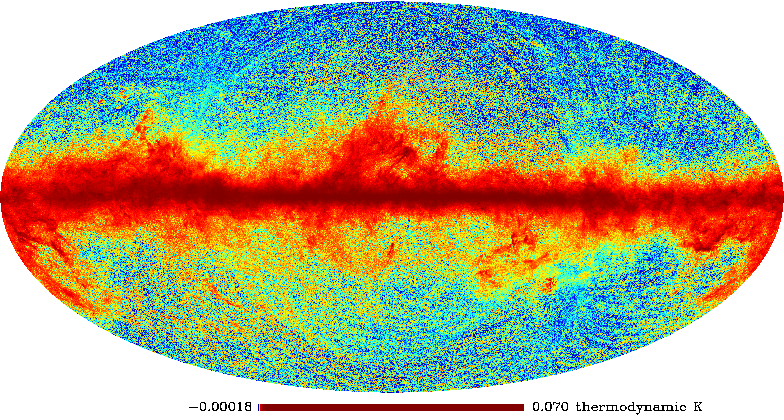} 
     \includegraphics[width=0.49\textwidth] {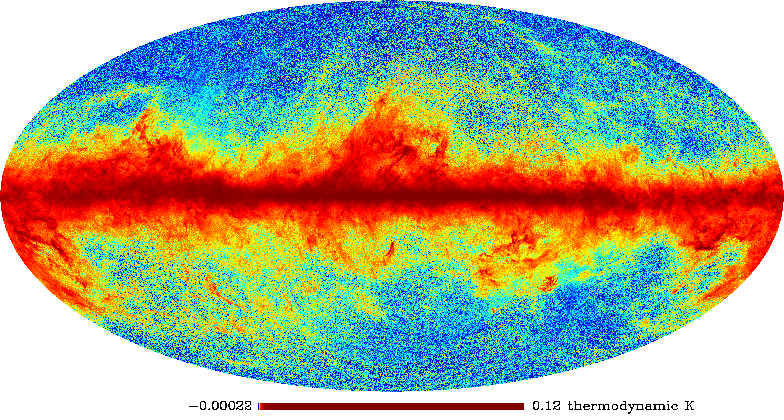} 
     \includegraphics[width=0.49\textwidth] {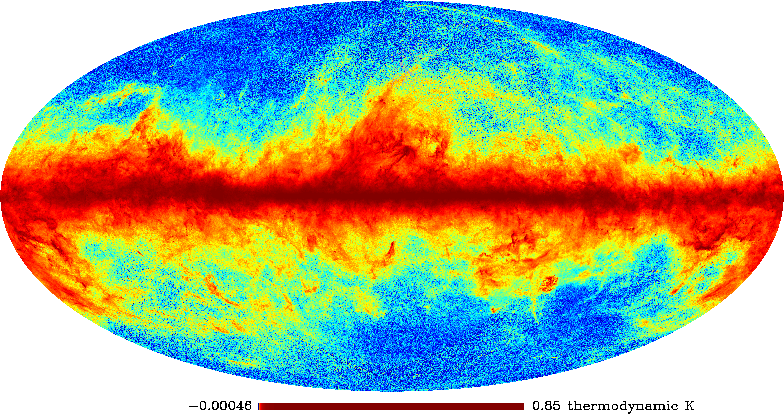} 
     \includegraphics[width=0.49\textwidth] {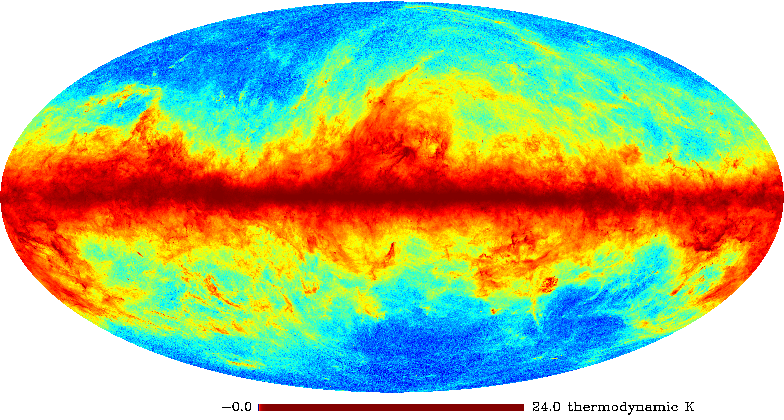} 
     \includegraphics[width=0.49\textwidth] {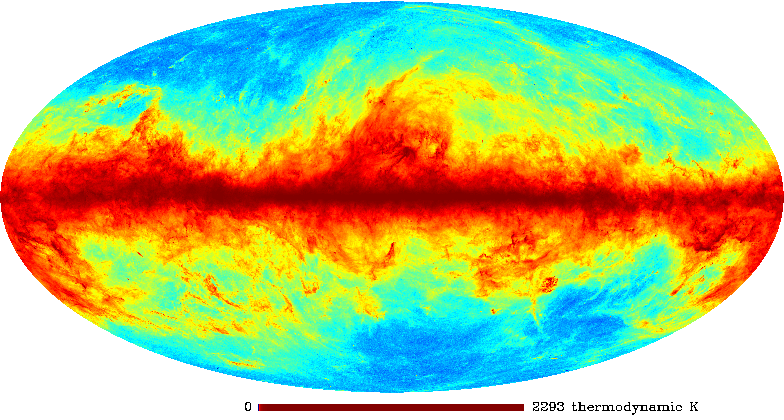}
     \caption{CMB-removed channel maps.  From left to right and top to bottom, 
      100, 143, 217, 350, 545, and 857\GHz. At high Galactic latitudes 
      of the the 100 and 143\GHz \ channel maps, the noise modulation caused by 
      varying integration times is clearly visible, cf. fig~\ref{fig:mapmaking_hits_maps}.   }
     \label{fig:sky-cmb_maps}
 \end{figure*}

 \section{Summary of processed data characteristics and conclusions} \label{sec:sum}

The data provided to the \Planck\ collaboration by the HFI DPC consisted of: \begin{enumerate}
\item channel (frequency) maps (\eg Fig.~\ref{fig:sky-cmb_maps}), 
\item hit count maps (Fig.~\ref{fig:mapmaking_hits_maps}), 
\item half-ring maps (Fig.~\ref{fig:mapmaking_noise_maps}), 
\item error map of the CMB template used for CMB removal (Figs.~\ref{fig:cmbdispersion} and \ref{fig:nilc_contamination}), and 
\item masks (\eg Fig.~\ref{fig:nilcgalmask}).
\end{enumerate} 
Channel maps were provided both with and without CMB removal. 
Table~\ref{tab:summary} summarizes their characteristics (prior to CMB removal). 
It gives most of the information needed to make use of the maps. Further details and comments are given below.  

 \begin{table*}[!htbp] 
 	\caption{Summary of the main characteristics of HFI early maps. The first column refers to following notes pertaining to the content of the line, while the units are between brackets [] at the right of column 2.} 
 	\label{tab:summary} 
 	\centering
	\scalebox{1}{
 \begin{tabular}{ll|rrrrrr} \hline \hline
 \multicolumn{8}{c}{HFI Early Maps - Main Characteristics } \\ 
 \hline 
\tablefootmark{a1} & $\nu$ \hfill [GHz] & 100 & 143 & 217 & 353 & 545 & 857 \\ 
\tablefootmark{a2} & $N_{Bolo}$ & 8 & 11 & 12 & 12 & 3 & 3 \\ 
\tablefootmark{a3} & $c_{WN}$ \hfill [$\mu$K\,degree] & 1.6 & 0.9 & 1.4 &  5.0 & 70 & $1180$\\  
 \hline 
\tablefootmark{b1} & $\theta_{S}$ \hfill [arcmin]& 9.53 & 7.08 & 4.71 & 4.50 & 4.72 & 4.42 \\ 
\tablefootmark{b2} & $\Delta \theta_{S}$ \hfill [arcmin] & 0.10 & 0.12 & 0.17 & 0.14 & 0.21 & 0.28 \\ 
\tablefootmark{b3} & $e_S$ & 1.20 & 1.03 & 1.13 & 1.10 & 1.17 & 1.35 \\
\tablefootmark{b4} & $\Delta\alpha_S$ \hfill [degree] & 0.80 & 2.08 &  0.28 & 0.28 & 0.13 &  0.07 \\   
\hline 
\tablefootmark{c1} & $\overline{\theta_{M}}$ \hfill [arcmin] & 9.88 & 7.18 & 4.87 & 4.65 & 4.72 & 4.39 \\  
\tablefootmark{c2} & $\sigma_{\theta_{M}}$ \hfill [arcmin] & 0.04 & 0.02 & 0.03 & 0.04 & 0.06 & 0.05 \\  
\tablefootmark{c3} & $\overline{e_{M}}$ & 1.15 & 1.01 & 1.06 & 1.05 & 1.14 & 1.19 \\  
\tablefootmark{c4} & $\sigma_{e_{M}}$ & 0.02 & 0.01 & 0.02 & 0.02 & 0.03 & 0.05 \\  
 \hline 
\tablefootmark{d1} & CMB relative calibration accuracy & $\simlt$ 1\% & $\simlt$ 1\%  & $\simlt$ 1\%  &  $\simlt$ 2\%  & & \\
\tablefootmark{d2} & CMB absolute calibration accuracy & $\simlt$ 2\% &  $\simlt$ 2\% & $\simlt$ 2\% & $\simlt$ 2\% & & \\
\tablefootmark{d3} & FIRAS gain calibration accuracy  & & & & & $\sim 7$\% & $\sim 7$\% \\  
\tablefootmark{d4} & FIRAS zero point uncertainty \hfill [MJy sr$^{-1}$]  & 0.8 & & & 1.4 & 2.2 & 1.7 \\  
 \hline 
\tablefootmark{e1} & $F_U$ \hfill  [MJy sr$^{-1}$/m$\mathrm{K_{CMB}}$] & $2.42\  10^{-1}$ & $3.69\ 10^{-1}$ & $4.81\ 10^{-1}$ & $2.88\ 10^{-1}$ & $5.83\ 10^{-2}$ & $2.24\ 10^{-3}$ \\  
\tablefootmark{e2} & $\Delta F_U$ & $\simlt$ 1\% & $\simlt 1$\%  & $\simlt 1$\% & $\simlt 1$\% & $\simlt 1$\% & $\sim 3$\%  \\  
\tablefootmark{e3} & $C(\alpha), \alpha=-2$ &  $1.011$ & $1.025$ & $0.999$ & $0.997$ & $0.998$ & $1.011$ \\   
 	& $C(\alpha), \alpha=0$ & $0.999$ & $0.985$ & $1.009$ & $1.011$ & $1.012$ & $0.999$ \\    
 	& $C(\alpha), \alpha=1$ & $1.008$ & $0.980$ & $1.027$ & $1.031$ & $1.035$ & $1.007$ \\  
 	& $C(\alpha), \alpha=2$ & $1.027$ & $0.985$ & $1.053$ & $1.060$ & $1.068$ & $1.024$ \\  
\tablefootmark{e4} & $F_{CO}$ \hfill [$\mu\mathrm{K_{CMB}} / \mathrm{K_{RJ}} \kms$] & $14.2 \pm 1.0$ & & $44.2 \pm 1.0$ & $171.0 \pm 6.0$ & & \\
 \hline 
 \end{tabular}
 } 
 \tablefoot{\newline
 \tablefoottext{a1}{Channel map reference frequency, and channel identifier.} \\
 \tablefoottext{a2}{Number of bolometers whose data was used in producing the channel map.} \\
 \tablefoottext{a3}{This estimate of the small scale noise in the maps
   comes from the average level between $\ell =100$ and $\ell=1000$ of
   the power spectra of the Jackknife map (1$^{st}$ versus 2$^{nd}$
   half of rings), with 40\% of it masked, of Fig.~\ref{fig:mapmaking_noise_spectra}. } \\   
 \newline
 \tablefoottext{b1}{Average FWHM of the \emph{scanning beam},
   $\theta_S$, determined on planets (Mars); it is obtained by 
   unweighed averaging the individual detectors FWHM. Each FWHM is
   that of the Gaussian beam which would have the same solid angle 
   ($\Omega_S = 2\pi(\theta_S/2\sqrt{2\ln 2})^2$) as that determined
   by using a full Gauss-Hermite expansion on destriped data. FWHM
   from straight Gaussian Elliptical fit would rather give 9.45, 7.01,
   4.68, 4.45, 4.48 \& 4.22 arcmin. } \\ 
 \tablefoottext{b2}{Uncertainty in determining the scanning beam FWHM,
   $\theta_S$. This conservative uncertainty is derived through the
   dispersion of results of several methods.}\\ 
 \tablefoottext{b3}{Ellipticity of the scanning beam. The formal
   uncertainty on these numbers is quite small, always smaller than
   1\% 
   , but it is likely  misleading. This formal uncertainty is defined
   as the square root of the second diagonal element of 
   $E_s$, the 3x3 covariance matrix for fitting ($\theta_S, e_S, \alpha_S$) to the data. } \\ 
  \tablefoottext{b4}{Typical uncertainty in determining the direction
    of the scanning beam elongation. (Square root of the third
    diagonal 
   element of the covariance matrix $E_S$ defined in note $b3$.)} \\
 \newline
 \tablefoottext{c1}{Average FWHM of the \emph{effective beam} at map level,
   $\theta_M$. This gives the typical width of the beam, as an average
 over 3000 locations in the map of the local effective beam resulting from 
 combining many measurements per pixel. It tends to increase the beam
 FWHM by a few percent with respect to the scanning one.} \\
 \tablefoottext{c2}{Standard deviation of the variation of the FWHM of the effective
   beam at the map level (at the location of the \Planck\ ERCSC\ sources), assuming the 
   input scanning beam is exact. This line
   shows that the variation of the effective beam FWHM from one
   location to the other is smaller than the uncertainty on the scanning
   beam FWHM quoted above. } \\
 \tablefoottext{c3}{Average ellipticity of the effective beam at the
   map level. }  \\
 \tablefoottext{c4}{Standard deviation of the ellipticity of the effective beam at the map
   level. } \\
 \newline
 \tablefoottext{d1}{Relative calibration accuracy between frequency channels. Estimate based on
   cross-correlation between maps of the CMB component, see details in 
  Sect.~\ref{sec:intercalib_cmb}. } \\
 \tablefoottext{d2}{Estimate based on simulations of the calibration
   procedure on the solar system kinematic dipole. This assumes \WMAP\
   determination is exact, and perfect data (\eg no LFER-induced
   systematics).} \\ 
 \tablefoottext{d3}{Estimate of the systematic error (which dominates the error budget) through the 
   dispersion of estimates obtained in different regions of the sky. } \\
 \tablefoottext{d4}{Overall error on the map zero point, for a 
   $\nu I_\nu =$ constant spectrum.} \\
 \newline
 \tablefoottext{e1}{This unit conversion factor, $F_U$, relies on the knowledge of the
   spectral bands.} \\
 \tablefoottext{e2}{Uncertainty estimate of the unit conversion factor.
 } \\
 \tablefoottext{e3}{Colour correction $C(\alpha)$, assuming a sky 
   emission of the form $I_\nu \propto \nu^\alpha$. This factor $C$ is the
   one by which the flux of the source need to be multiplied in order to
   compare with the HFI map or catalogue value. All colour corrections
   quoted are good at the 2\% level or better. 
 } \\ 
\tablefoottext{e4}{CO correction obtained as described in
  Sect.~\ref{sec:COcc}. These factors corresponds to the CO lines J\,1-0, 
  J\,2-1 and J\,3-2 (respectively) at 100, 217 and 353\GHz.}  
} 
\end{table*}

\subsection{Sensitivity}

The numbers given in line a3 of the table directly indicate the white noise
contribution to the \rms\ in pixels of 1 degree on a side. They show that, with
the current processing, the white noise level is slightly larger (or smaller) 
than the `Blue Book'  values \citep{planck2005-bluebook} (renormalised by duration \ie by
$\sqrt{14/10}$) by  20\% (100 \GHz), 11\% (143 \GHz), 8\%  (217 \GHz), 
25\% (353 \GHz), 16\% (545 \GHz)  and  -35\% (857 \GHz). 
These numbers pertain to the actual maps and do not include
data from the RTS bolometers ($\sim 8$\% of the 143\GHz\ data, and 25\% at 545 and 850\GHz), or from the 
depointing manoeuvres or glitch flagging  (another $\sim$ 20\% data loss in all bolometers).  
One should also bear in mind that any comparison in Kelvin at the highest frequencies is
very sensitive to spectral band assumptions.
     
\subsection{Angular response}

\paragraph{Scanning Beams:} are determined from planet scans (lines b1--b4). They include the effect of the optical beam, of the electronic detecting chain, and of the TOI processing pipeline. This is an important intermediate product, although it is usually not relevant for astrophysical applications.

\paragraph{Effective Beams:} provide the response of a map pixel to the sky (lines c1--c4). Compared to the
 scanning beams, they further account for the combined effect of the scanning strategy and additional data processing. 
Different applications need different levels of accuracy and detail, from the mean FWHM of a symmetrical Gaussian description (line c1) to an actual Point Spread Function at each specific map location.  For the time being, the dominant source of uncertainty remains the systematic uncertainties in the scanning beams. Mean ellipticities are given in line c3.

\subsection{Photometric accuracy}

The photometric calibration of the 100 to 353\GHz\ channels relies on
the Solar dipole. Comparing the common CMB component at these frequencies 
shows that the relative accuracy between these channels is better than
2\%, and is more probably at the per cent level. For the two highest
frequencies, we use a comparison with \FIRAS\ and find systematic
variations as large as 7\%, while the zero points of the maps are  better
than 2 MJy\,sr$^{-1}$ (Lines d1 --d4).

\subsection{Spectral response and conversions}

Line e1 provides a unit conversion coefficient, $U$, based on the
knowledge of HFI spectral transmissions, which is accurate to better
than one per cent. Other lines provide color correction factors for
different power-law emission. The last line provides an estimate of
the correction factor required  to perform CO corrections (line
e4). We now turn to details of its derivation and use.

\label{sec:CO}

\begin{figure*}[!htbp]
	\centering
	\includegraphics[width=\textwidth]{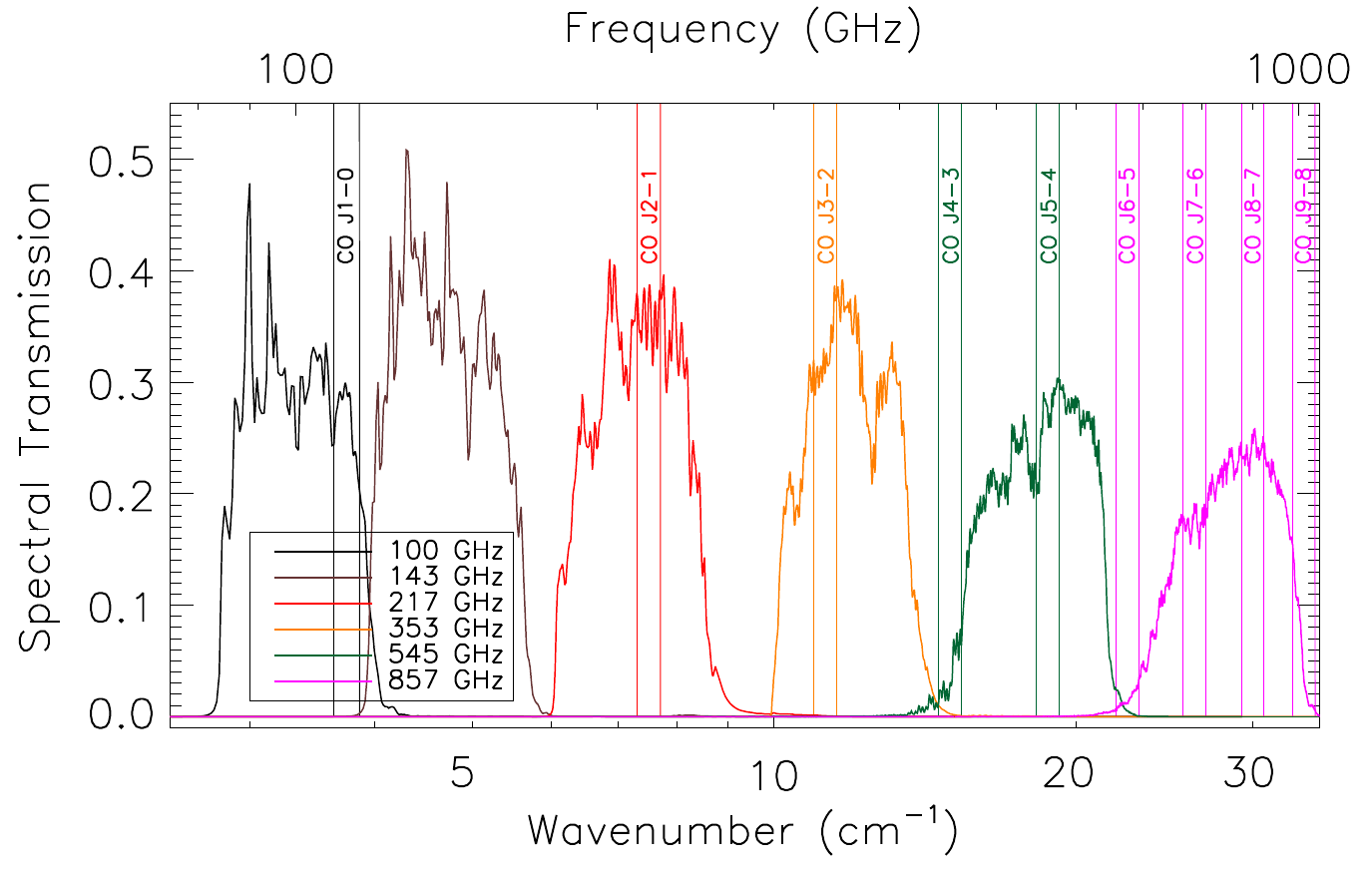}
	\caption{The average spectral response for each of the HFI frequency bands. The vertical
	 bars represent the spectral regions of CO transitions and are interpolated by a factor of
	 $\sim 10$.}
	\label{fig:AvgBPs} 
\end{figure*}

 Figure~\ref{fig:AvgBPs} shows the average band passes for each of the
 HFI frequency bands. The vertical bars show the rest
 frame frequencies of various CO-transitions. Appendix~\ref{Annex:CO}
 details how ground data has been reprocessed to allow
 computation of the CO correction factor described below.

A CO brightness temperature equivalent to a given CMB flux is determined by equating the integrated intensity of each.  For the CMB, the intensity is taken to be 
$\Delta T_{\mbox{\tiny{CMB}}} (\partial B_\nu /\partial T)|_{T_{\mbox{\tiny{CMB}}}}$, 
similar to the approach used in colour correction.  The CO flux is
expressed as the product of Rayleigh-Jeans temperature and the spectral
line width in velocity units, \ie\ K$_{RJ}$\kms. For CO regions of
interest to HFI, \ie the ones with excitation of the lowest rotational
CO transitions, the CO gas temperatures are such that a Doppler
broadening line profile may be assumed with $\nu \approx \nu_{CO}(1 -
v/c)$ ($v \ll c$). The CO brightness temperature is defined in terms
of an equivalent temperature for a Rayleigh-Jeans black body.  Thus,
the integrated CO flux may be expressed as $\int \Delta T_{CO}
\tau(\nu) (\nu_{CO}/c) (\partial B_\nu /\partial T )|_{[RJ,~T_{CO}]}
dv$.  Furthermore, as the CO transitions occur at discrete
frequencies, with a Doppler line-width much less than the transition
frequency ($\sim 10^6~$Hz versus \ $\sim 10^{11}~$Hz), and much
narrower than the available knowledge of the HFI detector spectral
response ($\sim 10^8~$Hz), the velocity distribution of CO intensity
may be approximated by a delta function at $\nu_{CO}$.  Therefore, the
conversion between CO brightness temperature and CMB temperature may
be expressed as follows 
\begin{equation}
\label{eq:COcc2}
F_{CO} = \displaystyle\frac{ \tau(\nu_{CO})(\partial B_\nu / \partial
  T)|_{[RJ,~T_{CO},~\nu_{CO}]}(\nu_{CO} / c)} 
{\int \tau(\nu) (\partial B_\nu / \partial T)|_{T_{\mbox{\tiny{CMB}}}} d\nu}\times 10^6 .
\end{equation}
Table~\ref{tab:summary} lists the average CO conversion factors (in $
\mu\mathrm{K_{CMB}}$ per $\mathrm{K_{RJ}} \kms$ unit) and the
associated uncertainties, determined using Eq.~\ref{eq:COcc2} for the
main CO transitions of concern; similar data are available for the
individual HFI detectors and the other transitions within HFI
bands. 

Statistical errors are computed for each of the detectors from the
respective spectral bandpass uncertainties assuming Gaussian errors.
The systematic errors account for spatial variations of those
coefficients due to both spectral mismatch between detectors and non
uniform weighting across the sky. Notice that the 
uncertainties are of the order of 10~\% and therefore of the same
order as the calibration errors on the currently available templates,
which are evaluated to be in the 10~--~20~\% range
\citep{dame2001,onishi2002}.   

Typically, it is found that in large molecular clouds the relative contribution
  of the CO line to the 100~GHz band is of the order of $10\,\%$ (up to
  $30\,\%$ in some locations).


 \subsection{Conclusions}

This paper has covered the processing completed to date to enable early scientific analysis 
and produce the \Planck\ Early Release Compact Source Catalogue (ERCSC).  
Many aspects will be improved in later 
data passes, using the  knowledge gained from this round of analysis and further rounds using
more data. Examples of possible improvements include an improved deglitching by further tuning
of the method we use (in particular for PSB-b and SWB bolometers), a better knowledge of the
effective beams through all of its components (in particular the instrument temporal transfer
function, and the scanning beams derived from more planet scans),  an improved pointing solution, 
a further reduction of 4K line
residuals, a further reduction of the low frequency noise, the use of the orbital dipole as a primary
calibrator, modelling the zodialal light contribution and far side
lobes of the beams, and quantifying polarization systematics. 

At this stage,
\Planck\ data appears to be of high quality and we expect that with further refinements
of the data processing we should be able to achieve, or exceed,  the science goals outlined in
the  `Blue Book' \cite{planck2005-bluebook}.


\begin{acknowledgements}
 
 A description of the \Planck\ Collaboration and a list of its members with the technical or scientific activities they have been involved into, can be found at
\url{http://www.rssd.esa.int/index.php?project=PLANCK&page=Planck_Collaboration}.

\end{acknowledgements}


\allearlypapers
\bibliographystyle{aa}
\bibliography{Planck_bib,HFIDPC_bib} 


\appendix

\section{Overview of the DPC Infrastructure} \label{Annex:infra}

The HFI Data Processing Centre can be thought of as a centralized
backbone providing hardware and software infrastructures to a
relatively large number of geographically distributed groups of
developers and other R\&D groups in the HFI and LFI core teams. This
appendix provides brief overviews of the main HFI infrastructures
elements.

\subsection{Code and configuration management} 
The modules in the pipeline have been programmed by the scientists and
engineers of the HFI DPC in C, C++, Fortran and Python. IDL is also available
and used for post-processing,  analysis, and  for interactive work.  All codes
are shared and tracked in a common versioning system (CVS), and are grouped
into \emph{packages} which can be tagged  with their version
number.  In addition to this versioning system, a set of scripts based on
CMT\footnote{\url{http://www.cmtsite.org/}} are used to define and build
(executable) code releases based on a specific tag of each CVS package
included.  Each code release includes basic libraries (\eg\ to access the
data), modules (basic executables), and pipelines (scripts that execute many
modules).  CMT provides a build system (based on a set of \emph{make} rules)
that ensure that a given code release is built in a coherent manner, \ie\
tracking the library dependencies and versions, including the versions of the
compiler and other external libraries.  Several code releases are maintained,
but old releases are deleted.  Developers can build  a local (edited)
version of any package but using the libraries from the release. They can thus
test their code before committing it to CVS, tagging it, and
finally \emph{announcing} it so that it will be included in the next release.
Code releases are built at irregular intervals (daily to weekly) depending on
needs (new features, corrections, bug fixes).

\subsection{Data management -- \HFIDMC} 
All data after L1 processing are stored in the HFI Data Management Component
(\DMC ) database, which is therefore the reference database. In fact, only the
metadata are effectively stored in a 
PostgreSQL\footnote{\url{http://www.postgresql.org/}} database. The data
themselves are stored on disk as binary files of a machine-dependent size
optimized for read and write speed.  Data access is possible only via queries
of the database. This induces a bottleneck in the system when
many modules need simultaneous access to a given data object, and thus query
the database at the same time.  Different technical solutions, including the
quick spawning of part of the database have been implemented to alleviate this
problem.  For massively parallel modules, a parallel query layer based on the MPI
library\footnote{\url{http://www.lam-mpi.org}} has been developed.   The
system is able to sustain up to 600 concurrent operations on the same database
object without noticeably slowing down due to the centralized database. 
Without our optimization, the centralized database model would develop a
bottleneck after a few concurrent operations on the same object (typically
the number of available cores on the server). With our
modifications, a central metadata handling system can provide
efficient  object management over a parallel system and also control of
concurrent access properties such as locking.

To optimise read and write rates, an \HFIDMC\ object can be stored in
different physical locations (node memory, node local disk, cluster disk array,
etc.). The use of local memory or disk is used intensively in the Monte-Carlo
pipeline  to avoid the overheads of transferring data to cluster
disks. In this way pipelines can read/write many TB of (temporary) data
very rapidly without using resources other than those of the nodes on which
they are running.
                                                   
The metadata stored in the database proper consists typically of generic file
access information (name of the creator, date of creation, data
rights, history) which is analogous to the fits file metadata (for example for
\healpix\ maps, the ordering, Nside...), as well as the list of all access
rights to the data, with a link to the pipeline, module or
interactive session that is responsible for that access.

Pipelines, modules, or interactive (IDL or Python) sessions that need access to
the data must use a dedicated library which interfaces with the database, and
stores a trace of the access as a dedicated, pure metadata object in the
database. Hence, module parameters, the version of the pipeline or module
as well as the release within which it has been compiled and run is stored in
the database. Log files are maintained external to the database, as for the data,
but are referenced therein. 

All of this information can easily be queried using command line
interfaces, or through a dedicated web interface.

Thus, at any stage, the full history of any data hosted in the
database can be obtained (using for example a dedicated web page).
Furthermore, computations applied to any data can be
reproduced simply by re-running the different pipelines or modules
(in the same version, with the same release) with their parameter
files,  provided a released version of the code has been used.  
    
\subsection{The \texorpdfstring{\IMO\ database}{IMO database} } \label{sec:IMO}
  
The purpose of the \IMO\ (Instrument Model) is to store the current
knowledge about the HFI behaviour in a unique location within the DPC. The
\IMO\ is necessarily restricted in complexity and is tailored specifically 
for use in processing the data. The \IMO\ is:
\begin{itemize} 

\item a database of the fixed parameters of subsystem models;

\item complementary to TOIs and Maps in the \DMC;

\item DPC oriented; \ie  a simplified (effective rather than physical) instrument model
      for the  data processing and simulations;

\item accessible via modules in C, Fortran, Python and IDL ;

\item browsable on the internet and can be exported to xml format; 

\item maintained as a consistent unit (global changes increment the 
major version counter). 

\end{itemize} 

Note that `transfer functions' (see below) in L1 use the
same \IMO\ than other procesing levels. This saves storage as only raw data
are stored and on-the-fly conversions are made to retrieve objects in physical
units. It also ensures that L2 and L1/QLA use the same functions and the same
parameters.

The \IMO\ contains either direct values (e.g. a detector gain, a FWHM) with
 uncertainties and units, or  links to \DMC\ objects
(e.g. bandpass tables, beams maps).  There is one distinct \IMO\ for each
`instrument' (CQM, PFM, Flight).  Although limited in scope, the
current reference version for the flight instrument contains about 17,000 values.

Finally, one should note that it is possible to create sets of \IMORealisation(s)
for Monte-Carlo purposes.  Each Monte-Carlo iteration use one version of
an \IMORealisation\ attached to the same \IMO\ object.  Simulations can use
\IMORealisation\ information if  specified, otherwise they use the 
\IMO.  An \IMORealisation\  therefore provides a way to redefine,  at each
Monte-Carlo iteration, some instrument properties based on the same
global instrument definition.

\subsection{Common libraries} \label{sec:comlib}

Apart from the IO library interfacing programs with the database, a
few common libraries are included in the HFI releases. Most of them are
well known mathematical libraries (FFTW\footnote{\url{http://www.fftw.org/}},
NAG\footnote{\url{http://www.nag.co.uk/}},
LAPACK\footnote{\url{http://www.netlib.org/lapack/}},
GSL\footnote{\url{http://www.gnu.org/software/gsl/}}). We are
also including in our releases classical IDL libraries such as the
Astronomy User's
Library\footnote{\url{http://idlastro.gsfc.nasa.gov/contents.html}},
as well as  Python packages such as
matplotlib\footnote{\url{http://matplotlib.sourceforge.net/}}, Cython
\footnote{\url{http://www.cython.org/}}, numpy and
scipy\footnote{\url{http://scipy.org/}}.  Finally, we are using \healpix\footnote{\url{http://\healpix.jpl.nasa.gov}} \citep{gorski2005}, 
its Python wrapper healpy\footnote{\url{http://code.google.com/p/healpy/}} 
as well as  libpsht\footnote{\url{http://sourceforge.net/projects/libpsht}}  \citep{reinecke2011}.

The HFI DPC also developed and maintains three specific libraries:   

The first  is a set of \emph{transfer functions}, interfaced with
the database, that can be automatically applied to data at read
time. Thus, many virtual views of the data can be built whilst saving disk
space. These different views can be simple unit changes, or more
complex combinations combining and processing several objects (\eg\ a
voltage from a stored resistance and a current);
  
The second  is a \emph{pointing library} which is used to
compute bolometer pointings. Only the attitude of the satellite is
stored in the database. Pointings for each individual bolometer are
computed on the fly, combining the attitude with the location of the
detectors in the focal plane reference frame. At low
computing cost, the library can also change the reference frame into
which the pointing is expressed, or the convention used to express it
(\eg\ Cartesian coordinates or RA and Dec).  The library is
 interfaced with the database so that the locations of the
detectors in the focal plane are retrieved from the \IMO\ directly. The
definition of this pointing (as a maximum of some analytical
approximation of the optical beam) can be easily changed;
   
Finally, the third library is dedicated to the computation of the
\emph{orbital dipole}. It is also interfaced with the database, and has
directly access to the data describing the orbit of the satellite.

\subsection{Data flow management -- \thinc} 

Pipelines represent an important part of the DPC code. They are
collections of modules (or other pipelines) linked by their dependency
relations. To pipe the execution of computational tasks,
modules and pipelines must be described by a `prototype', \ie\ the
definition of the content of their parameter files along with rules
to be imposed on those parameters (\eg\ that one parameter must be bigger
than another, or that a list must have the same number of elements
as another). These prototypes also allow for automatic documentation as
well as automatic production of template parameter files or pipelines.

The \thinc\ tool is a Python library interfaced with a runtime
environment that allows one to describe pipelines and execute
them. Pipeline descriptions are in fact Python scripts interfaced
with the \thinc\ runtime and the database. The script sets the content
of the parameter file of each executable that is part of the
pipeline. The runtime deduces the data flow relationship from those
parameters looking for corresponding output and inputs, takes care of
the submission of the executable to the batch system, and monitors the
correct execution of each computational task. Since pipelines are
Python scripts, direct access to the data is also possible within the
script (for example, reading a database object that has been produced
by a module and deciding on the execution of the rest of the pipeline
accordingly). The runtime also detects such access, and makes sure
that the data flow is still correct. Finally, all information related
to the pipeline (parameter files, log files...) is committed to the
database and can be queried later on, for example,  to check whether the
results of a given pipeline are consistent between releases.
   
The \thinc\ tool has several execution models that can be selected by
users. Indeed HFI-DPC infrastructure needs to and is used on various platform
types, massive computations need adaptations to the local hardware.
   
Finally, in one year (Oct-2009 to Oct-2010), using \MagiqueIII\ (one of
the dedicated computers of the HFI-DPC, see below) 559,000 processes were executed,
reading or writing 7.5 Peta-Bytes, using 450,000 CPU hours.
   
\subsection{Hardware} 

HFI-DPC has a dedicated computing facility hosted by the Institut d'Astrophysique de Paris
called \MagiqueIII. \MagiqueIII\ is a cluster consisting of 284 
Xeon processors/1128 cores built by IBM. The theoretical peak performance 
of the whole cluster
is around 13 TFlops. A GPFS parallel file system of around 100 TB 
usable storage is available and visible by all the login and compute nodes.

The nodes are interconnected with a high-speed (20 Gbit/s) point-to-point
InfiniBand switch. This medium-sized computing facility is used for operation
and most common analysis.  This hardware is used for the day-to-day analysis
and  provides enough processing power for preliminary MonteCarlo loops.

The HFI-DPC infrastructure is also available at several other computing
centre (CCIN2P3, Darwin at CPAC, NERSC).  These larger machines are available
for massive computations, in particular large-scale 
Monte-Carlo simulations.

\subsection{TOI simulations} \label{sec:sims}
%
The simulation methodology consists of two parts:  LevelS-Core, focused on the 
calculation of the sky power falling on detectors and LevelS-Desire, which simulates the
HFI detectors response. Both are described below. The simulated detector  response is 
specific to each instrument and requires access to information contained and
updated in the unique centralised database of parameters of the instrument
model, the \IMO.

\subsubsection{LevelS-Core: Sky power infalling on detector} \label{ssec:LSC}
The LevelS-Core is a set of modules (programs) developed 
jointly by  the HFI, LFI and MPA teams and integrated at MPA to compute  the sky power infalling
on Planck detectors \citep{reinecke2006}. 
Its HFI avatar takes advantage of the features of the HFI DMC.

The first pipeline, LSCmission, takes as input a Pre-programmed Pointing
List (PPL, a plain text file containing one fixed-length formatted line
per pointing period of the mission, typically generated by MOC) and translates
it into the HFI DMC
objects containing the mission parameters, namely:
\begin{itemize}

\item the satellite's position, rotation speed, spin axis orientation, start and
end time for each pointing period;
\item precise satellite position and orientation at 1Hz;
\item start and end sample number for each pointing period;
\item a catalogue of Solar System Objects (SSO) positions for each pointing
period.
\end{itemize}

\noindent
And optionally: 

\begin{itemize}

\item a quaternion representing the pointing and attitude of the satellite
at the bolometer's sampling frequency;
\item a vector of the satellite's spin axis in ecliptic spherical coordinates
for each pointing period.
\end{itemize}

The second pipeline, LSCorePipe, takes as inputs the results of LSCmission,
together with: \begin{itemize}

\item maps from the \Planck\ Sky Model;
\item a point sources catalogue from the PSM;
\item beam descriptions in GRASP format (from the \IMO);
\item the spectral response of the bolometers (from the \IMO).
\end{itemize}

LSCorePipe generates TOIs of antenna temperature per bolometer, containing the simulated 
power received from the user's selection of sky components (CMB, orbital dipole,
Sunyaev-Zeldovich effect, planets, point sources and Galactic emission).  We refer to
them as ``sky signal'' TOIs, and they are in units of $\mathrm{mK_{RJ}}$.

\subsubsection{LevelS-Desire: Detectors simulated response} \label{ssec:LSD}

The LevelS-Desire (Detectors simulated response) is a set of modules that
simulates the HFI instrument.  They convolve the sky signal TOIs from the
LevelS-Core with the instrument response to produce TOIs of modulated signal, in
ADUs, as given by the L1. These include: 
\begin{itemize}
\item scientific signal of the 52 bolometer channels;
\item fine thermometers -- 10 channels over the 4K, 1.6K and 0.1K stages;
\item housekeeping,  readout electronics unit.
\end{itemize}
It is divided into two parts that deal separately with the bolometers and
thermometers, and it accounts for the following instrumental effects: 
\begin{itemize}
\item temperature fluctuations of the telescope and  cryogenic stages;
\item non-linear response of the bolometers;
\item temporal response of the bolometers;
\item response of the electronic chain --  gain and filtering;
\item various components of the noise (see below).
\end{itemize}	

Models of HFI response to temperature fluctuations of the telescope and the
three cryogenic stages (4\,K, 1.6\,K and 0.1\,K) have been implemented.  
The behaviour of the detectors has been divided into a non-linear and
a linear time response. These parameters depend on the electronic
set-up of the REU and of the instrument (telescope
temperature, cryogenic stage temperatures, sampling frequency). The
non-linear response is modelled by a polynomial law and gives the
conversion factor from Watts received by the detector to ADU. The time
response effect is computed in Fourier space. The complex filters
contain the electronic time response and the bolometer time response
that is  modelled by two components: one short time constant
(about a few $ms$) and a long time constant (LTC) of a few
$100~ms$. These properties have been integrated into a complete model of
the instrument response SEB (Simulation of the Electronics and
Bolometer) that has been used to provide  the inputs to the Desire
module.

Finally one module is dedicated to the various components of the noise that are
added on the signal:  
\begin{itemize}
\item white noise (REU and JFET noise);
\item convolved noise (phonon and photon noise);
\item glitches;
\item 4K harmonics (electromagnetic coupling  between the 4K cooler and the detectors);
\item RTS noise.
\end{itemize}
These various components are optional and can be switched on or off to test the 
specific impact of each  systematic effect.
The parameters of these systematic effects have been estimated during the various
calibration runs of the instrument.

In addition, to ensure the completeness of the simulated dataset, the
housekeeping (HK) data are written to the \DMC\ database to be read subsequently  by
the TOI processing and L2 pipelines. This pipeline has been written to
work within the DPC environment (\eg\ DMC+ThinC), and uses the 
understanding of the instrument accrued in the  \IMO.

The performance of the LevelS-Desire pipeline was evaluated pre-launch.
Generating one year of  data  needs 2.5 cpu hours for each bolometer and 
2 cpu hours for thermometers with 37 GBytes of data generated for each. This
leads to 170 cpu and 2.7 TBytes storage to generate a simulation of
one year of data for the entire focal plane of the
\Planck\ HFI.

\section{Focal-Plane Measurements} \label{Annex:FPG}

In this appendix we describe the algorithms used to measure and monitor
the focal-plane geometry and the shape of the individual detector
beams. 

\subsection{The algorithm} \label{ssub:the_algorithm}

The pipeline for extracting beam and focal-plane measurements uses a series of
modules linked by the HFI DPC infrastructure (see
Appendix~\ref{Annex:infra}). As input data, the pipeline uses the product of
the TOI processing pipeline (cf.~Sect.~\ref{sec:toip}).  This data has been
calibrated, flagged for glitches, and has had various temporal filtering
applied as described in previous sections.  The pipeline is iterative,
requiring starting values for the focal-plane geometry which are then updated by
the pipeline.

To obtain a first in-flight check of the focal plane geometry prior to the
observation of the planets, we cross-correlated the TOIs of different detectors
to obtain the time lag of the easily identified crossings of the Galactic plane.
These lags were combined with the satellite rotation
rate to obtain the in-scan separations. The results proved consistent with
pre-launch values, taking into account that the measured lags are affected also
by the cross-scan displacement of the detectors and also by their observing
frequency. 

For a given object (planet or other compact source) we first find the rings
for which the focal plane intersects the object. This part of the pipeline is
also capable of removing constant offsets for each ring
(as calculated by the map-making pipeline). From this significantly reduced volume
of destriped data we make a `naive' map (\ie averaging into map pixels
assuming constant noise) around the source (for planets, the map is centred
around the known, moving, position) using the nominal focal-plane
geometry. This allows us to produce visualizations from each detector.

We then fit a parametric beam model either to  these maps, or 
to the destriped time-stream data. The two-dimensional Gaussian is parametrized as
\begin{equation}
	\label{eq:2dGaussian}
B(x_1, x_2) =  \frac{A}{|2\pi\mathbf{\Sigma}|^{1/2}}\exp\left[-\frac12\sum_{i,j=1}^{2}(x_i-{\bar x}_i)\Sigma^{-1}_{ij}(x_j-{\bar x}_j)\right], 
\end{equation}
where $A$ is an overall amplitude, $(x_1,x_2)$ are two-dimensional 
Cartesian coordinates (we project to a flat sky), $({\bar x}_1,{\bar x}_2)
 $ are the coordinates of the beam centre, and the correlation matrix is given by
\begin{equation}
	\mathbf{\Sigma} = \left( \begin{array}{cc}
	\sigma_1^2 & \rho\sigma_1\sigma_2 \\
	\rho\sigma_1\sigma_2 & \sigma_2^2 \end{array} \right)\;.
\end{equation}
Hence, for the Gaussian model, we fit for the parameters $A, {\bar
  x}_1, {\bar x}_2, \sigma_1, \sigma_2, \rho$. These can also be
expressed in terms of the ellipticity, $e$ (defined here as the ratio
between the major and minor axes), and rotation angle $\alpha$, of the
Gaussian ellipse.  

The \Planck\ beams are known to have non-Gaussian features. Hence, we
 model deviations from Gaussianity as a series of elliptical
Gauss-Hermite polynomials as discussed in the \Planck\ context by
\citet{huffenberger2010}. Note that we first fit an elliptical 
Gaussian as above and  then (\ie\ not simultaneously) solve for the
Gauss-Hermite coefficients. The basis functions are defined as 
\begin{equation}
    \Phi_{n_1n_2}(\mathbf{x}) \propto H_{n_1}({x_1'})  H_{n_2}({x_2'}) 
    \exp\left(-\mathbf{x'\cdot x'}/2\right), 
\end{equation}
where $H_n(x)$ is the order-$n$ Hermite polynomial and the primed coordinates
$\mathbf{x'}$ rotates into a system aligned with the axes of the Gaussian and
scaled to the major and minor axes $\sigma_i$ (\ie\ to the principle axes of
the correlation matrix $\mathbf{\Sigma}$).  Note also that the Gaussian
approximation, in general, underestimates the effective area of the beam compared
to the Gauss-Hermite parametrization.


In practice, we have found that the TOI-based code is better suited to
measuring the beam parameters, and the map-based code to the
focal-plane positions (using the result from the TOI code as a first
guess to a Levenberg-Marquardt search). Because the projection to a
two-dimensional coordinate system requires prior knowledge of the
detector positions, we can iterate this procedure to account for any
inaccuracies induced by differences with respect to the nominal
detector positions. In practice, this correction is negligible.

Within the DPC infrastructure, detector positions (and all other
directions relative to the spacecraft frame) are stored and
manipulated as quaternions \citep{Shuster1993}. The translation from
our two-dimensional fits into quaternions complicates the error
analysis somewhat. We represent the error (a difference between two
rotations) as  another rotation. However, because it is an
error, it is  convenient to represent it with quantities that are
both numerically small for small errors and are not subject to
constraints (\eg\ to represent a rotation, the four entries in a
quaternion must be normalized, and a rotation matrix must be
orthonormal); hence we store rotation errors as a vector pointing
along the rotation axis with magnitude given by the rotation angle. We
can then store a rotation error covariance matrix as a non-singular
three-by-three positive definite symmetric
matrix. 

To monitor the focal-plane geometry, we collect the results for all
detectors with a single source and calculate the offset of each
detector position with respect to the nominal (input) position. This
produces a diagnostic of any overall shift in the positions as well as
an  estimate of the noise on the measurement. The results from
individual detectors are then  combined to determine an
overall rotation (quaternion, with errors as explained above) and
scaling with respect to the nominal model.

\subsection{Validation on Simulated Data} \label{sec:results_on_simulated_data}

\begin{figure}[htbp]
	\centering
	\includegraphics[height=1.5in]{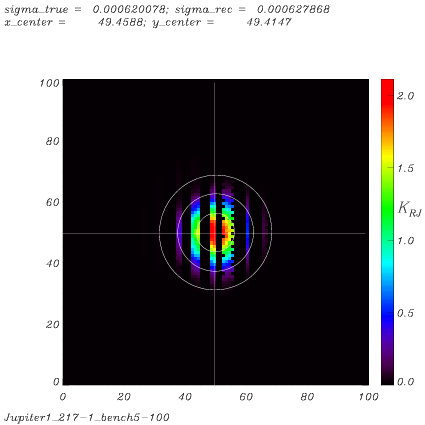}
	\includegraphics[height=1.5in]{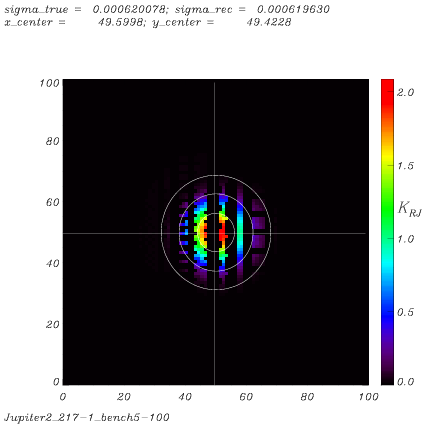} \\
	\caption{Simulated mini-map images of Jupiter for the 217-1 detector, 
         from the first scan (left) and  second scan (right).} 
	\label{fig:maps}
\end{figure}

In Fig.~\ref{fig:maps}, we show simulated maps
of scans of Jupiter for 
the 217-1 detector (see Sect.~\ref{sec:sims} for further information 
on the simulations). 
(Unlike the flight data, these simulations do not include a nonlinear response for Jupiter observations, \textbf{nor glitch residuals, 
but these are small effects compared to the other sources of uncertainties}.)
Vertical striping in the map is due to the
discrete jump of 2.5 arcmin between observation of rings. Hence,
measurements of detector positions and beam shapes in the scan
direction are considerably better than in the cross-scan direction. 

In Fig.~\ref{fig:FPRec-Saturn-bench6} we show simulation results of
scans of Saturn, including realistic noise. The results for the entire
HFI focal plane are plotted as offsets with respect to the actual
detector positions used in the simularions.  For the majority of
detectors, we find agreement to an accuracy of about 10 arcseconds,
except for several conspicuous outliers; these are the multi-moded
horns at 545\GHz, for which the input detector positions were defined
as the points of maximum response, as opposed to the centre of an
appropriately fitted beam shape. Discounting those points, we conclude that
we can  recover the detector  positions within 
the focal plane to an rms accuracy of $\sim4\;$arcsec.

\begin{figure}[\begin{figure}[htbp]
    \centering \includegraphics[width=3in]{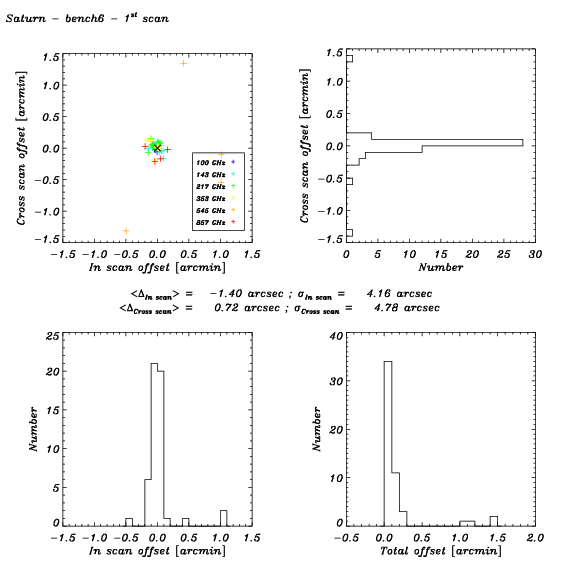} \caption{Recovery
      of the focal plane position in the in-scan and cross-scan
      directions using simulated scans of Saturn.  Upper left shows
      individual detectors, other three show histograms, as
      marked. } \label{fig:FPRec-Saturn-bench6}
  \end{figure}

\section{Detector noise measurements\label{Annex:detnoise}} 

  Our goal is to characterize the detector noise, as accurately as possible, starting from the time-streams produced by the TOI preprocessing pipeline.  We have therefore developed a diagnostic toolbox to monitor, among other things, localized auto and cross-power spectra of noise to provide information about the mean white noise level per detector, low frequency drift behaviour, narrow lines (caused, for example, by microphonic noise from the 4K cooler) and common noise modes between detectors.  To monitor time variable trends in the noise properties, the estimates are made on a ring-by-ring basis.  The noise estimation pipeline consists of three different modules: \begin{itemize} \item noise time-stream estimation from the input data\footnote {By `input data' we mean the output of the TOI processing pipeline, namely the total signal (sky + noise) time-stream calibrated in units of absorbed power} time-stream, on a ring-by-ring basis; \item empirical auto and cross-spectra estimation on predefined `stationary zones' (in practice, the rings); \item fitting to a parametric noise model to provide, \eg\ the Noise Equivalent Power, knee frequency, and the spectral index of the low-frequency noise component.  \end{itemize} 

\subsection{Theory}
%
\subsubsection{Noise power statistics} \label{sec:Noise-power-statistics} 

\cite{ferreira2000} show that the joint maximum likelihood estimation of the signal and noise power spectrum can be obtained iteratively.  Let us assume that the data time-stream, $d_{t}$ is related to the signal and the noise via \begin{equation} d_{t}=s_{t}+n_{t}=A_{tp}T_{p}+n_{t}, \end{equation} and $\mathbf{A}$ is the projection matrix relating the sky signal $\mathbf{T}$ in the map pixel $p$ to the signal detected in the time-stream at time $t$ (\ie\ the pointing matrix). The noise is assumed to be a Gaussian stationary process with covariance matrix $N_{tt'}=N(t-t')$.  If the signal and noise are assumed to be independent (which is a good approximation except perhaps for very strong signals where the linearity of the detector breaks down), then the joint posterior probability on signal and noise power can be factorized as follows: \begin{eqnarray} P(T_{p},\tilde{N}(\omega)|d_{t}) &\propto& P(\tilde{N}(\omega))
  \times P(T_{p}) \times \nonumber \\
  & & \qquad P(d_{t}|\tilde{N}(\omega),T_{p}), \end{eqnarray} where
the last term is just the joint likelihood $\mathcal{L}$ of the signal
and noise power spectrum $\tilde{N}(\omega)$, where \begin{equation}
  N(t-t') = \int \frac{\mathrm{d}\omega}{2\pi}\tilde{N}(\omega)
  \mathrm{e}^{-\mathrm{i}\omega(t-t')}.  \end{equation} The joint
likelihood can be expressed in the following way:\begin{eqnarray}
  -2\ln\mathcal{L} & = &
  \ln|\mathbf{N}|+(\mathbf{d}-\mathbf{s})^{T}\mathbf{N}^{-1}(\mathbf{d}-\mathbf{s})\ \nonumber\\
  & \simeq &
  \sum_{k}\ln\tilde{N}_{k}+|\tilde{d}_{k}-\tilde{s}_{k}|^{2}/\tilde{N}_{k},\end{eqnarray}
where the last line expresses the likelihood in terms of the discrete
Fourier modes and is approximate because the noise matrix $\mathbf{N}$
is Toeplitz rather than circulant. Let us now express the derivatives
of the joint posterior, where the noise power is assumed to be
constant within frequency band $\alpha$: \begin{eqnarray}
  \frac{\partial\ln
    P(T_{p},\bar{N}_{\alpha}|d_{t})}{\partial\mathbf{T}} & = &
  (\mathbf{d}-\mathbf{A}\mathbf{T})^{T}\mathbf{N}^{-1}\mathbf{A}, \\
  \frac{\partial\ln
    P(T_{p},\bar{N}_{\alpha}|d_{t})}{\partial\bar{N}_{\alpha}} & = &
  -\frac{1}{2\bar{N}_{\alpha}}\left[ (n_\alpha + 2\nu)-\frac{1}{\bar{N}_{\alpha}}\times \right. \ \nonumber\\
  & & \qquad \left. \ \sum_{k\in\alpha} \sum_p \vert
    \tilde{d}_{k}-\tilde{A}_{kp}T_{p} \vert^{2}
  \right], \end{eqnarray} where $n_\alpha$ is the number of
wavenumbers within the band $\alpha$, and we have assumed a noise
prior of the form \begin{equation} P(\bar{N}_{\alpha})\propto { 1
    \over \bar{N}_{\alpha}^{\nu}}.  \end{equation} Setting these
derivatives to zero, we obtain the following equations that can be
solved iteratively: \begin{eqnarray} \mathbf{T} & = &
  (\mathbf{A}^{T}\mathbf{N}^{-1}\mathbf{A})^{-1}\mathbf{A}^{T}\mathbf{N}^{-1}\mathbf{d}, \\
  \bar{N}_{\alpha} & = &
  \frac{1}{n_{\alpha}+2\nu}\sum_{k\in\alpha}\sum_{p}|\tilde{d}_{k}-\tilde{A}_{kp}T_{p}|^{2}.  \end{eqnarray}
However, here we are not interested in the signal estimate itself and
so we marginalize over all possible values of the signal to find the
posterior distribution on the noise power.  For an idealized ring
scanning strategy, in which the pointing matrix selects spin-harmonic
frequencies, the  a posteriori solution for the noise power
spectrum with signal marginalization can be expressed as a periodogram
estimate on the signal subtracted data time-stream, renormalized to
take into account the projection of the spin-synchronous noise modes.
For a realistic scanning strategy, the projection operator on
spin-synchronous signal modes is however more complicated and the
marginalized posterior is no longer factorized over frequencies. In
practice, we compute the joint posterior maximum, with the
simplification of replacing the maximum posterior estimate of the
signal at fixed noise spectrum with the `naive' (equal-weight) signal
estimate. This is justified because the noise covariance in phase
modes is almost diagonal (because of the specific scanning strategy on
a ring). For the pure diagonal case, the optimal signal estimator and
the naive ones are identical as the signal estimation can be done
independently for each phase mode. Nevertheless, as we discuss below,
the noise power estimates around the spin-synchronous frequencies are
pathological.  

\subsubsection{Unbiased periodogram  estimates} 

In practice, the data time-streams contain flagged data, identifying the presence of cosmic ray glitches, unstable pointing periods, etc. The missing data complicates the estimation of noise.  Our approach here is similar to the estimation of pseudo power spectra: we compute the `raw' periodogram on the estimated noise time-stream, with flagged data set to zero and we perform the same operation on the flagged time-stream (set equal to one in valid regions and zero otherwise). We then compute the Fourier transform of both periodograms, divide the first by the second, and take the inverse Fourier transform.  This gives an unbiased estimate of the underlying power spectrum.  Finally, in their equation (5), \cite{ferreira2000} make the approximation that the noise covariance matrix is circulant whereas it is, in fact, Toeplitz. To correct for this, we replace the simple periodogram estimates by windowed periodogram estimates (\cite{dalhaus1988}). The length of the periodograms is a parameter of the pipeline, as is the overlapping factor between adjacent periodograms. All periodogram estimates are averaged over a `stationary zone', chosen in practice to coincide with a given ring.  The procedure outlined above generalizes easily to the multi-channel case, where power spectra are complemented with cross-power spectra between detector time-streams.  

\subsubsection{Signal estimation procedure} \label{sec:FTR} 


As described in section \ref{sec:Noise-power-statistics}, the iterative optimal signal estimates required to maximize the joint likelihood of signal and noise power can be replaced by a `naive' (flat-weighted) estimate of the signal power to good accuracy.  This is because, to a first approximation, the only non-zero sub-block of the pointing matrix (when expressed as relating time frequencies to phase frequencies) on a given ring is equal to the identity matrix (for the subset of time-frequency modes that are spin-synchronous).  Small deviations from this pointing matrix model leads to second order corrections to the signal estimates.  \footnote{This reasoning implicitly assumes the circular stationarity of the noise on any given ring, whereas the noise covariance matrix is Toeplitz rather than circulant; therefore, we expect that a weighting scheme that takes into account edge effects would results in a slightly more optimal estimate of the signal.}.  In fact, the more significant problem is to find an \emph{unbiased }signal estimate (rather than a minimum variance estimate), especially in the regime where the signal-to-noise is very large (\eg\ on strong sources or when crossing the Galactic plane). A simple phase-binning procedure (corresponding to the idealized case where the sampling frequency is an integer multiple of the spin frequency) is not accurate enough for our purposes. Instead, we assume that the beam smoothing allows us to use a band-limited signal parametrization in phase space, and that the pointing matrix acts as an (irregular) sampler in phase space. The problem, then, is to compute these irregular samples to high accuracy in a reasonable time.  The solution that we have adopted is to use a Fourier-Taylor \citep[see][and references therein]{colombi2009} expansion of the phase Fourier modes to achieve the irregular sampling. Through simulations of different frequency channels, we found that the signal bias using this solution can be made arbitrarily small by increasing the order of the Taylor expansion. In practice, a $4{th}$ order expansion was sufficient to leave no signal residuals in the estimated noise time-streams.

\subsubsection{Parametric estimation of the noise power}

The noise power spectrum estimates in different frequency bins are
expected to be mildly correlated.  However, for wide frequency bins
this correlation is expected to be small. We thus model the likelihood
of a given parametrized model power spectrum as factorized over the
frequency bins. Each (averaged) periodogram estimate is therefore
modelled as $\chi^2$ distributed, where the number of degrees of
freedom is given by twice the number of periodograms used in the
averaging (accounting for the cosine and sine mode for each
periodogram frequency). The model power spectrum is then a simple
scaling factor in the distribution.

The joint likelihood over all frequency bins is then maximized with
respect to the parameters using a Levenberg-Marquardt algorithm.  The
user can choose to include or exclude any given number of frequency
bins from the fit, which is useful if one wants to
compute a simple noise power amplitude (\eg\ a white noise model)
by restricting the fitting procedure to the specific spectral range.

\subsection{Implementation}

The \emph{detnoise} pipeline is implemented in Python using HFI DPC's
\thinc\ library, and it is maintained under CVS in the HL2\_FMpipes
package. A block diagram of the pipeline for a single detector is
shown in Fig. \ref{fig:scheme}. The pipeline consists of three modules:

\begin{figure}[htb]
	\centering
	\includegraphics[width=.5\textwidth]{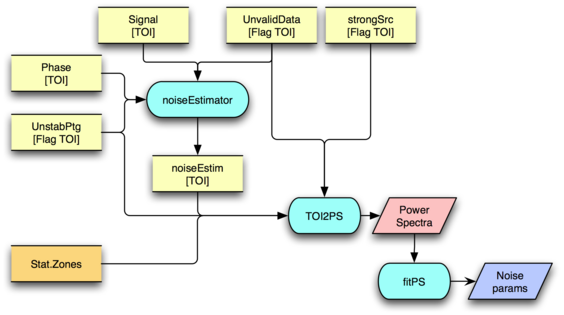}
	\caption{Schematic diagram of the noise estimation pipeline. }
	\label{fig:scheme}
\end{figure}
\begin{itemize}

\item{the \emph{noiseEstim} module produces the TOIs of noise. For each ring,
  the valid (\ie excluding glitched data and data obtained during periods of
  unstable pointing) input data are binned in phase and combined as described
  above to produce a phase-binned-ring, or PBR, containing the best estimate
  of the sky signal component.  The PBR is defined to contain $10\,822$ bins,
  to match the detector sampling time\footnote{Note 
  that since the satellite's spin rate varies by $\pm \sim
  0.05\,sec$ peak-to-peak around the nominal rate of 1 rpm, there is not a
  perfect match between the sampling time, which is fixed in time, and the
  bin size, which is fixed in space.}.  The signal PBR is then subtracted
from the original input TOI by first interpolating it back onto the TOI, and
then subtracting.  This results in  a timeline containing mostly noise;  }

\item{the \emph{TOI2PS} module computes, for each ring, the power
  spectrum of the noise timeline.  Within each ring, power spectra are
  computed over successive chunks of $2^{p}$ samples with, depending on the
  window function, some overlap between successive chunks.  The power spectra
  are averaged, divided by the sampling frequency, and the square root of the
  result is returned as the spectrogram for the ring, which is in units of
  [units of the input noise TOI]$/\sqrt{{\rm Hz}}$.  The length of the final
  spectrograms and the associated frequency resolution as a function of $p$
  is given in Table (\ref{tab:SpLength}).  The value  $p=18$ is used for
  the sample spectra shown below.  A flag TOI is used to specify which data
  samples to skip. This same flag is then used to correct the empirical
  autocorrelation function (Fourier transform of the reversed periodogram)
  by the autocorrelation of the gaps; }

\item{The \emph{fitPS} module fits one of several noise models to the
  ring spectra.  The simplest model is  pure white noise within a given
  spectral range, 
\begin{equation}P(f;\ \sigma)=\sigma^{2}.
\end{equation}  
A more complex model includes low frequency  $1/f$ noise, 
\begin{equation}
P(f;\ \sigma,\ f_{knee},\ \alpha)=\sigma{}^{2}(1+(f_{knee}/f)^{\alpha}), \label{E1}
\end{equation}
  where  $\alpha$ is the slope of the $1/f$ component,
  and $f_{knee}$ is an effective `knee-frequency'. Finally, we have also 
  implemented  a three parameter
  model including photon and phonon noise
   \begin{equation}
   P(f;\,\sigma_{b},\,\sigma_{p},\,\tau)=\sigma_{b}^{2}+\sigma_{p}^{2}/[1+(2\pi\tau
    f)^{2}], 
   \end{equation}  where $\sigma_{b}$ and $\sigma_{p}$ are the white noise
  components due to photons and phonons, respectively, and  $\tau$ is a 
  phonon time constant.

}
\end{itemize}

\begin{table}
	\centering
	\begin{tabular}{l|ll} \hline \hline
		$p$ & length  & $F_{min}$ \\ \hline
		10 & 5.68 sec & 176.2 mHz \\
		12 & 22.7 sec & 44.04 mHz \\
		14 & 1.51 min & 11.01 mHz \\
		16 & 6.06 min & 2.752 mHz \\
		17 & 12.1 min & 1.376 mHz \\
		18 & 24.2 min & 0.688 mHz \\ \hline
	\end{tabular}
	\caption{Length of the data chunks ($=2^p$) on which the power spectra 
  		are computed and  the associated minimum frequency,  $F_{min}$.}
  	\label{tab:SpLength} 
\end{table}

\subsection{Results on simulated data}

The pipeline was tested extensively on simulations that include the full sky
signal (CMB, Galaxy, point sources). The noise of the  HFI instrument 
is modelled by equation \ref{E1}.  The details are
presented in \cite{planck2011-1.10sup}, but a brief
description is given here for completeness.

The results given below were obtained from simulations that used a
mission-like scanning strategy, in which the ring length varied with
time, as shown in Fig~\ref{fig:RingLength}.  The analysis was
performed over rings 3000--7000, indicated by the blue vertical lines
in the figure.  Four detectors at different frequencies were studied, 
all of which showed the behaviour reported below.

\begin{figure}[htb]
	\noindent \centering
	\includegraphics[width=0.45\textwidth]{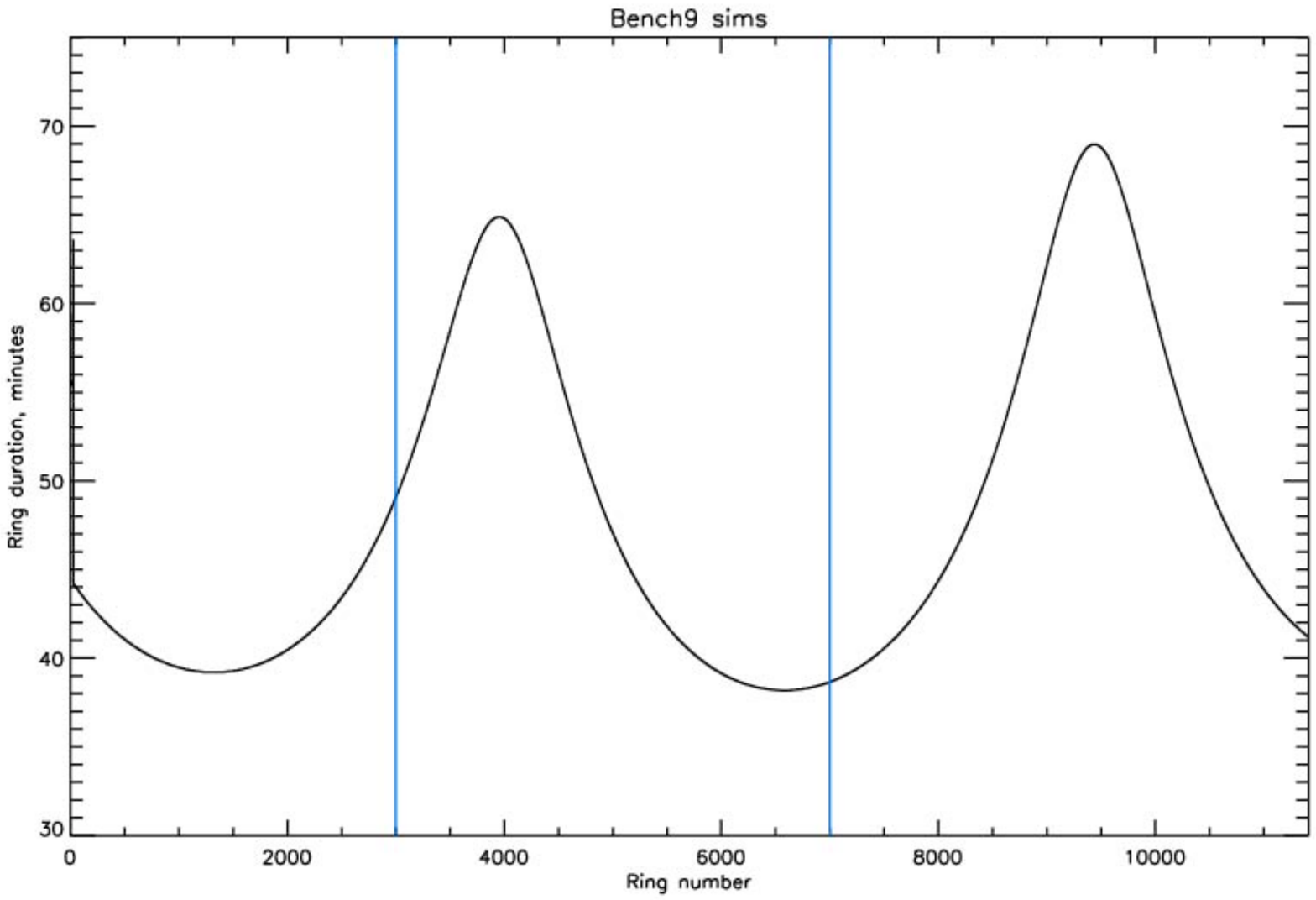}
	\caption{Length of pointing periods in the simulations. The vertical blue lines indicate the range 
		of rings used. }
	\label{fig:RingLength}
\end{figure}

\subsubsection{Effects on the spectra}

The noise estimator can be applied to  the sky+noise timeline or to the pure
noise timeline.  The differences between the two are negligible except when
the timeline includes very strong sources (planets or, at high frequencies, the
central regions of  the Galactic plane). Below we report results on the noise estimator
obtained from the pure noise timeline.

\begin{figure}[htb]
	\noindent \centering{}
	\includegraphics[width=0.5\textwidth]{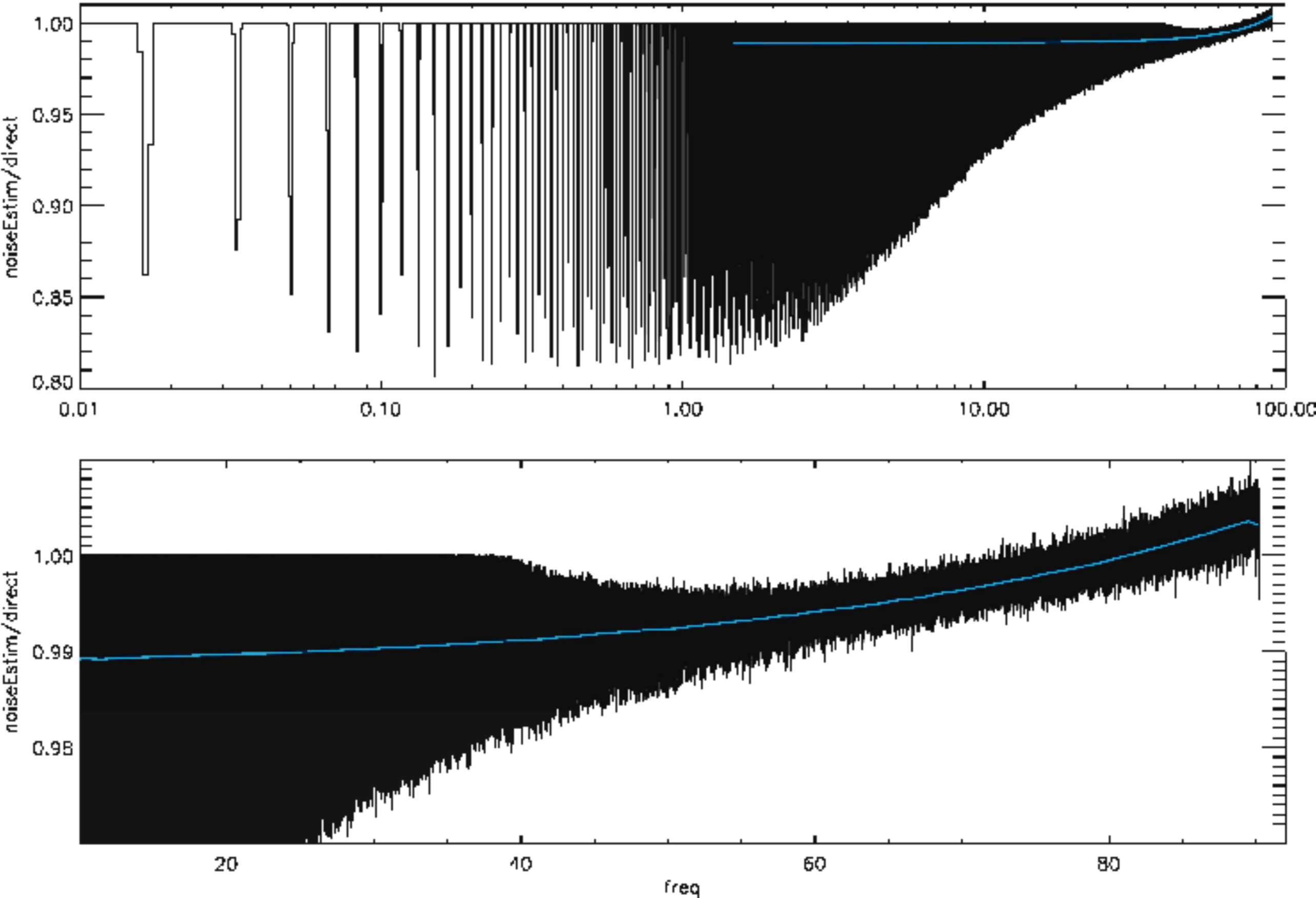}
	\caption{Ratio of the power spectra obtained on the noise
          estimator timeline to that from   
		the input noise timeline. Top: on a logarithmic
                frequency scale , to emphasise the  
		effect at low frequency. Bottom: on a linear frequency
                scale . The blue line is the  
		same spectrum smoothed to 1\Hz\ resolution to show
                the mean value of the bias.} 
	\label{fig:detnoiseBias} 
\end{figure}

This bias is best determined by comparing the spectrum of the noise
estimator to that of the input noise TOI.  To accentuate the
difference, we compare the average (in quadrature) of 4000 ring
spectra. The ratio of the two power spectra is shown in
Fig.~\ref{fig:detnoiseBias} and displays dips at multiples of the spin
frequency.  These dips arise from the particular ring-based scanning
strategy of \Planck.  In the case of an idealized scanning where the
sampling frequency is an integer multiple of the spin frequency, the
signal subtraction operation would remove all power at spin-harmonic
frequencies, and leave other frequencies of the timeline
untouched.  Since this idealized situation is not realized in
practice, the spin-harmonic frequencies do not coincide exactly with
the discrete (time) frequencies of the periodogram, and the dips
(decrement of power around spin harmonics) acquire wings.

The white noise level is measured, in practice, in the 1--3 \Hz\
region of the spectra to avoid the rise due to the $1/f$ component at
lower frequencies and the rise caused by the time constant
deconvolution at higher frequencies. The bias in the white noise level
measured from the 1--3 \Hz\ range is $\sim 1\%$.  This bias can also
be determined by comparing the mean and the median value of the
spectrum in the desired frequency range.  Provided the spectral resolution is
sufficiently high, as is the case in the region selected, the median
will reflect that level of the `continuum' which is identical to the
noise level measured directly on the pure noise timelines.

\subsubsection{Comparison to expected values}

The input noise timeline is computed using detector-specific input NET and
$\epsilon$ (cross-polar leakage) values that are stored in an instrument
database.  Given those input parameters, the  NET measured from the 
noise estimation pipeline  is expected to be:
\begin{equation}
NET_{\rm out} = 0.5 \sqrt{2} \; (1+\epsilon) \; NET_{\rm in}.
\end{equation}
The simulations agree with this expectation to much better than $1\%$ once the bias 
in the noise estimator described above is corrected.

\section{Determination of the Effective Beams} \label{Annex:FEBECOP}

Pure analytical approaches to the effect of asymmetrical beams, as in \cite{fosalba2002} are
illuminating, but are usually of limited accuracy due to a number of simplifying assumptions. This appendix provides some details on the results with the two approach which we have developed to deal in practice with \Planck\ data, FEBeCoP and  FICSBell .

\subsection{The FEBeCoP method}

The `Fast Effective Beam Convolution in Pixel space' algorithm, referred to as FEBeCoP, and associated software to derive the effective beams appropriate to a fixed observing period of the
\Planck\ mission is described in detail in \cite{mitra2010}. FEBeCoP uses a
record of detector pointings and a mathematical model of the scanning
beam (either as an analytical function or a numerical table of the
scanning beam values) to produces a data object of effective beams
computed over a  few hundred pixels centred at every pixel of the sky
map produced from the detector pointings. After this large data object
is computed, a convolution of the full sky input signal with the
spatially varying effective beam becomes a  numerically fast
contraction of the stored beam with the sky model. Further, once the
effective beam data object is computed, the point spread function in
any specific direction on the sky can easily be extracted and utilized
in the analysis of the compact sources detected in the \Planck\ sky
maps.

We present two sets of  results for the effective
beams. The first set is based upon the 
elliptical Gaussian scanning beams and the second 
uses a truncated  Gauss-Hermite series, as
described in section~\ref{ssub:beams}.

\begin{figure*}[!htbp]
	\centering
	\includegraphics[height=0.9\textheight]{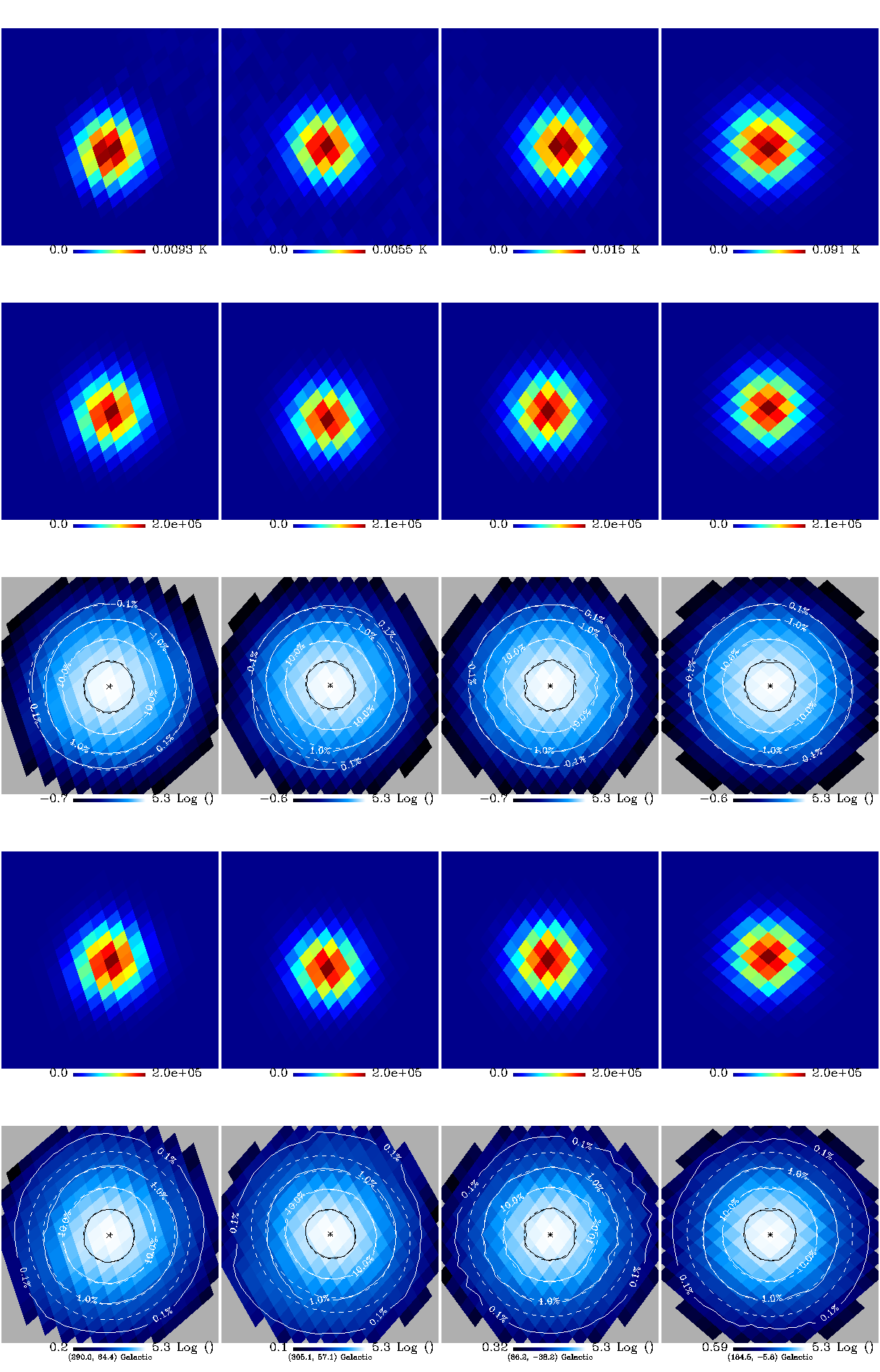}
	\caption{Images obtained at 143\GHz. Each column corresponds to one of
          four sources. a) The first row 
		displays a zoom of the 143\GHz\ map around each source. 
		b) The second row shows the FEBeCoP beam at that
                location, as computed when using 
		the elliptical Gaussian description of the scanning
                beam derived from Mars observations. 
		c) The third row shows on a logarithmic scale the same
                Point Spread Functions (PSF) with superimposed  
		iso-contours shown in solid line, to be compared with
                elliptical Gaussian fit iso-contours 
		shown in broken line.  
		d) The fourth row shows the FEBeCoP beam at that location, as in b),   when 
		using instead the Gauss-Hermite description of the scanning beam in input.
		e) The fifth row shows as in c) the PSF on a
                logarithmic scale, for the Gauss-Hermite 
		input.}
	\label{fig:stamps@143} 
\end{figure*}

\begin{figure*}[!htbp]
	\centering
	\includegraphics[height=0.9\textheight]{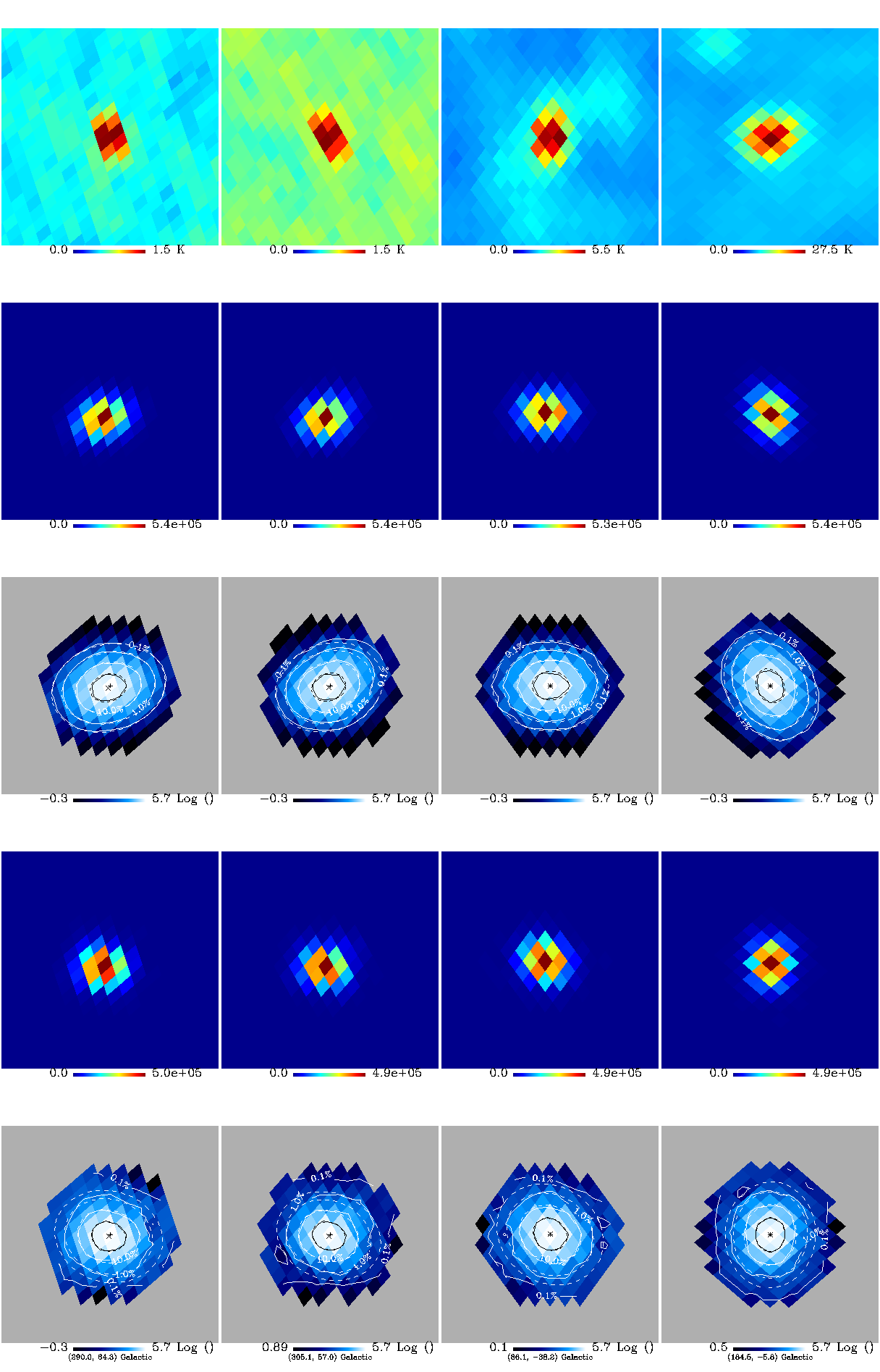}
	\caption{Images obtained at 857\GHz. The arrangement is the same than for the four sources at
		143\GHz\ displayed in fig.~\ref{fig:stamps@143}.}
	\label{fig:stamps@857} 
\end{figure*}

Figs.~\ref{fig:stamps@143} and ~\ref{fig:stamps@857} shows images at
143 and 857\GHz\ of four sources from the ERCSC (detected at all six
frequencies of the HFI) together with the FEBeCoP derived Point Spread Functions
evaluated at the same locations on the sky. At frequencies of 353\GHz\
or less, the main errors in the FEBeCop representation of the scanning
beams are in the low-amplitude outer parts of the effective PSF. The
143\GHz\ PSFs plotted in in Fig.~\ref{fig:stamps@143} represent
\Planck's most sensitive frequency channel. These detectors also have
the most symmetric beams. The multi-moded nature of the 857\GHz\
and 545\GHz\ channels leads to flat-topped beam profiles that are
poorly described by an elliptical Gaussian. This is apparent in
Fig.~\ref{fig:stamps@857} which show that the differences between the
FEBeCop results based on elliptical Gaussian versus Gauss-Hermite fits
are more pronounced than at 143\GHz.

\begin{figure*}[!htbp]
	\centering
	\includegraphics[width=\textwidth]{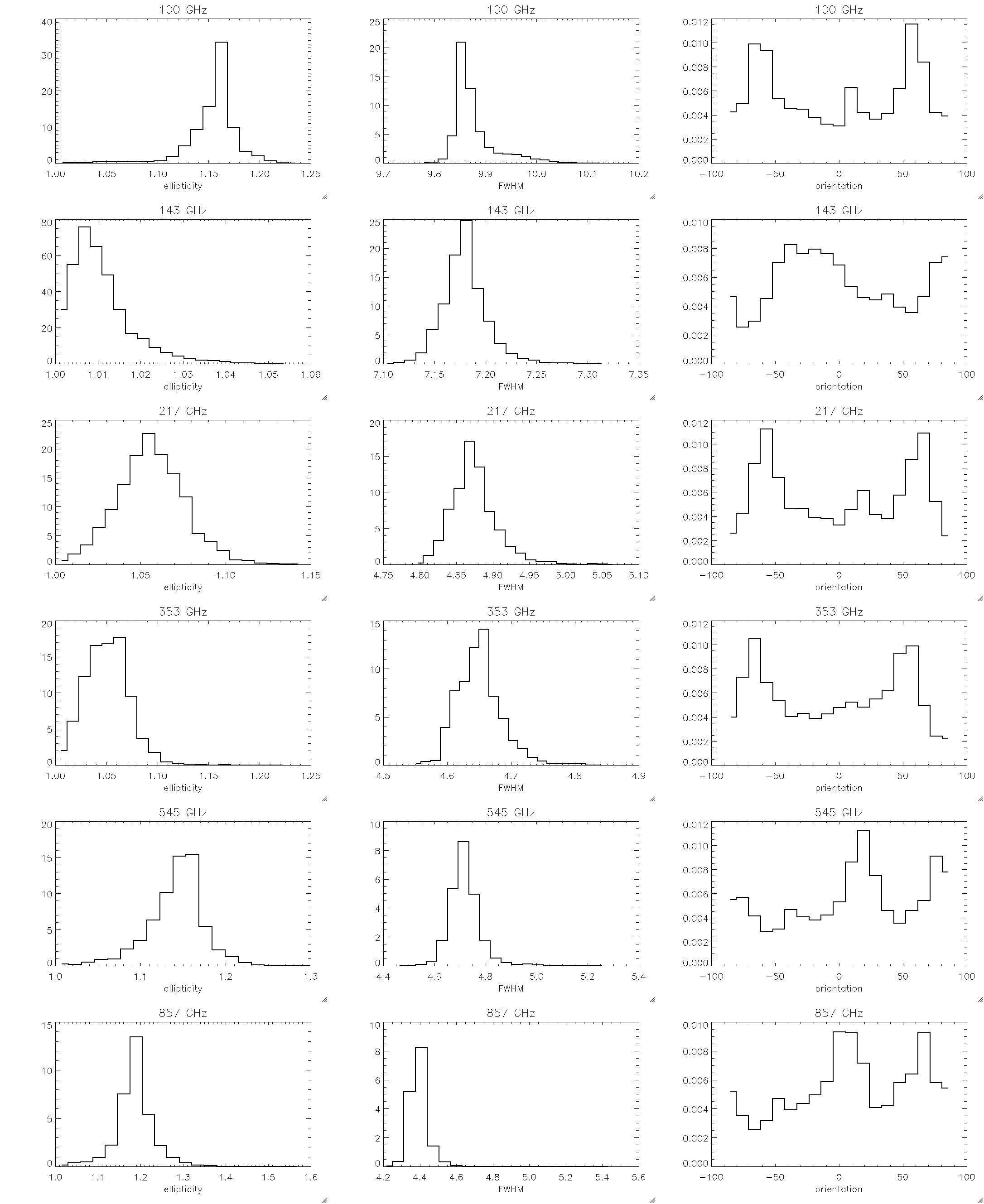}
	\caption{Statististics of the variation in ellipticity, effective
	beam width and orientation of elliptical Gaussian fits to the FEBeCoP beam pattern
	around the sky. These plots are for the Gauss-Hermite fits to the scanning beams.}
	\label{fig:febecop_GHstat}
\end{figure*}

The \Planck\ scanning strategy leads to large variations in the number
of observations of a given map pixel and in the range of scanning beam
orientations (see, for example, the `hit-count' maps plotted in
Fig.~\ref{fig:mapmaking_hits_maps}). As a result, effective beams vary
in size, shape, and orientation depending on location with no
azimuthal or latitudinal symmetry.  To give an impression of the
range of these variations, we sampled the effective beams and PSFs at
3000 uniformly distributed locations on the sky. At each location we
 fitted an  elliptical Gaussian to the FEBeCoP beam. We then constructed
histograms of the fitted values of the FWHM (the geometric
mean of the major and minor axes), ellipticity, and orientation with
respect to the local meridian of the central pixel of the beam. 
Fig.~\ref{fig:febecop_GHstat} shows those histograms for the Gauss-Hermite
fits to the scanning beams.  Overall, we find a dispersion of few percent 
in the FWHM and ellipticity around the sky. Numerical values are given
in the summary table~\ref{tab:summary}. 

\subsection{The FICSBell method}

The FICSBell method \citep{hivon2011} generalizes to
polarization and to include other  
sources of systematics
the approach used for temperature power spectrum  estimation in WMAP-3yr \citep{hinshaw2007} and
by \cite{smith2007} in the detection of CMB lensing in WMAP maps.
The different steps of the method can be summarized as follows:
\begin{enumerate}
	
\item The scanning related information (i.e., statistics of the orientation of
	each detector within each pixel) is computed first, and only once for a given
	observation campaign. Those hit moments are only computed up to degree 4, for
	reasons described below.

\item The (Mars based) scanning beam beam map (or any beam model) of each detector $d$ is 
	 analyzed into its Spherical Harmonics coefficients
	\begin{equation}
	        b^d_{ls} = \int d{\bf r} B_d({\bf r}) Y_{ls}({\bf r}),
	\end{equation}
	where $B_d(\bf{r})$ is the beam map centred on the North pole, and
	$Y_{ls}(\bf{r})$ is the Spherical Harmonics basis function.
	Higher $s$ indexes describes higher degrees of departure from azimuthal symmetry
	and, for HFI beams, the coefficients $b^d_{ls}$ are decreasing functions of $s$ at
	most multipoles. It also appears that, for $l<3000$, the coefficients with
	$|s| > 4$ account for $1\%$ or less of the beam throughput. 
	For this reason, only modes with $|s| \le 4$ are considered in the present analysis.
	\citet*{armitage-caplan2009} reached a similar conclusion in their deconvolution of 
	Planck-LFI beams.

\item  The $b^d_{ls}$ coefficients computed above are used to generate $s$-spin weighted maps for a 	given CMB sky realization. 

\item The spin weighted maps and hit moments of the same order $s$ are
	combined for all detectors involved, to provide an ``observed'' map. 

\item The power spectrum of this map can then be computed, and compared to the
	input CMB power spectrum to estimate the effective beam window function over 
	the whole sky, or over a given region of the sky. 
\end{enumerate}
Monte-Carlo (MC) simulations in which the sky realisations 
are changed can be performed quickly by repeating steps 3, 4 and 5 only. The impact of beam
model uncertainties can be studied by including step 2 into the MC simulations.

\begin{figure}[!htbp]
	\centering
	\includegraphics[width=\columnwidth]{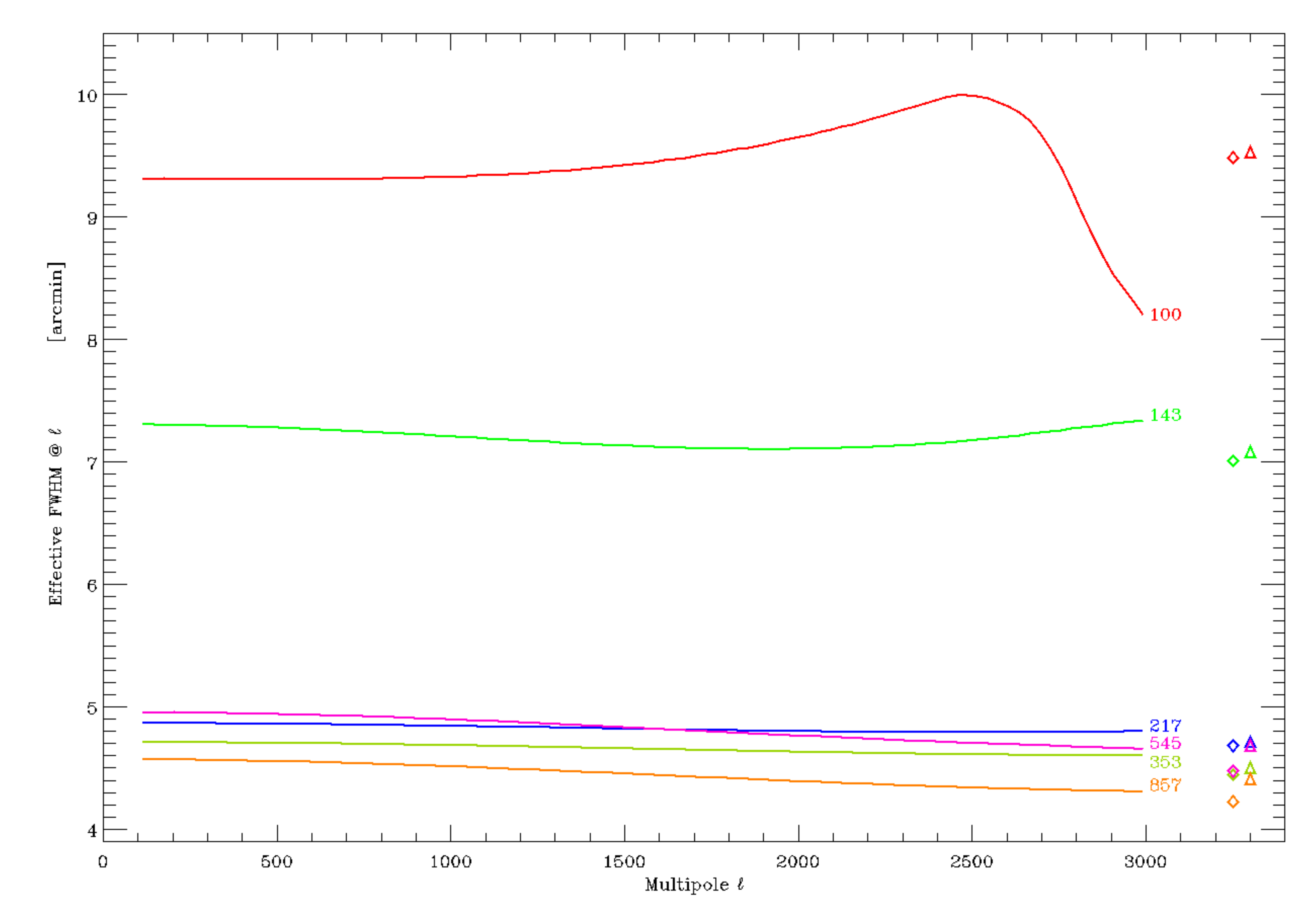}
	\caption{FWHM versus multipole of the Gaussian which would give the same value as
		effective beam transfer function at that multipole.The diamond at right gives the scanning 	
		FWHM obtained from Mars observations, and the triangle the value derived from our best
		estimate of the beam solid angle given in line b1 of the summary
		table (Table~\ref{tab:summary}). }
	\label{fig:ficsbell}
\end{figure}

Figure~\ref{fig:ficsbell} shows that a Gaussian beam is a reasonably
accurate approximation of the effective beam. Indeed one sees that on
average the effective beam derived from the Mars scanning beam differs
only little from a Gaussian, except for  100\GHz for which the scanning beam
is quite elliptical. The 857\GHz\ scanning beams are also rather
elliptical  but this  would show up at higher multipoles than those  shown in the
plot. In most cases the deviations are only at the 0.1\arcmin\ level,
which is negligible for  the analyses of the early results.

\section{Band-pass measurements at the rest-frame frequency of CO lines}\label{Annex:CO}

\begin{table*}[!htbp]
	\caption{Rotational CO transitions within the HFI bands.}
	\label{tab:CO}
	\centering
	\begin{tabular}{|c|c|c|c|}
	\hline
	\begin{tabular}{c} Band \\ (GHz) \end{tabular} & \begin{tabular}{c} CO transition \\
	($J_{\mbox{\tiny{upper}}} - J_{\mbox{\tiny{lower}}}$) \end{tabular} & 
	\begin{tabular}{c} $\nu_{o~\mbox{\tiny{C}}}$\tiny{12}$_{\mbox{\tiny{O}}}$\tiny{16} \\ 
	(GHz) \end{tabular} 
	& \begin{tabular}{c} over-sampled region \\ 
	(GHz) \end{tabular} \\
	\hline
	100 & 1 -- 0 & 115.2712018 & 109.67 --  115.39 \\
	\hline
	217 & 2 -- 1 & 230.5380000 & 219.34 --  230.77 \\
	\hline
	353 & 3 -- 2 & 345.7959899 & 329.00 --  346.15 \\
	\hline
	545 & 4 -- 3 & 461.0407682 & 438.64 --  461.51 \\
	545 & 5 -- 4 & 576.2679305 & 548.28 --  576.85 \\
	\hline
	857 & 6 -- 5 & 691.4730763 & 657.89 --  692.17 \\
	857 & 7 -- 6 & 806.6518060 & 767.48 --  807.46 \\
	857 & 8 -- 7 & 921.7997000 & 877.04 --  922.73 \\
	857 & 9 -- 8 &1036.9123930 & 986.57 -- 1037.95 \\
	\hline
	\end{tabular}
\end{table*}

After launch it became apparent that the contribution of CO rotational
transitions to the HFI measurements was greater than anticipated,
especially for the 100 \GHz\ band.  To isolate the narrow CO features
from the rest of the signal components (CMB, dust, etc.), precise
knowledge of the instrument spectral response as a function of
frequency is required.  The original spectral resolution requirement
for the spectral response measurements of a given HFI detector was $\sim$3\GHz;
this corresponds to a velocity resolution of $\sim$8000 km/s for the
CO J\,1-0 line.  As \Planck\ does not have the ability to measure absolute
spectral response within a frequency band during flight, the
ground-based FTS measurements may provide the most accurate data on the
HFI spectral transmission.  With good S/N, spectral information can be
inferred to a fraction (\ie\ $\sim$1/10th) of a spectral resolution
element \citep{spencer2010}.  Thus, the Saturne FTS measurements
\citep{Pajot2010}, which were carried out at a spectral resolution 
of $\sim$0.6\GHz, may be used to estimate the spectral response at a 
resolution equivalent to $\sim$150 km/s 
for the CO J\,1-0 line.

\begin{figure}[!htbp]
	\centering
	\includegraphics[width=\columnwidth]{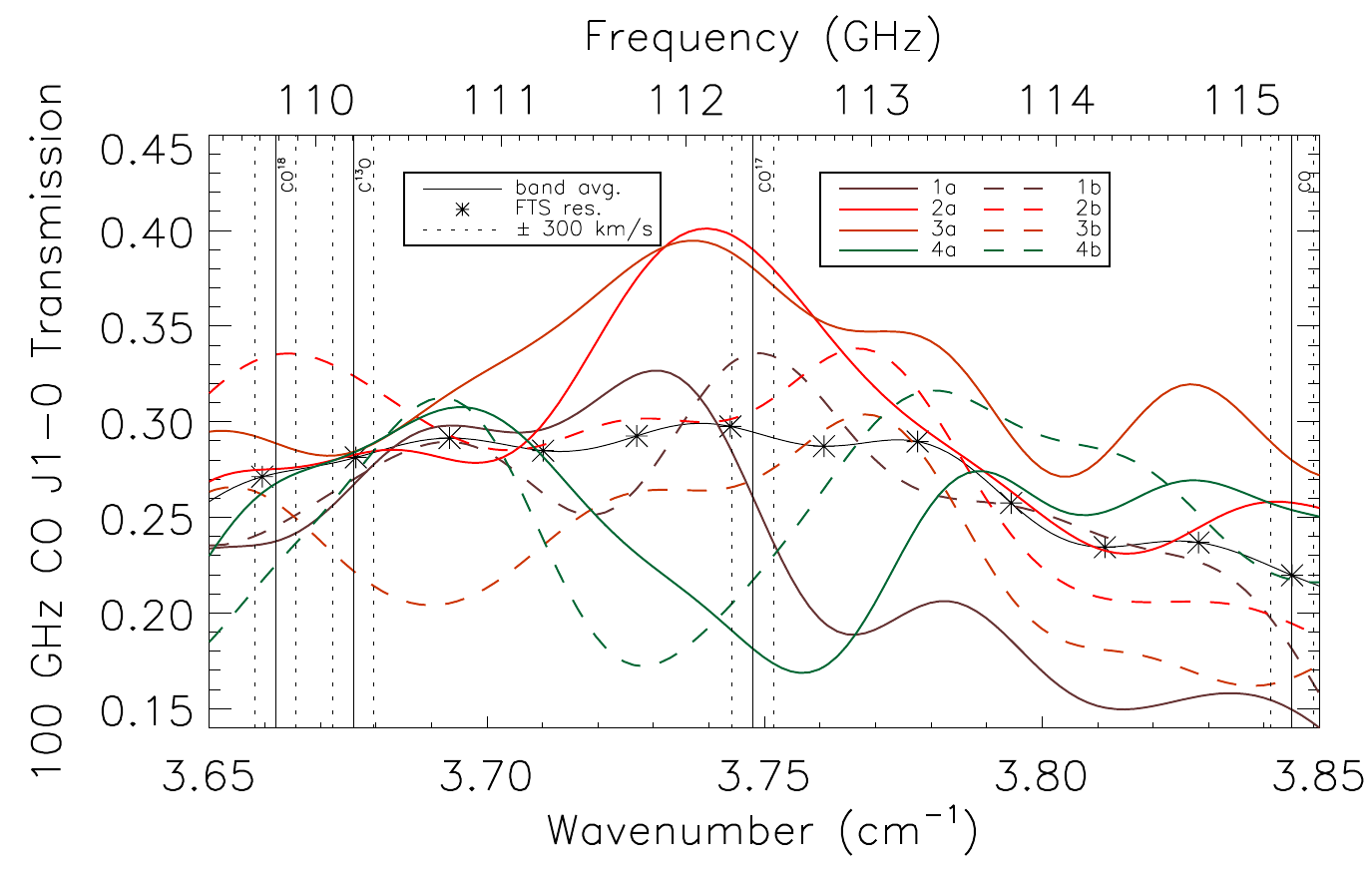} 
	\caption{The spectral response for the CO J\,1-0 transition region.  
		The spectral resolution has been oversampled by a
                factor of $\sim$10 for a window 
		surrounding each of the CO transitions listed in Table
                \ref{tab:CO}, the asterisks show  
		the spacing of independent spectral data points. This
                region was selected to include  
		the rest frequency ($\pm$300 km/s -- vertical dotted
                lines) for each of the CO,  
		C$^{13}$O, CO$^{17}$, and CO$^{18}$ isotopes.  Also shown is the band average
		spectrum for each region.}
	\label{fig:COzoom} 
\end{figure}

For spectral regions near CO rotational transitions, therefore, the
spectral response was oversampled by a factor of $\sim$10 using an
interpolation based on the instrument line shape of the FTS.  Table
\ref{tab:CO} lists the relevant CO transitions, and the frequency
ranges over which the HFI spectral response was oversampled.  The
oversampling ranges were extended to include all of the common CO
isotopes.  The vertical bars above the spectral response curves of
Figure \ref{fig:AvgBPs} illustrate these oversampled regions.  The
spectral responses for the CO J\,1-0 frequency range are shown in 
Fig.~\ref{fig:COzoom}; similar data is available for all of the CO
transitions within HFI bands.

\end{document}